\documentclass[bottom,nosig]{mythesis}

\title{Probabilistic and Flux Landscapes of \\
       the Phage $\lambda$ Genetic Switch}
\author{Nathan Borggren}

\month{December} \year{2011}%

\program{Physics}%

\director{Jin Wang}{Associate Professor, Department of
 Chemistry}%
\chairman{Jacobus Verbaarschot}{Professor, Department of Physics and
Astronomy}%
\fstmember{Dmitri Averin}{Professor, Department of Physics and
Astronomy}%
\outmember{Nuggehalli Ajitanand}{Senior Research Associate}{
 Department of Chemistry}%
\dean{Lawrence Martin}%

\begin{document}

\singlespacing %
\pagenumbering{roman} %



\begin{abstract}
    
The phage $\lambda$ infection of an \textit{E. coli} cell has become a paradigm for understanding the molecular processes involved in gene expression and signaling within a cell.  This system provides an example of a genetic switch, as cells with identical DNA choose either of two cell cycles: a lysogenic cycle, in which the phage genome is incorporated into the host and copied by the host; or a lytic cycle, resulting in the death of the cell and a burst of viruses. The robustness of this switch is remarkable; although the first stages of the lysogenic and lytic cycles are identical, a lysogen virtually never spontaneously flips, and external stressors or instantaneous cell conditions are required to induce flipping.  In particular, the cell fate decision can depend on the populations of two proteins, cI and Cro, as well as their oligomerization and subsequent binding affinities to three DNA sites.  These processes in turn govern the rates at which RNAp transcribes the cI and Cro genes to produce more of their respective proteins. 

Although the biology in this case is well understood, the fundamental chemistry and physics underlying the bistability remains elusive.  In this work, a dynamical model of the non-equilibrium statistical mechanics is revisited, generalized, and explored.  The low number of proteins and other sources of noise are non-negligble and corrections to the kinetics are essential to understanding the stability.  To this end, general integral forms for advection-diffusion equations appropriate for finite element methods have been developed and numerically solved for a variety of mutants and assumptions about the state of the cells.  These solutions quantify the probabilistic and flux landscapes of the ensembles' evolution in concentration space and are used to predict the populations of the cell states, entropy production, passage times, and potential barriers of wild type and mutant bacteria to illuminate some structure of the configuration space from which Nature naturally selects.         

\end{abstract}

\tableofcontents
\listoffigures %
\listoftables %



\begin{acknowledgements}
\chapter*{Acknowledgements}

The original title of this thesis, \textit{One Continuous Mistake}, seemed to have already been taken, though time has now evolved sufficiently long to have absorbed enough contribution from elsewhere to break this continuity sufficiently to warrant a new name and all of the residual mistake is entirely my own.    

I would like to thank Jin Wang for this opportunity, guidance in looking universe, and for the space and time to diffuse through many ideas and approaches to the questions he had asked of me.  His grand canonical ensemble: Wei Wu, Weixin XuBo Han, Zhiqiang Yan, Yuan Yao , Xiaosheng Luo, Haedong Fong, Jeremy Adler, Qiang Lu, Zaizhi Lai, Ruonan Lin, Ronaldo Olviera, and to those whose bodies have fluctuated elsewhere but kept their influence close: Liu Fang, David Lepzelter, Saul Lapidus, Keun Young Kim, and Amber Carr have been endlessly knowledgeable, kind, and helpful and add some reason to the rhymes and rhyme to the reasons to a vast array of problems in the field.  

I would like to thank my committee for everything, and in particular it is quite a privilege that I have been able to ponder deeply questions with Ajit Nuggenhali about nuclear matter and Dmitri Averin with condensed matter.  This thanks extends to Jiangyong Jia, Roy Lacey, Rui Wei, Arkadij Taranenko, John Anderson, Damian Reynolds and Alex Mwai as well as the rest of the Phenix collaboration, Vasili Semenov, Jie Ren, and Supradeep Narayana have been generous with their time, expertise, laboratories as well.  My other FEniCS collaboration has been fast, thorough, and efficient with their help and expertise on the rare occasions direct mimicry of their methods didn't suffice.  The scientific influence of Chris Tartamella, Stanislav Srednyak, and John Schreck are also deeply ingrained in this work, but that influence is trivial next to their friendship.  

I would like to thank Mark Ptashne for an early version of \cite{ptashne_principle} and Xue Lei and Ping Ao for helpful discussion of units.   

I have never stopped counting in densities since leaving Roger Yelle's company though his teachings and insight still seem to anticipate me and welcome me in every new direction I turn.      

My mom, dad and $E^3$ 

and Amber, la luz, \textit{Let simple and old fashioned myself stay with you, while ordinary things go disappearing from this world.}

\end{acknowledgements}
\pagenumbering{arabic} %
\chapter{\textit{Introduction}}
\vspace{-0.4in}
\begin{quotation}
  \textit{Life is an entanglement of lies to hide its basic mechanisms.}

\hspace{0.1in}William S. Burroughs, \textit{The Place of Dead Roads}
\end{quotation}

\section{Computation, Noise and Limit Cycles in Biophysics}

It is a season in which the porousness of the borders delineating the subjects that comprise Knowledge are growing more so, or altogether vacuous.  The overlap is revealing some truths to be quite universal.  It has been said that \textit{more is different} \cite{anderson_science} and \textit{information is physical} \cite{landauer1961irreversibility}, and concepts involving emergence and entropy are proving to be fundamental to understanding the physical basis of biological systems.  The notion of a complete and clockwork deterministic universe has been left behind, and the necessity of dealing with probabilities, partial information, measurement, and open systems has become essential to all of the major sciences.  

Among these disciplines, the common goals have become those of elucidating which states are possible, determining the relative probabilities or amplitudes of those states, and examining their stability and robustness.  In the subject of physics, the need for these particular goals is most easily evidenced by the existence and necessity of quantum mechanics, which assigns to each system a wavefunction from a spectrum of permitted states; entanglement amongst these states; and, by measurement, results ultimately in the collapse of the wavefunction to a single eigenstate of the spectrum.  In chemistry, it has become increasingly necessary to delineate the single pathway that a molecule may take, as it selects a seemingly random walk in a solvent, or geometrically aligns with a catalyst in a reaction.  In biology, evolution provides a framework by which we understand how each member in a myriad of diverse species becomes finely adapted to their single, specific niche, and how these populations are interrelated.    

Although the underlying theories are generally accepted, there remain open questions and challenges in the implementation of these theories that are similar across disciplines.  Analogies can even be present in the mathematical formalisms that are often developed in parallel for these subjects.  For example consider time evolution in quantum mechanics, Brownian motion, and population dynamics.  The integral of a specified initial condition is commonly implemented iteratively as a matrix operator acting on a vector which represents the quantum state in the case of quantum mechanics, the population densities in the case of chemical dynamics, or the population density in the case of an ecosystem.

The limit cycles are of considerable interest.  Although teleological explanations in evolution are largely vanquished \cite{Mayr_Evolution}, at least some of the properties of living matter can be predestined by the highly non-negligible physical and chemical constraints placed on it by the supporting planet.  Understanding physically how systems acquire their properties through gene expression and feedback from an environment leads to fundamental physics.  However large the number of unknown degrees of freedom, they serve as a source of energy and entropy for the system, or as a place to deposit energy and entropy.  In some shape or form, this exchange introduces a mechanism for choice, chance, or randomness into the time evolution of the system, which is commonly implemented as a matrix acting on a vector.\footnote{Consider, for example, $\dot{x}=Ax$, $i\hbar \partial_t \Psi = H\Psi$, $\partial_t P = -\partial_i (D_{ij}\partial_j P - F_iP)$ with solutions, $x(t)=e^{At}x(0)$, $\Psi(t)=e^{-\frac{i}{\hbar}\int_0^tHdt}\Psi(0)$, and $P(t)=e^{\int_0^t(D_{ij}\partial_j - F_i)dt}P(0)$ respectively.  The essential problem, be it a continuous Markov process described by A, Schr\"{o}dinger's equation, or the Fokker-Planck equation, the equation boils down to the evaluation of an exponential of an operator.}  Be it the decoherence processes that select an eigenfunction out of entangled quantum states or the natural selection processes that select the fittest of a species, it is \textit{time} that selects an actuality out of the sea of possibilities that could potentially occur.  That is to say, whatever the equations, it is the asymptotics of those equations that are of interest.  When the system is open, the nature of this selection speaks equally as much of the measurement device used or the planet underfoot as it can of the system measured or the species selected.  

Assuming that we have equations that describe these phenomena, is it feasible to imagine finding solutions to them?  The question we ask is what the best models are for the physics we can develop given the computational resources that are available, and how we can improve the precision of the physics as our algorithms and computing power improve.  When we point our computers at the abstractions we use for these systems, the polynomial algorithms seem nowhere to be found.  While for biological problems, even by counting only the emergent degrees of freedom (proteins) of the emergent degrees of freedom (RNA) of the emergent degrees of freedom (DNA) of the emergent degrees of freedom (the nucleotides) of the emergent degrees of freedom (the periodic table) of the emergent degrees of freedom (the protons and the neutrons), our ability to directly probe the relevant scales of length and time is limited by our current information technology. 

Because many questions in one field are categorically identical to questions in another, the achievements and methods in one science are non-negligible to the others.  Often they are the same achievement or require cooperation.  The contribution of physics thus far to biology is largely that of engineering: it is the microscopes, spectrometers and transistors that are mainly of service.  There is no evidence suggesting that the physics of biology requires more then quantum mechanics, statistical mechanics, and classical mechanics to operate, but the extent to which the solutions to our equations lag behind our equations provides ample room for doubt and countless theorems remain to be demonstrated to \textit{show} this.  As an example, biology has provided us with the ingenuity through experiments to deduce the relevant underlying networks for the $\lambda$ phage \cite{ptashne1992genetic}\footnote{and references therein}, the flexibility of those networks \cite{atsumi_synthetic_2006,little_robust}, and how to introduce and parameterize the empirical force fields that are introduced to study the dynamics \cite{shea_or_1985}.  These numbers are improved every day as better microarray data is accumulated and understood \footnote{In a just universe, the models such as discussed in this thesis would be coupled with data.  The rates used in the partition functions should be fit simultaneously with the numbers used in the diffusion tensor as well.  We do not take such a course here but provide the reference \cite{Kinney09012007}, in addition to \cite{shea_or_1985} to illustrate how $\Delta G$ numbers are accumulated}.        

Evolution, like gravity and thermodynamics, is so seamlessly entangled into our lives as to often go unnoticed.  Once, however, the concept is understood it is quite self-evident, and is as readily seen as the selection of an item off of a menu or the selection of students or employees from a pool of applicants.  This interplay between chance and determinism, the arbitrary and the necessary, and populations and their individuals provides more then enough room for complexity to emerge, even before granting a capacity for volition to the individuals.  

The key mechanism again is emergence - be it quarks to form a proton, nucleotides to form DNA, or cells to form an organism, the salient necessity for evolution to occur at any hierarchy of a universe is a mechanism to build up systems of arbitrary complexity out of simpler and readily available resources from the nearby hierarchical levels \cite{Jacob,McCoy}.  Unsurprisingly, we find ourselves present in such a universe, and Nature has no shortage of mechanisms to build complex systems, and clever algorithms for their improvement \cite{darwin,tol}\footnote{The notion of species-as-algorithm is evident in \cite{toffoli_physics_1982}, where he gives the example of a typical control function: the octopus.}.     

It has been said that the universe \textit{is} a quantum computation \cite{Lloyd_universe}, and it will not resist being perceived as one, but other modes of computation the universe readily performs, those using the counting of molecules, are worthy of deep reflection \cite{bennett1982thermodynamics}.  These classical modes can also supply the speedup to which quantum computers esteem.   This Brownian computer, as Bennett calls it, is not dissimilar to a quantum computer, in the sense that it relies on interference to decrease the probability of certain possibilities and increase the probabilities of others.  The fundamental difference is in the nature of the interference, which for a quantum computation occurs at the amplitude level, where numbers can be complex, and for a Brownian computer this occurs at the probability level, where the numbers are strictly positive.  The fundamental similarity between the two modes of computation is that they occur essentially in parallel.   

To illustrate, consider an impenetrable door, nay, there is an alphabet with four letters and one password in $4^n$ words that will let us in were we to enter that word.  The keepers of the door also granted us a classical, quantum, and chemical keypad.  The classical keypad takes too long, qubits have their own agenda at those numbers, so how then will we ever open the door?

What is a chemical keypad?  Say we have n sites and we enter our code by binding 1 of our 4 molecules onto each site, and if we get the arrangement just right, then the door opens.  A molecule will simply go away after some time if it did not fit.

Our only hope then without any more prior information is Bennett’s Brownian computer.  We simply gas the keypad with concentrations of all the molecules and wait.  Sooner or later the necessary configuration will occur and our key is found and the door is opened.

This is still a classical computation in the sense that it is no different than our fingers punching numbers into a keypad as far as the information processing is concerned.  Indeed the speedup here is simply of scale, as molecules can typically bind and unbind to sites and each other faster then we can push buttons.  The probabilistic nature of this process comes only as a result of our description of the scenario.  We have only macroscopic information, the thermodynamic variables, and may be able to tune them to some degree, but the goal is a particular conformation. This information is not accessible to thermodynamics proper, but we know the answer once the goal has been achieved, as we read it out of the keypad only after the door is opened.  As is often the case, the key itself might fall out of the door and this information too is lost, but sometimes just an open door is sufficient to proceed.      
     
Evolution \textit{is} such a mode of computation: a wide variety of possibilities are attempted, and if any open doors, or are built to last, then doors open and things last.  The rate and mechanisms it achieves to do this provides curious insight into new computers and algorithms for achieving and dealing with computational complexity, typically involving non-binary modes of thinking.   

Granting a billion years from the birth of Earth to the birth of the first prokaryotes, say $4*10^6$ bp (base pairs) \cite{Mayr_Evolution}, it can be infered that the effective rate at which natural selection processed the $4^{4*10^6}$ configurations is $4^{4*10^6-27}$ Hz.  This is a rephrasing of Levinthal's paradox in protein folding for the problem of evolution.  It is thus safe to infer that natural selection is not picking numbers at random from a flat distribution.        

Genes are an example of an algorithm that has provided exponential speedup to this sampling.  Genes might be a transcript for a protein and take up a thousand nucleotides of space, and at this stage Nature's trial and error processes of evolution revolve around a sampling of the space of gene combinations; this is a considerably truncated space from the space of permissible DNA chains.  Sexual reproduction is one example exhibiting the benefits of such a sampling, and it emerged as a means of speeding up the quest for fitness about 2.7 billion years ago.  The DNA of the resulting offspring is far removed in a symbol space from either parent's DNA, yet the genes persist.  It is quite a luxury that a child should take after their parents.  

With a configuration space so vast and the demands of a niche so extreme, one can hardly suspect a few random bits flipping to send one to a more favorable potential well.  Quite the contrary is fact and indeed the structure of DNA, with its Watson-Crick pairing and strength of its bonds, is largely preventative of just this situation.  Unsurprisingly, there exist proteins and mechanisms that specialize in DNA repair to further parry these crude attempts to debase the bases.  In eukaryotes, DNA is further protected via a profound capacity for packing and elaborate protection from histones and proteins.

So it is genes that drive development and the space of gene configurations that DNA prefers to play trial and error with, for exactly the reason to exponentially speedup the search for fitness.  There is then a very strong and productive motive to treat the genes and their functions as the relevant degrees of freedom, and indeed that is the approach and success of systems biology. \cite{ptashne2002genes,comingtolife}   

It can be seen that evolution can proceed with large steps.  There are mechanisms in which the step nature of evolution of DNA itself is evident.  This revolves around the discussion of genes, not nucleotides, where very specific and relatively long transcripts can be inserted irreversibly into a genome.  Such is the case we study now, the phage $\lambda$ infection of an E. coli.   The biological content of this thesis is well known and actively extended, \cite{St.Pierre,ptashne1992genetic}.    

\begin{figure}[h]
\begin{center}
\includegraphics[width=0.7\textwidth]{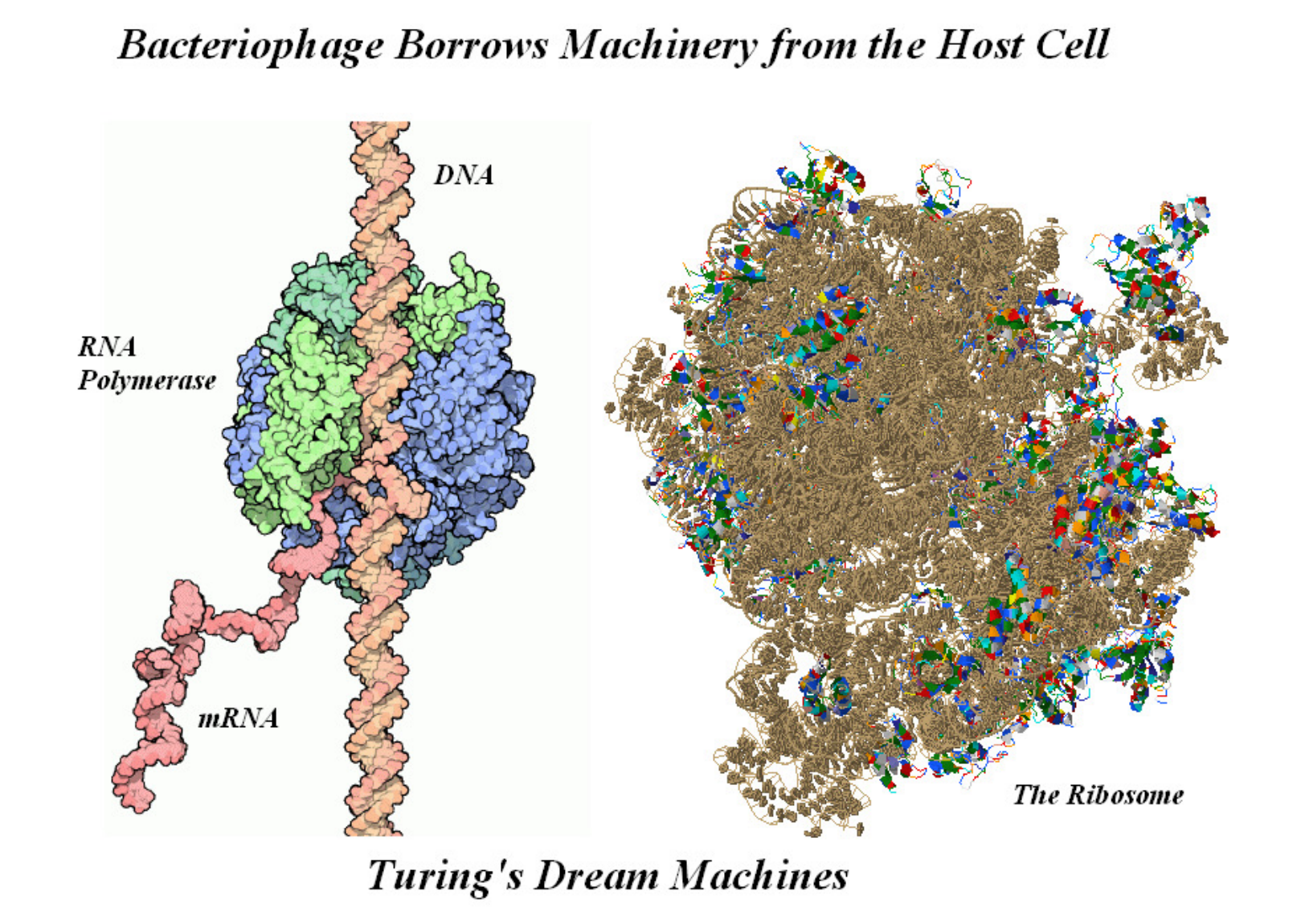}
\caption{{\bf $\lambda$ phage requires a host for RNAp and ribosome to express its genes.}}
\label{RNAp}
\end{center}
\end{figure}

$\lambda$ is not alive and requires an E. coli host cell to provide the habitat and ingredients, in particular RNA polymerase and the ribosome, to express and replicate the genes present in that genome.  A head provides protection to the DNA outside of a cell as the virus uses a tail to seek out a host upon which it pierces the cell membrane and injects the DNA into the cell.  A special feature of the genome is the six bases on each end; they are unpaired 'sticky ends,' and are dual to each other.  These sticky ends stick to to each other upon entering a cell and form a ring.  The viral DNA then waits for the cell contents to regain their closest approach to equilibrium.  This tricks the RNA polymerase and ribosome in the host into expression and synthesis \ref{RNAp} of its genes. 

\begin{figure}[h]
\begin{center}
\includegraphics[width=0.7\textwidth]{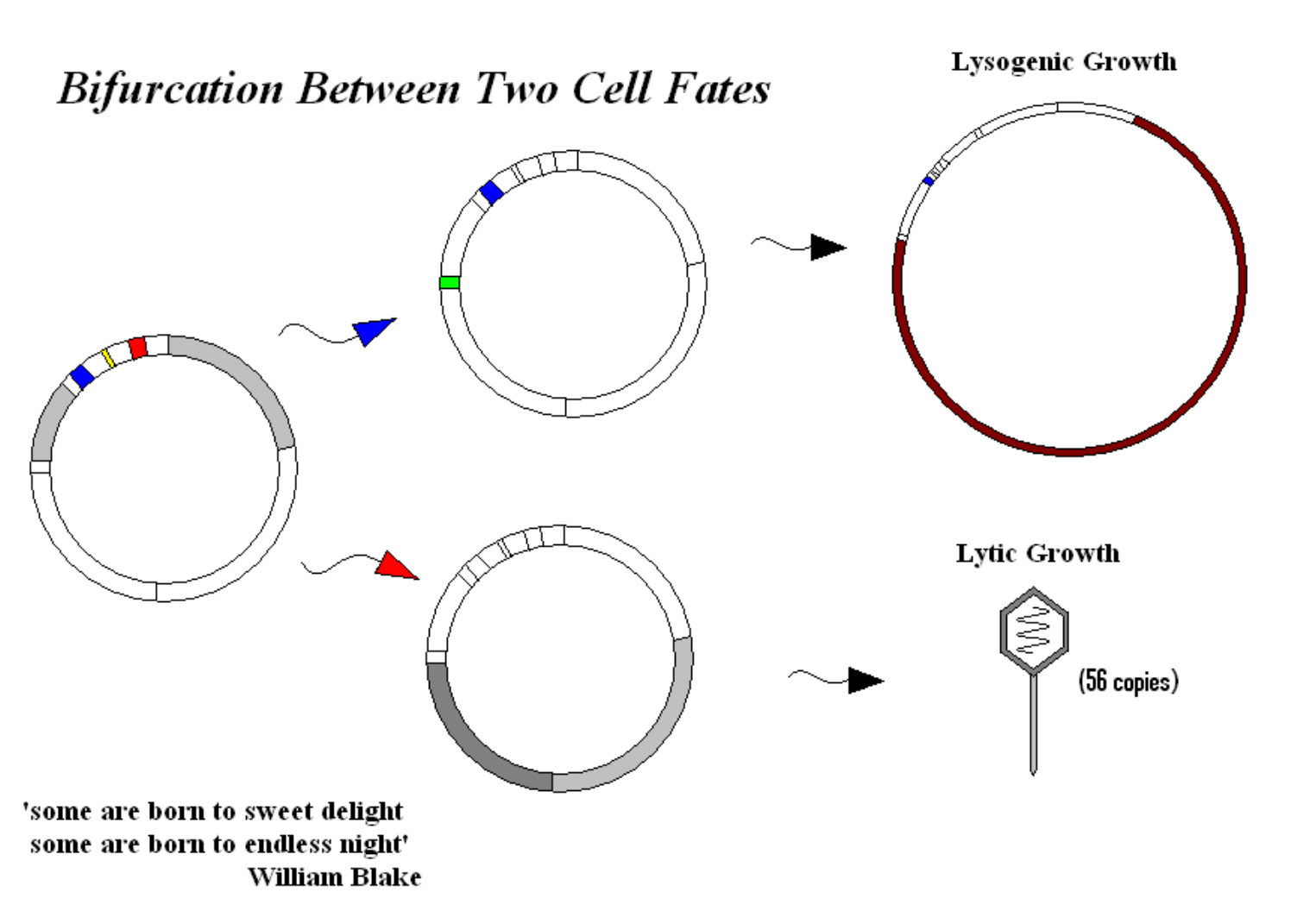}
\caption{{\bf The two $\textit{E. coli}$ outcomes, lysogenic above and lytic below, of a $\lambda$ phage infection.  The $O_R$ is yellow, the active cI gene is blue, the active Cro gene is red, and grey are other active genes in that stage of the cell cycle.  The Att gene in green is the gene responsible for incorporating the virus into the bacteria.  White regions are DNA segments inactive during that stage.}}
\label{activity}
\end{center}
\end{figure}

This splitting of a cell population into different possible fates can be seen in a variety of circumstances.  Some cells burst into a multiplicity of phages immediately (lysis), while the bacteria that do survive are irreversibly transformed into a new genome, now nearly 48,502 bp longer, and henceforth replicating along with its old identity and the entire phage (lysogeny), see \ref{activity}.  This behavior seems more akin to a jewel heist then a thermal, stochastic, or a quantum fluctuation.  It is believed that noise influences the decision at the level of selection between fates.  This noise has been studied elsewhere, \cite{Ao,wang_kinetic_2010,arkin}, but the exact nature of this noise is not understood, and is typically presumed Gaussian for computational and analytic convenience.  The methods developed to study this noise can easily implement more complicated noise, for example, as known functions of space and time.  This switch is a perfect example of a switch component of a Brownian computer, but in the end just one component of a larger $\lambda$ gene network, presented in \cite{arkin,St.Pierre}.        

The rates at which these transcriptions occur enter the model's equation of motion as simply numbers and set the time scale of the computation, i.e. the expression of the gene, so let us revisit RNAp, and the ribosome, to first marvel at two of the most amazing molecules this side of the big bang, and chieftains among the hyperspecialized computational hardware inside a cell.  

Indeed it was RNA polymerase that was the source of inspiration for the Brownian computer in the first place, and its functioning had been described in information theoretic terms already in 1982 \cite{bennett1982thermodynamics}:

\begin{quotation}
To see how a molecular Brownian computer might work, we first consider a simpler apparatus: a Brownian tape-copying machine.  Such an apparatus already exists in nature, in the form of RNA polymerase, the enzyme that synthesizes a complementary RNA copy of one or more genes of a DNA molecule.
\end{quotation}

Bennett goes on to describe the RNA polymerase action in terms of a one-dimensional random walk along a DNA chain.  In the forward reaction RNA polymerase opens the DNA to read the content and append to the growing RNA strand the nucleotide part of the relevant nucleotide pyrophosphate ATP, GTP, CTP, or UTP and subsequently releasing the pyrophosphate molecule into the cell, advancing a notch on the DNA chain.  In the backward reaction RNA polymerase takes a free pyrophosphate (PP) from the cell and combines it with the last RNA bit and releases into the cell the ATP, GTP, CTP, or UTP.  In equilibrium this reaction would balance and RNA and protein synthesis would not occur.  However, the metabolic processes are such that a bias in the direction of RNA synthesis occurs by removing free PP and supplying ATP, GTP, UTP, and CTP to the vicinity of RNA polymerase.  The numbers used in his analysis are RNA polymerase advancing at 30 nucleotides/s (330 forward reactions -300 backward reactions), and dissipates 20kT per nucleotide all while making less than one mistake per ten thousand nucleotides. \cite{bennett1982thermodynamics}

By 1999, this one-dimensional random walk was directly observed in single molecule experiments \cite{guthold1999direct}.  The rates therein were much lower (1.5 nucleotides/s $\pm$ 0.8 nucleotides/s), but these processes were likely slowed down by the experimental conditions which confined the DNA to diffuse along a 2D substrate, restricting the motions of many degrees of freedom, to facilitate the observation of RNA polymerase moving along the DNA as if it were a 1D chain.  It had been argued that effectively diffusing in lower dimensions enables RNA polymerase to find the promoters much more easily.        

$\lambda$ has developed an additional strategy that speeds up this transcription involving cI protein \cite{ptashne_principle}.  It involves a relatively small cooperativity (-2kcal/mol) binding between a cI dimer and RNAp at the promoter site, which also shortens the time RNAp requires to diffuse to that point.  This is a small number and can not alone describe the observed factor of 10-100 increase in transcription rate when cI is present.   An explanation I provide here is that this speedup could be a result of another complex binding process, where cI octamers short the DNA loop, as has been observed.  That this results in a speedup of transcription is a result of the simple fact that, if RNAp can maintain a 30 nucleotide/s rate, this rate is increased in units of m/s if the chain is shortened and there is simply less distance to diffuse to the end of the transcript and back to the promoter.  Indeed the looping between the $O_R$ and the $O_L$ leaves RNAp again approaching in physical space the promoter as it is finishing the transcription of cI.  This increase does not occur for the Cro gene since the shorting of the loop does not contain the Cro gene, only the cI gene, and requires cI to be occupying the promoter site for Cro anyway.             

The subsequent RNA strand is fed to the ribosome which implements a many-to-one dictionary, converting three consecutive nucleotides into an amino acid.  The 64 4-letter words of length 3 get mapped to 20 amino acids, aiding in the formation of the protein.  This allows for a large multiplicity of DNA words that can create the same protein and thus allow for a finer resolved selection process to occur amid the words that produce a given protein.

cI is a chain of 236 amino acids and cro is 66.  The rates at which these processes occur are actively measured and correspond to the value of $\nu_{xi}$ for cI and Cro in this situation.  The rates for the lambda phage production of cI and Cro used in these simulations are 0.115 cI monomers/second for states with cI  assisting RNAp, and 0.30  Cro monomers/second.  Other impacts on the rates are the 0.01045 cI monomers/second contribuition from certain configurations, and that the ribosome produces 20 Cro for each reading of the Cro transcript.  These numbers are chosen to correspond to the values use in \cite{Ao,wang_kinetic_2010} where this dynamical system is investigated.  The complete set of parameters used in this simulation will be shown in the dynamics section. 
    
There are two sources of non-linearity in the equations of motion, first from there simply being multiple binding sites, but the main complication, and source of error in this work, is that both cI and Cro first dimerize in the cell and bind in dimer form to DNA with a helix-bend-helix motif.  Different configurations have slightly different affinities and thus occur with different frequencies.  All the configurations however have negative $\Delta G$ and as such are spontaneous reactions; they are energy favorable in comparison with an empty strand of DNA.  In addition, cI dimers are found to bind cooperatively on adjacent sites.  This added energy bonus from cooperation aids in the competition with Cro.  Cro is also limited by the fact that both $O_{R1}$ and $O_{R2}$ site need to be unoccupied for RNAp to bind to the Cro promotion site.  The partition function is dominated by the triply liganded species and the faster rates and large bursts for Cro get largely repressed by the relative scarcity of the singly liganded states in comparison to the doubly liganded states that promote cI.  

The phage genome in a lysogen patiently waits for cell conditions to appear, a process known as induction, that switch the genes being expressed to phage-bearing ones and result in bacteria death and phage progeny.  Ultraviolet light, small cell volume, and elevated Cro or depleted cI protein levels are known to be able to induce the switch.  It is the object of this dissertation to further quantify the nature of this swith and compare with the words nearby in simple space.  These are very specific mutations in the $O_R$, chosen so that the binding properties can be predicted, and switch properties studied.                          A result of this thesis, and a testable hypothesis, is that energetically the use of ultraviolet light for induction is a little extreme, as radiation targeted at the $O_R$ equivalent to ~15kT should be sufficient to provide the energy for some of the lysogens to flip.  Indeed, NMR techniques specialized to control cI to the extent that it can be removed from $O_{R1}$ and $O_{R2}$ should also be sufficient for induction even at larger concentrations.    
         
\subsection{Chemical Oscillations}

That classical chemical reactions used to describe the balancing of morphogens within an embryo is sufficient to give rise to non-trivial spatial and temporal structures was noted carefully in the last works of Alan Turing \cite{turing1952chemical}.  See \cite{kondo_2010} for a modern review. 

Classical treatment of chemical reactions is sufficient to understand many rhythms of the body, like a heart beating, and it is wise to push the utility of such treatments as far as possible. 

Reactions are not even necessary for oscillations to occur, indeed mixtures of perfect gases oscillate on grand scales in atmospheres in an attempt to maintain hydrostatic equilibrium for the mean density despite tidal or other forces \cite{borggrenA, borggrenB}.       

The variety of chemical oscillations considered in this thesis are particular to those believed to be relevant in gene expression, which involve the binding of ligands to operator sites  resulting in the creation of more ligands.

The chemistry section of this work states quantitatively our basic assumptions: namely the partition functions used to describe the equilibrium statistical mechanics of the protein/DNA reactions.  It is stated to a reasonable generality what shapes these partition functions can make, facts that we will need to motivate the extent of accessible dynamics to which these binding polynomials along with transcription factors can give rise.

At low numbers, chemical oscillations take on much greater complexity, where noise can play dominating roles.  Most importantly this work relies on the discussions in \cite{wang_potential_2008}.    

\subsection{Dynamics}

Universal computation takes on a whole new meaning in chemical and quantum computation.  One need not seek only classical answers to classical questions, as analog circuitry and quantum computation can greatly enrich the classes of problems that can be understood.  The dynamics resulting from gene expression could be of interest to the purest of mathematicians, and in some sense are the n-body generalization of Hilbert's 16th problem, which remains unsolved.  We applaud chemistry and biology for their courage in asking such questions.  Hence it is quite appropriate to begin with the most general equations for dynamical systems.  The qualitative and quantitative theories of dynamical systems are old and advanced \cite{Lia93,volterra,arnold_ordinary_1978, Perko}\footnote{a study of dynamical systems, to which noise is largely a perturbation, can on no accounts be trivialized or neglected.  Many texts on the fundamentals are available, in particular \cite{Perko} illustrates the linearization methods to understand the local, qualitative dynamics of a system.} and the need for general methods to general equations in simulating gene expressions is noted in \cite{Ao,wang_kinetic_2010,aurell_epigenetics_2002}.

The codes used for this work were written to maintain this generality and the software can be used to study dynamical systems over real or complex fields.

\subsection{Diffusion}

Of the few equations Schr\"{o}dinger reluctantly allows appear in \textit{What is Life?} \cite{whatislife} the first is

\begin{equation}
 \partial_tP=D\Delta P
\label{diffusion}
\end{equation}

\noindent Which he need not trouble you to explain.  The gradients and diffusion of which he speaks are in physical space, whereas we will be concentrating on concentration space.  The motive and solutions are, however, similar.  

The robustness of the phage $\lambda$ is of the type Schr\"{o}dinger would have claimed at the time to be impossible.  He deliberates extensively on the $\sqrt(n)$ law and expected that where there is order, such as ferromagnetic order, there will be a requirement of a large number of particles involved.  Yet the order one would expect from large concentrations is still present at the small numbers of 100-300 proteins and has given rise to the so-called stability puzzle \cite{aurell_stability_2002}.     

The analogous equation of (\ref{diffusion}) for the case of concentration space, to which numerical solutions herein are devoted to address, is

\begin{equation}
\partial_t P = \partial_i (D_{ij}\partial_j P - F_i P)  .
\label{diffusion2}
\end{equation}

\noindent The codes allow for $F_i$ to take on the generality prescribed in the dynamics section and $D_{ij}$ can be functions of space and time.  Our mathematical analysis will proceed with the simplest case of a positive and diagonal diffusion tensor. 

A main contribution of this thesis is the adaptation of (\ref{diffusion2}) for use with finite elements.

\section{Computers and Phage $\lambda$}

Before we posit equations and automate solutions let us note that biological systems are not dissimilar from computers and their behavior is not dissimilar from a program.   There is no hope of putting it more eloquently then \cite{rothstein1982physics}:

\begin{quotation}
‘Computers and living systems share many characteristics: they behave selectively, show tremendous differences in response to similar inputs, and are nonequilibrium systems, generally metastable.  They usually require an ongoing dissipation to maintain their characteristic behaviors, which depend on stored information.  Both involve loosely coupled subsystems, use elaborate internal communication and control systems, and key interactions within them and with their environments are typically irreversible and \textit{all or nothing} (nonlinear).'
\end{quotation}

One may also add that they have already been selected from a vast space.         

The phage $\lambda$ is simply one of the $4^{48502}$ molecules DNA could have chosen from this vast space.  Yet, this sequence warrants its own name, for it is uniquely adaptated to secure the fidelity of that same sequence through the infection of a host cell.  To do so, it chooses selectively between lysis and lysogeny and requires an ongoing interaction with RNA polymerase and the ribosome to achieve its steady state protein concentrations through the transcription of the cI and Cro genes.  This involves the coupled subsystems, the cI and Cro proteins and the $O_R$, with elaborate internal communication through oligomerization processes, and key interactions within them are indeed nonlinear.  The fate of the decision is thermodynamically irreversible.  Let us elaborate.

\section{phage $\lambda$}


The enterobacteria phage $\lambda$ infection of $\textit{Escherichia Coli}$ is proving to be an insightful example in biology of how the small-scale molecular interactions involved in gene signaling and expression lead to distinct outcomes \cite{ptashne_principle,ptashne2002genes}.  In this case, these outcomes are the very life and death of the cell.  A history of ingenious experiments has revealed the interactions involved in this determination of cell fate in exacting detail \cite{ptashne1992genetic}, and the $\lambda$ phage's strategy for reproduction is a clear example where the molecular strategy involved in the evolution and natural selection of the organism are well understood.  The questions then remaining are the chemical and physical questions in the cellular environment, particularly, what the interactions between these macromolecules are, and the type of model that is best used to quantify these interactions. 
The timescales involved in the lambda phage genetic switch are on the order of $10^2$ s to $10^4$ s, thereby placing the system beyond the limits of an all-atom molecular simulation with an empirical forcefield. Phenomenology has been introduced, and the origin of the model herein is the formulation of \cite{shea_or_1985} with subsequent revision in \cite{aurell_stability_2002} and propagated by \cite{wang_kinetic_2010,ao_potential_2004} and in the framework of \cite{wang_potential_2008}: the diffusion equation.  This has been adapted herein for the finite element method. 

\begin{figure}[h]
\begin{center}
\includegraphics[width=0.7\textwidth]{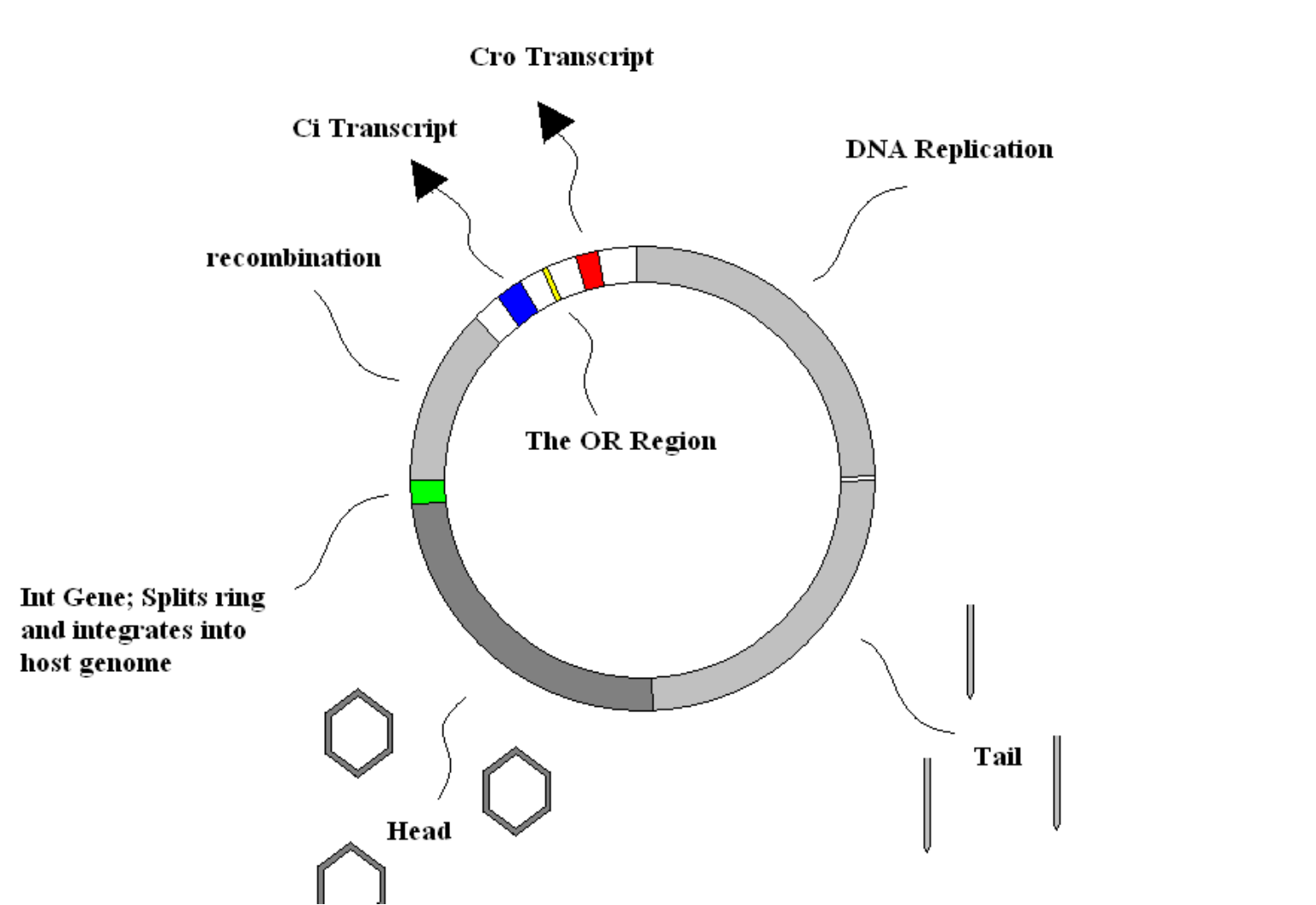}
\caption{{\bf Some basic features of the $\lambda$ phage genome.}}
\label{genome}
\end{center}
\end{figure}

The first description of the $\lambda$ phage system as a two-state switch was given by the biologists Lwoff, Jacob, and Monod \cite{Lwoff}. The $\lambda$ phage is a virus that recognizes and binds to its host, the bacterium \textit{Escherichia Coli}, and subsequently injects its genetic material into the bacterium in the form of double-stranded DNA, forming a ring.  After the virus has infected the cell, there are two main mechanisms by which the viral DNA may be replicated: the lytic cycle, or the lysogenic cycle. In the lytic cycle, the bacterial DNA is destroyed, and the virus takes control of the metabolic activity of the cell with the goal of replicating its own DNA. The host cell is quickly overcrowded with new viruses, and the cell bursts, releasing the new progeny viruses. In the lysogenic cycle, the viral DNA is incorporated into the DNA of the host, and cellular replication occurs normally. The host cell is not destroyed, and the virus does not produce progeny.  A summary of the genome is depicted in Fig. \ref{genome}.  
The goal of the experiment of Lwoff et al. was to understand the genesis and predictability of the lytic cycle. In their work, a collection of \textit{E. coli} cells infected with the $\lambda$ phage was irradiated by a moderate dose of ultraviolet light. In contrast to the expected scenario observed without radiation, in which many generations of cells underwent uninterrupted reproduction, the irradiated cells stopped growing and 90 minutes later a large fraction of the cells burst into a crop of $\lambda$ viruses. Repetition of the experiment led to reproducibility of the result, and the phenomenon was understood to be an example of the switching on or off of a gene controlling the lytic or lysogenic fate of the cell.

\begin{figure}[h]
\begin{center}
\includegraphics[width=0.7\textwidth]{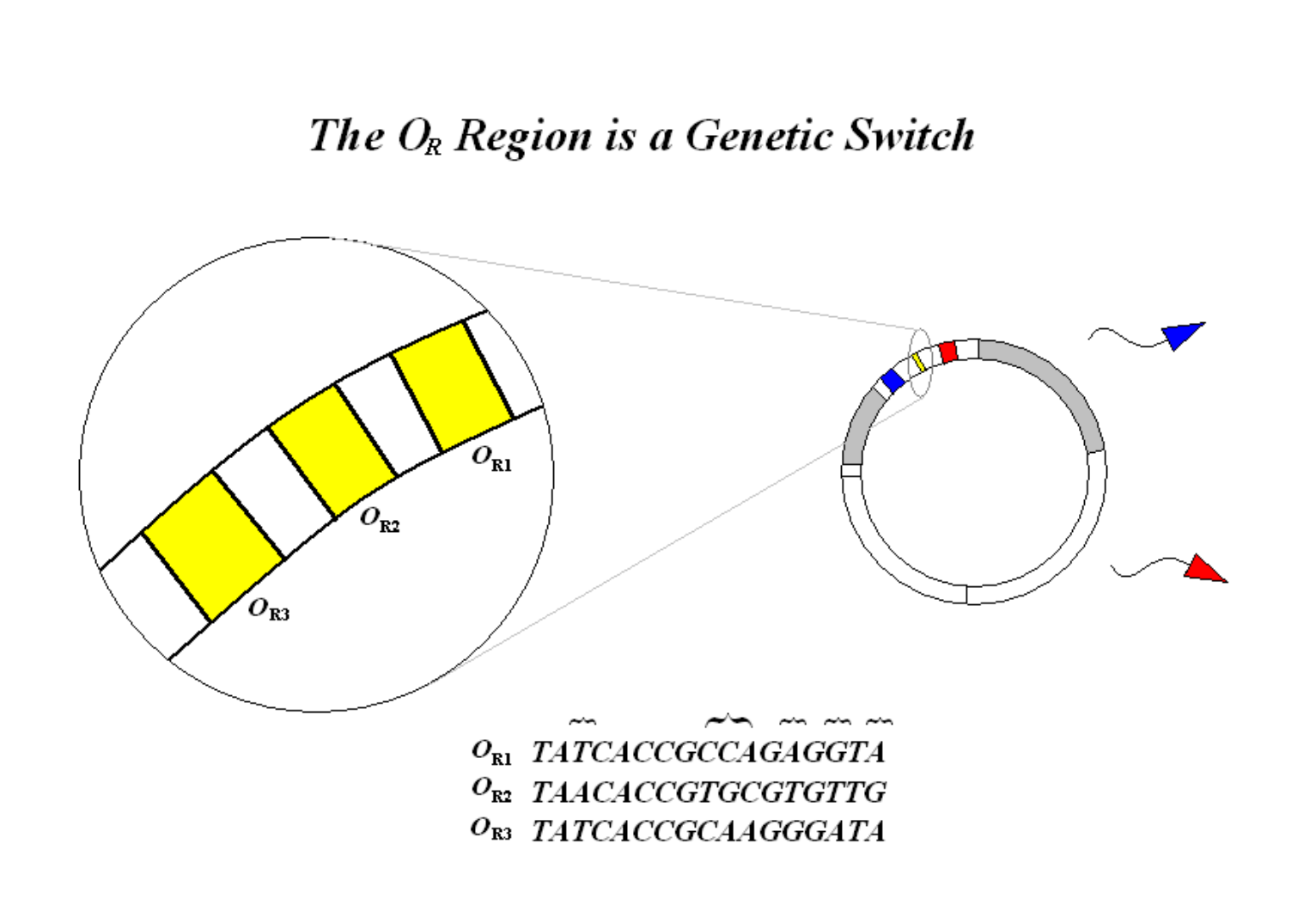}
\caption{{\bf Zooming into the $O_R$ site reveals more structure, the $P_R$ site overlaps with $O_{R1}$ and $O_{R2}$ and $P_{RM}$ overlaps with $O_{R3}$.}}
\label{switch}
\end{center}
\end{figure}

The next major step was the discovery of the $O_R$ region \cite{ptashne1992genetic}.  The 48502 base pairs of the genome had been sequenced, and experiments showed that between basepairs 37940 to 38020 there existed a region, the $O_R$, between the cI and Cro genes where the cI and Cro proteins compete on binding sites for their transcription.  This region overlaps with two RNA polymerase promoters, $P_{RM}$, $P_R$, for promoting cI and Cro, respectively.  $P_{RM}$ overlaps with $O_{R3}$ and $O_{R2}$, while $P_R$ overlaps with $O_{R1}$, Fig. \ref{switch}.  This has given the cI protein the name $\lambda$ repressor, since in lysogeny the cI protein actively represses the expression of the Cro gene by blocking the $P_R$ binding site from RNA polymerase.  When Cro is present the affinity is hightest to the $P_{RM}$ site, blocking cI promotion. 

\begin{figure}[h]
\begin{center}
\includegraphics[width=0.7\textwidth]{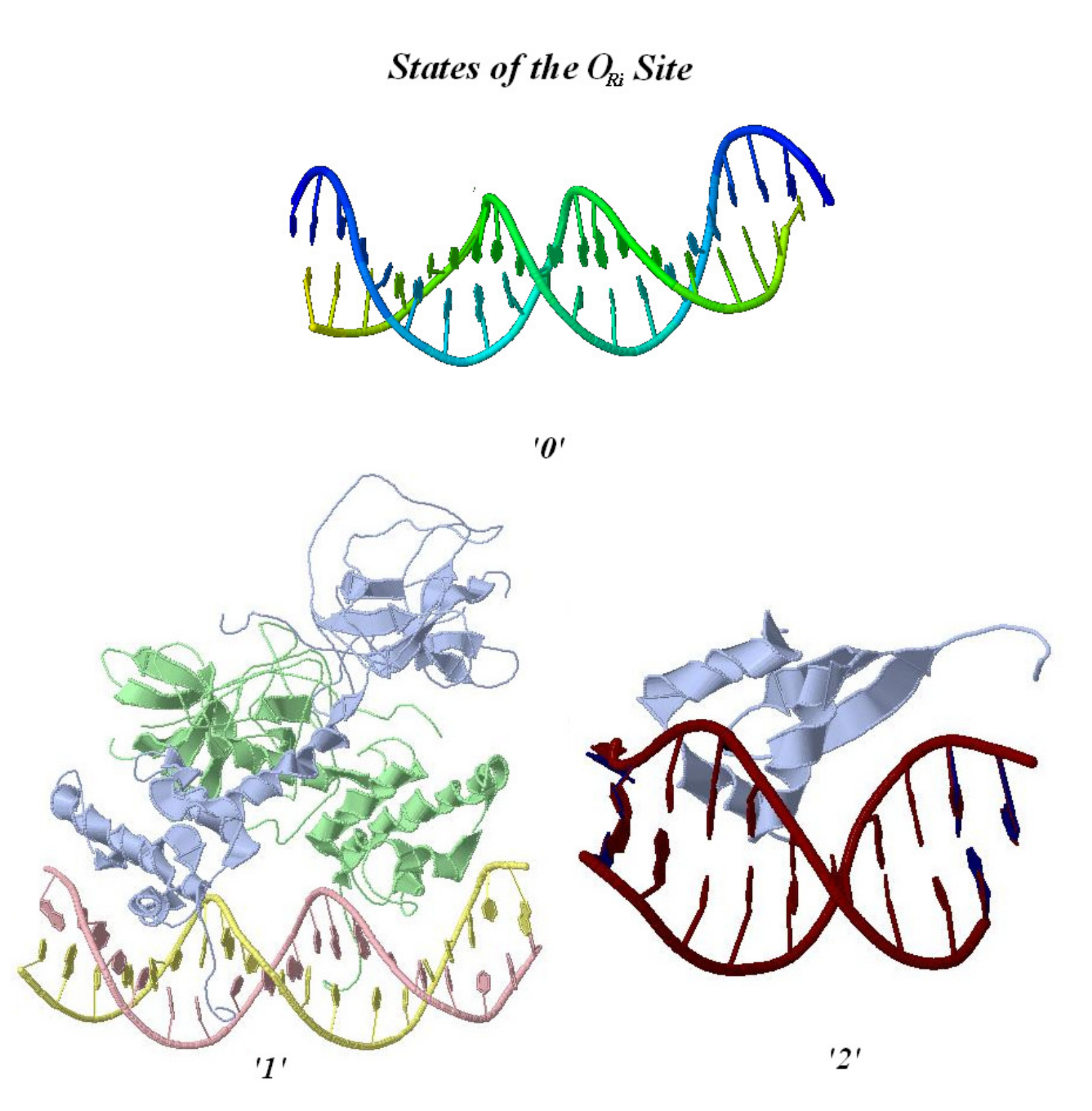}
\caption{{\bf Depicted are three possible states of the $O_{Ri}$ site. Each of the three DNA sites can be either empty, occupied by a cI dimer, or occupied by a Cro dimer.}}
\label{regions}
\end{center}
\end{figure}

In this study, we focus on the fate of the \textit{E. coli} bacterial cell at the moment that infection with the $\lambda$ phage occurs.

In the chemistry section of this work, we introduce the equilibrium assumptions used in the model and tabulate the rates involved in some of the system kinetics, namely the binding of dimers to DNA, and the dimerization of the proteins in the cell.  The $\Delta G$ values are the result of fits to experimental studies and are largely the focus of experimental work on which this model depends.  All numbers have been chosen to correspond with the numbers implicit to \cite{wang_kinetic_2010} which we repeat here.  With modern crystallographic studies of these interactions and the state of empirical force fields, these values could in principle begin to be given a basis without the biological experiments.  This avenue will not be pursued here, alhough it is representative of the type of hurdle involved in getting clinically useful information out of molecular structure-based studies and we stress the importance of such calculations to provide some atomic and molecular justification of this phenomenology. 

In the dynamics section of this work, we review the phenomenology that has been introduced.   This phenomenology is interesting in its own right as a method for exploring non-equilibrium statistical mechanics behavior over large time scales given some understanding of the equilibrium picture.  The success with the $\lambda$ phage is suggestive of rather general and powerful methods for exploring other gene regulatory circuits and complex systems as well.  The essential content of the dynamical system of \cite{shea_or_1985} is contained in the rate equation that follows

\begin{equation} \dot{x_i} = \nu_{x_i}\frac{Z_{x_i}(x_1,...,x_n)}{Z(x_1,...,x_n)}-\frac{x_i}{\tau_{x_i}}. 
\label{sanda}
\end{equation}  

\noindent Here, the time derivative of the $x_i$ protein is written in terms of the rate, $\nu_{x_i}$, at which RNAp and ribosome can actively create more $x_i$.  $Z_{x_i}$ is itself the sum over states that allow a creation process and is weighted by a sum over all states.  In the cell the protein decays at a rate of $\tau_{x_i}$.  No summation convention is employed with this equation.  

The inevitable limitation of Eq. (\ref{sanda}) is the presence of noise, and corrections are necessary.  To this end, the diffusion section of this work introduces an integral method for solving a class of partial differential equations including the Fokker-Planck equation $\partial_t P = -\partial_i(F_iP-D_{ij}\partial_j P)$.  This is to provide a numerical method of solution to the general formalism for using diffusion equations to describe noise in chemical equations, \cite{wang_potential_2008}.  This is an analogy to stochastic methods for chemical equations \cite{ge_physical_2010,qian_concentration_2002}, and path integral methods for chemical equations \cite{wang_kinetic_2010}.  The class is then restricted to the case of flows induced by the dynamic assumptions of the phage.  This method is typical of a finite element method and its origins even predate the existence of the computer that has made it so practical \cite{fenicsbook}.  The following two equations are the main result of this thesis.

\begin{equation}
a  =  \int_\Omega (v u_{t+\delta t}+\frac{\delta t}{2}\partial_i v [D_{ij}\partial_j u_{t+\delta t} - F_i u_{t+\delta t}]) dx, 
\end{equation}

\begin{equation}
L  = \int_\Omega (v u_{t}-\frac{\delta t}{2}\partial_i v [D_{ij}\partial_j u_{t} - F_i u_{t}]) dx.
\end{equation}

\noindent The solutions $u_t$ and $J=D_{ij}\partial_j u_{t} - F_i u_{t}$ are the probabilistic and flux landscapes, respectively, that appear in the title of this work.  

a = L is a linear matrix equation in a function space when one specifies an initial condition.  The dimension of the space depends on the computational resources at our disposal and the flexibility in this direction is just one luxury of FEniCS software.\footnote{Various mesh sizes were used in this thesis, a 600 by 600 space of quadratic elements, corresponding to a vector space of $\approx 720000$ dimensions was about the limit of a desktop computer.}  The expression a is bilinear, in $u_{t+\delta t}$ and in v, when used for propagating time in the positive direction.  The right hand side in this equation is only linear, depends only on v, i.e. $u_t$ is fixed by the previous time step and thus by the initial condition.   

 This is done by casting all the functions into a function space, or vector function space as needed.  When these expressions hold for all v in the function space, $u_{t+dt}$ is a solution to the related Fokker-Planck equation if $u_t$ is.   Integral nth passage time equations are similarly introduced and first passage time equations are solved for all the mutants.  The integral equation corresponding to $F_i \partial_i t_1(x)- D\partial_i\partial_i t_1(x)=-1$ follows      

\begin{equation}
\int_\Omega (t(x)(F_i \partial_i t_1(x))- D\partial_i t(x)\partial_i t_1(x)) dx = -\int_\Omega t(x)dx,
\end{equation}

\begin{equation}\int_\Omega (t(x)(F_i \partial_i t_n(x))- D\partial_i t(x)\partial_i t_n(x)) dx = -n\int_\Omega t(x)T_{n-1}dx.
\end{equation}

    The passage times proceed recursively.  After the first passage times are found and inserted for $T_0$ the second equation is rendered linear and the higher moments follow similarly.  The situation in the $O_R$ requires only the first passage time; the extreme value of these numbers compared to the cell cycle is how the robustness is quantified here.  There is some dispersion in the literature as to what that value is, as it is difficult to measure directly, but an estimate of the upper bound is that 1 in $10^{7 \pm 2}$  flip in a cell cycle, see \cite{little_robust}.        

The integral equations are discussed in the \textit{diffusion} section and further tabulation of the results occurs in chapter 6.  The methods were kept general and modular to quickly facilitate deeper exploration as more computational resources are employed or parameters changed.  It is taken for granted that the models of systems that evolve should likewise undergo evolution.

The $\lambda$ phage also meets the criterion for a selective system to be worthy of intensive study; the $\lambda$ phage is a modular system, a system of interacting components that can be systematically permuted \cite{rothstein1982physics}.  

He also motivates purely computational approaches to theoretical biology.

\begin{quotation}
‘Progress in theoretical biology, we believe, needs help from more than physics and biology.  It needs help from computer science, where any [selective system] that can be described by either a statistical, mathematical, or logical theory can be modeled.  In particular, reconfigurable systems consisting of interacting subsystems should be studied intensively.’
\end{quotation}

The $\lambda$ phage has proven to be a case in point of the benefits of such intensive study.  The effects of some mutations in the $O_R$ have already been experimentally realized, \cite{little_robust}, and $\lambda_{121}$, $\lambda_{323}$ and $\lambda_{323'}$ have been experimentally synthesised and studied. The most successful exploration in this varietal has been by far directly from the biologists, where DNA and proteins are directly manipulated and observed.  It is the authors hope that in time the contributions from physics, mathematics, computer science, and chemistry will provide the means to observe much more quickly and cost effectively some of the effects possible in this vast configuration space, and direct biological experiments towards interesting possibilities.  It is largely in the direction of this ideal that this thesis explores, and we will simulate those and similar mutations, proposing theoretical predictions to feasible experiments which could further constrain the models.

\section{Evolution}

Evolution, like gravity and thermodynamics, is so seamlessly entangled into our lives as to often go unnoticed.  Once, however, the concept is understood it is quite self-evident, and is as readily seen as the selection of an item off of a menu or the selection of students or employees from a pool of applicants.  This interplay between chance and determinism, the arbitrary and the necessary, and populations and their individuals provides more then enough room for complexity to emerge, even before granting a capacity for volition to the individuals.  
Much confusion about evolution can be avoided by clearly seperating what one is selecting from and selecting for, \cite{Mayr_Evolution}.  This thesis takes leave from the \textit{noblesse oblige}, in speaking of matters in which one is not an expert,\footnote{somebody has to bridge the gaps...\cite{whatislife}} but the motivation of such an investigation as this is largely rooted in Evolution, and such a discovery warrants discussion in a physics thesis.  Indeed in \cite{ao_borges_2008} it is speculated that Evolution \textit{is} the theory of everything physicists have vainly sought after.  In addition, the main methodology of this thesis is one of tinkering; to do lots of trials and make a lot of errors.  This process is in correspondence with how natural selection naturally selects, see \cite{Jacob}.    

\subsection{Natural Selection from...}

We will be considering natural selection from the set of DNA configurations.  We choose 27 of these configurations corresponding to mutations in the $O_R$ region for which present knowledge enables a prediction of what the binding affinities would be.  It is demonstrated that just this restricted space is enough to sample both sides of a bifurcation.

\subsection{Natural Selection for...}

If biological systems are like other physical systems they seek to minimize energy and maximize entropy.  We will rank our mutants from fittest to least fit based on passage times, energy, and entropy.  Natural selection is a fine balancing of many considerations.

\section{Three modes of Computation}

Many modes of computing with molecules have been proposed and these can be split into three overlapping categories.

Although the number of genes in the human genome has likely been constant for a few million years, the number in the refereed literature steadily decreases with time.  In the human genome there are believed to be around 25000 protein coding genes, and much more of the genome is understood to be dedicated to regulation and signaling processes than initially thought.  

Cooperativity between regulatory proteins and RNAp for transcription is seeming to be the rule rather then the exception \cite{ptashne_principle}.  The signaling involved masterfully involves the time domain and is the major port of departure between DNA type computation and a Turing tape.  Whereas a Turing maching requires a head to physically transform the bit strings along the tape, DNA strictly avoids this phenomenon and instead prefers to modify the tape through binding of ligands.  This is the tragic flaw of Bennet's proposed DNA computer in 1982 in which a ribosome like structure alters base pairs.  It may be more easily implemented on RNA then DNA, but still the lesson to be learned is to program with the time domain as much as possible.  The paradigms of causality are being replaced by the paradigms of synchronicity; GPU, DuoCore, QuadCore, quantum computation, and chemical computation are all parallel computations that rely on a balancing of rates in order for theircomponents to sync and perform properly.  

\subsection{Binding Sites}
Anything that can count can compute, and RNAp transcribing a gene into a protein is such a means of computing.  This is the type of computation discussed quantitatively in this thesis, using the phage $\lambda$ as a concrete example.  In the dynamics section we will discuss the rich space of analog computation, which computation with binding sites can simulate.    

\subsection{Polymerization}
A type of computation DNA can employ, in which a large configuration space is rapidly sampled through polymerization processes has demonstrated working, fast algorithm for NP problems.  \cite{adleman1994molecular,adleman1998computing,braich2002solution}

\subsection{Solitons}
A method of computation that macromolecules could employ is through solitons.  The effective field theories for some of the empirical force fields used to describe the molecular dynamics are known to emit soliton solutions, \cite{yakushevich_nonlinear_2002}.  The presence of soliton solutions is sufficient to define a bit, and the propagation of the soliton is a sufficient means to transfer a signal.  The empirical model used in protein folding is similar in structure to the empirical models used in superconducting circuits.  Cascading logic implemented through solitons in superconducting circuits is the most promising approach to superconducting-based computation, \cite{likharev_rsfq_1991,semenov_negative-inductance_2003,ren_progress_2009,jie_thesis}.      
It will be found that biological systems have the same physical limitations as engines or other computational devices and deep understanding of those limitations are required, \cite{likharev_classical_1982}.

To specify, consider the Hamiltonian used in \cite{yakushevich_nonlinear_2002},

\begin{equation}
\begin{split}
H &=  \sum_n (\frac{1}{2}I_{n,1}\dot{\phi}_{n,1}^2+\frac{1}{2}I_{n,2}\dot{\phi}_{n,2}^2+ \epsilon_{n,1}sin^2(\frac{\phi_{n+1,1}-\phi_{n,1}}{2}) \\ &  +\epsilon_{n,2}sin^2(\frac{\phi_{n+1,1}-\phi_{n,2}}{2})+V_{\alpha\beta}(\phi_{n,1},\phi_{n,2})). 
\end{split}
\end{equation}

Now consider a Hamiltonian used in superconducting circuits \cite{averin_rapid_2006},

\begin{equation}
H = \sum_n[\frac{Q_n^2}{2C}+2E_jsin^2(\frac{\phi_n}{2})+E_L(\phi_{n+1}-\phi_n-\phi_n^{(e)})^2].
\end{equation}.

We can see that the correspondence is not just a matter of romance.  Letting the nucleotides correspond to junctions, and the phase of the superconducting order parameter correspond to physical angles, one can proceed quite formally.

The solitons would then be of the sine-Gordon varietal and approachable through standard methods, \cite{instantons_solitons}.  

It is a mystery of gene expression and cell behaviour how genes turn on and off, how several distinct and large regions of DNA, genes, can be quiet for the majority of a cell cycle and then be triggered for expression of certain proteins at critical moments.  Although these processes are often known to be caused by chemical triggers, mechanical triggers are not ruled out, and anything that can induce long-range spatial and temporal correlations over a chain of DNA should be considered deeply.  Solitons typify long-range spatial coordination, and are at least an entertaining possibility of signaling modes available in DNA.       

The analogy is suggestive also of modeling systems via direct mimicry.  Junctions could be chosen with capacitance and junction energies to directly simulate on a transmission line the dynamics of a DNA chain.  Similar circuits can be designed for protein folding with a corresponce between the junctions and the amino acids.  Although DNA and proteins are brilliant, they are not particularly fast with that brilliance, and such circuits could boast protein folding faster than proteins can fold a protein.  The lesson to be learned in superconducting circuits from DNA is a promise of diverse effects emerging from the aperiodicity of a crystal.  One might be capable of developing dramatically sophisticated circuits, without a large number of components, by using similar-but-different junctions.  The richness of the protein functions themselves are believed to be derivative of their structure, which itself is yet another chain, this time of geometric motifs, helices, hairpins.  One is left curious to find analogues of these structures in the circuits.        
 
\section{Methods of Numerical Analysis}

The rich diversity intrinsic to biological systems lends necessity towards mathematics and computation that is equally rich and diverse.  In computational terms, biophysics necessitates modularity, as code needs to be useful and applicable in a wide variety of settings.  Emphasis has been placed on exactly this point throughout the design of numerical investigations reported in this thesis.  This outlook has allowed for scanning large regimes of parameters of particular systems as well as investigating many systems and models.  It is a well known problem of computer science that generality comes at the cost of requiring large amounts of code to consider all possible cases.  This problem has largely been circumvented through the use of automated form compilation, in which the code written is designed to write more specific code when a particular instance is requested, \cite{fenicsbook}\footnote{One cannot overstate the importance of the software and contributors therein.  Their achievement is momentous.}.  Such a method lends to an elegant correspondence between the mathematics describing a problem and the corresponding computer instruction.

The mathematics employed for this analysis is a modern incarnation of finite element methods developed before computers existed to provide a general method to solve or approximate large classes of partial differential equations from theories of hydrodynamics, electromagnetism, and elasticity, and is sufficient to solve the partial differential equations arising from biophysics as well.  We find that even in circumstances where analytic solutions exist and can be found, numerics are more illuminating then the resulting expression and provide greater accesibility to the particular values of a function at given points in the domain.  An analytic expression may require just as much numerical work to, say, evaluate and sum the necessary hypergeometric functions when a particular value at a particular point for a particular set of parameters.  Thus, even in instances where symbolic analysis of solutions are accesible we insist on numerics.  

Another important necessity of computational techniques used to study biological systems is scalability.  Computer instructions, if not the computer itself, should easily adapt as numbers of parameters, degrees of freedom, and resolution of meshes increase.  Equally as diverse as the biological systems themselves are the assertions found in the literature about the most relevant degrees of the system, the rate of changes exhibited by the system, the parameter values the system exhibits, etc.  Scalability ensures that the choices made in approximating a system can be based on physics or other considerations and not depend exclusively on access to computational power.  For example, better computers should result in better approximations but not be a prerequisite for the utility of the method; indeed, the use of these methods predates the existence of the computer.  The desire for scalability also motivated the use of FEniCS software which can be parallelized for larger systems and meshes.  The scaling that these methods utilize are suboptimal in that as the number of system degrees of freedom are increased, the computational resources required for the same resolution increases exponentially.  For an optiplex duo core, a computer of modest price and performance, we found this to limit the size of mesh to two dimensions with 600 quadratic Lagrange elements in each direction.  Three-dimensional queries required the use of adaptive mesh techniques to locate the active regions of interest and localize the resolution to these areas.  The computer instructions, the human interface to the computer, scale ideally.  That is to say, the only instructions to change as the systems analyzed change are the definitions of the system. 
 
This partial scaling manifests in the code simply crashing when more resources are requested then available, but would continue to work on machines or clusters where the resources are indeed available.  A pleasant corrolary of the active development of FEniCS is that code already written ages like a fine wine.  The benchmarks stated have changed over time, constantly speeding up, and one only needs patience for the physical interface to be as fast as one desires, as the human interface is nearly optimized.  If you want a solution, it is now as easy as a well-defined question.      

Another important aspect of the computation, to the author, is that all results and methods are reproducible and extendable from open-source utilities.  This facilitates the transfer of knowledge from the more fortunate to the less fortunate and simply evolves faster than closed source analogues.  That is not to say that open-source is better than closed-source so much as that it will be better, and contributions in the direction of automated form compilation are quickly eliminating the gap between the quality of freely available symbolic mathematics code and the privatized software for the same tasks.  

\subsection{FEniCS}

The diffusion equations herein would never have been solved without the expertise and their freely available software.  If the codes used throughout this thesis are to be used succesfully a basic understanding of not only object-oriented programming is necessary, but also understanding and installation of the software available at www.fenicsproject.org is necessary. 

\textit{Data aequatione}, spoketh Newton, \textit{quotcunque fluentes quantitae involvente fluxiones invenire et vice versa.}  Which Arnold put into modern mathematical prose: \textit{It is useful to solve differential equations}.  Finite elements do just that. 

\subsection{ROOT}

ROOT is an open source utility developed at CERN and the standard analysis framework used at all major particle accelerators of the world, \cite{CERN}.  The object-oriented tools like graphing and histogramming utilities largely extend the utility and accesibility of data generated in this thesis.  The tools long since built for analysing nuclear data are readily available for the analysis of other data as well and provide automatable procedures for problems such as fitting peaks and simple accounting.

In particular, a pyROOT paradigm was embraced to combine this functionality with the ease of python and integrate seamlessly with the python tools built by FEniCS.     

\subsection{Aleph}

It is taken for granted that models of systems that evolve should evolve as well, and Launchpad documents the code and the evolution of this code, \cite{Aleph}.

\section{Overview}

This work is largely a survey; it is the study of a biological system with the techniques of applied mathematics, engineering, computer science, and chemistry, directed at an audience of physicists and philosophers.  As such it runs the risk of being neither here nor there, to shoot three bears while only infuriating them.  To this end offense is inevitable.  An ISI Web of Knowledge query reveals 170,303 results for Finite Elements, 11,445 for $\lambda$ Phage, and 3,151,834 for genes.  To contrast, Quantum Mechanics appears 31,673 times and Quantum Hall Effect 6,557 times.  Thus, the references are largely a matter of taste, and do not attempt a reflection of history, so much as what I felt to be instructive accounts of the systems and useful ports of entry to the historical literature.  

A goal is maintained to provide a perspective on generality while keeping the $\lambda$ phage as a concrete example.  This work seeks to explain to a physical audience the details of a biological system.

The philosophy of physics and information took inspiration from many places: \cite{anderson_science,anderson_brainwashed_2000}, \cite{bennett1982thermodynamics,toffoli_physics_1982}, \cite{likharev_dynamics_1996, likharev_classical_1982,landauer1961irreversibility}, and \cite{rothstein1982physics}.

In matters of solitons there is \cite{yakushevich_nonlinear_2002,averin_rapid_2006,likharev_rsfq_1991,likharev_dynamics_1996}.

For background and data in the $\lambda$ phage, see \cite{ptashne_principle,ptashne1992genetic}, \cite{ptashne2002genes}, \cite{little_robust,atsumi_synthetic_2006}.  

For evolution there are good introductions, see \cite{Mayr_Evolution,Jacob}.

For mathematics, \cite{arnold_ordinary_1978,Perko}.

For early modeling and the head node for the taxonomy tree of the $\Delta G$ values, \cite{ackers_quantitative_1982,shea_or_1985}.

Later modeling, \cite{aurell_epigenetics_2002,aurell_stability_2002,Ao,wang_kinetic_2010}.

The methods and ideas herein are largely in correspondence to: \cite{wang_potential_2008}, \cite{Ao,wang_potential_2008,wang_kinetic_2010,ge_physical_2010}, \cite{feng_potential_2011} and have benefited from numerous conversations with Jin Wang.

An accomplishment of this thesis is one of method; it is the computational framework in \cite{Aleph}, this is but a modest interface and synthesis of tools \cite{fenicsbook, CERN}.

\chapter{\textit{Binding Polynomials and the Chemistry of the $O_R$ Region}}
\begin{quotation}
\textit{Pray be under no constraint in this house.  This is Liberty-hall, gentlemen. You may do just as you please here.}

\hspace{0.1in}Oliver Goldsmith, \textit{She Stoops to Conquer}
\end{quotation}

\section{$\Delta G$ Measurements}

In this section we consider the partition functions used to describe the chemical binding interactions in the $O_R$ region that must be posited before we proceed to introduce dynamics.  There is some choice so we explore multiple possibilities.  
    
For one, the original partition functions include explicitly RNA polymerase at a fixed concentration, whereas many new pictures also include bindings to the $O_L$ region.  A complication is that both cI and Cro proteins bind to any of the 3 sites, on $O_R$ or $O_L$ in their dimer form.  It is also now established that up to octamers can be present and, amazingly, cI octamers can bind at once to the $O_R$ as the $O_L$ and short the DNA loop exactly across the cI gene.  We will continue to use the equilibrium relationship between dimer and monomer concentrations in our partition function, \cite{Reinitz,wang_kinetic_2010,aurell_epigenetics_2002,Ao}.

 Again, x, y are cI, Cro protein numbers and [cI],[Cro] denote the dimer concentrations.     
The simplest partition function used does not include RNAp binding affinities or concentration, and consists of the 27 states permitted by having nothing, cI, or Cro on each of the three sites.  Of these states, 1 is zero liganded, 6 are singly liganded, 12 are doubly liganded, and 8 are triply liganded \footnote{Since $\Delta G$ values are negative and the individual processes occur at similar rates to each other, this establishes a sort of band structure; the triply liganded configurations occur the most often, then the doubly liganded, then the singly liganded, and then the scarcest event is to find an open DNA configuration.}  The nature of the partition function is perhaps most evident when written in the form

\begin{equation}
  \begin{split}
Z &=1+\\
  &  [cI](e^{-\frac{\Delta G_{001}}{RT}}+ e^{-\frac{\Delta G_{010}}{RT}}+ e^{-\frac{\Delta G_{100}}{RT}})+\\
  &  [Cro](e^{-\frac{\Delta G_{002}}{RT}}+ e^{-\frac{\Delta G_{020}}{RT}}+ e^{-\frac{\Delta G_{200}}{RT}})+\\
  &  [cI]^2(e^{-\frac{\Delta G_{011}}{RT}}+ e^{-\frac{\Delta G_{101}}{RT}}+ e^{-\frac{\Delta G_{110}}{RT}})+\\
  &  [Cro]^2(e^{-\frac{\Delta G_{022}}{RT}}+ e^{-\frac{\Delta G_{202}}{RT}}+ e^{-\frac{\Delta G_{220}}{RT}})+\\
  &  [cI][Cro](e^{-\frac{\Delta G_{012}}{RT}}+ e^{-\frac{\Delta G_{021}}{RT}}+ e^{-\frac{\Delta G_{102}}{RT}}+\\
  &      e^{-\frac{\Delta G_{201}}{RT}}+ e^{-\frac{\Delta G_{210}}{RT}}+ e^{-\frac{\Delta G_{120}}{RT}})+\\
  &  [cI]^3(e^{-\frac{\Delta G_{111}}{RT}})+\\
  &  [Cro]^3(e^{-\frac{\Delta G_{222}}{RT}})+\\
  &  [cI]^2[Cro](e^{-\frac{\Delta G_{112}}{RT}}+ e^{-\frac{\Delta G_{121}}{RT}}+ e^{-\frac{\Delta G_{211}}{RT}})+ \\
  &  [cI][Cro]^2(e^{-\frac{\Delta G_{122}}{RT}}+ e^{-\frac{\Delta G_{212}}{RT}}+ e^{-\frac{\Delta G_{221}}{RT}})
\end{split}
\label{partition}
\end{equation}

This form makes it clear that it is symmetric with respect to permutation of the sites.  The argument of \cite{aurell_epigenetics_2002} is that RNAp needs an open $P_R$ site to promote Cro and an open $P_{RM}$ site to promote cI, and if the site is open, RNAp will be overwhelmingly first in line to bind and promote such states.  It is known that the $P_R$ site overlaps with $O_{R_1}$ and $O_{R_2}$ and $P_{RM}$ overlaps with $O_{R_3}$.  This symmetry is not present for the subspaces; the sum over the substates with a free $P_{RM}$ can promote x, a cI monomer, and the sum over the substates with a free $P_R$ site can promote y for a Cro monomer  

\begin{equation}
  \begin{split}
Z_x &=1+\\
  &  [cI](e^{-\frac{\Delta G_{010}}{RT}}+ e^{-\frac{\Delta G_{100}}{RT}})+\\
  &  [Cro](e^{-\frac{\Delta G_{020}}{RT}}+ e^{-\frac{\Delta G_{200}}{RT}})+\\
  &  [cI]^2e^{-\frac{\Delta G_{110}}{RT}}+\\
  &  [Cro]^2e^{-\frac{\Delta G_{220}}{RT}}+\\
  &  [cI][Cro](e^{-\frac{\Delta G_{210}}{RT}}+ e^{-\frac{\Delta G_{120}}{RT}}),
\end{split}
  \begin{split}
    Z_y &=1+\\
    &  [cI]e^{-\frac{\Delta G_{001}}{RT}}+\\
    &  [Cro]e^{-\frac{\Delta G_{002}}{RT}}.
  \end{split}
\label{zxandzy}
\end{equation}

with

\begin{equation}
\begin{split}
[cI] &= [x*pnc]/2+e^{\Delta G_{cI}/RT)}/8\\
     & -([x*pnc]\frac{e^{(\Delta G_{cI}/RT)}}{8}+\frac{e^{2 \Delta G_{cI}/RT}}{64})^{1/2}
\end{split}
\label{dimerizationcI}
\end{equation}

\begin{equation}
\begin{split}
[Cro] & = [y*pnc]/2+e^{\Delta G_{Cro}/RT}/8\\
     & -([y*pnc]\frac{e^{(\Delta G_{Cro}/RT)}}{8}+\frac{e^{2 \Delta G_{Cro}/RT}}{64})^{1/2}
\end{split}
\label{dimerizationCro}
\end{equation} Interlude: On dimerization.\footnote{

\bf{\Large{\textit{On Dimerization}}}

\begin{quotation}

$1+1=1$
  
\textit{Una goccia pi\'u una goccia, fanno una goccia pi\'u grande, non due!} 

  \hspace{0.1in}the Tarkovsky equation and explanation, \textit{Nostalghia}
\end{quotation}

\begin{quotation}
\textit{Todos los fuegos el fuego}

  \hspace{0.1in}Julio Cortazar, \textit{Todos los fuegos el fuego}
\end{quotation}

 The dimerization equation used in this work originates from a quasi-steady state assumption for the dimerization reaction.

\begin{equation}
  \begin{split}   
   [cI_1]^2&= [cI_2]e^{\frac{\Delta G_{cI}}{RT}}\\
    \end{split}, \qquad \begin{split} 
   [N_{cI}]&= [cI_1]+2[cI_2]\\ 
   \end{split} 
 \end{equation}

This gives a quadratic equation with roots giving the cI dimer concentration, $[cI_2]$ in terms of the total protein concentration, $[N_{cI}]$.  

\begin{equation}
[cI_2]=\frac{[N_{cI}]}{2} + \frac{e^{\frac{\Delta G_{cI}}{RT}} \pm \sqrt{e^{\frac{2\Delta G_{cI}}{RT}}+8e^{\frac{\Delta G_{cI}}{RT}}[N_{cI}]}}{8}
\end{equation}

The root with the positive sign is the expression stated in \cite{wang_kinetic_2010,ao_potential_2004} and the negative sign corresponds to the expression used in the calculations of \cite{Reinitz,ao_potential_2004, wang_kinetic_2010}.  The negative sign seems to be the better physical choice in that, in the limit of zero monomer concentration, one also has zero dimer concentration.      
}

\begin{table}[ht]
 \footnotesize
  \centering
     \begin{tabular}{|c|c|c|c|c||c|c|c|}
          \hline
          \multicolumn{5}{|c|}{$\lambda_{123}$, $wild-type$ $O_R$ affinities (kcal/mol)}
          & \multicolumn{3}{|c|}{$\Delta G_{ijk}$}\\ \hline \hline
          protein & $O_{R1}$&$O_{R2}$&$O_{R3}$&ref.& equation & monomial & premotes\\ \hline
       $\Delta G_{001}$,$\Delta G_{010}$,$\Delta G_{100}$   & -15 & -13 & -12 & \cite{Ao} & $\Delta G_{000} = 0.0 $ & 1 & $Z_x$,$Z_y$ \\ \hline
       $\Delta G_{002}$,$\Delta G_{020}$,$\Delta G_{200}$   & -18.4 & -17.1 & -19.5 & \cite{Ao} & $\Delta G_{001} = -12.0 $ & $[cI]$ & $Z_y$ \\ \hline
       \multicolumn{5}{|c||}{} & $\Delta G_{010} = -13.0 $ & $[cI]$ & $Z_x$ \\ \hline
       \multicolumn{5}{|c||}{$\Delta G_{coop} = -6.9 $} & $\Delta G_{100} = -15.0 $ & $[cI]$ & $Z_x$ \\ \hline
       \multicolumn{5}{|c||}{$pnc = 1.5*10^{-11}$} & $\Delta G_{002} = -19.5 $ & $[Cro]$ & $Z_y$ \\ \hline
       \multicolumn{5}{|c||}{$mpi = 3$\footnote{mpi is phages per cell.}} & $\Delta G_{020} = -17.1 $ & $[Cro]$ & $Z_x$ \\ \hline
       \multicolumn{5}{|c||}{} & $\Delta G_{200} = -18.4 $ & $[Cro]$ & $Z_x$ \\ \hline
	\hline
       \multicolumn{3}{|c|}{$\Delta G_{ijk} (cont)$} & \multicolumn{3}{|c|}{$\Delta G_{011} = \Delta G_{010}+\Delta G_{001}+\Delta G_{coop} $} & $[cI]^2$ & \\ \hline
       \multicolumn{3}{|c|}{} & \multicolumn{3}{|c|}{$\Delta G_{110} = \Delta G_{010}+\Delta G_{100}+\Delta G_{coop} $} & $[cI]^2$ & $Z_x$ \\ \hline
       \multicolumn{3}{|c|}{} & \multicolumn{3}{|c|}{$\Delta G_{101} = \Delta G_{100}+\Delta G_{001}$} & $[cI]^2$ & \\ \hline
       \multicolumn{3}{|c|}{} & \multicolumn{3}{|c|}{$\Delta G_{022} = \Delta G_{020}+\Delta G_{002}$} & $[Cro]^2$ & \\ \hline
       \multicolumn{3}{|c|}{} & \multicolumn{3}{|c|}{$\Delta G_{220} = \Delta G_{020}+\Delta G_{200}$} & $[Cro]^2$ & $Z_x$ \\ \hline
       \multicolumn{3}{|c|}{} & \multicolumn{3}{|c|}{$\Delta G_{202} = \Delta G_{200}+\Delta G_{002}$} & $[Cro]^2$ & \\ \hline
       \multicolumn{3}{|c|}{} & \multicolumn{3}{|c|}{$\Delta G_{120} = \Delta G_{100}+\Delta G_{020}$} & $[cI][Cro]$ &$Z_x$ \\ \hline
       \multicolumn{3}{|c|}{} & \multicolumn{3}{|c|}{$\Delta G_{210} = \Delta G_{200}+\Delta G_{010}$} & $[cI][Cro]$ &$Z_x$ \\ \hline
       \multicolumn{3}{|c|}{} & \multicolumn{3}{|c|}{$\Delta G_{102} = \Delta G_{100}+\Delta G_{002}$} & $[cI][Cro]$ &\\ \hline
       \multicolumn{3}{|c|}{} & \multicolumn{3}{|c|}{$\Delta G_{201} = \Delta G_{200}+\Delta G_{001}$} & $[cI][Cro]$ &\\ \hline
       \multicolumn{3}{|c|}{} & \multicolumn{3}{|c|}{$\Delta G_{012} = \Delta G_{010}+\Delta G_{002}$} & $[cI][Cro]$ &\\ \hline
       \multicolumn{3}{|c|}{} & \multicolumn{3}{|c|}{$\Delta G_{021} = \Delta G_{020}+\Delta G_{001}$} & $[cI][Cro]$ &\\ \hline
\hline
       \multicolumn{3}{|c|}{} & \multicolumn{3}{|c|}{$\Delta G_{111} = \Delta G_{100}+\Delta G_{010}+\Delta G_{001}+\Delta G_{coop}$} & $[cI]^3$ &\\ \hline
       \multicolumn{3}{|c|}{} & \multicolumn{3}{|c|}{$\Delta G_{222} = \Delta G_{200}+\Delta G_{020}+\Delta G_{002}$} & $[Cro]^3$ &\\ \hline
       \multicolumn{3}{|c|}{} & \multicolumn{3}{|c|}{$\Delta G_{112} = \Delta G_{100}+\Delta G_{010}+\Delta G_{coop}+\Delta G_{002}$} & $[cI]^2[Cro]$ &\\ \hline 
       \multicolumn{3}{|c|}{} & \multicolumn{3}{|c|}{$\Delta G_{121} = \Delta G_{100}+\Delta G_{020}+\Delta G_{001}$} & $[cI]^2[Cro]$ &\\ \hline
       \multicolumn{3}{|c|}{} & \multicolumn{3}{|c|}{$\Delta G_{211} =\Delta G_{200}+\Delta G_{001}+\Delta G_{010}+\Delta G_{coop}$} & $[cI]^2[Cro]$ &\\ \hline    
       \multicolumn{3}{|c|}{} & \multicolumn{3}{|c|}{$\Delta G_{221} = \Delta G_{200}+\Delta G_{020}+\Delta G_{001}$} & $[cI][Cro]^2$ &\\ \hline
       \multicolumn{3}{|c|}{} & \multicolumn{3}{|c|}{$\Delta G_{212} = \Delta G_{200}+\Delta G_{010}+\Delta G_{002}$} & $[cI][Cro]^2$ &\\ \hline
       \multicolumn{3}{|c|}{} & \multicolumn{3}{|c|}{$\Delta G_{122} = \Delta G_{100}+\Delta G_{020}+\Delta G_{002}$} & $[cI][Cro]^2$ &\\ \hline

        \hline
        \end{tabular}
        \caption{Values of variables referenced by the partition function.  }
        \label{wtgval2}
    \end{table}%

\normalsize

The key feature to notice here is that, since all the binding affinities are negative, the sites are dominated by the triply liganded terms.  The occurrence of an x promoting state is already rare, and a y promotion state exponentially more scarce then those.  The partition function here is seen to be a sum over monomials of the forms, $1$, $[cI]$, $[Cro]$, $[cI]^2$, $[Cro]^2$, $[cI][Cro]$,$[cI]^3$,$[Cro]^3$,$[cI]^2[Cro]$, and $[Cro][cI]^2$.  In the original Shea and Ackers contribution we note that additional monomials of the form $[RNAp]$, $[RNAp][cI]$, $[RNAp][Cro]$, $[RNAP][cI]^2$, $[RNAP][Cro]^2$, $[RNAP][cI][Cro]$ and $[RNAp]^2$ were also included.  This contributed to 13 more states included in the partition function for a total of 40.  Only 40 of the $3^4$ states are permissible since an RNAp at $P_R$ precludes any binding to $O_{R_1}$ and $O_{R_2}$.   The $-\Delta G_s$ values for these states are also actively measured by the experimental community and could be included in the partition function insofar as RNAp does indeed bind in the $O_R$, so that in any given interval of time there is indeed the probability that the $O_R$ is in a state not considered in the above partition function.  The effect is said that eliminating the RNAp is \textit{constant}, though there is a continuation in the literature that prefers to keep RNA polymerase explicitly in the partition function and no mention to what the value of that \textit{constant}.  The situation is a little subtler then that and the value of this effect, as shown here, relates to the RNAp binding energies to the $P_R$, and $P_{RM}$ as well as absolute concentrations.  The benefit is still a simpler partition function and since RNAp is fixed for the simulation, the additional monomials do not alter the algebraic structure of the equations but only append to the weights of the already present monomials, but this effect can still induce bifurcations and one should be careful.   

It is then necessary to write the dimer concentrations in terms of the monomer number.  Here we use the expression implicit to the results of \cite{Ao, wang_kinetic_2010}, where we explicitly show how the number pnc, labeled as a conversion factor from number to concentration, enters into their equation. 

We note that the above equations appear with the mistake of a positive sign, in \cite{wang_kinetic_2010} and the arxiv version of \cite{Ao}, and use the expression from \cite{Reinitz} with the insertion of the pnc, a converting factor from protein number to concentration, and implicit assumptions about RNAp (see the note on dimerization).  The bacterial volume used in \cite{Ao} is stated to be $.7*10^{-15}L$ and pnc is in inverse proportion to the volume.  The value used for pnc is $1.5*10^{-11}$.  A calculation reveals and is corroborated by a private communication that in actuality $pnc = (volume*N_A)^{-1} = 2.37*10^{-9} Moles/L$ and the listed value is a typo.  Use of that number reveals unrealistic protein concentrations and it is this authors belief that the number pnc includes implicit assumptions about the RNAp concentrations and affinities.  Provided RNAp is removed from the partition function, a pnc value of $1.5*10^{-11}$ is indeed more consistent with the numbers obtained from experiment, and was likely tuned ad hoc to the present value for that reason.  No reference is made about the value of the number or conversion in \cite{wang_kinetic_2010}, although the calculations present suggest the use of the conversion from Ao with the equation from Reinitz despite the errors in their written equation.  Many calculations should be performed twice, once with the values similar to \cite{Ao} and again with RNAp in the partition function and $pnc = 2.37*10^{-9}$ which is postponed for future work.  As noted in the introduction this equation should be modified in future work anyway as a result of the higher order polymerization processes that occur.

\section{A Partition Function; RNAp or no RNAp?}

The partition function in \cite{shea_or_1985} reads, with $[RNAp]$ the concentration and assigning the index $'3'$ to denote the RNAp occupancy.    

\begin{equation}
  \begin{split}
Z &=1+([RNAp](e^{-\Delta G_{03}}+ e^{-\Delta G_{30}})+[RNAp]^2e^{-\Delta G_{33}})+\\
  &  [cI](e^{-\Delta G_{001}}+ e^{-\Delta G_{010}}+ e^{-\Delta G_{100}}+[RNAp](e^{-\Delta G_{31}}+ e^{-\Delta G_{103}}+ e^{-\Delta G_{130}}))+\\
  &  [Cro](e^{-\Delta G_{002}}+ e^{-\Delta G_{020}}+ e^{-\Delta G_{200}}+[RNAp](e^{-\Delta G_{32}}+ e^{-\Delta G_{203}}+ e^{-\Delta G_{230}}))+\\
  &  [cI]^2(e^{-\Delta G_{011}}+ e^{-\Delta G_{101}}+ e^{-\Delta G_{110}}+[RNAp]e^{-\Delta G_{113}})+\\
  &  [Cro]^2(e^{-\Delta G_{022}}+ e^{-\Delta G_{202}}+ e^{-\Delta G_{220}}+[RNAp]e^{-\Delta G_{223}})+\\
  &  [cI][Cro](e^{-\Delta G_{012}}+ e^{-\Delta G_{021}}+ e^{-\Delta G_{102}}+e^{-\Delta G_{201}}+ e^{-\Delta G_{210}}+ e^{-\Delta G_{120}}+ \\
  & +[RNAp](e^{-\Delta G_{123}}+e^{-\Delta G_{213}}))+\\
  &  [cI]^3(e^{-\Delta G_{111}})+\\
  &  [Cro]^3(e^{-\Delta G_{222}})+\\
  &  [cI]^2[Cro](e^{-\Delta G_{112}}+ e^{-\Delta G_{121}}+ e^{-\Delta G_{211}})+ \\
  &  [cI][Cro]^2(e^{-\Delta G_{122}}+ e^{-\Delta G_{212}}+ e^{-\Delta G_{221}})
 \end{split}
\label{RNApartition}
\end{equation}

It is clear that the effect of [RNAp] is not constant; [RNAp] does not factor out of the partition function.  Three experimental numbers, the RNAp concentration, the RNAp binding to the cI premoter, and an RNAp binding to the Cro premoter are entangled into this partition function.  Removing RNA polymerase from the partition function also has the effect of disturbing the time scales, in that equation \ref{partition} only accounts for the fraction of a dt in which no RNA polymerase is bound.     

We note that this form of the partition function would require corrections for low numbers in that it uses the approximations $n^2 \approx n*(n-1)$, $n^3 \approx n(n-1)(n-2)$.  It is likely that some of the complexity of these systems near the axis is residual of such an approximation.  This becomes especially more important if multiple phages have invaded the cell or the $O_L$ is included in the partition function.  Implicit to the figures in Zhu is the $T_r$, $T_{rm}$, and $T_{rr}$ values reported are 1/3 of the values used in the calculation.  This is due to the inclusion of 3 phages in the cell as in \cite{aurell_epigenetics_2002}.  The calculation left out presumes that $Z'=Z^3$, and the states that promote a ci, for example, are enhanced by the number of phages present $\frac{3*Z_{x}*Z^2}{Z^3}$.  Also, it is often presumed that for the composite system of the left and right operator regions $Z'$=$Z_{OL}Z_{OR}$.  For a partition function to be the product of the subsystems, the subsystems must not interact.  This is not the case here, and especially for the low numbers involved.  One need only plot, for example, $x^9$ next to $x*(x-1)*(x-2)*(x-3)*(x-4)*(x-5)*(x-6)*(x-7)*(x-8)$ to see that the difference is considerable.  This error is further increased with monomers present as well, since this too reduces the number of dimers and increases the fluctuation.

We use $\Delta G_{cI}=-11.1$, $\Delta G_{Cro}=-7.0$, and $\Delta G_{coop}=-6.9$.  The states and their values are collected in Table \ref{wtgval2}

The values for the multiply liganded states are found by summing up the values of the constituent singly liganded states.  Cooperativity gives an additional sum, $\Delta G_{coop}=-6.9$, for the states with cI and cI on adjacent sites.  These are the $G_{*11}$, $G_{1*1}$ configurations where $*$ can denote 0,1, or 2.  $RT=.617 kcal/mol$ is the Boltzmann factor for the 310 Kelvin laboratory conditions which these were accumulated.   
When we speak of the mutant $\lambda_{ijk}$ we mean that the affinity for the ith site is used in place of $O_{R_1}$, jth in place of $O_{R_2}$, and kth in place of $O_{R_3}$.  Some of these are likely biologically unstable but it is inevitable that different strings for the operator configurations may have similar $\Delta G$ and thus similar dynamics.

\chapter{\textit{Dynamics Near Equilibrium}} 

\begin{quotation}
  \textit{The key to the treasure is the treasure.}

\hspace{0.1in}John Barth, \textit{Chimera}
\end{quotation}  

\section{Gene Expression and General Dynamical Systems}

Although this is a simplified model, the need for general dynamical systems is already evident.                             

 The number of stationary points is complicated by dimerization but in the basic case this number is found to be in proportion to the $(number of binding sites+1)\times(number of protein species)$.  The positivity of the partition function and concentration ensures that some of these are physically realizable.  The dimerization used here doubles the number of stationary points.  Thus it is reasonable to start from very general mathematical considerations \cite{Lia93,Perko} and well over a century of elegant mathematics that would be required to understand the mathematics of gene expression. One is compelled to introduce a modified model that is more general and symmetric in the form in which creation and annihilation processes enter.  For example, consider the n equations to replace for the discussion equation 1.3.  
 
\begin{equation}
\small{\dot{x_i} = \frac{\nu_{x_i}Z_{x_i}(x_1,...,x_n;H_2O,RNAp,Ribosomes, ...)-\nu_{x^\dagger_i}Z_{x^\dagger_i}(x_1,...,x_n;H_2O,RNAp,Ribosomes, ...)}{Z(x_1,...,x_n;H_2O,RNAp,Ribosomes, ...)}}
\end{equation}

Here one can begin with a partition function for the n proteins and let quantities like $H_20$, RNAp, all the things that contribute non-negligible partial pressures in the cell, enter as control parameters.  Say, $Z(x_1,...,x_n;H_2O,RNAp,Ribosomes, ...)$.  Again $\nu_{x_i}$ is the rate at which transcription in the $Z_{x_i}$ states occur.  Introduced is an expression for annihilation, $\nu_{x^\dagger_i}Z_{x^\dagger_i}$, to account for the specific events that destroy a protein.  For example, in the $\lambda$ phage situation, the cI protein is destroyed by another protein, RecA, which destroys a cI monomer by cutting it in half.  In such a situation one would include a term $Z_{cI^\dagger}\propto [cI][RecA]$ governed by a rate, $\nu_{cI^\dagger_i}$ of this destruction process. 

The motivation is that the stationary points would appear at the points $x_1,...,x_n$ where \small{$\nu_{x_i}Z_{x_i}(x_1,...,x_n;H_2O,RNAp,Ribosomes, ...)-\nu_{x^\dagger_i}Z_{x^\dagger_i}(x_1,...,x_n;H_2O,RNAp,Ribosomes, ...)=0$} for all i.  This is a set of polynomial equations in the protein variables $x_i$.  This allows one to either fit to known systems, or construct others.  Also since, $Z_{x,y,z}$ is positive, and any term appearing in the numerator appears with a positive sign in the denominator, many qualitative features persist. 
Say $\dot{x}=f(x)$ and $\dot{y}=f(y)/g(y)$.  One can see that if $\dot{x}=0$, then $\dot{y}=0$, since g(y) is non-negative that the stationary points in the first system are the same as in the latter.  Next, in the neighborhood of a stationary point the dynamics are qualitatively the same as the linear system, 
\begin{equation}  
\dot{x}\approx \frac{\partial f}{\partial x}\bigg|_{x=x_{stationary}}(x-x_{stationary}),
\end{equation}

 and  

\begin{equation}
\dot{y}\approx \frac{\frac{\partial f}{\partial y}g(y)-\frac{\partial g}{\partial y}f(y)}{g(y)^2}\bigg|_{y=y_{stationary}}(y-y_{stationary}).
\end{equation}

\noindent Since at a stationary point $f(y)=0$, evidently, 

\begin{equation}
\dot{y}\approx \frac{\frac{\partial f}{\partial y}}{g(y)}\bigg|_{y=y_{stationary}}(y-y_{stationary}).
\end{equation}

 \noindent The two systems have similar quantitative behavior then too, with the dynamics simply scaled: 

\begin{equation}
t \leadsto t*g(y)\bigg|_{y=y_{stationary}}.
\end{equation}

\noindent For an overview of such methods in dynamical systems many texts are available, \cite{Perko}.   

This interlude is to motivate some of the extent to which these kinetics serve as analog circuits.  Restricting the numerators to polynomials, $f_i(x_1,...,x_n)$, in which $f_i$ is non-negative when $x_i = 0$, that is to say we require proteins to exist before we can destroy them.  Then we can always \textit{program a set of binding-expression interactions} to simulate such dynamics.    
 
For example, say one would like to construct a circuit with dynamics similar to $\dot{x}=x+x^2-x*z$, $\dot{y}=y+y^2-x*y$, $\dot{z}=z+z^2-y*z$.  Weighting by $Z_{x,y,z}=1+x+x^2+y+y^2+z+z^2+x*y+x*z+y*z$ does not change the locations of the stationary points.   That is to say, one can accomplish this circuit with three proteins, with two binding sites each, in which the occupation of either would enable their own transcription; also the z protein destroys x, y protein destroys z, and x destroys y.

To be able to fully fit the experiments of \cite{St.Pierre} the phase space would partition into more than just two states, but four states, in addition to a splitting of a cell population into lysis and lysogeny, some cells (refractory) were immune to the phage and a small fraction of the lysogens had daughter cells that lyse.  The number and properties of lysogens that had daughters that underwent lysis (5/100) could be useful in additional studies of robustness \footnote{An interesting explanation without requiring more basins of attraction would be, if, for example, as a cell divides the daughter cells have each taken half the cI dimers from the parent lysogen, and in so doing switch to the lytic basin of attraction, causing the cells to lyse.  (It is imaginable that whereas $x$, $y$ may sit in one basin of attraction $x/2$, $y/2$ may reside in the other)}.   We continue the analysis of \cite{Ao,wang_kinetic_2010}.  Including the $O_L$, RNAp, tetramers, etc.  could contribute this additional complexity, codes are provided \cite{Aleph} if one was inclined towards such study.  

At different times in the cell cycle different genes are turned on or off.  Often these genes are transcripts for proteins that are necessary for functions in that particular interval.  The concern here is with the lysogeny decision and in this interval we have only the $O_R$, \textit{cI} gene, and \textit{Cro} gene on.  
    
Letting $x$, $y$ denote the cI and Cro monomer numbers respectively we write for the dynamics 

\begin{equation}
\dot{x} = \nu_xZ_x(x,y)/Z(x,y)-x/\tau_x,
\end{equation}

\begin{equation}
\dot{y} = \nu_yZ_y(x,y)/Z(x,y)-y/\tau_y.
\end{equation}

\noindent The creation term in the dynamics of the associated protein populations within a cell are governed by the rates at which RNA polymerase binds and translates the DNA into mRNA, together with the rate at which the ribosome translates the mRNA into the chain of amino acids that comprise the protein.  In this way a cell seems to possess three different molecules encrypting the same message.  Categorically speaking, the fundamental biological information processing is governed by the physics of $RNAp:DNA \rightarrow RNA$, $Ribosome:RNA^3 \rightarrow \textit{amino acid}$.  The net effective rate of these processes enters directly into the equations of motion through the $\nu_{x,y}$ variables.  

One can say much about these questions without specifying numbers of binding sites or which promotional states enter into $Z_i(x,y)$ if we grant a few specifications.  Firstly, we assert the positivity of $\tau_x$ and $\nu_x$.  This is also largely natural selections choice, though alternatives are imaginable.  For example, RNAp can indeed unscribe a transcript, the backward reaction is permissible, and it is said that for any 11 forward reactions, there were 10 steps back \cite{bennett1982thermodynamics}.  The partition functions are positive by virtue of being well-behaved non-negative partition functions or a subset thereof.  To avoid singularities and to insure $\dot{x_i}>0$ we include at least one state that does not depend on $x$ and $y$ in each of $Z$, $Z_x$, and $Z_y$.  In the case of $\lambda$ this is provided by the open DNA configuration.  This is the reference state $\Delta G_{000}=0$.  In the original formulation, \cite{shea_or_1985}, security against singularities was also provided by the explicit inclusion of $[RNAp]$ and $[RNAp]^2$ states.  For two-dimensional systems, for example two genes expressing two proteins, this is sufficient to prove the existence of any non-negative number of stationary points provided a sufficient supply of binding sites.  

Stationary states occur when the derivatives vanish:  

\begin{equation}
x Z(x,y) - \nu_{x}\tau_{x}Z_{x}(x,y) = 0,     
\end{equation}

\begin{equation}
y Z(x,y) - \nu_{y}\tau_{y}Z_{y}(x,y) = 0,     
\end{equation}
This is to illustrate some of the most important aspects of the equation and how these states interplay to create the desired states.  


When $x$, $y$ are zero, $\dot{x}$,$\dot{y}$ are positive since $\nu_{x,y}$ and $Z_x(x,y)/Z(x,y)$ are.  As $x$, $y$ go to $\inf$ the $-x/\tau_x$ and -$y/\tau_y$ terms dominate and since $\tau_{x,y}$ are positive, the dynamics are confined to the positive quadrant.  It follows similarly for $y$ as well as $x$ that there exists a contour for which $\dot{x}=0$ on their respective axis since  $0<Z_x(x,y)/Z(x,y)\le 1$;  thus $\dot{x}$ switches eventually to negative and stays negative and this crossing requires at most $\nu_x\tau_x$ proteins.  


This provides an experimental test, or at least an avenue into parameterizing the models.  For example, if one were to measure only $\nu_{x,y}$, $\tau_{x,y}$, and just the expectation value of $Z_{x,y}(x,y)/Z(x,y)$, one already has means to predict the average number of proteins $x$ and $y$.  Since the expectation value of $Z_{x,y}(x,y)/Z(x,y)$ is just the fraction of time RNAp is promoting, it is a much easier process than discovering all possible states, measuring their relative affinities and then simulating.       
%

 It is the fitting of empirical models to experimental data that comprises the bulk of the quantitative features of biophysics.  
 
With cI at 236 amino acids, requiring at minimum 708 base pairs and Cro, 66 amino acids for 198 pairs, it is not surprising that Cro can be transcribed faster.  The decay time in seconds is $\tau_{cI}=2943$ and $\tau_{Cro}=5194$.   

In this model the states that contribute to the numerator for x are those in which $O_{R3}$ is unoccupied and thus open for RNAp binding.  Of these states those with a cI dimer bound at $O_{R2}$ are believed to further enhance the rates.  Thus,  

\begin{equation}
  \begin{split}
\frac{cI}{dt} = & \frac{1}{Z}(\\
  & mpi*Trm*Eci([cI]e^{-\frac{\Delta G_{010}}{RT}}+[cI]^2e^{-\frac{\Delta G_{110}}{RT}}+[cI][Cro]e^{-\frac{\Delta G_{210}}{RT}}) \\
  & +Trmu*Eci(1+[cI]e^{-\frac{\Delta G_{100}}{RT}}+[Cro](e^{-\frac{\Delta G_{020}}{RT}}+ e^{-\frac{\Delta G_{200}}{RT}})+[cI][Cro]e^{-\frac{\Delta G_{120}}{RT}}+[Cro]^2e^{-\frac{\Delta G_{220}}{RT}})\\
  & )-\frac{cI}{\tau_{cI}}
\end{split}
\end{equation}

\begin{equation}
\begin{split}
\frac{Cro}{dt} = &\frac{1}{Z}(\\
 &  mpi*Trr*Ecro*(1+[cI]e^{-\frac{\Delta G_{001}}{RT}}+[Cro]e^{-\frac{\Delta G_{002}}{RT}}\\
 & )- \frac{Cro}{\tau_{Cro}}
\end{split}
\end{equation}

Here $T_{RM}$,$T_{RMU}$, and $E_{cI}$ are 0.115 cI monomers/sec, 0.01045 cI monomers/sec and 1 copy per mRNA synthesized.  Next, RNAp is assumed to require both $O_{R1}$ and $O_{R2}$ free to bind to the Cro promotion site.  This occurrence is repressed in comparison with the states that promote cI but promotions proceed with a higher burst.  $T_{RR}$,$E_{Cro}$ are 0.30 Cro monomers/sec and 20 copies per mRNA synthesized, respectively.  $mpi$ denotes the number of phage $O_R$ sites per infected cell.  The mpi factor curiously seems to only have been multiplied by the premotional states with cI dimer at $O_{R2}$ and has been given the value 3.  Here, Z is the partition function from \ref{partition}, while [cI] and [Cro] are the assumed relations in \ref{dimerizationcI} and \ref{dimerizationCro} respectively.  Again, these are the assumptions used to correspond to \cite{Ao,wang_kinetic_2010} and to references therein.  We will posit different equations of motion, including the correction to the $pnc$ confusion, and the effect of RNA polymerase in \textit{variations on a theme}.  Experimental collaboration, references, or data to analyze that can pin down any to all parameters are welcome, appreciated, and daresay necessary. \footnote{nborggren@gmail.com}    

The situation is more complicated in that cI promotion occurs at 2 different rates, the states with a cI dimer on $O_{R2}$ enhance the actual rate which RNAp transcribes, as well as binding more favorably, \cite{ptashne_principle}, and how this assistance translates physically into the enhanced rate is an interesting open question.  Here it is expressed by using the two different transcription rates, $T_{RM}$,$T_{RMU}$.  

To approximate the effects of specific mutations we consider mutations, $\lambda_{ijk}$, by which it is meant that new cells have been synthesized with $O_{R1} \rightarrow O_{Ri}$ ,$O_{R2} \rightarrow O_{Rj}$ and $O_{R3} \rightarrow O_{Rk}$, the chemical mutations correspond to changes in the strings shown in Fig. \ref{switch} which lead to changes in the binding affinities in Table 2.1.

\section{Stationary Points}

The stationary points of the system are found empirically and tabulated herein.    

To inquire into the nature of these stabilities it is fruitful to consider the linearized equations of motion at a stable point which approximates the flow in the neighborhood.  Stable, unstable, and saddle points have negative, positive, and mixed eigenvalues for their respective linearized equations.  An imaginary part suggests the dynamics around the point have non vanishing circulation.

Although we will maintain a global perspective in the Diffusion section to follow, in correspondence with the main utility of the methods in \cite{wang_potential_2008} or \cite{jin_chapter_2011} where a local analysis is not necessary before proceeding to find global features, dynamics near stationary points is mathematically very well understood \cite{Perko} and even with stochastic features \cite{kwon_structure_2005}, and any such analysis would begin with an enumeration of the fixed points provided here.  It is also noticed that the peaks of the distributions to be found are displaced from the fixed points of the underlying dynamical system.  This displacement is purely the result of the inclusion of noise and has been observed in theory and numerical studies before \cite{kwon_nonequilibrium_2011}.  An experimental situation where the classical dynamics can be fixed and the noise part can be scanned independently of these dynamics could greatly facilitate the observations and interpretations of the non-equilibrium steady states and the breaking of detailed balance suggested in \cite{wang_potential_2008, jin_chapter_2011, qian_vector_1998,ao_potential_2004} by directly measuring this displacement.  \footnote{Doing this in a minimally biased fashion could be difficult.  For example, scanning noise by simply letting the number of particles decrease can easily put one in a domain where the classical kinetics are no longer a good approximation \cite{arkin_cme_2006}.  In such an instance the measured displacement, in addition to the contribution from noise, would also include a contribution from reality deviating from the kinetic assumptions of the model.}

%

For these systems the relevant matrix components are, writing $x_1,...,x_n$ as x,

\begin{equation}
\begin{split}
\mathcal{D}f(x_1,...,x_n)_{ij} & = \frac{\nu_{x_i}}{Z(x_1,...,x_n)^2}[Z(x_1,...,x_n)\frac{\partial}{\partial x_i}Z_{x_j}(x_1,...,x_n) \\
                    &   -Z_{x_j}(x_1,...,x_n)\frac{\partial}{\partial x_i}Z(x_1,...,x_n)] \\ 
                    &  -\frac{\delta_{ij}}{\tau_{x_i}}, 
\end{split}
\end{equation}
where $\delta_{ij}$ is one when $i=j$ and is otherwise zero while the partition functions are evaluated at the stationary point under query.   
 
For the two by two case the eigensystem is 

\begin{equation}
\alpha_1 = \frac{1}{2}(-\sqrt{a^2-2ad+4bc+d^2}+a+d),
\end{equation}

\begin{equation}
e_1 = (-\frac{ \sqrt(a^2-2ad+4bc+d^2)-a+d}{2c},1),
\end{equation}
  
\begin{equation}
\alpha_2 = \frac{1}{2}(\sqrt{a^2-2ad+4bc+d^2}+a+d,
\end{equation}

\begin{equation}
e_2 = (\frac{ \sqrt(a^2-2ad+4bc+d^2)+a-d}{2c},1),
\end{equation}

with 

\begin{equation}
a = \frac{\nu_{x}}{Z^2}[Z\frac{\partial}{\partial x}Z_{x}-Z_x\frac{\partial}{\partial x}Z]-\frac{1}{\tau_x},
\end{equation}

\begin{equation}
b = \frac{\nu_{x}}{Z^2}[Z\frac{\partial}{\partial y}Z_{x}-Z_x\frac{\partial}{\partial y}Z],
\end{equation}

\begin{equation}
c = \frac{\nu_{y}}{Z^2}[Z\frac{\partial}{\partial x}Z_{y}-Z_y\frac{\partial}{\partial x}Z],
\end{equation}

\begin{equation}
d = \frac{\nu_{y}}{Z^2}[Z\frac{\partial}{\partial y}Z_{y}-Z_y\frac{\partial}{\partial y}Z]-\frac{1}{\tau_y}.
\end{equation}

In two dimensions the quantity $(ad-bc)$ is the determinant.  The fact that this term need not vanish is why the notion of energy landscape must be generalized into the flux framework to include fields that are not the gradient of a potential, see \cite{wang_potential_2008}.   

Understanding these equations can tell us a great deal about the type of forces at work since near a particular point, $x\prime=(x_1\prime,...,x_n\prime)$ through an expansion \cite{Perko}, we know that the system is qualitatively similar to the linearized equations

\begin{equation}
\dot{x}\approx f(x\prime)+\mathcal{D}f(x\prime)x.
\end{equation}

If $\alpha_1 <0$, $\alpha_2<0$ then the point serves as a probability sink, if $\alpha_1 >0$, $\alpha_2>0$, then it is a probability source.  Mixed eigenvalues are transition states.  For example, one can verify that any of the following conditions lend to stability:

\begin{description}
\item[$a < 0$, $b=0$, $d<0$]
\item[$a = 0$, $b>0$, $c<0$, $d \le -2 \sqrt{-bc}$] 
\item[$a = 0$, $b<0$, $c>0$, $d \le -2 \sqrt{-bc}$] 
\item[$a > 0$, $b>0$, $c<-\frac{a^2}{b}$, $\frac{bc}{a} < d \le a -2 \sqrt{-bc}$]
\item[$a > 0$, $b<0$, $c>-\frac{a^2}{b}$, $\frac{bc}{a} < d \le a -2 \sqrt{-bc}$]
\item[$a < 0$, $b>0$, $c \ge 0 $, $d<\frac{bc}{a}$]
\item[$a < 0$, $b>0$, $-\frac{a^2}{b}<c<0$, $a +2 \sqrt{-bc} \le d <\frac{bc}{a}$]
\item[$a < 0$, $b>0$, $-\frac{a^2}{b}<c<0$, $d \le a - 2 \sqrt{-bc}$.]
\end{description}

Of the 27 mutations considered, 15 had underlying dynamical systems with 3 stationary points, 11 had one stationary point, and one had two stationary points.  The locations are tabulated in the tables \ref{tab:zeroes}, \ref{tab:twozeroes}, \ref{tab:threezeroes1} and \ref{tab:threezeroes2}.  In particular we note that the permutation group $S_3$ acting on $O_{R1}$,$O_{R2}$, and $O_{R3}$, only six mutations, is sufficient to provide mutants with a single lytic peak $\lambda_{231}$, $\lambda_{321}$ a single lysogenic peak $\lambda_{132}$ while $\lambda_{123}$,$\lambda_{312}$, and $\lambda_{213}$ retain the switch properties.

The 27 phase portraits of this model can be seen in Tables \ref{phase1}, \ref{phase2}, \ref{phase3}.  Dynamics are also combined with the results of the diffusion section to demonstrate the importance of statistical, trajectory based approaches.  Those results are in the tables \ref{tab:ltraj1}, \ref{tab:ltraj2}, \ref{tab:ltraj3} and demonstrate that the different limit cycles correspond to regions of different entropy or energy density.  

This model, however, does not probe what the downstream reactions do to the phase space distribution given to it; it is quite likely that in the case of, for example, $\lambda_{231}$ lysogens will still attempt to form but with a loss of stability.  It does, however probe transient phenomena and the initial impulse of the system is, even in the monostable case, towards two states.



 \begin{equation}
  \begin{split}   
   \frac{\partial Z_x}{\partial x} &= \frac{\partial [cI]}{\partial x}(e^{-\Delta G_{010}}+ e^{-\Delta G_{100}})+\\
   & 2*[cI]*\frac{\partial [cI]}{\partial x}e^{-\Delta G_{110}}+\\
   & \frac{\partial [cI]}{\partial x}[Cro](e^{-\Delta G_{210}}+ e^{-\Delta G_{120}})
    \end{split}, \qquad \begin{split} 
    \frac{\partial Z_x}{\partial y} &=\frac{\partial [Cro]}{\partial y}(e^{-\Delta G_{020}}+ e^{-\Delta G_{200}})+\\
   & [cI]^2e^{-\Delta G_{110}}+\\
   & 2[Cro]\frac{\partial [Cro]}{\partial y}e^{-\Delta G_{220}}+\\
   & [cI]\frac{\partial [Cro]}{\partial x}(e^{-\Delta G_{210}}+ e^{-\Delta G_{120}})
   \end{split} 
 \end{equation}

  \begin{equation}
    \begin{split}
      \frac{\partial Z_y}{\partial x} &=\frac{\partial [cI]}{\partial x}e^{-\Delta G_{001}}
    \end{split}, \qquad \begin{split}
      \frac{\partial Z_y}{\partial y} &=\frac{\partial [Cro]}{\partial y}e^{-\Delta G_{002}}
    \end{split}
   \end{equation}
%
%

The robustness of the lysogenic cycle would be unsurprising if the system were wholly deterministic.  Dynamical systems in general can partition into multiple basins of attraction and thus so too can those arising from chemical kinetics.  In two dimensions, like that presumed here, it can often be the case that there exists some contour that partitions the phase space \cite{Perko}.  This contour is unstable in the perpendicular direction and deterministic trajectories beginning on this contour take the steepest descent to a saddle point if the system is bounded.  Any crossing from one region to the other would require an intersection with this line and is forbidden with the rules for ordinary differential equations.  Thus, hopping would require a stochastic process to change this fact for this model.  The puzzle of this stability, which we consider in the next section, is why there is not enough noise.

\chapter{\textit{Diffusion; Evolution of Ensembles}}

\begin{quotation}
  \textit{Lo otro no existe: tal es la fe racional, la incurable creencia de la raz\'{o}n humana.  Identidad=realidad, como si, a fin de cuentas, todo hubiera de ser, absoluta y necesariamente, uno y lo mismo.  Pero lo otro no se deja eliminar; subsiste, persiste; es el hueso duro de roer en que la raz\'{o}n se deja los dientes.}

\hspace{0.1in}Antonio Machado, \textit{Juan de Mairena}, quoted in \textit{El laberinto de la soledad, Octavio Paz}
\end{quotation}

When Galileo first assisted his gaze with the telescope and directed it towards Saturn, he discovered its ring, and Titan, one of its moons.  The point-like orb of the ancients revealed more structure still when Cassini subsequently fixed his gaze upon this ring, revealing a large division; the ring was not one, but two.  Further query revealed yet more structure, courtesy of an \textit{in situ} spacecraft, also named Cassini, and one is now disinclined to the naming of rings, which appear with the abundance of grooves in a phonograph.

Indeed the rings of Saturn are not rings at all, but a collection of granules of various sizes caught in circular potential wells; the only rings are in the energy landscapes, our abstractions, and only appear in the data in that there was a sufficient supply of granules to sample the space so that the typical distance between stones was shorter than the resolution of our instruments.

With that said let us revisit the diffusion equation   
\begin{equation}
\partial_t u = \partial_i (D_{ij}\partial_j u - F_i u).
\label{fplanck}
\end{equation}

For, example if one were to construct a Hamiltonian of the Saturn system and fill a vector, $F_i$, with each of the equations of motion, then $u$ would denote the density of granules, and in the case of $D_{ij}$ being zero, the equation would describe the evolution of that initial configuration as it samples the Saturn force field; depending on the details some fraction fall into Saturn's core, some might fall into Titan's surface, others escape, and some get trapped into the resonances of the system, the rings.  Letting the $D>0$ would allow our granules to interact with each other, although they are small enough so as not to perturb $F_i$, they are of comparable sizes to each other and collisions lend them to diffuse through space.   

In the case of $\lambda$ phage, the numbers of proteins involved are small, and intrinsic noise, shot noise, is indeed a real source of diffusion for the system; the system kinetics do not behave the same with every trial, but spread throughout the phase space.  However, Eq. (\ref{fplanck}), can equally describe such a situation.  For references and the general state of affairs of Fokker-Planck equations used as a \textit{theory of everything}, query \cite{friston_free_2012}.     

\section{Master Equations, Langevin Dynamics, or Diffusion Equation}

Many approaches are used to capture the chemical kinetics or gene expression in a noisy environment, all of which are phenomenological and have their advantages and disadvantages.  The proper approach can vary from system to system and depend on resources available.  See for example
\cite{wang_potential_2008}, which this thesis follows and extends with the finite elements, and also the potential approaches in \cite{ao_potential_2004,Ao}; for field-theoretic approaches there is \cite{wang_instantons_1996,wang_kinetic_2010,feng_potential_2011}, and for Langevin dynamics and master equations see \cite{wang_funneled_2006,ge_physical_2010, qian_concentration_2002}.

\subsection{An Integral Approach for Diffusion Equations and Passage Times}

The finite element method is emerging as a universal method for the solution of differential equations of arbitrary complexity from all walks of science, see \cite{fenicsbook}.  The following derivation is typical of the procedures implemented throughout the FEniCS book (\cite{fenicsbook}).   In order to use a finite element method to solve partial differential equations it is necessary to first translate the equations into an integral equation.  This integral equation can be seen as setting a bilinear functional equal to a linear functional and solving the corresponding system by standard linear algebraic methods.  It will be described in detail for the Fokker-Planck equation and corresponding passage time equations.       

\subsubsection{Diffusion Equations}
First we seek to solve: 

\begin{equation}
\partial_t u = \partial_i (D_{ij}\partial_j u - F_i u).
\end{equation}

For numerics the function space is specified by defining a mesh over $\Omega$ and the type of element associated with each point in the space.  These specifications alter the analysis mainly in how long it takes for a desired level of accuracy to be reached.  We continue thinking of the function space as the space of twice differentiable functions over $\Omega$.  The functional form is found by multiplying a test function v and integrating over the mesh  

\begin{equation}
\int_\Omega v \partial_t u dx =\int_\Omega v  \partial_i (D_{ij}\partial_j 
u - F_i u) dx .
\end{equation}

\noindent
 Continuing with integration by parts yields 
\begin{equation}
\int_{\partial\Omega} v (D_{ij}\partial_j u - F_i u) n_i ds - \int_\Omega \partial_i v (D_{ij}\partial_j u - F
_i u) dx.
\end{equation}

\noindent
 Here $D_{ij}$ denotes the spatial dependance of the Diffusion, $F_{i}$
, is the drift force, $n_i$, a normal vector to the mesh.  The mesh is denoted $\Omega$ with boundary $\partial\Omega$.

 We approximate to find an expression for $u_{t+\delta t}$ 
given the solution at $u_t$

\begin{equation}
\begin{split}
\int_\Omega v (u_{t+\delta t} -  u_t) dx \approx &\delta t [\int_{\partial \Omega} v (D_{ij}\partial_j u - F_i u) n_i ds \\
 &- \int_\Omega \partial_i v (D_{ij}\partial_j u - F_i u) dx ].
\end{split}
\end{equation}

\noindent
We take the average of the expression over time $\delta t$ by asserting $u \approx \frac{u_{t+\delta t} +  u_t}{2}$. Continuing gives 

\begin{equation}
\begin{split}
&\frac{\delta t}{2} [\int_{\partial\Omega} v (D_{ij}\partial_j u_{t+\delta t}- F_i u_{t+\delta t}) n_i ds \\
&- \int_\Omega \partial_i v (D_{ij}\partial_j u_{t+\delta t} - F_i u_{t+\delta t}) dx \\
&+ \int_{\partial \Omega} v (D_{ij}\partial_j u_t - F_i u_t) n_i ds \\
&- \int_\Omega \partial_i v (D_{ij}\partial_j u_t-F_i u_t) dx ].
\end{split}
\end{equation}

\noindent
Collecting terms with $u_{t+\delta t}$ on one side of equality and $u_t$ on the other gives the variational forms $a$ and $L$ that we will use

\begin{equation}
a  =  \int_\Omega (v u_{t+\delta t}+\frac{\delta t}{2}\partial_i v [D_{ij}\partial_j u_{t+\delta t} - F_i u_{t+\delta t}]) dx ,
\end{equation}

\begin{equation}
L  = \int_\Omega (v u_{t}-\frac{\delta t}{2}\partial_i v [D_{ij}\partial_j u_{t} - F_i u_{t}]) dx .
\end{equation}

\noindent
$a(v,u) = L(v)$ then defines a linear system.  To see this note that if $v$, $v'$ satisfy $a$ then so does $\alpha(v+v')$ and similarly for $L$ for all scalars $\alpha$.  Since it also holds that $a$ is linear in $u_{t+\delta t}$, a is bilinear.  The $u_{t}$ in $L$ is specified by propagating $u_0$ and thus explicitly appears as a known function on the right side of the equations.  Thus, finding the $u$ that solves $a(v,u)=L(v)$ for all $v$ advances our solution a timestep.  Eq. (\ref{varform}) shows the simplicity of the lines of code to perform these calculations.  Indeed the computation is more general and concise then the notation here.  
          
The passage time equations are much simpler.  The integral equation corresponding to $F_i \partial_i t_1(x)- D\partial_i\partial_i t_1(x)=-1$ follows by simply multiplying through by the test function and integrating      

\begin{equation}
\int_\Omega (t(x)(F_i \partial_i t_1(x))- D\partial_i t(x)\partial_i t_1(x)) dx 
= -\int_\Omega t(x)dx ,
\end{equation}

\begin{equation}
\int_\Omega (t(x)(F_i \partial_i t_n(x))- D\partial_i t(x)\partial_i t_n(x)) dx 
= -n\int_\Omega t(x)T_{n-1}dx .
\end{equation}

\noindent
One can then find the passage times recursively.  The induction step is bilinear provided one has first solved for $T_{n-1}(x)$ before proceeding to find $T_n(x)$.

\begin{figure}[h]
\begin{center}
\includegraphics[width=0.75\textwidth]{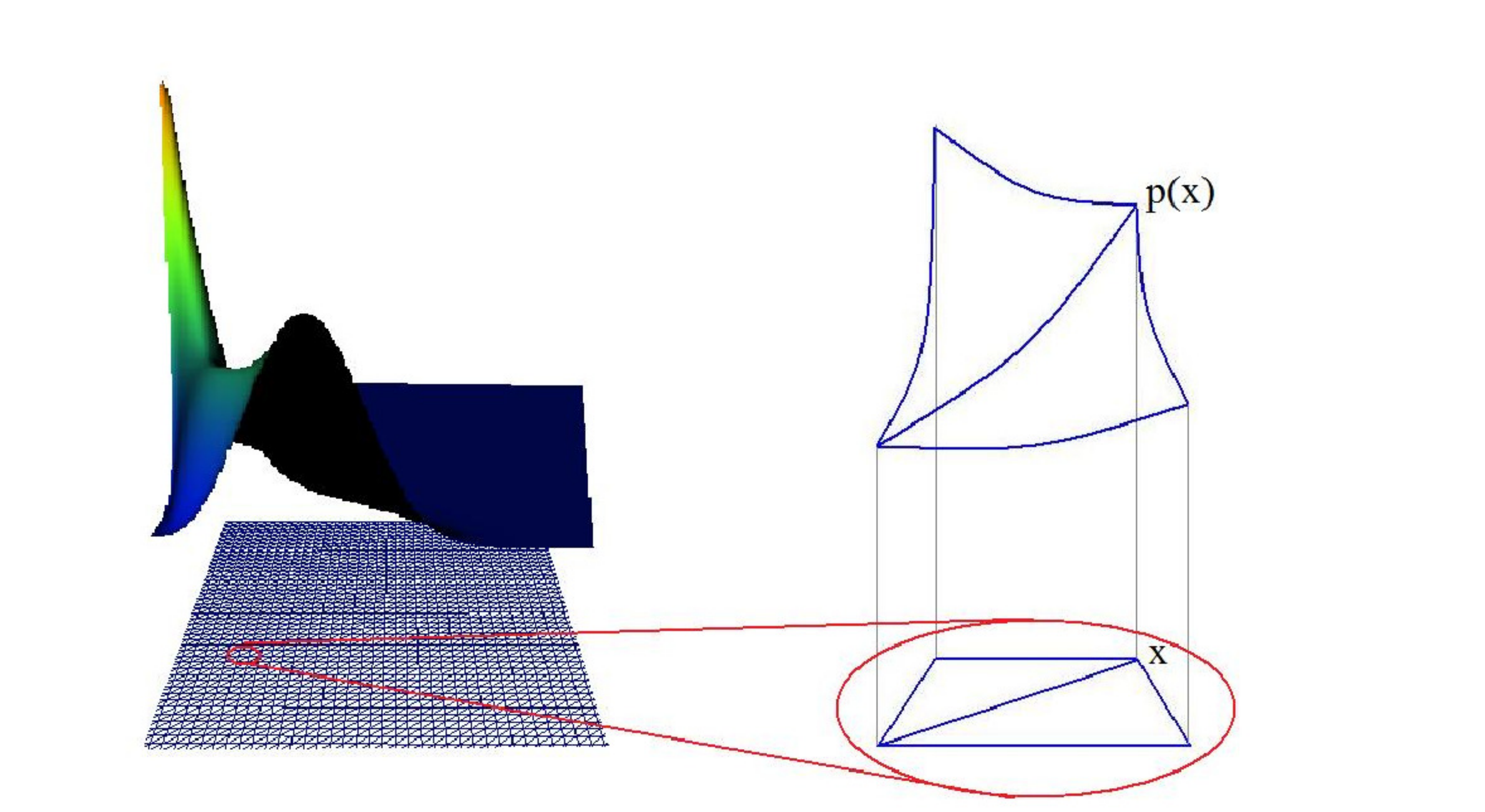}
\caption{{\bf Finite elements construct a surface by combining simplices together.  The quadratic elements used fit the function inbetween the mesh points with a quadratic formula.}}
\label{finelem}
\end{center}
\end{figure}

To perform the simulations in this work we need still to specify a mesh and type of finite element.  We choose for $\Omega$ the domain in protein number space $0<cI<1000$ and $0<Cro<1000$.  We discretize this into at least a 50 by 50 grid and use a quadratic Lagrange element.  The quadratic Lagrange elements span the surface by connecting the adjacent mesh points with a quadratic function (see Fig. \ref{finelem}).  The function space is the set of all such functions.  One benefit of a finite element method over a finite difference is the functions in the function space are continuous over $\Omega$ and we can evaluate any function, and in particular our solutions, anywhere in the domain.  We see this smoothness manifest in the plots which evaluate the functions along the trajectories.    

\subsection{On Convergence}

Technically, the asymptotics of the Fokker-Planck equation take infinitely long to converge.  Such a program, see Fig. \ref{traj}, can still be written, albeit not technically admissible into von Neumann architectures.  Taking the $a$ and $L$ as bilinear and linear forms, respectively, defined anywhere, we have an infinite trajectory pursuing eternally its asymptotic states.  A break can be entered by establishing a criterion for convergence.  The distributions present in the Appendix were found using 53, 7.5 second time steps and well approximate the steady state solutions.

\begin{figure}
\begin{python}

def TimeStep(u0,u1):

    A = assemble(a)

    while 'no your solutions have not converged':
        b = assemble(L)
        bc.apply(A,b)  
        solve(A, u1.vector(), b)
        u0.interpolate(u1)
        yield u1

while InfiniteTraj:InfiniteTraj.Next()


\end{python}
\caption{{\bf This would be an example of a legitimate python code that runs indefinitely.  Computationally, the asymptotics of the equations require a function that does not terminate.}}
\label{traj}
\end{figure}

\noindent

This loop was carried out 53 times before I choose to break it in the states labeled steady in this work, with $k = 7.5$ 'seconds' the time step.

\subsection{Choice of Gauge}

It is noticed that equation (\ref{fplanck}) presupposes a choice of gauge and the curl of any vector field can be added to the flux landscape, $D_{ij}\partial_j P -F_iP+\epsilon_{ijk}\partial_jA_k$, and still satisfy the basic equation.  We choose a gauge such that $\epsilon_{ijk}\partial_jA_k=0$.  We notice that what has been called the entropy production in \cite{wang_potential_2008} is a gauge-dependent quantity and how it translates into entropy or heat in absolute units is open.  We note that the different kinetic paths will have different thermodynamic signatures, and understanding the connection can enable direct measurement of some of the quantities that have been proposed.
  
A recent paper has elaborated on the nature of the relationship between related Fokker-Planck equations and gauge theories, \cite{feng_potential_2011}, this is likely to prove a rather pregnant approach to chemical kinetics.   

\subsection{$\Delta x \Delta J$}
After having computed the solutions to the equations we can calculate some moments of the distribution.  So far the diffusion has been picked to correspond to the numbers believed to quantify robustness, see \cite{Ao,wang_kinetic_2010}, however the boundary condition can be tuned to make this number vary over many orders of magnitude and it may be more prudent to use moments of the distribution to parameterize $ D$.  For example when cI proteins are reported in numbers 150-250 in a cell this can suggest an expectation value of, say, 200 for $x$ and a spread of 50 for the second moment.  These numbers are more easily measureable then the robustness, which has been the subject of amendment over time. 

The figure \ref{expectationvals} illustrates how FEniCS software can compute expectation values from the command line, and the type of calculations that can be performed once one has solutions in hand.  Let us compute $\Delta x$ and $\Delta y$ with the distributions from \ref{tab:l123} and recompute the related expectation values. 

\begin{figure}
\begin{python}
>>> exprs = [Expression("x[0]"),Expression("x[1]"),
... Expression("x[0]*x[0]"),Expression("x[1]*x[1]")]
>>> ans=[
... assemble(i*u*dx,mesh=mesh)/assemble(u*dx,mesh=mesh) for i in exprs]
>>> ans
[93.076476131804114, 169.77672009751805, 13624.244707344435, 36974.455745161322]
>>> sqrt(abs(pow(ans[0],2)-ans[2]))
70.434468112069482
>>> sqrt(abs(pow(ans[1],2)-ans[3]))
90.279128585129428

\end{python}
\caption{{\bf Calculating expectation values is straightforward, even on a command line.}}
\label{expectationvals}
\end{figure}

It is important to realize that it is not the mass of an electron that biologists measure.  The errors do not necessarily go away with better measurements.  The numbers are supposed to fluctuate and this can actually be interpreted to mean that the higher moments of the distribution are relevant.  A spread of 70 in the number of $cI$ proteins and 90 $Cro$ proteins is predicted by this model.    

\section{Passage Times}

The equations again for the first passage time is 

\begin{equation}
\int_\Omega (t(x)(F_i \partial_i t_1(x))- D\partial_i t(x)\partial_i t_1(x)) dx 
= -\int_\Omega t(x)dx,
\end{equation}

\noindent
where we force a boundary condition of zero, in the present case a circle of radius ten is placed around the critical point, where we want to calculate the passage time to and following we tabulate the results for the mutants we can construct from the binding values available.  First, we explore the properties of the 15 of 27 mutants that maintain the switch properties in this model.  We notice however that the illusion of bistability is manifest in the transient phenomena even for the monostable states.  That is to say, if the switch behavior is a non-equilibrium phenomenon, monostable mutations may be relevant.

\begin{figure}[h]
\begin{center}

\begin{python}
...
Assembling matrix over cells [======================================] 100.0%
Applying boundary conditions to linear system.
Solving linear system of size 1442401 x 1442401 (PETSc LU solver, umfpack).
*** Warning: Using LU solver, ignoring preconditioner "default".
evaluate at wells,  [0.14175299999999999, 462.57299999999998] 112474076.373
evaluate at wells,  [644.05200000000002, 0.0061165799999999999] 112473547.501
evaluate at wells,  [29.979800000000001, 178.22999999999999] 112469292.141
\end{python}

\includegraphics[scale=.72]{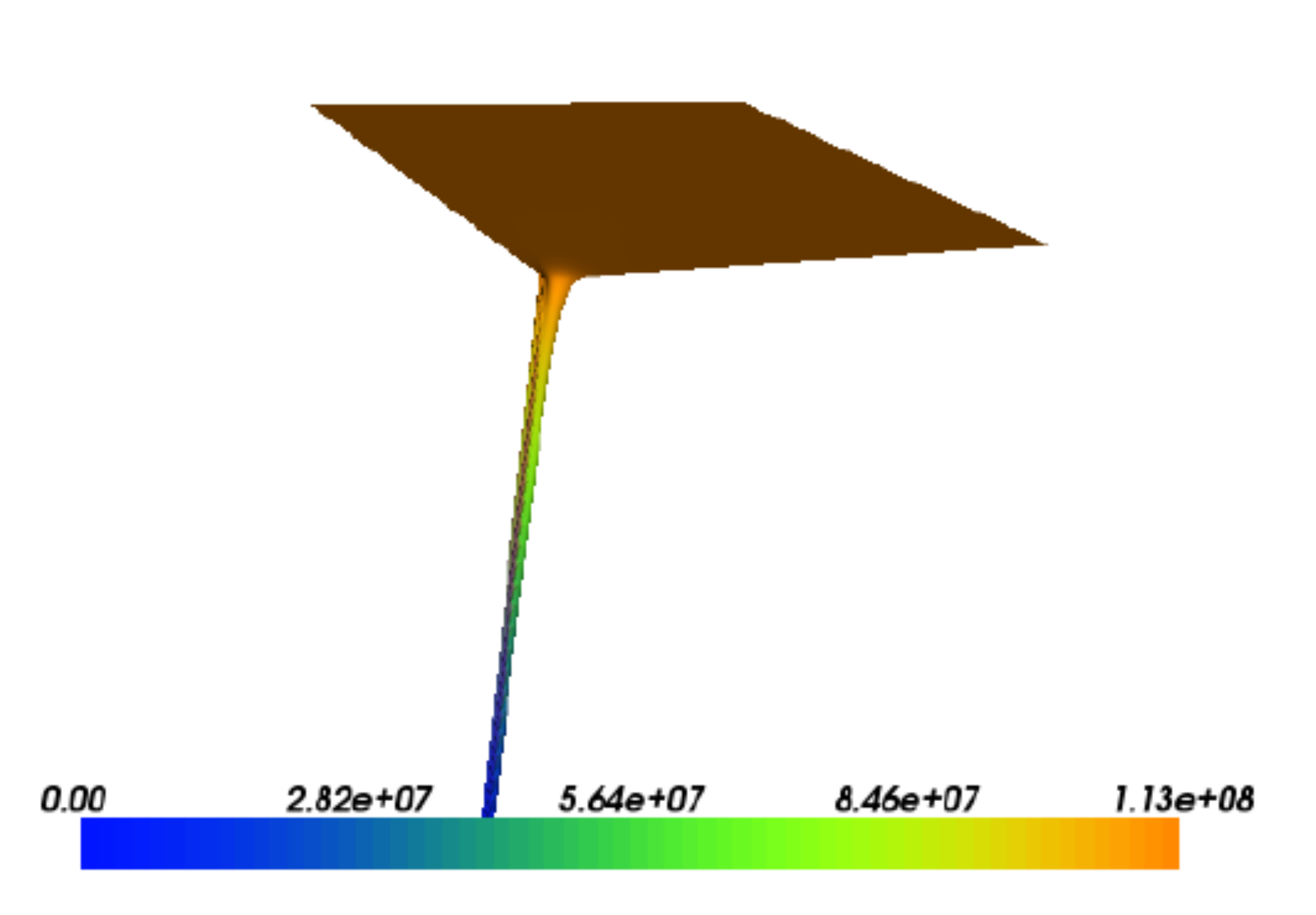}
\caption{{\bf The first passage time with an absorbing boundary condition within 10 units of the origin. The origin is extremely unstable; The steepness of the funnel of the first passage time landscape illustrates the overwhelming propensity for the system to be driven towards some non-trivial state.  The numbers, $\approx 10^8$, appear in seconds.}}
\label{fpt_origin}
\end{center}
\end{figure}

\section{Noise-induced stability}

The dynamics picture of the previous chapter is quite idealized.  Even if all the parameters were perfectly known there is still the issue that we are discussing mesoscopic regimes and our large $N$ intuitions and classical assumptions are certain to break down in this limit.  The constant rate assumptions would need to be replaced with discrete events and the manner in which a unit of time splits into subunits becomes important, as well as the annihilation processes.  The changes to the partition function in this model are again relevant.  

It is found here that the noise added displaces to a large extent the locations of the peak from the corresponding stable points of the underlying dynamical system.  Viewing the noise as diffusion through concentration space and developing effective field theories and path integrals has provided interesting and useful constructions, see \cite{wang_kinetic_2010}.  The fact that there even exist corrections to be made is interesting.  This is the type of functionality that Schrodinger deliberates on being prohibited for an organism, in his discussion of the $\sqrt(n)$ law.

\begin{quotation}
  ... an organism must have a comparatively gross structure in order to enjoy the benefit of fairly accurate laws, both for its internal life and for its interplay with the external world.  For otherwise the number of co-operating particles would be too small, the 'law' too inaccurate.  The particularly exigent demand is the square root.  For though a million is a reasonably large number, an accuracy of just 1 in 1,000 is not overwhelmingly good, if a thing claims the dignity of being a 'Law of Nature.' 

Erwin Schr{\"o}dinger 'What is Life'
\end{quotation}  

For $\lambda$ phage this is just a restatement of the stability puzzle: how then can the numbers of proteins be within the range 0-1000, a far worse scenario then Schrodinger deemed possible?  
\textit{Why} does a deterministic dynamical analysis in the small concentration regime \textit{work at all}?  Do we retain the observed robustness when we consider noise?  Is the noise \textit{really} $\sqrt(n)$ noise?  What paths are permissible and what are their relative probabilities?  General questions and methods of this variety are the subject of intense debate and muse to much creativity, see \cite{wang_instantons_1996,wang_kinetic_2010,stock_maximum_2008,ghosh_teaching_2006,qian_concentration_2002}, and have been applied to the $\lambda$ system, see \cite{Ao,wang_kinetic_2010}. 
This debate has led to legitimate concerns and elegant measurements to seek explanations from other variables, like volume, where the variation across an ensemble are less severe or can be experimentally minimized, see \cite{St.Pierre}.

It is written that $D_{cI,Cro}=const*\tau_{cI,Cro}/N_{CI,Cro}$ then $\Lambda=D^{-1}$, \cite{Ao,wang_kinetic_2010}.  However, their expression for $\Lambda$ is the same as their expression for $D$.  The calculations require 

\begin{equation}
D = \left( 
  \begin{array}{cc}
    17.85*\frac{N_{cI}}{\tau_{cI}} & 0 \\
    0 & 25.0*\frac{N_{Cro}}{\tau_{Cro}}
  \end{array}
  \right) .
\label{diff}
\end{equation}

\noindent
An $N_{cI}=644.052$ corresponds to the value at the lysogen zero and $N_{Cro}=462.573$ the lytic zero.  $\tau_{cI}=2943 s$ and $\tau_{Cro}=5194 s$ giving 

\begin{equation}
D_{wt} \approx \left(
  \begin{array}{cc}
    3.91 & 0 \\
    0 & 2.2271
  \end{array}
  \right) .
\label{diff2}
\end{equation}

\noindent
The numerics proceed with this diffusion tensor.  If the analogy is to extend from diffusion in ordinary space to diffusion in concentration space, then the mean squared displacement is a displacement in concentration space and the units of this $D$ should be ${dN_{CI}dN_{Cro}}/{ds}$.     

Let us attempt to motivate the form of Eq. (\ref{diff}) and the scale of Eq. (\ref{diff2}).
If 
\begin{equation}
x\Rightarrow x \pm \sqrt{x}\quad  {\rm then} \quad \dot{x} \Rightarrow \dot{x} \pm \frac{\dot{x}}{2\sqrt{x}} \approx  f(x) \pm \frac{f(x)}{2\sqrt{x}}.
\end{equation}  
\noindent
For large $x$, we have seen that the term $-x/\tau$ dominates and recover fluctuations of order $\sqrt{x}$.  Since $D$ goes as the square of these fluctuations we get the proportionality to $x$ as appears in \ref{diff}.  The situation is more subtle when $x$ is small.  For Shea and Acker type models the $D_{xx}$ term, say, would be dominated by terms of the form ${Z_x^2}/{xZ^2}$ which can diverge.  We expect corrections from this end to increase the average diffusion and therefore facilitate the crossing of a barrier, thereby reducing the robustness of the switch more than Eq. (\ref{diff2}) may suggest.         

The cI protein is found in numbers typically less than 180-350 for a healthy cell, and error from counting statistics alone contributes $\pm \sqrt(N)$ to the intrinsic noise.  The uncertainty in $\Delta G$ and the dimerization equation of state also induces a spread of trajectories.  A convenient way to catch these and other sources of stochasticity is to assert a diffusion tensor, $D_{ij}$, with $i, j$ twice indexing over the protein species.  This is equivalent to considering some Langevin-like equation but allows us to use partial differential equations to explore ensembles of trajectories.  Noise of greater complexity then Gaussian noise can be readily studied with the software developed.

As with the other parameters in this model they are left determined by experiments and have been given a simple form to understand the effects that chance events could have on the dynamical picture.  In particular, it can wash away higher resolution features which can, for example, remove the gap between two nearby stabilities, facilitating perhaps the appearance of bistability when the dynamical picture may suggest more structure.  From the modeling perspective it can be useful to use large, perhaps unphysically large (analagous to a high temperature limit), diffusion to speed up simulations and see some of the more basic underlying features of the phase space.  As diffusion gets smaller, the steady states take exponentially longer to converge and can quickly become computationally prohibitive.  In these scenarios, where converging to a steady state from any arbitrary initial condition is computationally prohibitive, the steady states from larger diffusion cases can serve as useful initial conditions since they have already undergone a large degree of this convergence process and the remaining time to equilibrium is for the finer resolution features to develop since the coarse features were already present at the beginning of the simulation.

\section{Little's Mutants}

Mutations in the $O_R$ region have been performed, see \cite{little_robust}, and we would like to check the performance of our model and suggest means to calibrate accordingly.  The qualitative features compare to experiment but quantitative comparison is premature; in particular, the time and concentration units need to be understood better.  As is, the first passage times predict a smaller barrier then observed.  

\begin{table}[ht]

  \centering
     \begin{tabular}{|c|c|c|}
          \hline
          \multicolumn{3}{|c|}{Diffusion Tensor}
          \\ \hline \hline
          mutant & $D_{xx}$ & $D_{yy}$ \\ \hline
          $\lambda_{123} \textit{wild-type}$ &3.90778 &2.22668 \\ \hline
          $\lambda_{121}$ &0.775855 &2.22668\\ \hline
          $\lambda_{323}$ &2.6462 &1.5024\\ \hline
          $\lambda_{323'}$ &2.0016&2.86862\\ \hline
        \hline
        \end{tabular}
        \caption{Values for diffusion tensors.  $\lambda_{323p}$ in \cite{little_robust} is nearly $\lambda_{222}$, the $cI$ affinities are identical and $Cro$ values differ by -.6 kcal/mol in the first and third site.}
        \label{tab:wtgval}
    \end{table}%

\subsection{Wild-Type Evolution}

Four trajectories of distributions are depicted in Tables \ref{tab:gt4}, \ref{tab:gt5}, \ref{tab:gt6} and \ref{tab:gt7}, corresponding to the wild-type and mutants $\lambda_{121}$, $\lambda_{323}$, and $\lambda_{323'}$.  It is evident that in the first 200 seconds, an initially peaked distribution has developed a second peak and after about 600 seconds the distributions have reached their steady states.    

\subsection{Steady state properties for $O_R$ Mutations}

The Tables \ref{tab:l121}, \ref{tab:l123}, \ref{tab:l131}, \ref{tab:l133}, \ref{tab:l211}, \ref{tab:l213}, \ref{tab:l222},
 \ref{tab:l223}, \ref{tab:l232}, \ref{tab:l233}, \ref{tab:l312}, \ref{tab:l313}, and \ref{tab:l323} archive the results.  Depicted are the phase space portraits alongside the Probability and Flux Landscapes and above the passage time distributions to each of the three stationary points.  Assorted expectation values are calculated and can be compared with the wild-type.    

In particular, we notice that with diffusion the peaks of the solutions are considerably displaced towards smaller numbers from where one would expect the peaks from the deterministic equations.  This is required to satisfy the boundary condition, and it would be interesting to see how a Langevin-based approach can reproduce this ensemble property in that different trajectories would require correlation.

Noise in this model seems to destabilize at least the mutants $\lambda_{111}$, $\lambda_{131}$, and $\lambda_{312}$; either no gap, or a small gap is present.  The phage seems most robust in response to mutations in $O_{R1}$ in that $\lambda_{*23}$ all have a sizeable gap.        

\section{Hybrid particle and ensemble approaches}

It is important to see how different paths contribute differently to the overall energy and entropy profiles, see \cite{seifert_paths,stock_maximum_2008}.  To illustrate the shape of these landscapes it is sufficient to evaluate the steady state ensembles values along determinstic trajectories.  These calculations are depicted in Tables \ref{tab:ltraj1}, \ref{tab:ltraj2}, and \ref{tab:ltraj3}.  \footnote{\cite{seifert_paths} suggests a definition of path entropy as $s(\tau)=-ln((x(\tau),\tau))$, with $x(\tau)$ some stochastic trajectory.  The definition is sufficient to define a path entropy for any function $x(\tau)$, regardless of the motivation of such functions.  The principle of maximum caliber, see e.g. \cite{stock_maximum_2008}, is an effort to define a dynamic partition function, over all such paths.  There has been some success with this method for low number of states, but in the discrete case, or the continuous case here, the implications are largely unknown.}  It is evident that different paths can converge to different locations and demonstrates the motivation for a (more fundamental) dynamic partition function.  

\section{Variations on a theme}

We revisit the equations of motion with RNAp still included in the partition function.  Here, the promotional states are the natural ones, the binding of RNAp to $prm$ for cI promotion and the binding of RNAp to $pr$ for Cro promotion.  These promotional states enter the dynamics as follows.   

\begin{equation}
  \begin{split}
\frac{d(cI)}{dt} = & \frac{1}{Z}(\\
  & mpi*Trm*Eci([RNAp][cI]e^{-\frac{\Delta G_{013}}{RT}}+[RNAp][cI]^2e^{-\frac{\Delta G_{113}}{RT}}+[RNAp][cI][Cro]e^{-\frac{\Delta G_{213}}{RT}}) \\
  & +Trmu*Eci([RNAp]^2e^{-\frac{\Delta G_{33}}{RT}}+[RNAp][cI]e^{-\frac{\Delta G_{103}}{RT}}+[RNAp][Cro](e^{-\frac{\Delta G_{203}}{RT}}+ e^{-\frac{\Delta G_{023}}{RT}})\\
  & +[RNAp][cI][Cro]e^{-\frac{\Delta G_{123}}{RT}} +[RNAp][Cro]^2e^{-\frac{\Delta G_{223}}{RT}}+[RNAp]e^{-\frac{\Delta G_{03}}{RT}})\\
  & )-\frac{cI}{\tau_{cI}}
\end{split}
\end{equation}

\begin{equation}
\begin{split}
\frac{d(Cro)}{dt} = &\frac{1}{Z}(\\
 &  mpi*Trr*Ecro*([RNAp]e^{-\frac{\Delta G_{30}}{RT}}+[RNAp][cI]e^{-\frac{\Delta G_{31}}{RT}}+[RNAp][Cro]e^{-\frac{\Delta G_{32}}{RT}}+[RNAp]^2e^{-\frac{\Delta G_{33}}{RT}})\\
 & )- \frac{Cro}{\tau_{Cro}}
\end{split}
\end{equation}

The Z in these equations is now from \ref{RNApartition}, with values listed in table \ref{wtgval3}.  The inclusion of these changes leads to much higher numbers for cI, up to 1500 proteins, though the lysogen peak still dominates.  The time scale is slowed down in comparison to the dynamics without RNAp.  The $[RNAp] = 600 nM$ was chosen to be in the range of the experiment \cite{klump_rnap} and is presumed to be a fixed number for the simulations.  The results are displayed in tables \ref{rnapdiff} and \ref{rnapdiff2}.

\begin{table}[ht]
 \footnotesize
  \centering
     \begin{tabular}{|c|c|c|c|c||c|c|c|}
          \hline
          \multicolumn{5}{|c|}{$\lambda_{123}$, $wild-type$ $O_R$ affinities (kcal/mol)}
          & \multicolumn{3}{|c|}{$\Delta G_{ijk}$}\\ \hline \hline
          protein & $O_{R1}$&$O_{R2}$&$O_{R3}$&ref.& equation & monomial & premotes\\ \hline
       $\Delta G_{001}$,$\Delta G_{010}$,$\Delta G_{100}$   & -15 & -13 & -12 & \cite{Ao} & $\Delta G_{000} = 0.0 $ & 1 &  \\ \hline
       $\Delta G_{002}$,$\Delta G_{020}$,$\Delta G_{200}$   & -18.4 & -17.1 & -19.5 & \cite{Ao} & $\Delta G_{001} = -12.0 $ & $[Ci]$ &  \\ \hline
       \multicolumn{5}{|c||}{$\Delta G_{coop} = -6.9 $} & $\Delta G_{010} = -13.0 $ & $[Ci]$ & \\ \hline
       \multicolumn{5}{|c||}{$pnc = 2.5*10^{-9}$} & $\Delta G_{100} = -15.0 $ & $[Ci]$ & \\ \hline
       \multicolumn{5}{|c||}{$[RNAp] = 300,600 nM$ \cite{klump_rnap}} & $\Delta G_{002} = -19.5 $ & $[Cro]$ &  \\ \hline
       \multicolumn{5}{|c||}{$\Delta G_{30} = -11.5,-10.5$, RNAp binding to Cro promoter.} & $\Delta G_{020} = -17.1 $ & $[Cro]$ &  \\ \hline
       \multicolumn{5}{|c||}{$\Delta G_{03} = -11.5,-13.5$, RNAp binding to cI promoter.} & $\Delta G_{200} = -18.4 $ & $[Cro]$ &  \\ \hline
	\hline
       \multicolumn{3}{|c|}{$\Delta G_{ijk} (cont)$} & \multicolumn{3}{|c|}{$\Delta G_{011} = \Delta G_{010}+\Delta G_{001}+\Delta G_{coop} $} & $[Ci]^2$ & \\ \hline
       \multicolumn{3}{|c|}{} & \multicolumn{3}{|c|}{$\Delta G_{110} = \Delta G_{010}+\Delta G_{100}+\Delta G_{coop} $} & $[Ci]^2$ & \\ \hline
       \multicolumn{3}{|c|}{} & \multicolumn{3}{|c|}{$\Delta G_{101} = \Delta G_{100}+\Delta G_{001}$} & $[Ci]^2$ & \\ \hline
       \multicolumn{3}{|c|}{} & \multicolumn{3}{|c|}{$\Delta G_{022} = \Delta G_{020}+\Delta G_{002}$} & $[Cro]^2$ & \\ \hline
       \multicolumn{3}{|c|}{} & \multicolumn{3}{|c|}{$\Delta G_{220} = \Delta G_{020}+\Delta G_{200}$} & $[Cro]^2$ &  \\ \hline
       \multicolumn{3}{|c|}{} & \multicolumn{3}{|c|}{$\Delta G_{202} = \Delta G_{200}+\Delta G_{002}$} & $[Cro]^2$ & \\ \hline
       \multicolumn{3}{|c|}{} & \multicolumn{3}{|c|}{$\Delta G_{120} = \Delta G_{100}+\Delta G_{020}$} & $[Ci][Cro]$ & \\ \hline
       \multicolumn{3}{|c|}{} & \multicolumn{3}{|c|}{$\Delta G_{210} = \Delta G_{200}+\Delta G_{010}$} & $[Ci][Cro]$ & \\ \hline
       \multicolumn{3}{|c|}{} & \multicolumn{3}{|c|}{$\Delta G_{102} = \Delta G_{100}+\Delta G_{002}$} & $[Ci][Cro]$ &\\ \hline
       \multicolumn{3}{|c|}{} & \multicolumn{3}{|c|}{$\Delta G_{201} = \Delta G_{200}+\Delta G_{001}$} & $[Ci][Cro]$ &\\ \hline
       \multicolumn{3}{|c|}{} & \multicolumn{3}{|c|}{$\Delta G_{012} = \Delta G_{010}+\Delta G_{002}$} & $[Ci][Cro]$ &\\ \hline
       \multicolumn{3}{|c|}{} & \multicolumn{3}{|c|}{$\Delta G_{021} = \Delta G_{020}+\Delta G_{001}$} & $[Ci][Cro]$ &\\ \hline
\hline
       \multicolumn{3}{|c|}{} & \multicolumn{3}{|c|}{$\Delta G_{111} = \Delta G_{100}+\Delta G_{010}+\Delta G_{001}+\Delta G_{coop}$} & $[Ci]^3$ &\\ \hline
       \multicolumn{3}{|c|}{} & \multicolumn{3}{|c|}{$\Delta G_{222} = \Delta G_{200}+\Delta G_{020}+\Delta G_{002}$} & $[Cro]^3$ &\\ \hline
       \multicolumn{3}{|c|}{} & \multicolumn{3}{|c|}{$\Delta G_{112} = \Delta G_{100}+\Delta G_{010}+\Delta G_{coop}+\Delta G_{002}$} & $[Ci]^2[Cro]$ &\\ \hline 
       \multicolumn{3}{|c|}{} & \multicolumn{3}{|c|}{$\Delta G_{121} = \Delta G_{100}+\Delta G_{020}+\Delta G_{001}$} & $[Ci]^2[Cro]$ &\\ \hline
       \multicolumn{3}{|c|}{} & \multicolumn{3}{|c|}{$\Delta G_{211} =\Delta G_{200}+\Delta G_{001}+\Delta G_{010}+\Delta G_{coop}$} & $[Ci]^2[Cro]$ &\\ \hline    
       \multicolumn{3}{|c|}{} & \multicolumn{3}{|c|}{$\Delta G_{221} = \Delta G_{200}+\Delta G_{020}+\Delta G_{001}$} & $[Ci][Cro]^2$ &\\ \hline
       \multicolumn{3}{|c|}{} & \multicolumn{3}{|c|}{$\Delta G_{212} = \Delta G_{200}+\Delta G_{010}+\Delta G_{002}$} & $[Ci][Cro]^2$ &\\ \hline
       \multicolumn{3}{|c|}{} & \multicolumn{3}{|c|}{$\Delta G_{122} = \Delta G_{100}+\Delta G_{020}+\Delta G_{002}$} & $[Ci][Cro]^2$ &\\ \hline
\hline
\multicolumn{3}{|c|}{} & \multicolumn{3}{|c|}{$\Delta G_{30} = -10.5$} & 1 &$Z_y$\\ \hline
\multicolumn{3}{|c|}{} & \multicolumn{3}{|c|}{$\Delta G_{03} = -13.5$} & 1 &$Z_x$\\ \hline
\multicolumn{3}{|c|}{} & \multicolumn{3}{|c|}{$\Delta G_{33} = \Delta G_{30}+\Delta G_{03}$} & 1 &$Z_x,Z_y$\\ \hline
\multicolumn{3}{|c|}{} & \multicolumn{3}{|c|}{$\Delta G_{31} = \Delta G_{30}+\Delta G_{001}$} & $[cI]$ &$Z_y$\\ \hline
\multicolumn{3}{|c|}{} & \multicolumn{3}{|c|}{$\Delta G_{32} = \Delta G_{30}+\Delta G_{002}$} & $[Cro]$ &$Z_y$\\ \hline
\multicolumn{3}{|c|}{} & \multicolumn{3}{|c|}{$\Delta G_{103} = \Delta G_{100}+\Delta G_{03}$} & $[cI]$ &$Z_x$\\ \hline
\multicolumn{3}{|c|}{} & \multicolumn{3}{|c|}{$\Delta G_{203} = \Delta G_{200}+\Delta G_{03}$} & $[Cro]$ &$Z_x$\\ \hline
\multicolumn{3}{|c|}{} & \multicolumn{3}{|c|}{$\Delta G_{013} = \Delta G_{010}+\Delta G_{03}$} & $[cI]$ &$Z_x$\\ \hline
\multicolumn{3}{|c|}{} & \multicolumn{3}{|c|}{$\Delta G_{023} = \Delta G_{020}+\Delta G_{03}$} & $[Cro]$ &$Z_x$\\ \hline
\multicolumn{3}{|c|}{} & \multicolumn{3}{|c|}{$\Delta G_{113} = \Delta G_{100}+\Delta G_{010}+\Delta G_{03}$} & $[cI]^2$ &$Z_x$\\ \hline
\multicolumn{3}{|c|}{} & \multicolumn{3}{|c|}{$\Delta G_{123} = \Delta G_{100}+\Delta G_{020}+\Delta G_{03}$} & $[cI][Cro]$ &$Z_x$\\ \hline
\multicolumn{3}{|c|}{} & \multicolumn{3}{|c|}{$\Delta G_{213} = \Delta G_{200}+\Delta G_{010}+\Delta G_{03}$} & $[cI][Cro]$ &$Z_x$\\ \hline
\multicolumn{3}{|c|}{} & \multicolumn{3}{|c|}{$\Delta G_{223} = \Delta G_{200}+\Delta G_{020}+\Delta G_{03}$} & $[Cro]^2$ &$Z_x$\\ \hline

        \hline
        \end{tabular}
        \caption{Values of variables referenced by the partition function with RNA polymerase states included.  \ref{rnapdiff} uses the values of  $[RNAp]$,$\Delta G_{30}$, and $\Delta G_{03}$ listed first, whereas \ref{rnapdiff2} and \ref{rnapspacediff} use the values listed second.}
        \label{wtgval3}
    \end{table}%

An important contribution of the codes written in this thesis is the capacity for noise with spatial and temporal noise.  As an example, a spatially dependant noise that distorts locally the classical kinetics is given in \ref{diff2}.  A multiplier of 1000 is used to control the numerics.  \footnote{Computationally, diffusion equations with finite elements requires a very fine grid, or a rotating coordinate system when the noise is smaller or comparable to the underlying flow.  The factor of a thousand is an artificial enhancement to control the numerics. (i.e. a high temperature limit of sorts)  In general, it is best to fit all parameters simultaneously with experiments.}  

\begin{equation}
D(cI,Cro) = \left( 
  \begin{array}{cc}
    1000*cI^2*\frac{(\frac{dcI}{dT})^2}{1+cI^2+Cro^2} & 0 \\
    0 & 1000*Cro^2*\frac{(\frac{dCro}{dT})^2}{1+cI^2+Cro^2}
  \end{array}
  \right) .
\label{diff2}
\end{equation}

\normalsize

\chapter{\textit{Conclusions and Future Directions}}

\begin{quotation}
  \textit{Da\ss{} es mir - oder Allen - so \textit{scheint}, daraus folgt nicht, da\ss{} es so \textit{ist}.}

\hspace{0.1in}Ludwig Wittgenstein \textit{On Certainty}
\end{quotation}

\begin{quotation}
  \textit{We have compared the probability that an unknown scientist should have found out what has been vainly sought for so long, with the probability that there is one madman the more on the Earth, and the latter has appeared to us the greater.}

\hspace{0.1in}Henri Poincar\'e \textit{Science and Hypothesis}
\end{quotation}

  The particular contributions of this author towards this thesis, and towards other dynamical systems not discussed in this thesis, are archived in \cite{Aleph}.  This has allowed for much deeper quantitative inspection of the dynamical systems with noise then hitherto discussed with the phage $\lambda$.  This has allowed for the prediction of behaviour of phenotypes not yet synthesized and further comparison with those that have been.  Some subtlety involved with the removal of RNA polymerase from the partition function has been discussed.  The $\lambda$ phage has been discussed in physical and computational terms in hopes to present the system to a physics, mathematics, and computational audience.  The $\lambda$ phage is proving to be to biophysics what hydrogen was for quantum mechanics \footnote{A point Jin Wang has made to motivate this study.} and as such is a perfect testing ground for empirical and ab initio physical, computational and mathematical methods for gene expression.  

Efforts should be made to link the connection with the decision portion of the cell cycle to the bigger picture involving the upstream and downstream reactions and understand the connections between all the parameters and variables, see \cite{St.Pierre,arkin}.  Solutions are only initial conditions for the next portion of the cell cycle and our simple models are in danger of being too simple, as we are not left with much room under the rug to sweep degrees of freedom we choose to ignore.  

In particular, volume should be connected into the modeling with greater care.  The binding and unbinding are in the end very specific bond types occurring at very specific times, and proceeding in terms of idealized concentrations tends to wash out these subtleties.  I cannot say with confidence if the model here can or cannot corroborate the findings of \cite{St.Pierre} with respect to the sensitivity of the lysis and lysogeny decision being governed by volume.  It can be said though, that the decision being sensitive to volume is not mutually exclusive to the decision also being controlled by protein concentrations.  In the mathematical sense, we expect the spaces to be connected; volume is in the end the denominator of these densities.  Much of the fluctuation can be considered fluctuation of specific volume of cI.  For example, if one were to carefully wrap a cylinder around the helix-bend-helix motif of the protein and associated dimerizations and tabulate the volume, that volume would vary greatly over the course of, say, a binding and unbinding process.  This model lacks any specific protein configuration and in general, specific volume can vary a great deal over this space.  Discrete jumps in that space would appear as noise in this model and would need to be tabulated as such in the diffusion tensors, for example.  
From a phenomenological perspective it is a useful construct that is parameterized, ideally, by direct measurement and is a highly intuitive way to reproduce the general features of the circuit.               

\subsection{On Computation and Information}

It is important to realize that computers evolve simultaneously in two directions.  There is the physical interface, the its that make the bits and the its that control them, but there is also the human interface that establishes how we control the its that control the bits, and how the consequent bits are presented back to us in such a way usable to our central nervous system. 

\subsubsection{The Physical Interface}

The physical interface evolves in the direction of greater complexity.  Moore's law illustrates famously that rates of (serial and electronic) computation are limited by size and heat production.  
Parallelization, reversible logic, and quantum computation have all been under continual development to circumvent these issues and with considerable success.  Where a decade ago one would speak of how fast their computer was, now one speaks of how many cores it has.  This is parallelization's contribution.  $\aleph_0$ core would be the limiting case if this trend of the processor to $duo-core$ to $quad-core$ to $2^n$ core persists, but appended to the technical difficulties involved with space are now the technical difficulties involved with time, the synchronization of the bits from all the its.  The limitation that won't go away here is the speed of light; if A needs a signal from B before time t, the situation is hopeless unless B is spatially located within ct of A.  Multiple processors on a chip enable a reduction in heat dissipated by providing channels for which heat can be carried away from a processor.  Cells to cells exhibit analogous, although much more sophisticated signaling, are far closer to the $\aleph_0-core$ limit, and being waterproof would certainly be convenient for transporting heat elsewhere; 98.6 degrees is certainly quite a clue that bodies are quite adept at managing this housekeeping heat.  
Reversible logic evades the heat issue while exacerbating the spatial issue since intermediate steps are kept. 

\subsubsection{The Human Interface}

The human interface evolves in the direction of greatest simplicity.  Quantum Computer, classical computer, topological quantum computer, chemical computer,...is trivia far from the concern of the user who ultimately just wants to push a button and get a reliable answer.  Contrast the Fortran punch codes of old with the following lines of python code that defines our variational forms:

\begin{figure}
\begin{python}
a = v*u*dx + 0.5*k*(inner(grad(v),-velocity*u)*dx + inner(grad(v), D*grad(u))*dx)
L = v*u0*dx - 0.5*k*(inner(grad(v),-velocity*u0)*dx + inner(grad(v), D*grad(u0))*dx)

\end{python}
\caption{Variational forms used in this work.}
\label{varform}
\end{figure}      

These lines of code are the same no matter what mesh we choose, velocity we choose, time step, initial condition, and diffusion tensor.  When the code is called we better have defined what they mean, but until then they are freely transferrable.  See more examples in \cite{fenicsbook}.

A Gui was built to provide buttons to push to generate dynamical trajectories, Langevin trajectories, mesh trajectories (in which an entire domain is propagated in accordance with the flow field), and diffusion trajectories.  Systems built are the $\lambda$ phage, $\lambda$ phage with RNA polymerase, $\lambda$ phage approximating $O_L$ states, the cyclin system (see \cite{wang_potential_2008}), the Lorenz attractor, and can operate with imaginary numbers for the ode case.

Here is a screenshot showing multiple instances of the Gui from \cite{Aleph}.

\begin{figure}[h]
\begin{center}
\includegraphics[width=0.7\textwidth]{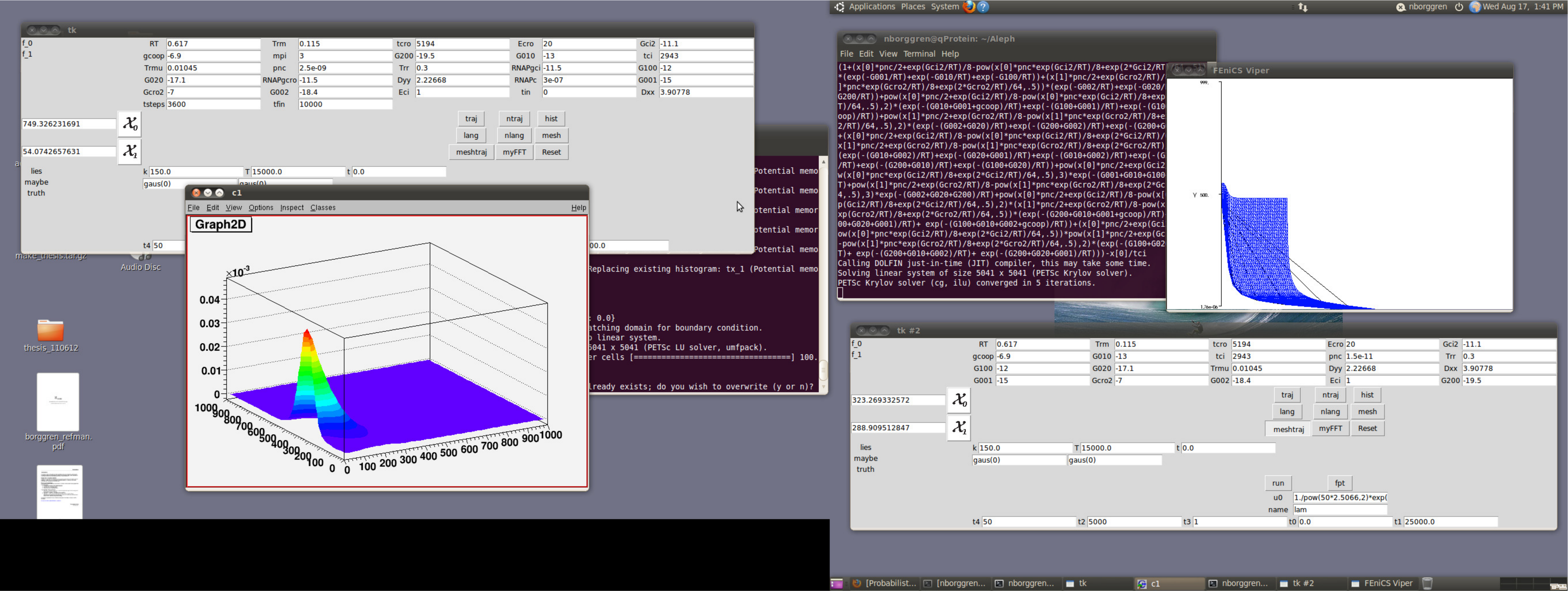}
\caption{{\bf This is the $\aleph_{00000}$ Gui demonstrating user interaction with equations of motion.  It reads equations of motion of a system and builds a Gui to interact with that system}}
\label{aleph}
\end{center}
\end{figure}

\section{On the importance of generality and the abstract}

It is important that methods and calculations be \textit{reproducible} and transferrable to different systems.  The study of dynamics in the most general settings cannot be neglected for proper understanding of the circuitry and signaling of cells.

Sun Ra says that it is laughable when knowledge is attributed to a man, \cite{JoyfulNoise}, and this rings all the more true when the discussion is one that seeks to synthesize elements from the great discoveries of physics: quantum mechanics and statistical mechanics, with the great discoveries of biology, evolution, genes, and DNA to just begin.  Chemistry has found itself central to the discussions of biology as a result of being best positioned and adept at comprehending the molecular basis and even the purest of mathematics at this stage cannot escape its insights from being applied to the problems of life.  At this interface; engineers, medical doctors, biologists, physicists, mathematicians, chemists, logicians, theologians and everybody else with hands and a mind are trying to contribute to this story.  The vacuousness of the borders of our disciplines has become evident and we find Nature ready and willing to adapt to any of our biases.  The wisdom of Poincar\'e on mathematics equally applies to the whole of science as well:

\begin{quotation}
We know very well that mathematics will continue to develop, but we have to find out in what direction. I shall be told "in all directions," and that is partly true; but if it were altogether true, it would become somewhat alarming. Our riches would soon become embarrassing, and their accumulation would soon produce a mass just as impenetrable as the unknown truth was to the ignorant. 
\end{quotation}

\noindent
So it is the case that Science and Method begins with a selection of facts, see \cite{poincare}.  For Poincar\'{e} this consisted in those with the greatest chance of recurrence and to each of us our lives come to be mirrors of our particular criterion of this selection process.  Wittgenstein in his youth would further require that this selection be from all the facts, see \cite{tractatus}, but by the end of his term he would have brushed that off to some of the symbolic games that we play, see \cite{certainty}.  The world is more akin to a collection of statements then a collection of facts and a statement can be true, false, or meaningless.  Perhaps one may want to continue this line of thinking to get a world that is a collection of words, but words require letters and something to denote nothing to tell where one word ends and the next word begins.  The statistical mechanics of random words and strings of letters may be equally important, see \cite{four,nofour}.    

What sets the subjects apart on this stage is no longer the content of their subject matter - it is one and the same - but the psychological predispositions of the practitioners.  What are the relevant degrees of freedom?  No two agree, but nor is it necessary for them to do so.  A mutual respect is in order, and as the tricks of noise and dynamics that can crash markets are now being applied to the problems of life, let us walk gently into this night, none of these days have been promised to us.  Recall a giant meteor from outerspace was required to wipe out the dinosaurs and open up our niche.  Nobody can look a three-headed pterodactyl with flourescent wings in the eyes and call themself the pinnacle of creation, and as far as I can tell, their is nothing prohibiting DNA to make such a genome.  Science is still new, our ignorance will be infinite, even while our knowledge can be infinite too.  For want of a common goal let us choose, at least, a moral one.     

\begin{quotation}
 \textit{The only thing that could unite the planet is a united space program [health center]... the earth becomes a space station [health center] and war is simply out, irrelevant, flatly insane in a context of research centers, spaceports, and the exhilaration of working with people you like and respect toward an agreed-upon objective, an objective from which all workers will gain.  Happiness is a by-product of function.  The planetary space station [health center] will give all participants an opportunity to function.}

\hspace{0.1in}William S. Burroughs, \textit{The Place of Dead Roads} [NAB]
\end{quotation}
  
Toss in some clean water, money, and the cure for pain and we may asymptotically approach a civil society.  When the subject matter is infection and disease and the consequences are pain and joy it would be imprudent for physics to wait for a topological quantum computer and all orders of precision worked out to chime in to this story, which is still caught up in tabulating everything.  From the looks of the White House, and the dazzling noise of stock markets apparently stuck between radio stations, the stalemate seem to persist straight up to the top.  

The challenges of this century can provide an incomparable opportunity towards global scientific collaboration as we learn to harness the resources we are enveloped in, or we can watch ourselves fumble the oil into sea, and another mysterious earthquake, tornado and tsunami wipe out our cities, as we choose the right color for our data points, and argue over what data points to delete.  

The human genome is at least the magnitude of mystery it was a decade ago and bioinformatics has been more successful at deciphering the code than electricity, magnetism, statistical mechanics, quantum mechanics, and all the other tools physicists pridefully declare universality with.  Some basic assumptions clearly need to be relinquished.   

\section{On the importance of the particular and the measurable; Case Studies}

A model of the phage $\lambda$ has been analyzed to illustrate how a particular system is queried.
Given that these crystal structures have been measured and good empirical force fields exist, these values of $\Delta G_s$ can in principle begin to be predicted and compared to the values deduced from experiments.  Ab initio calculations of the $O_R$ region from first principle Hamiltonians may take some time but until then it is exactly these types of mechanisms that motivate experiments like NSLS to do such crystallographic measurements that empirical protein and DNA force fields are developed to illuminate.  Understanding the relation of these nanoscopic structures and dynamics to the macroscopic flows of energy and entropy is the basic question necessary to address in order to appreciate the full value of the accumulated data and begin to use it for clinically relevant purposes.  Nuclear magnetic resonance, NMR, is also regularly used to investigate more directly these ensemble properties in vivo, near equilibrium, and can also be used to constrain these numbers.  The difference between in vitro and in vivo values for the weights has already been demonstrated to induce topological changes. These numbers have a taxonomy tree of their own as measurement techniques get refined.  Our phenomenology here continues to rely on fits to the primary datum from the biological measurements.  Physics will learn a lot from the networks that make up biological systems and the manners in which they give rise to emergent degrees of freedom.  In order to address the concerns of \cite{little_robust}, and perhaps many other scientists who are typically more concerned about large scale qualitative differences, like the difference of lysogeny and lysis, we will scan parameters as much as we can in this volume of spacetime.  The particular values of $\Delta G$ are dependent in some complicated way to evolution, volume, temperature, pH, doping, in vitro, in vivo and a host of other unknowns, can and do change on large scales.  Schrodinger’s anthropomorphic remark can be applied here: it is exactly robustness to these types of changes that allow it to exist at 300K for so long in the first place.  Here we note a difference in the use of the terminology ‘robust’ between biologists and the use in subsequent theoretical work.  For Little, robustness is the property that two states, lytic and lysogenic, are still present after large scale mutations, changes in $\Delta G$ due to switching, say, site 3 with a copy of site 1.  This is a transformation on the level of the partition function in that the $\Delta G$ values change.  This differs from the use of robustness in the sense of \cite{Ao, wang_kinetic_2010,aurell_epigenetics_2002}, which uses this word to describe the small rate at which a cell state switches spontaneously as a result of noise.  The phage can be robust in both senses.  It is even robust against our attempts to model it.

Binding studies for drug docking often involve phenomena of this sort and computational drug design as well.  This is a largely trial and error process, and perhaps some topological information is due.  Biologists are often skeptical of the latest grand unification scheme from the physicists and the applied mathematicians, and their skepticism is mimed here.  The main point of separation, as I see it, between the two subjects is in a sense a question of topology.  The mass of an electron is a number with small error bars that with hard work can be made even smaller.  It is repeatable and exists on other planets too, the numbers of biology are scarce when present.  The facts they prefer to select on are topological in nature: to lyse or not to lyse?     

\subsection{Robustness, robustness, stability, and stability}

\begin{quotation}
 \textit{..., I am referring to something that mere words will never be able to express, relative, absolute, full, empty, still alive and no longer alive, because, sir, in case you don't know it, words move, they change from one day to the next, they are as unstable as shadows, are themselves shadows,...}

\hspace{0.1in}Jos\'{e} Saramago, \textit{Death with interruptions}
\end{quotation}

In the overlap of subjects it is quite expected for there to be ambiguity between the different uses of the same word.  This is all too commonplace and has been a perennial source of confusion in this study.  Two examples are worthy of note, the notion robust in the title of \cite{little_robust} is different than the use of the same word in the subsequent work of \cite{wang_potential_2008,Ao}.  The first use is to denote the propensity of phage $\lambda$ to maintain a working switch after particular mutations in the $O_R$ sites.  This is a change at the level of the partition function; the genome has fundamentally changed, yet the bistability persists, it is robust.  The second use refers to something fundamentally different: the capacity of a lysogen to stay a lysogen by continuous maintenance by $cI$ protein to keep, in particular, the cro gene from expression.  The use of the word stability in \cite{little_robust} is exactly what \cite{wang_potential_2008,Ao} refer to as robustness, reserving their use of the word stability to refer to the mathematical use of describing the nature of the stationary points.  Stability and robustness are used interchangeably in \cite{aurell_epigenetics_2002}.

With a four order magnitude spread in the refereed values of the numbers quantifying robustness, it is possible that there is some misunderstanding here.  Indeed the numbers reported as experimental in \cite{wang_kinetic_2010} are not strictly speaking experimental numbers at all, they involve simulation as well, and Little has now decreased those numbers to $10^{-8}$ or $10^{-9}$, in units of lysogens flipping per generation, but has not published these numbers in a refereed journal and caveats the measurement as very difficult when it arises.  Finding a more realistic diffusion tensor and better oligomerization equations of state are likely to help better quantify the robustness in the model.  We note that in the large concentration limit, where the dynamics picture is sufficient, the phase space completely partitions into two basins of attraction and the number is simply 0.  The stability puzzle is still puzzling.    
      
\section{Algebraic, Analytical, or Computational Methods?}
A computational approach, the finite elements, were eventually decided upon for the practical solution of the differential equations herein \cite{fenicsbook}.  Other methods were available and attempted.  In particular, an algebraic approach was also investigated (see \cite{marcus_1960,figueiredo_algebraic_1998}), and a brute-force analytic approach was used in \cite{snider}.  These are briefly discussed but were found not as immediately illuminating as the numerical approach throughout.  \subsection{Non-Associative Commutative Algebra}  A differential equation is associated with a non-associative algebra such that invariant manifolds of the differential equation correspond to invariant sub-algebras of the algebra.  In particular, given a dynamical system, expand in a form,
\begin{equation}
\dot{x_i}= \sum_{j=1}^mA_{ij}\prod_{k=1}^nx_k^{B_{jk}}, 
\end{equation}
we can then define a matrix $M = BA$ and a system $\dot{U_i}=U_iM_{ij}U_j$.  This is then a quadratic system and the methods of Marcus apply; we can define an algebra by introducing a product $u_i*u_j=\frac{1}{2}(\delta_{ik}M_{kj}+\delta_{ij}M_{jk})$. 

 \subsection{Recursion Equations for Expansion coefficients } In particular, a dynamical system with some noise is concisely written adopting the summation convention

\begin{equation}
\partial_t P = \partial_iD_{ij}\partial_jP - \partial_i{(F_iP)},
\label{diffusionagain}
\end{equation}
\noindent
Roman indices run from 1 to $K$ over the generalized coordinates of the system.  This relates the time change of the surface $P$ with a diffusion term and the flow field of the dynamical system.  The noise has effectively been smoothed and encoded in the surface $P$ through the diffusion tensor $D_{ij}$. To be explicit, by $\partial_iD_{ij}\partial_jP$ we mean $\partial_i(D_{ij}(\partial_jP))$.

The zero is used to denote the time index and thus $K+1$ indices over the non-negative integers are summed over.  Ansatz $P$, a polynomial expansion of the form

\begin{equation}
P(x_0,x_1,...,x_K) = p_{i_0,i_1,...,i_K}\prod_{j=0,...,K}x_j^{i_j},
\end{equation}   
and similarly for the components of $F$

\begin{equation}
F_m(x_0,x_1,...,x_K) = f_{m,i_0,i_1,...,i_K}\prod_{j=0,...,K}x_j^{i_j}.
\end{equation} 
\noindent
We have allowed the index corresponding to powers of $x_i$ to run over a countably infinite set if need be, however the $ m$ counts only the $K$ indices of the coordinates.   In general we need an analogue expansion for $D_{mn}$.  We will soon specify to the constant diffusion case, which is lowest order of the general case

\begin{equation}
D_{mn}(x_0,x_1,...,x_K) = D_{m,n,i_0,i_1,...,i_K}\prod_{j=0,...,K}x_j^{i_j} .
\end{equation}  

Term by term we will decipher equation (\ref{diffusionagain}),  

\begin{equation}
\partial_t P(x_0,x_1,...,x_K) = p_{i_0,i_1,...,i_K}(i_0x^{i_0-1})\prod_{j=1,...,K}x_j^{i_j}= p_{i_0+1,i_1,...,i_K}(i_0+1)\prod_{j=0,...,K}x_j^{i_j}.
\end{equation}
\noindent
Here we specify constant diffusion tensors.  For the next term in equation (\ref{diffusionagain}), the coefficients on the diagonal are related by

\begin{equation}
D_{mm}\partial_m\partial_mP(x_0,x_1,...,x_K) = D_{mm}p_{i_0,i_1,...,i_m+2,...,i_K}(i_m+2)(i_m+1)\prod_{j=1,...,K}x_j^{i_j},
\end{equation}
\noindent
and off diagonal by

\begin{equation}
D_{mn}\partial_m\partial_nP(x_0,x_1,...,x_K) = p_{i_0,i_1,...,i_m+1,...,i_n+1,...,i_K}(i_m+1)(i_n+1)\prod_{j=1,...,K}x_j^{i_j}.
\end{equation}
\noindent
The last term in equation (\ref{diffusionagain}) has two contributions through the product rule,

 \begin{equation}
(\partial_mF_m)P(x_0,x_1,...,x_K) = (f_{m,i'_0,i'_1,...,i'_m+1,...,i'_K}(i'_m+1)\prod_{j=0,...,K}x_j^{i'_j})p_{i_0,i_1,...,i_m,...,i_K}\prod_{j=1,...,K}x_j^{i_j},
\end{equation}

 \begin{equation}
F_m\partial_mP(x_0,x_1,...,x_K) = (f_{m,i'_0,i'_1,...,i'_K}\prod_{j=0,...,K}x_j^{i'_j})p_{i_0,i_1,...,i_m+1,...,i_K}(i_m+1)\prod_{j=1,...,K}x_j^{i_j}.
\end{equation}
\noindent
These become 

\begin{equation}
(\partial_mF_m)P(x_0,x_1,...,x_K) = f_{m,i'_0,i'_1,...,i'_m+1,...,i'_K}(i'_m+1)p_{i_0,i_1,...,i_m,...,,i_K}\prod_{j=1,...,K}x_j^{i'_j+i_j},
\end{equation}

 \begin{equation}
F_m\partial_mP(x_0,x_1,...,x_K) = f_{m,i'_0,i'_1,...,i'_K}p_{i_0,i_1,...,i_m+1,...,i_K}(i_m+1)\prod_{j=1,...,K}x_j^{i'_j+i_j}.
\end{equation}

Let us simplify by saying the diffusion tensor is diagonal.  Substituting into the differential equation yields 

\begin{eqnarray}
p_{i_0+1,i_1,...,i_K}(i_0+1)\prod_{j=0,...,K}x_j^{i_j} & = & D_{mm}p_{i_0,i_1,...,i_m+2,...,i_K}(i_m+2)(i_m+1)\prod_{j=1,...,K}x_j^{i_j} \nonumber\\
 & & {} -f_{m,i'_0,i'_1,...,i'_m+1,...,i'_K}p_{i_0'',i_1'',...,i_m'',...,,i_K''}(i'_m+1)\prod_{j=1,...,K}x_j^{i'_j+i_j''}\\
 & & {} -f_{m,i'_0,i'_1,...,i'_K}p_{i_0'',i_1'',...,i_m+1'',...,i_K''}(i_m''+1)\prod_{j=1,...,K}x_j^{i'_j+i_j''} .\nonumber
\end{eqnarray}
\noindent
Setting the coefficients of $\prod_{j=0,...,K}x_j^{i_j}$ equal on both sides gives the recursion relation

\begin{eqnarray}
p_{i_0+1,i_1,...,i_K}(i_0+1) & = & D_{mm}p_{i_0,i_1,...,i_m+2,...,i_K}(i_m+2)(i_m+1) \nonumber \\
 & & {} -\sum_{i_j'+i_j''=i_j}f_{m,i'_0,i'_1,...,i'_m+1,...,i'_K}p_{i_0'',i_1'',...,i_m'',...,,i_K''}(i'_m+1)\\
 & & {} -\sum_{i_j'+i_j''=i_j}f_{m,i'_0,i'_1,...,i'_K}p_{i_0'',i_1'',...,i_m+1'',...,i_K''}(i_m''+1) .\nonumber
\end{eqnarray}
\noindent
A steady state solution in particular requires $p_{i_0}=0$ for all i.  So that the index can be dropped altogether. 

\begin{eqnarray}
0 & = & D_{mm}p_{i_1,...,i_m+2,...,i_K}(i_m+2)(i_m+1) \nonumber \\
 & & {} -\sum_{i_j'+i_j''=i_j}f_{m,i'_1,...,i'_m+1,...,i'_K}p_{i_1'',...,i_m'',...,,i_K''}(i'_m+1)\\
 & & {} -\sum_{i_j'+i_j''=i_j}f_{m,i'_1,...,i'_K}p_{i_1'',...,i_m+1'',...,i_K''}(i_m''+1). \nonumber
\end{eqnarray}

The software of \cite{Aleph}, can also perform calculations based on these approaches. 

\section{From biology to physics to biology}

Physics has always relied heavily on progress from biology for its own progress.  The lung predates the steam engine and it is unlikely the latter would have emerged without the scientific inspection of the former.  The central nervous system predates the understanding of electricity and again the latter would not have emerged without the scrutiny of the former.  

Consider the timelime of the study of metabolism, see \cite{GreatExperiments}.
In around 1500, Leonardo da Vinci likened the burning of a candle to animal nutrition, from which he inferred that an atmosphere that could not support combustion would also not support animals.  In 1648, Van Helmont coined the word 'gas' in a journal for medicine under the title \textit{Ortus Medicinae}.  The first gas described as such was $CO_2$, the biproduct of animal respiration.  From 1660-1678 Robert Boyle developed experiments that studied the changes in a volume of air due to respiration and combustion and similarly for Mayow's \textit{Treatise on Respiration}.  In 1754 Black discovered \textit{fixed air}, $CO_2$.  In 1766, Cavendish discovered \textit{inflammable air}, $H_2$.  In 1772, Rutherford described \textit{residual air}, $N_2$.  In 1780, with their new invention, \textit{the calorimeter}, Lavoisier and Laplace concluded that da Vinci's hypothesis of respiration being a form of combustion is indeed the case in their beautiful paper \textit{Memoir on Heat}.  Thus the groundwork had been laid for thermodynamics and the subsequent industrial revolution, the driving force of which having been the study of respiration.

One only need to read the bibliography of Hermann Helmholtz to see that physiology was a driving force in the study of electricity and acoustics, see \cite{OnSensations}.

Progress in physics in the 20th century, given the equipment and good mathematical abstractions of the 19th century, managed on its own for a century without much aid from biology and chemistry, but now it is time to revisit the old phenomena, with our new lights, and actually solve some problems.  

Biologists and chemists have handed us the structures of their molecules, the wirings of their systems, and the phase diagrams of their compounds.  If those are the answers and statistical mechanics and quantum mechanics provide the questions, what is left but the proof?

Bibliography
\bibliography{bibliography}
\bibliographystyle{unsrtnat}
\appendix
\chapter{Appendix; Tabulation of Results}

\begin{quotation}
  \textit{Ineluctable modality of the visible: at least that if no more, thought through my eyes. Signatures of all things I am here to read, seaspawn and seawrack, the nearing tide, that rusty boot. Snotgreen, bluesilver, rust: coloured signs. Limits of the diaphane. But he adds: in bodies. Then he was aware of them bodies before of them coloured. How? By knocking his sconce against them, sure. Go easy. Bald he was and a millionaire, maestro di color che sanno. Limit of the diaphane in. Why in? Diaphane, adiaphane. If you can put your five fingers through it, it is a gate, if not a door. Shut your eyes and see.}

\hspace{0.1in}James Joyce, \textit{Ulysses}
\end{quotation}

\begin{quotation}
\textit{We live, I regret to say, in an age of surfaces.}

\hspace{0.1in}Oscar Wilde, \textit{The importance of being earnest}
\end{quotation}

\section{Results}

\begin{figure}[h]
\begin{center}
\includegraphics[width=0.7\textwidth]{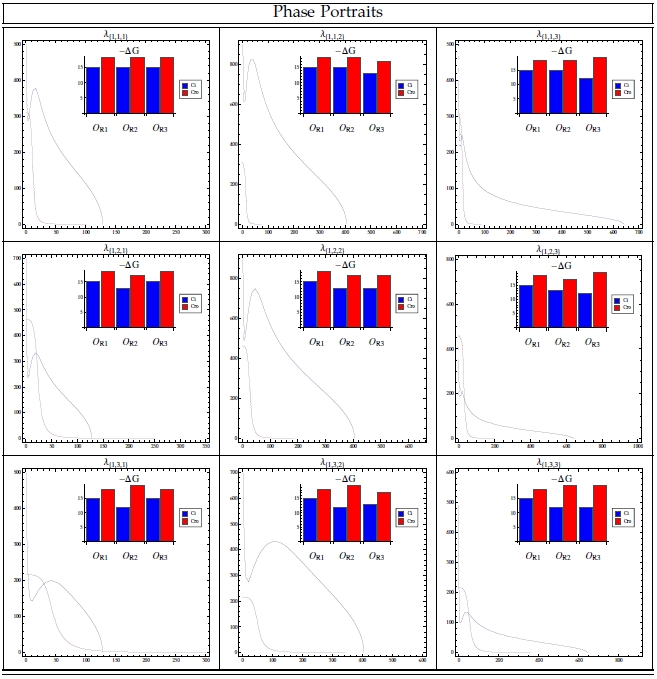}
\caption{\bf {Contours with stationary Ci and Cro production with $O_{R_1}$ fixed.  The axis are in protein numbers.  3 of the 9 mutations are destabilized: $\lambda_{112}$,$\lambda_{122}$, and $\lambda_{132}$}}
\label{phase1}
\end{center}
\end{figure}

\begin{figure}[h]
\begin{center}
\includegraphics[width=0.7\textwidth]{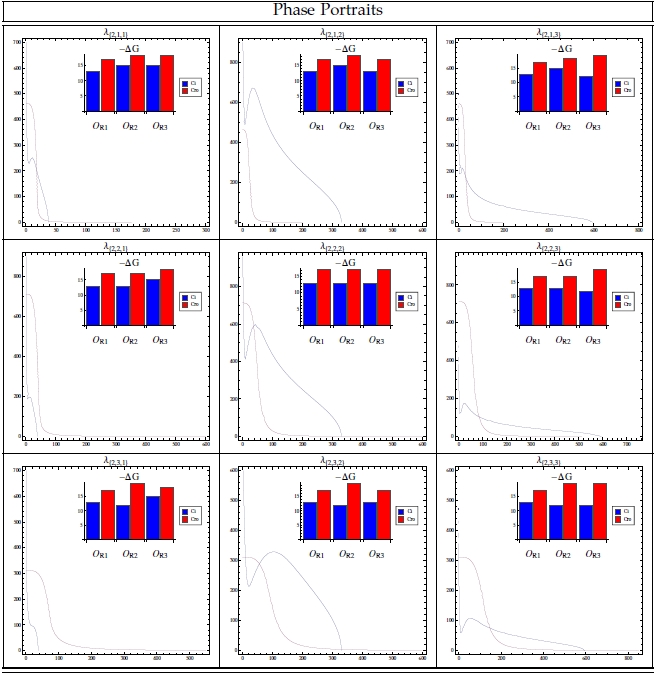}
\caption{\bf {Contours with stationary Ci and Cro production with $O_{R_1}$ fixed.  The axis are in protein numbers.  3 of the 9 mutations are destabilized:$\lambda_{212}$,$\lambda_{221}$, and $\lambda_{231}$. }}
\label{phase2}
\end{center}
\end{figure}

\begin{figure}[h]
\begin{center}
\includegraphics[width=0.7\textwidth]{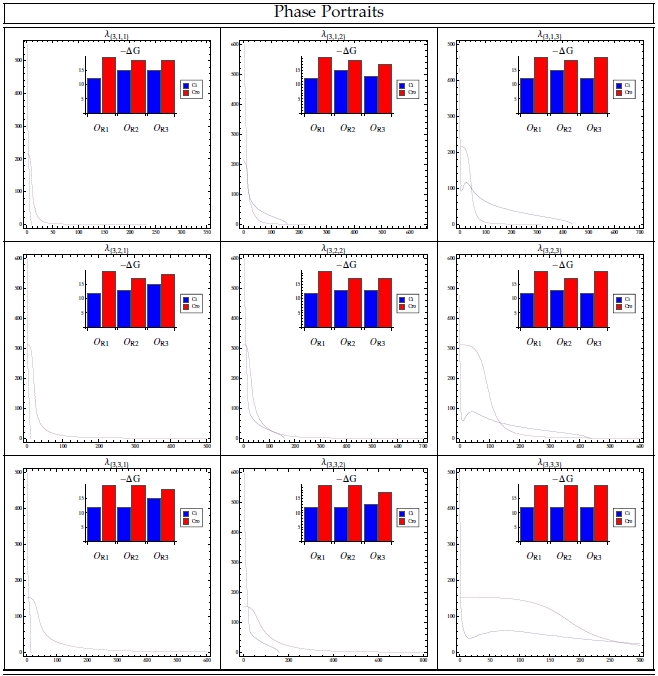}
    \caption{\bf {Contours with stationary Ci and Cro production with $O_{R_1}$ mutated to $O_{R_3}$.  5 of the 9 mutations are destabilized: $\lambda_{311}$,$\lambda_{321}$, $\lambda_{322}$, $\lambda_{331}$ and $\lambda_{332}$. }}    
\label{phase3}
\end{center}
\end{figure}

The phase portraits for the 27 mutations are shown in Tables \ref{phase1}, \ref{phase2} and \ref{phase3}.

The switch component of 3 of the 9 mutations in Table \ref{phase1} are destabilized: $\lambda_{112}$,$\lambda_{122}$, and $\lambda_{132}$.  Transients however are still largely important for the out of equilibrium switch properties.
3 of the 9 mutations in Table \ref{phase2} are destabilized: $\lambda_{212}$,$\lambda_{221}$, and $\lambda_{231}$.  Transients however are still largely important for the out of equilibrium switch properties.  
5 of the 9 mutations in Table \ref{phase3} are destabilized: $\lambda_{311}$,$\lambda_{321}$, $\lambda_{322}$, $\lambda_{331}$ and $\lambda_{332}$.  Transients however are still largely important for the out of equilibrium switch properties.

The locations of the stationary points are tabulated in Tables 
\ref{tab:zeroes}, \ref{tab:twozeroes}, \ref{tab:threezeroes1}, and \ref{tab:threezeroes2}.

Time evolution of distributions is given in Tables 
\ref{tab:gt4}, \ref{tab:gt5}, \ref{tab:gt6} and \ref{tab:gt7}, corresponding to the wild-type and mutants $\lambda_{121}$, $\lambda_{323}$, and $\lambda_{323'}$.  

The Tables \ref{tab:l121}, \ref{tab:l123}, \ref{tab:l131}, \ref{tab:l133}, 
\ref{tab:l211}, \ref{tab:l213}, \ref{tab:l222}, \ref{tab:l223}, 
\ref{tab:l232}, \ref{tab:l233}, \ref{tab:l312}, \ref{tab:l313}, and \ref{tab:l323} archive the results of mutants with three stationary points.  Depicted are the phase space portraits alongside the Probability and Flux Landscapes and above the passage time distributions to each of the three stationary points.  Assorted expectation values are calculated and can be compared with the wild-type.    

Representative trajectories, encoding global features, are depicted in Tables \ref{tab:ltraj1}, \ref{tab:ltraj2}, and \ref{tab:ltraj3}.

\begin{table}[ht]

  \centering
     \begin{tabular}{|c|c|c|}
          \hline
          \multicolumn{3}{|c|}{Properties of the Stationary Point} 
          \\ \hline \hline
          Mutant &\multicolumn{1}{|c|}{x (Ci proteins)}&\multicolumn{1}{|c|}{y (Cro proteins)} \\ \hline
          \multicolumn{3}{|l|}{$\lambda_{112}$} \\ \hline
          &404.90 & 0.0009 \\ \hline

          \multicolumn{3}{|l|}{$\lambda_{122}$}\\ \hline 
          &404.89 & 0.0232 \\ \hline

          \multicolumn{3}{|l|}{$\lambda_{132}$}\\ \hline 
          &404.89 & 0.1173 \\ \hline

          \multicolumn{3}{|l|}{$\lambda_{212}$}\\ \hline 
          &329.88& 0.02954\\ \hline

          \multicolumn{3}{|l|}{$\lambda_{221}$}\\ \hline 
          &0.3535& 712.24\\ \hline
          
          \multicolumn{3}{|l|}{$\lambda_{231}$ }\\ \hline
          &1.7522& 312.15\\ \hline

          \multicolumn{3}{|l|}{ $\lambda_{311}$}\\ \hline 
          &3.4415& 214.70\\ \hline

          \multicolumn{3}{|l|}{ $\lambda_{321}$ }\\ \hline
          &1.7532& 312.14 \\ \hline

          \multicolumn{3}{|l|}{ $\lambda_{322}$ }\\ \hline
          &10.253& 309.57 \\ \hline

          \multicolumn{3}{|l|}{$\lambda_{331}$ }\\ \hline
          &6.1379& 153.09 \\ \hline

          \multicolumn{3}{|l|}{$\lambda_{332}$ }\\ \hline
          &20.403& 152.35 \\ \hline
          \hline
     \end{tabular}
     \caption[Mutants with one stationary point.]{Mutants with one stationary point.  }
\label{tab:zeroes}
    \end{table}%

\begin{table}[ht]

  \centering
     \begin{tabular}{|c|c|c|c|}
          \hline
          \multicolumn{4}{|c|}{Properties of Stationary Points}
          \\ \hline \hline
          Mutant & \multicolumn{1}{|c|}{index}&\multicolumn{1}{|c|}{x (Ci Proteins)}&\multicolumn{1}{|c|}{y (Cro Proteins)}\\ \hline
          \multicolumn{4}{|l|}{$\lambda_{113}$}\\ \hline 
          & 0 & 644.05 & 0.00023\\ \hline
          & 1 & 0.31299 & 310.32500 \\ \hline

        \hline
        \end{tabular}
        \caption[Mutants with two stationary points]{Mutants with two stationary points. }
        \label{tab:twozeroes}
    \end{table}%

\begin{table}[ht]
  \centering
     \begin{tabular}{|c|c|c|c|}
          \hline
          \multicolumn{4}{|c|}{Properties of Stationary Points}
          \\ \hline \hline
          Mutant & \multicolumn{1}{|c|}{index}&\multicolumn{1}{|c|}{x (Ci Proteins)}&\multicolumn{1}{|c|}{y (Cro Proteins)}\\ \hline 

          \multicolumn{4}{|l|}{$\lambda_{111}$}\\ \hline 
          & 0 & 5.6833& 301.287\\ \hline 
          & 1 & 1.8180& 310.218\\ \hline 
          & 2 & 127.87& 0.03220\\ \hline

          \multicolumn{4}{|l|}{$\lambda_{121}$}\\ \hline 
          & 0 & 19.9875& 332.348\\ \hline 
          & 1 & 0.82427& 462.571\\ \hline
          & 2 & 127.86& 0.8235\\ \hline 

          \multicolumn{4}{|l|}{$\lambda_{123}$ wild-type }\\ \hline 
          & 0 & 0.141753& 462.57273 \\ \hline
          & 1 & 644.05& 0.00611 \\ \hline 
          & 2 & 29.979& 178.229 \\ \hline 

          \multicolumn{4}{|l|}{$\lambda_{131}$ }\\ \hline 
          & 0 & 30.0039& 191.3784 \\ \hline 
          & 1 & 3.43968& 216.27658 \\ \hline 
          & 2 & 127.812& 4.168042 \\ \hline 

          \multicolumn{4}{|l|}{$\lambda_{133}$ }\\ \hline 
          & 0 & 48.6464& 130.6941 \\ \hline 
          & 1 & 0.6369& 216.311 \\ \hline 
          & 2 & 644.043& 0.0309 \\ \hline 

          \multicolumn{4}{|l|}{$\lambda_{211}$ }\\ \hline 
          & 0 & 17.505& 206.876 \\ \hline 
          & 1 & 37.3015& 4.8364 \\ \hline 
          & 2 & 0.8244& 462.570 \\ \hline

          \multicolumn{4}{|l|}{$\lambda_{213}$ }\\ \hline 
          & 0 & 0.14175& 462.572 \\ \hline 
          & 1 & 595.05& 0.00679 \\ \hline 
          & 2 & 29.938& 177.711 \\ \hline 

          \multicolumn{4}{|l|}{$\lambda_{222}$ }\\ \hline 
          & 0 & 37.234& 590.517 \\ \hline
          &1 & 2.6864& 712.240 \\ \hline 
          &2 & 329.884& 0.7555 \\ \hline 

        \hline
        \end{tabular}
        \caption[Mutants with three stationary points]{Mutants with three stationary points.  }
        \label{tab:threezeroes1}
    \end{table}%

\begin{table}[ht]
  \centering
     \begin{tabular}{|c|c|c|c|}
          \hline
          \multicolumn{4}{|c|}{Stationary Points}
          \\ \hline \hline
          Mutant & \multicolumn{1}{|c|}{index}&\multicolumn{1}{|c|}{x (Ci Proteins)}&\multicolumn{1}{|c|}{y (Cro Proteins)}\\ \hline 
          \multicolumn{4}{|l|}{ $\lambda_{223}$ }\\ \hline 
          &0 & 0.06002& 712.247\\ \hline 
          &1 & 595.04& 0.17382\\ \hline 
          &2 & 83.624& 103.62\\ \hline 

          \multicolumn{4}{|l|}{$\lambda_{232}$ }\\ \hline 
          &0 & 56.762& 289.227\\ \hline 
          &1 & 10.272& 312.072\\ \hline 
          &2 & 329.829& 3.82207\\ \hline 

          \multicolumn{4}{|l|}{ $\lambda_{233}$ }\\ \hline 
          &0 & 0.30934& 312.153\\ \hline
          &1 & 146.0568& 74.57208\\ \hline 
          &2 & 594.8508& 0.87984\\ \hline 

          \multicolumn{4}{|l|}{$\lambda_{312}$ }\\ \hline 
          &0 & 19.08756& 102.4368\\ \hline
          &1 & 13.3744& 163.4361\\ \hline 
          &2 & 158.8459& 0.47835\\ \hline

\multicolumn{4}{|l|}{ $\lambda_{313}$ }\\ \hline 
&0 & 44.4330& 100.41068\\ \hline 
&1 & 0.63744& 216.3115\\ \hline
&2 & 436.73205& 0.054279\\ \hline

\multicolumn{4}{|l|}{ $\lambda_{323}$ }\\ \hline 
&0 & 0.30935& 312.15312\\ \hline 
&1 & 133.92& 52.23903\\ \hline 
&2 & 436.115& 1.393207\\ \hline 

\multicolumn{4}{|l|}{ $\lambda_{333}$} \\ \hline 
&0 & 1.24472& 153.1449\\ \hline 
&1 & 267.603& 28.44186\\ \hline 
&2 & 417.693& 7.90558\\ \hline 

        \hline
        \end{tabular}
        \caption{Mutants with three stationary points (continued)}
        \label{tab:threezeroes2}
    \end{table}%


\clearpage

\begin{table}[ht]
        \centering
        \begin{tabular}{|p{0.11\textwidth}|p{0.29\textwidth}|p{0.29\textwidth}|p{0.29\textwidth}|}
          \hline
          \multicolumn{4}{|c|}{Time Evolution}
          \\ \hline \hline
          time&$+0$&$+100$&$+200$ \\ \hline
          $t=1$ & \includegraphics[scale=.22]{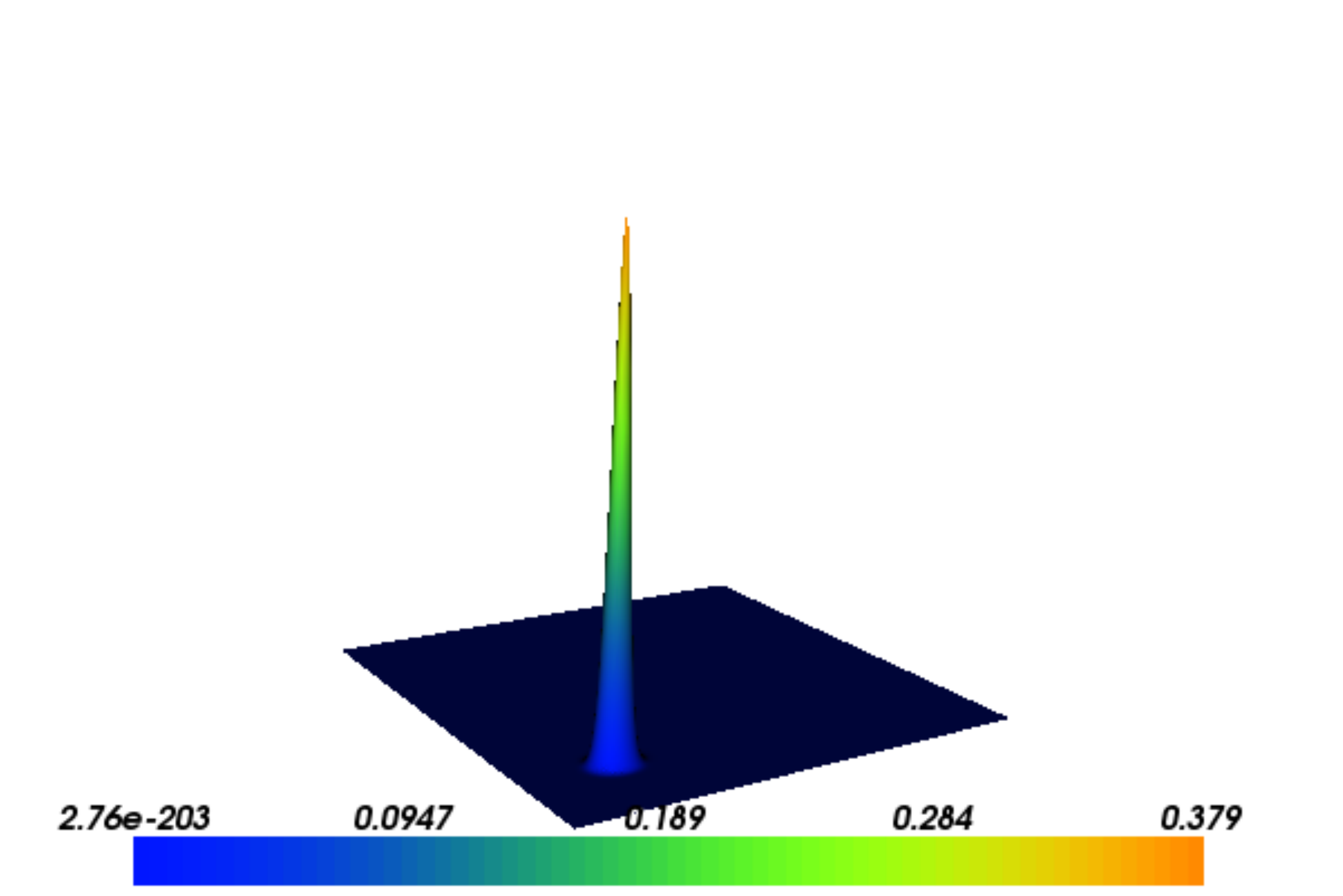}& \includegraphics[scale=.22]{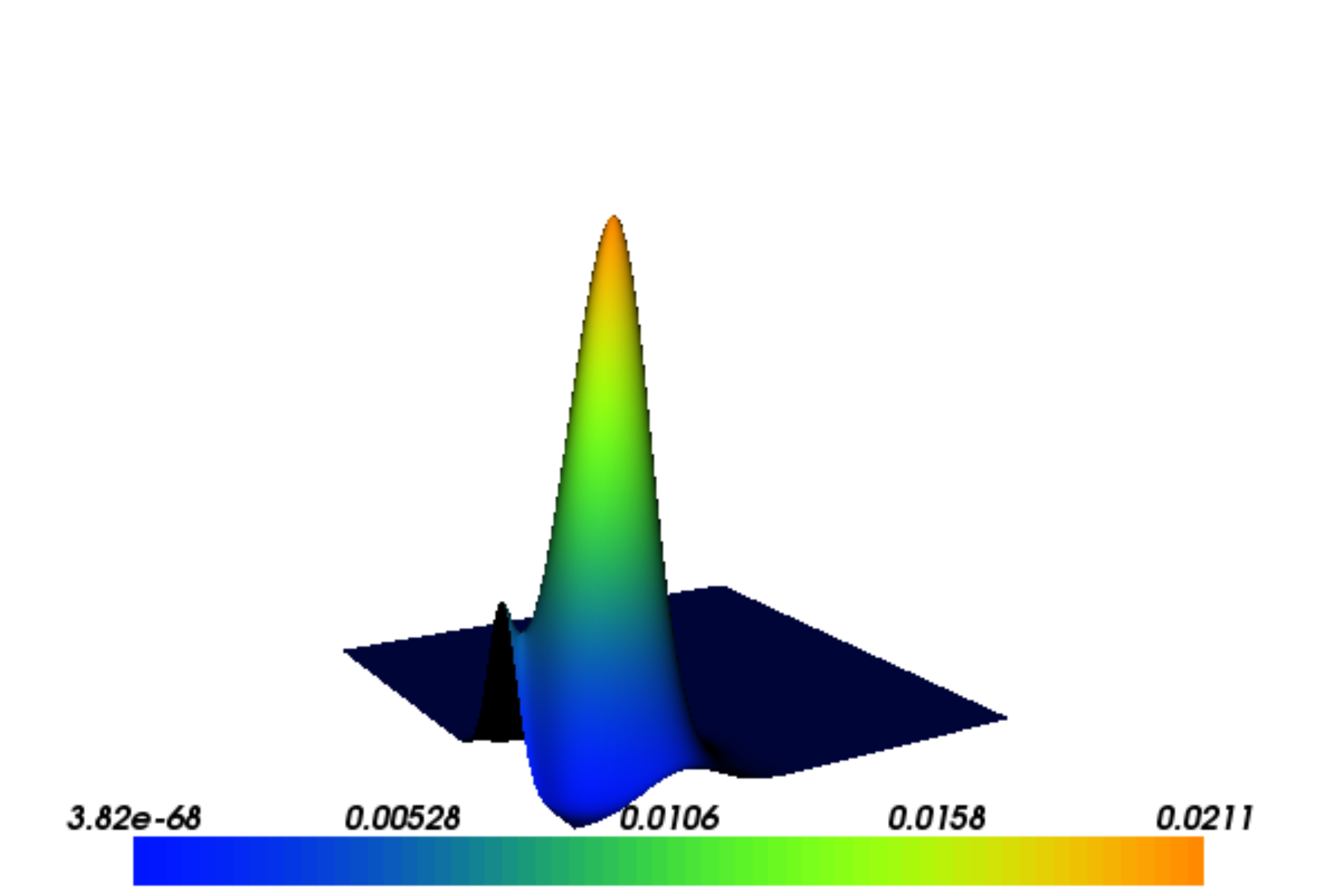}&\includegraphics[scale=.22]{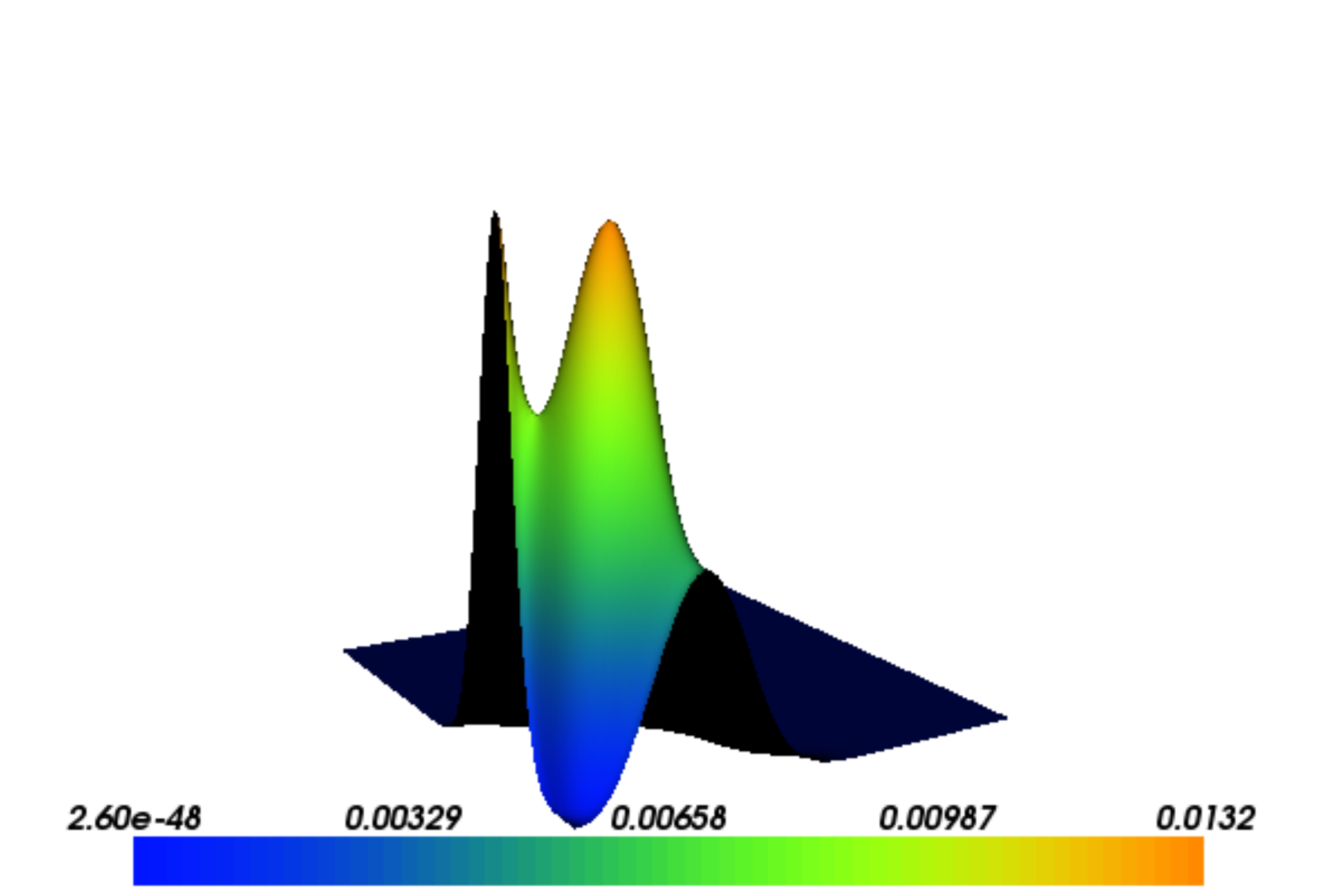}\\ \hline
          $t=301$ & \includegraphics[scale=.22]{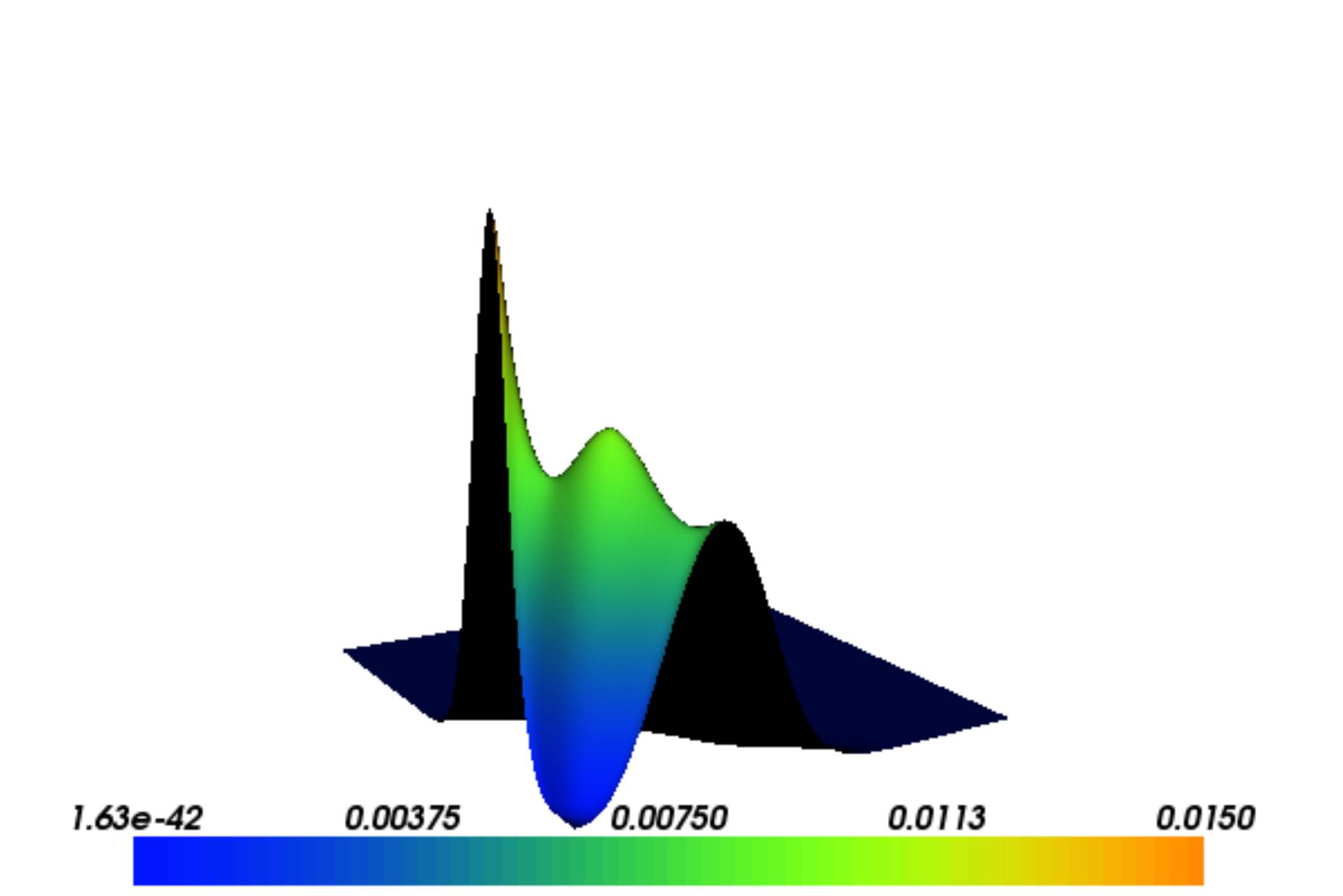}& \includegraphics[scale=.22]{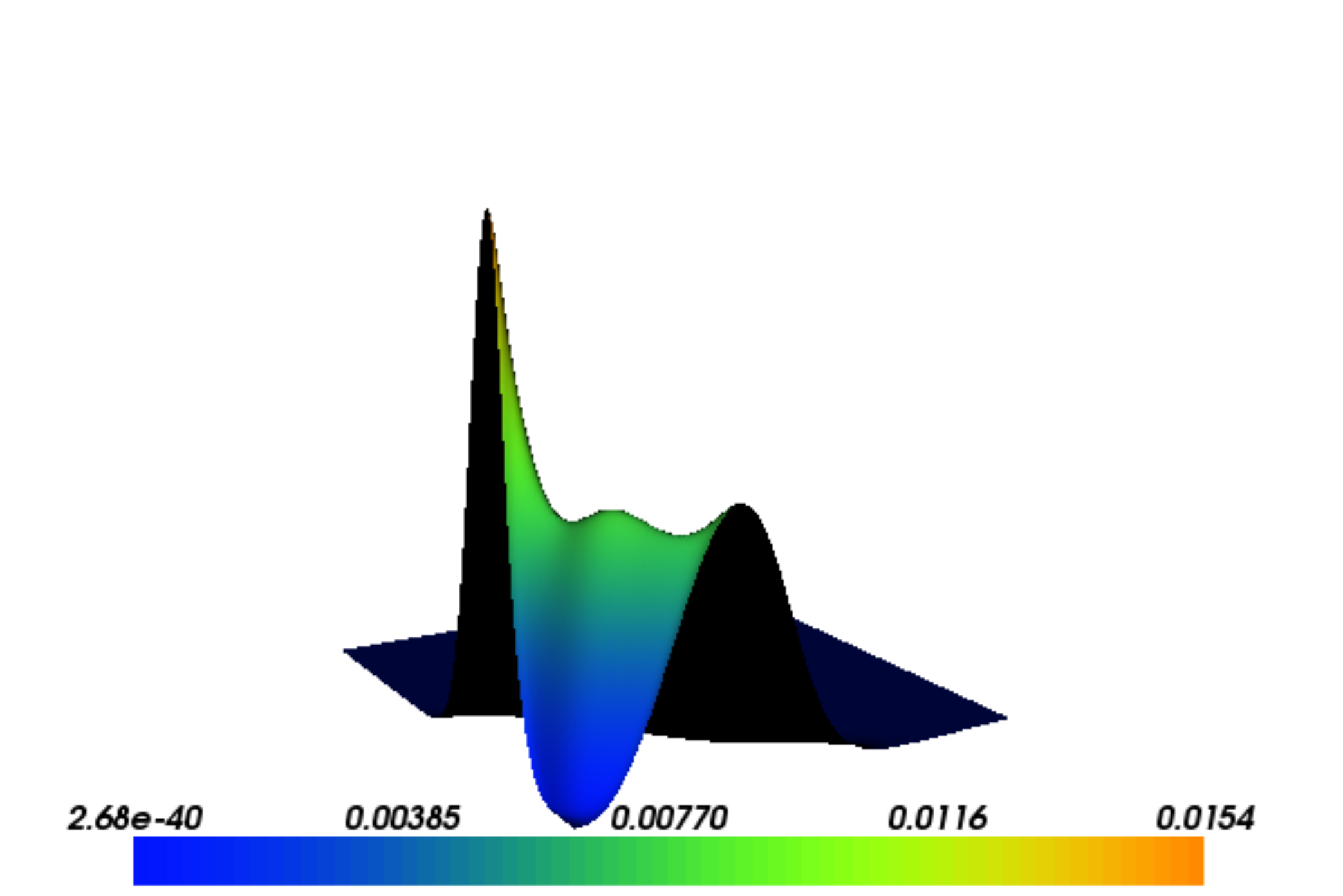}&\includegraphics[scale=.22]{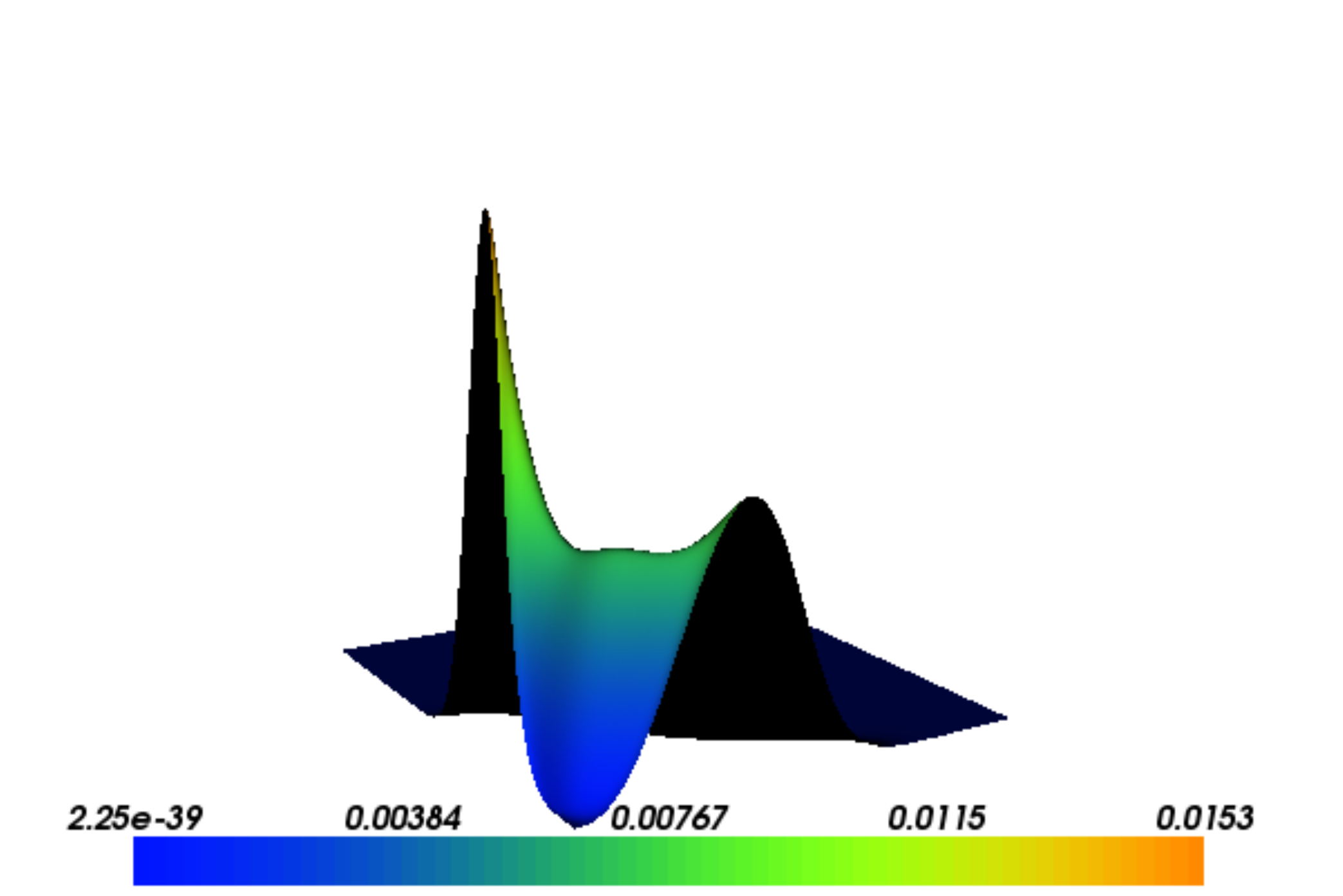}\\ \hline
          $t=601$ & \includegraphics[scale=.22]{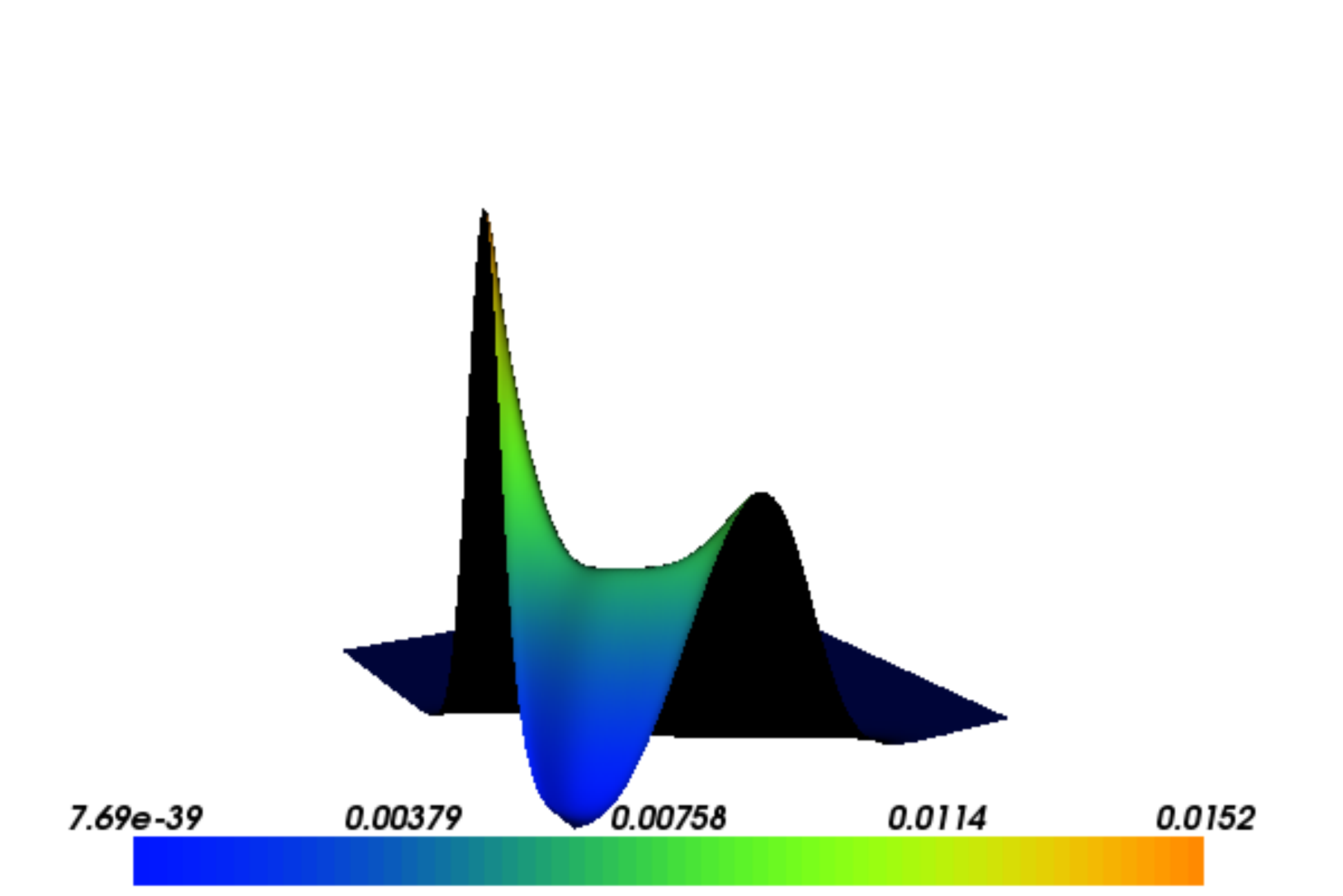}& \includegraphics[scale=.22]{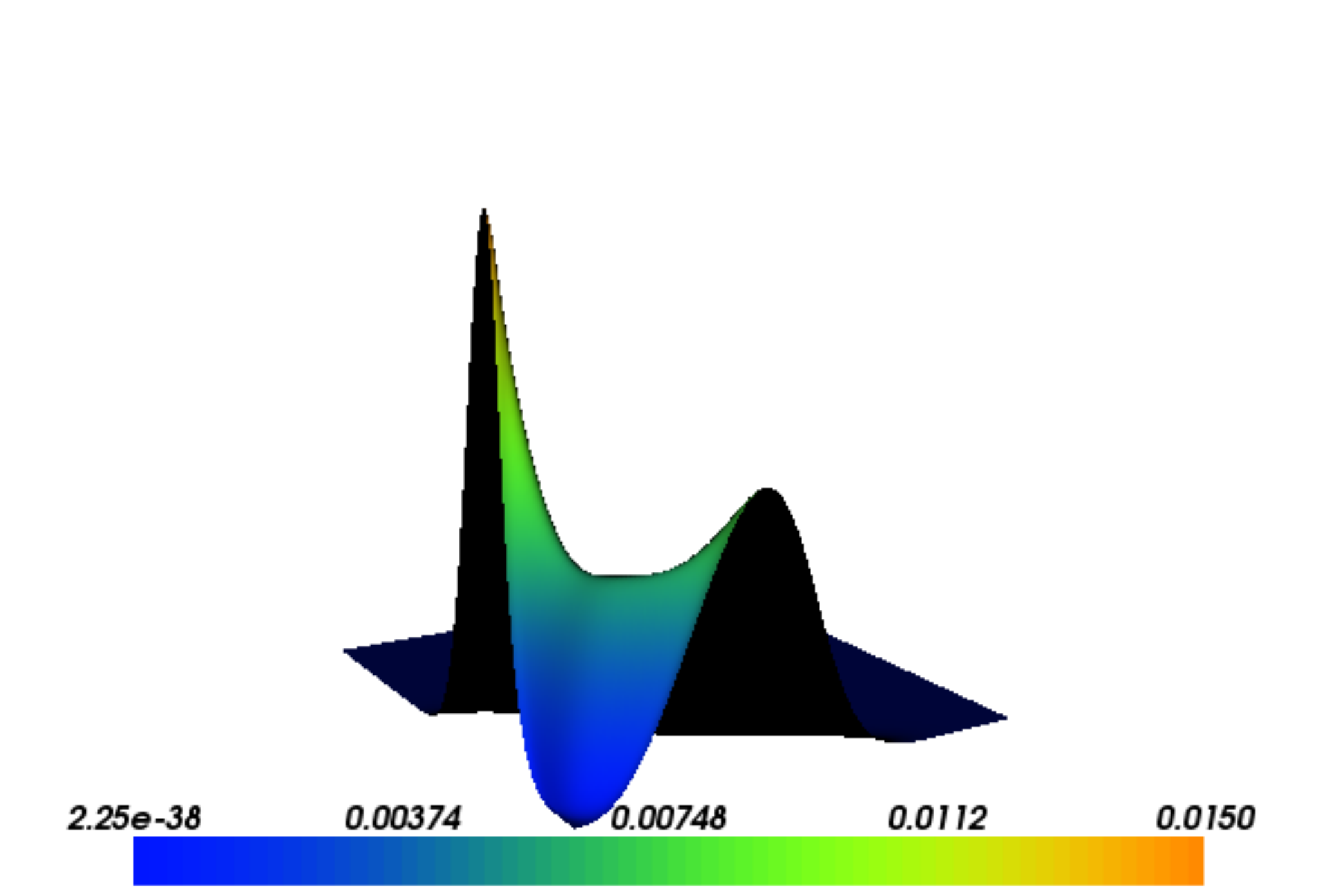}&\includegraphics[scale=.22]{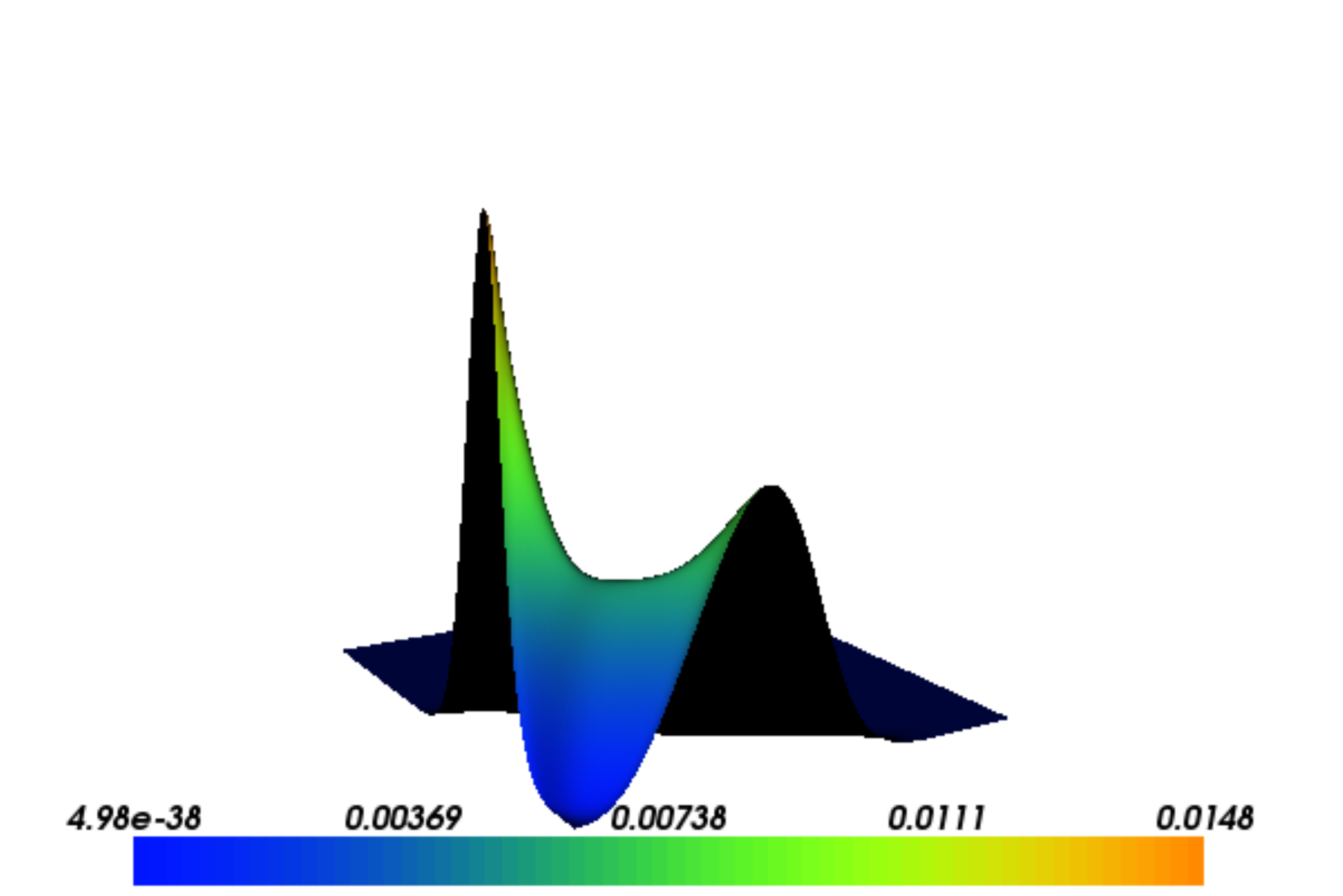}\\ \hline
          $t=901$ & \includegraphics[scale=.22]{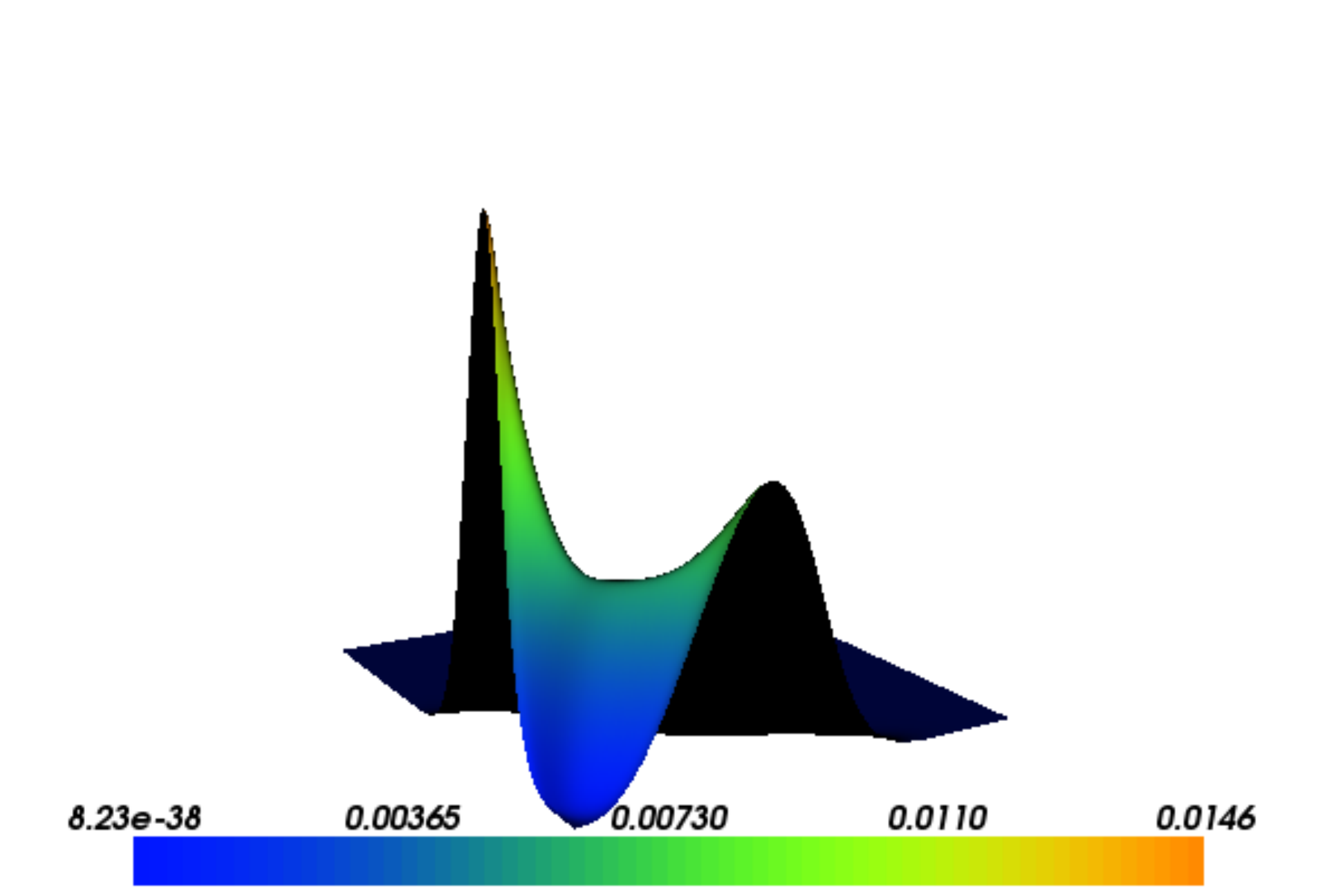}& \includegraphics[scale=.22]{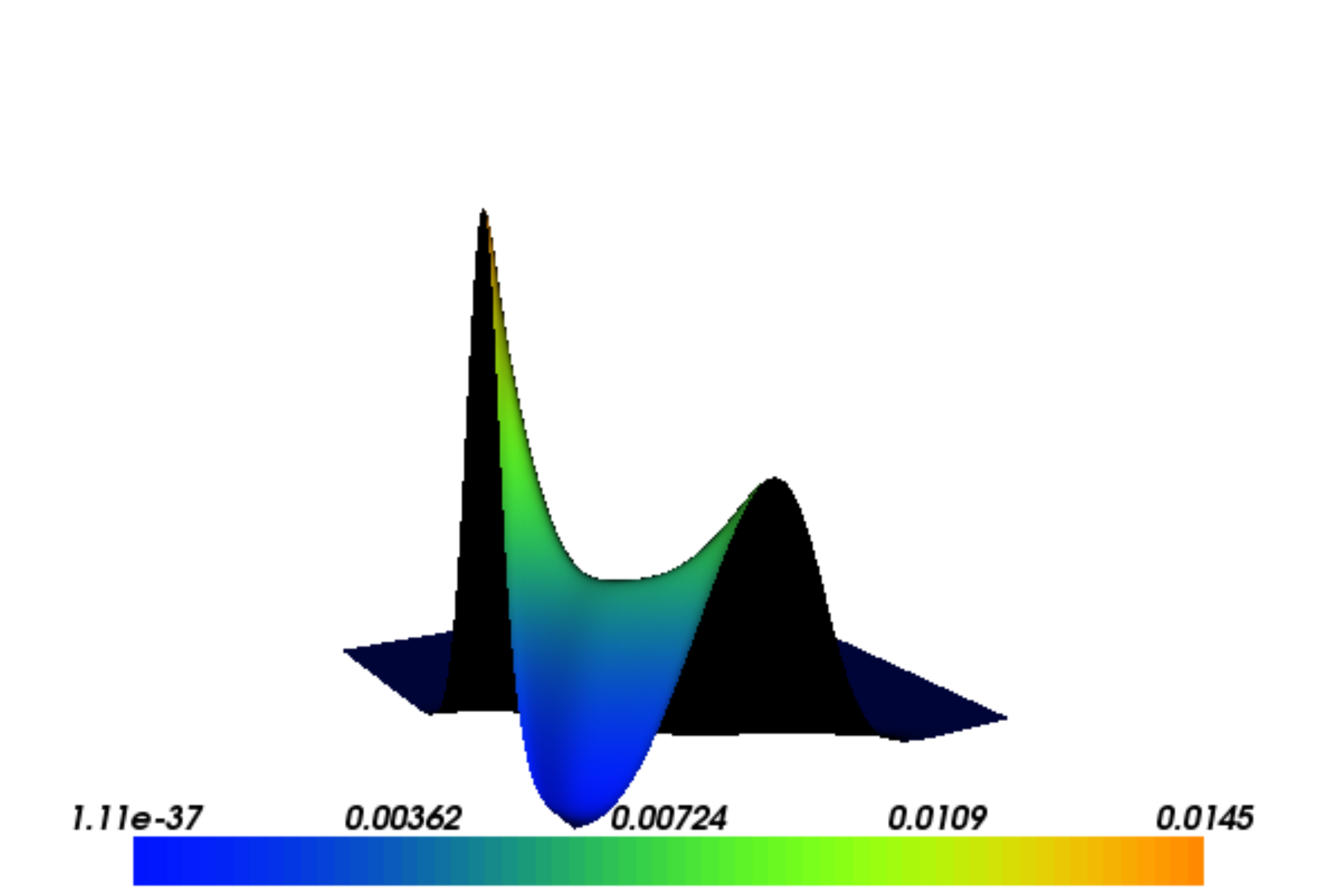}&\includegraphics[scale=.22]{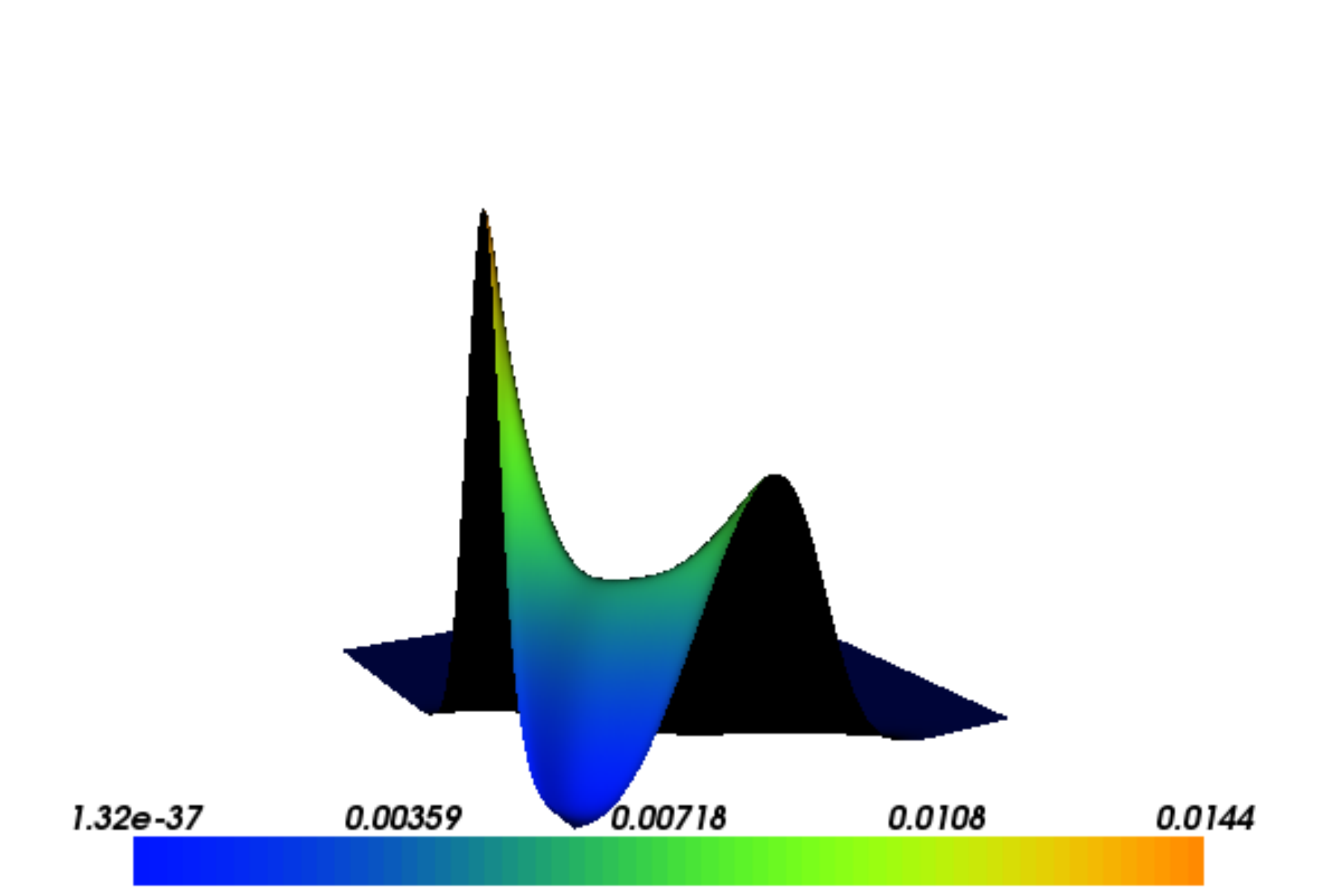}\\

          \hline
        \hline
        \end{tabular}
    \caption{\bf {The time evolution of a gaussian initial condition relaxing into the steady state for the wild type. The higher peak corresponds to lytic while the wider peak is lysogenic.}}    
        \label{tab:gt4}
    \end{table}%

\begin{table}[ht]

        \centering
        \begin{tabular}{|p{0.11\textwidth}|p{0.29\textwidth}|p{0.29\textwidth}|p{0.29\textwidth}|} 
          \hline
          \multicolumn{4}{|c|}{Time Evolution}
          \\ \hline \hline
          time&$+0$&$+100$&$+200$\\ \hline
          $t=1$ & \includegraphics[scale=.22]{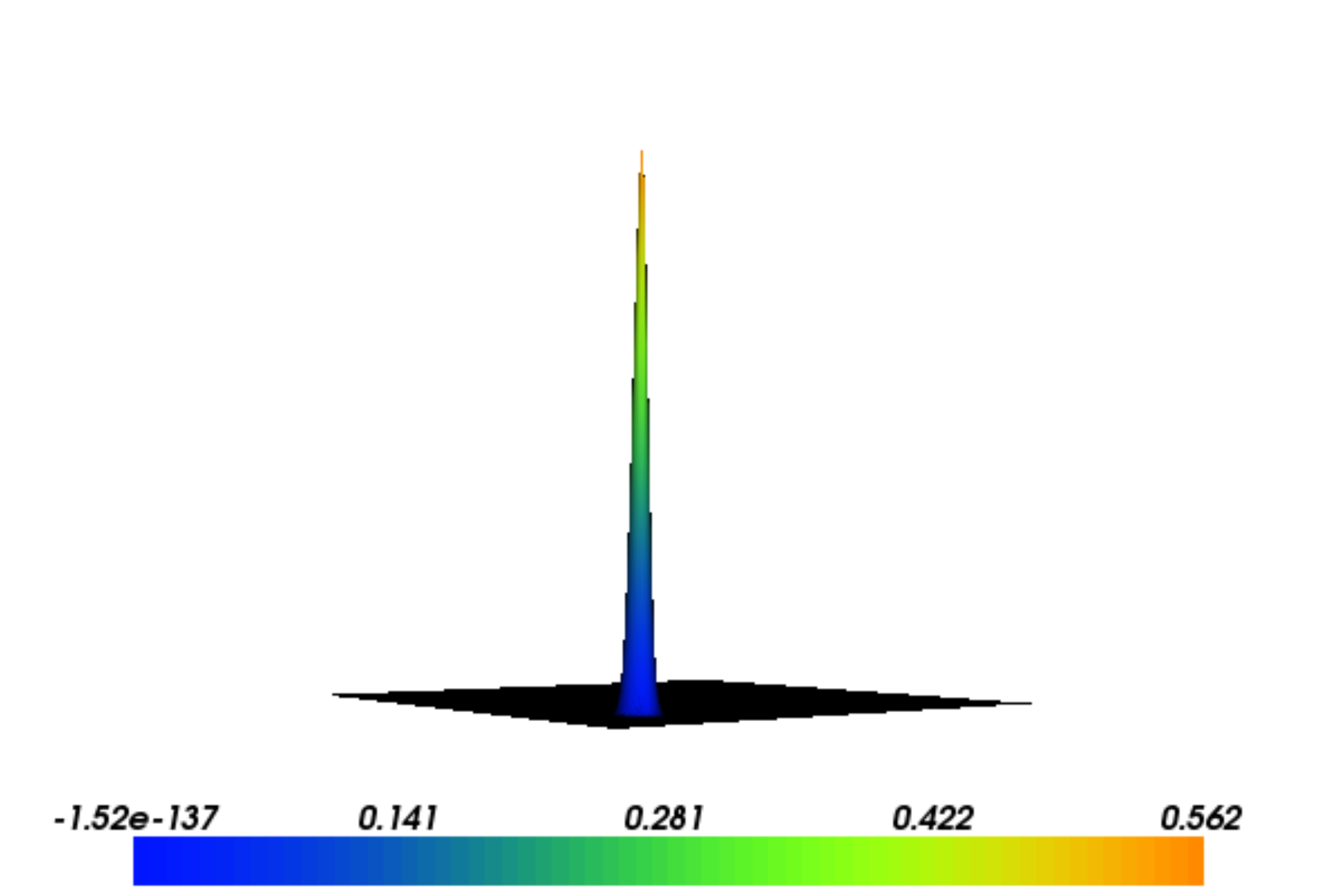}& \includegraphics[scale=.22]{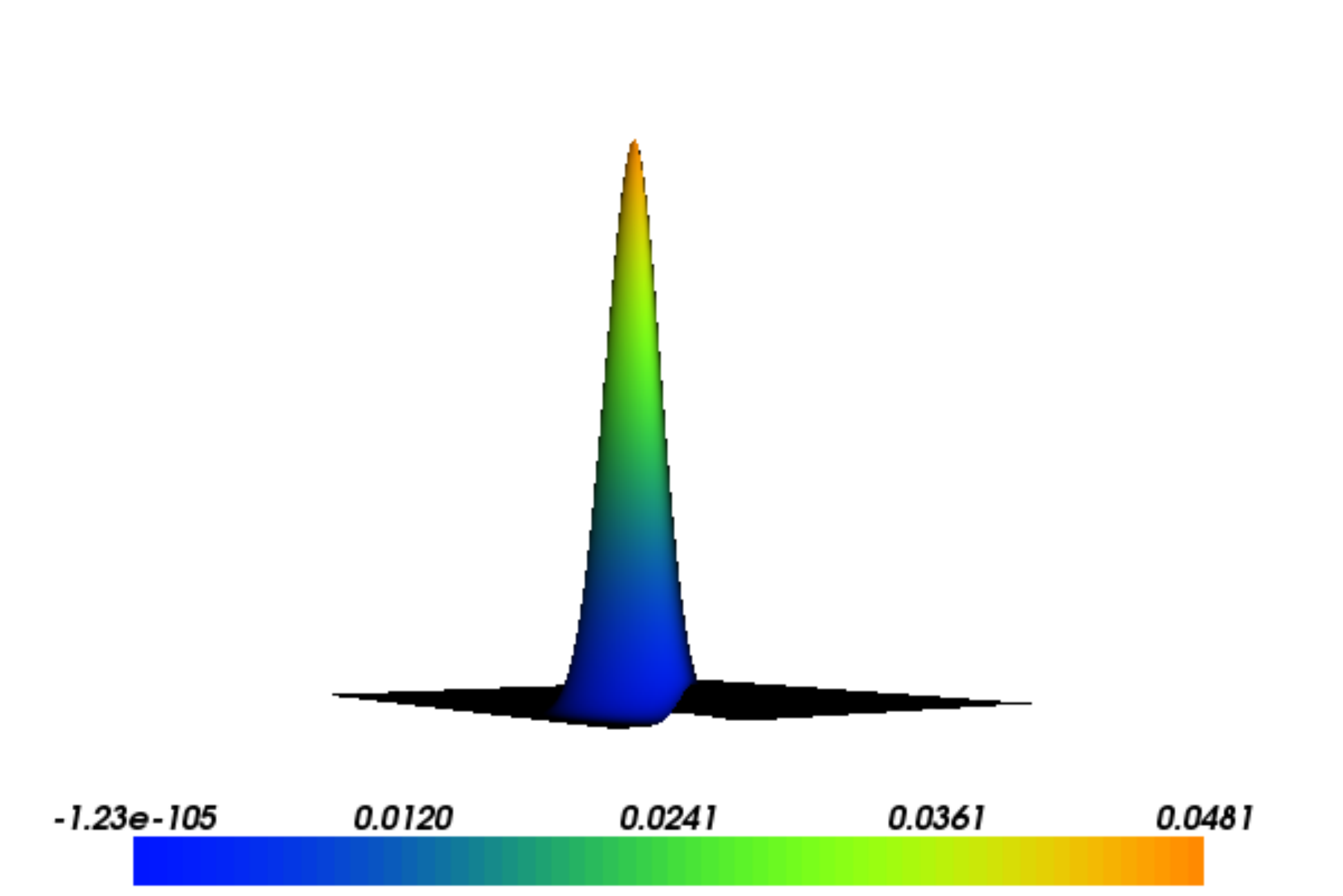}&\includegraphics[scale=.22]{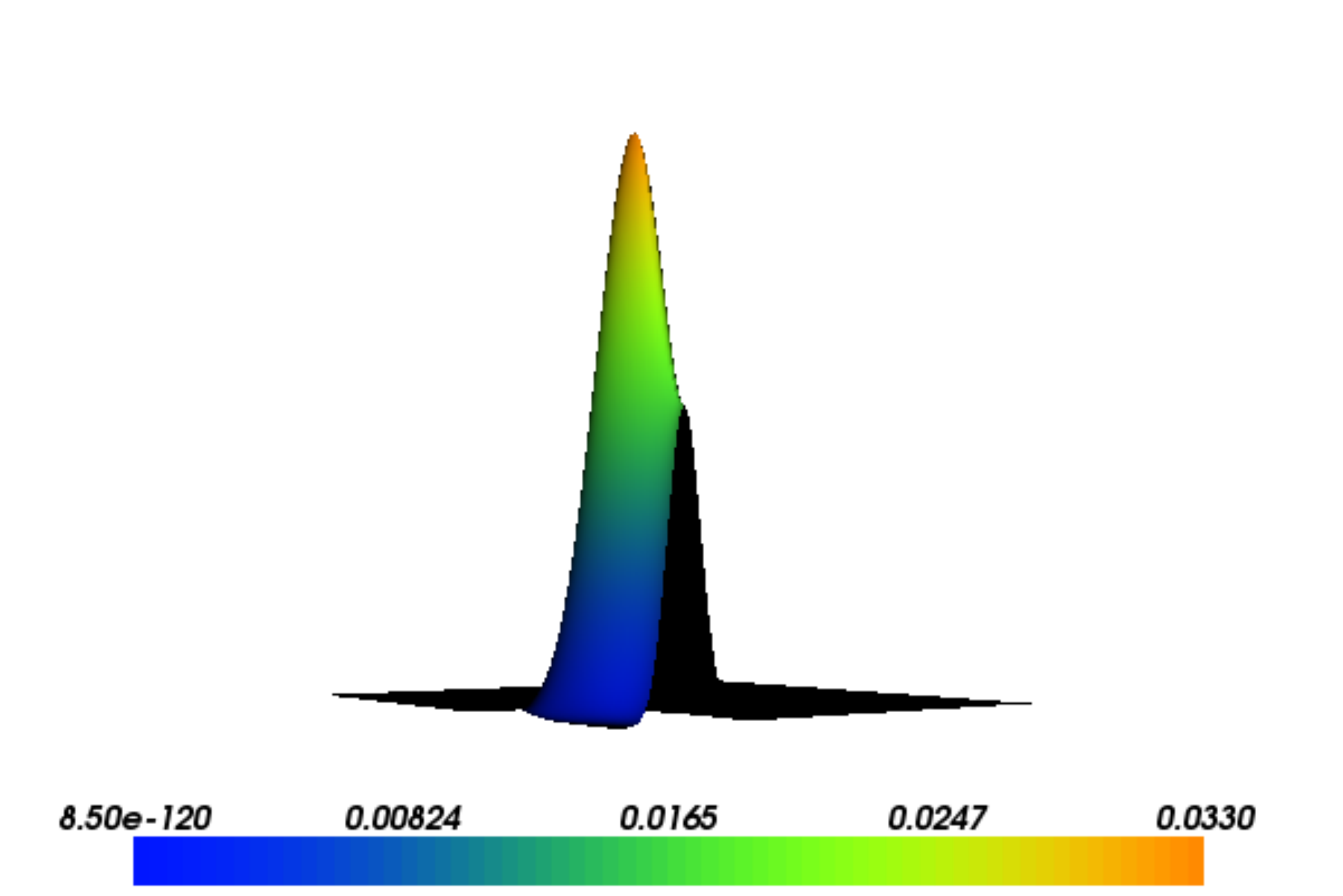}\\ \hline
          $t=301$ & \includegraphics[scale=.22]{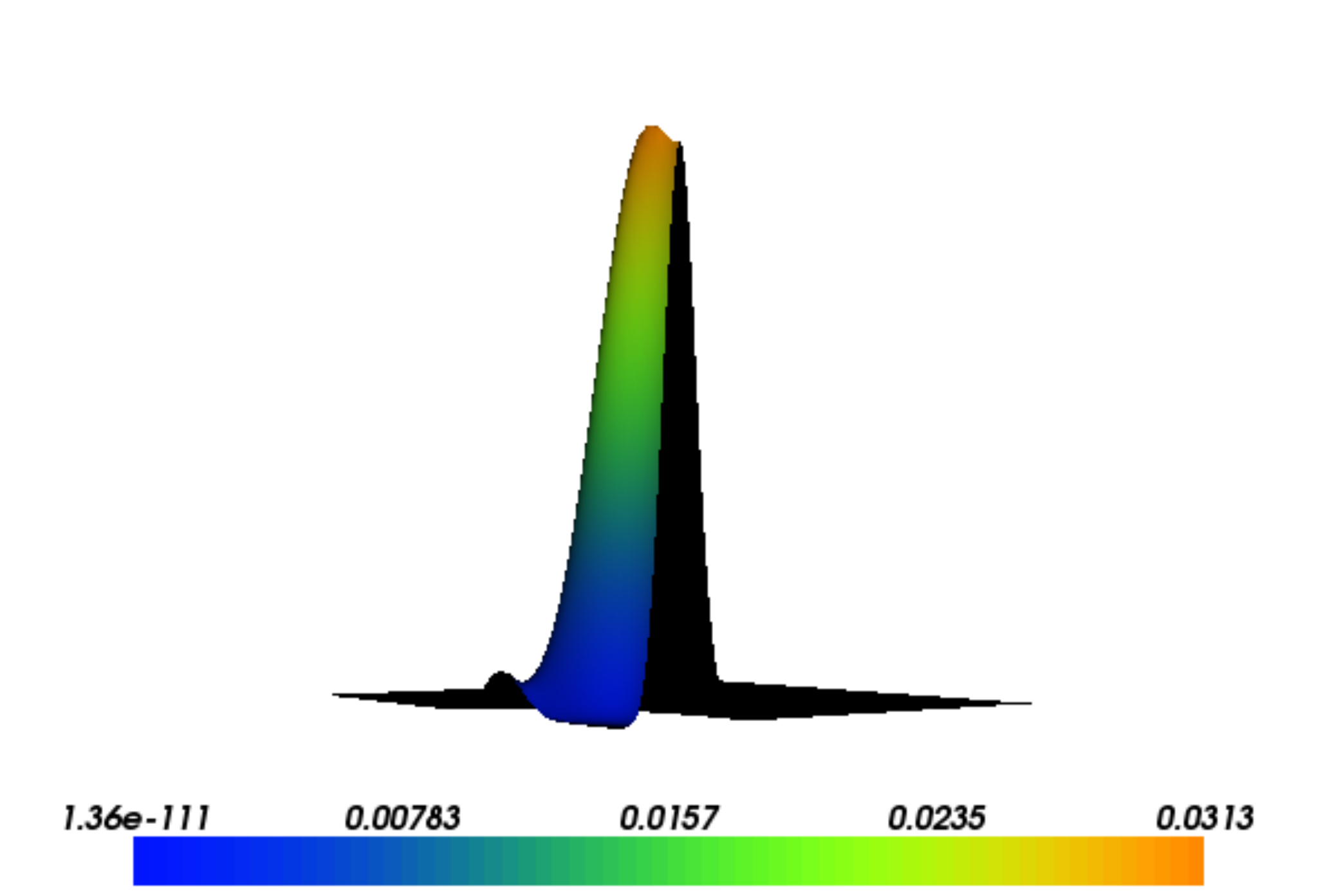}& \includegraphics[scale=.22]{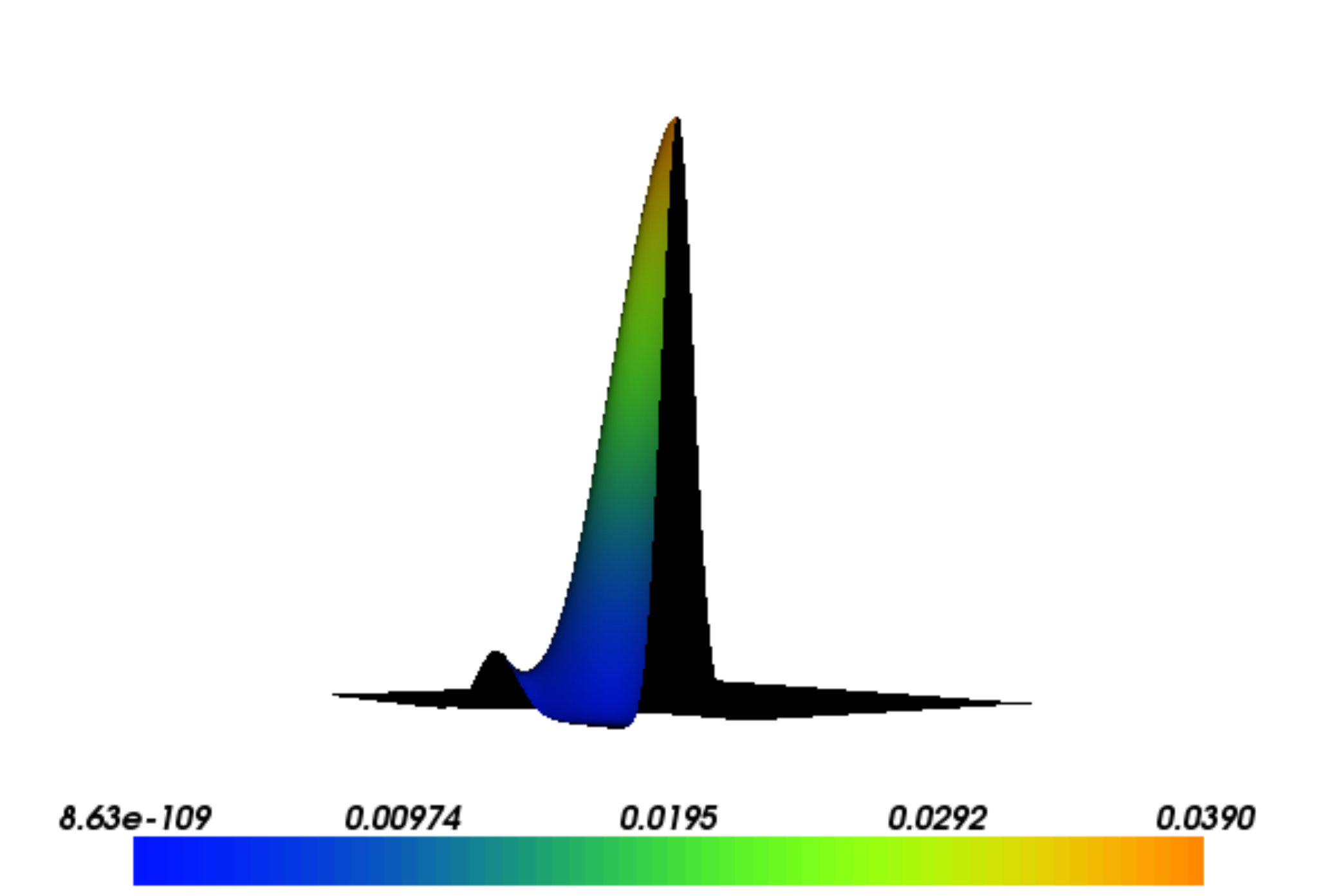}&\includegraphics[scale=.22]{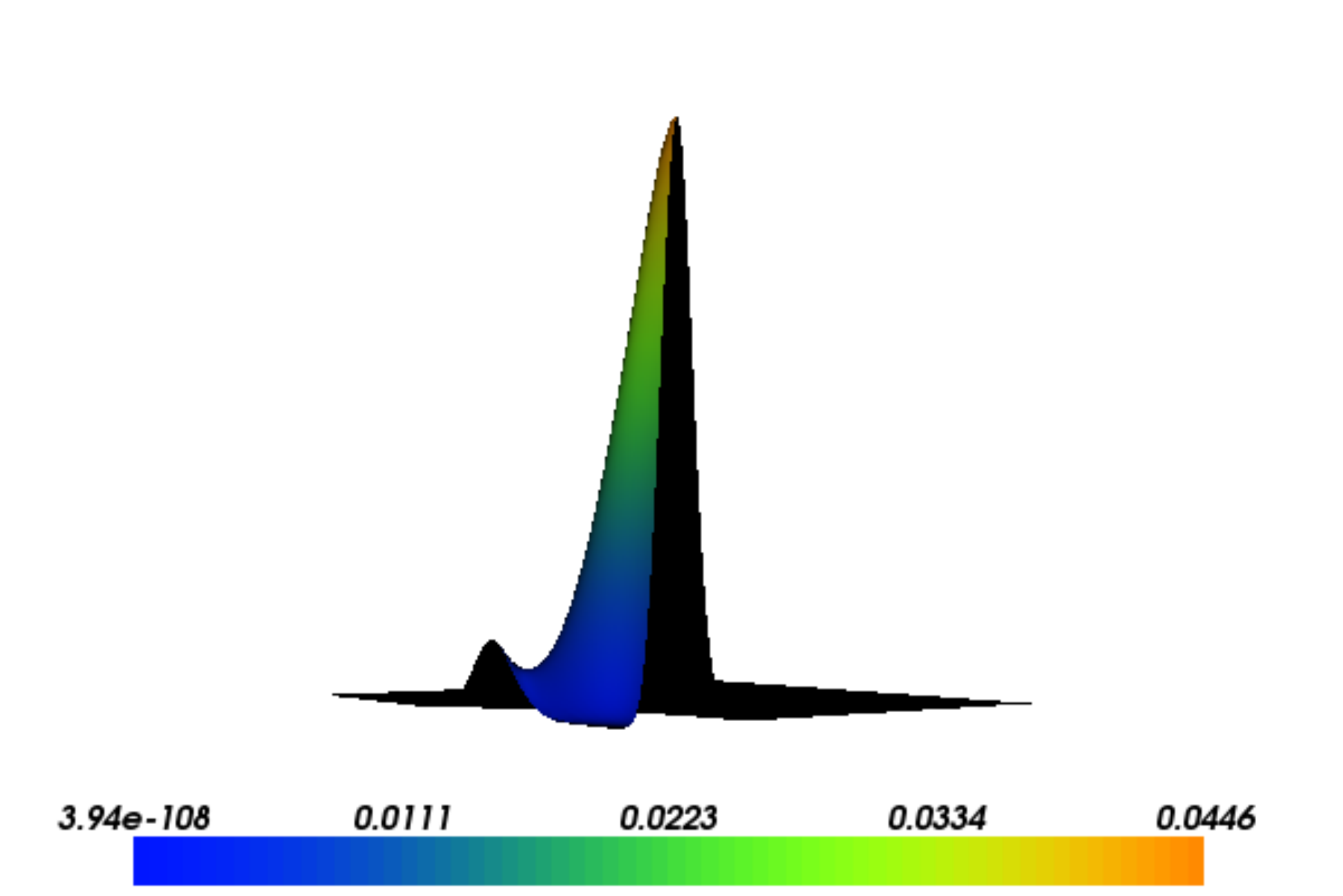}\\ \hline
          $t=601$ & \includegraphics[scale=.22]{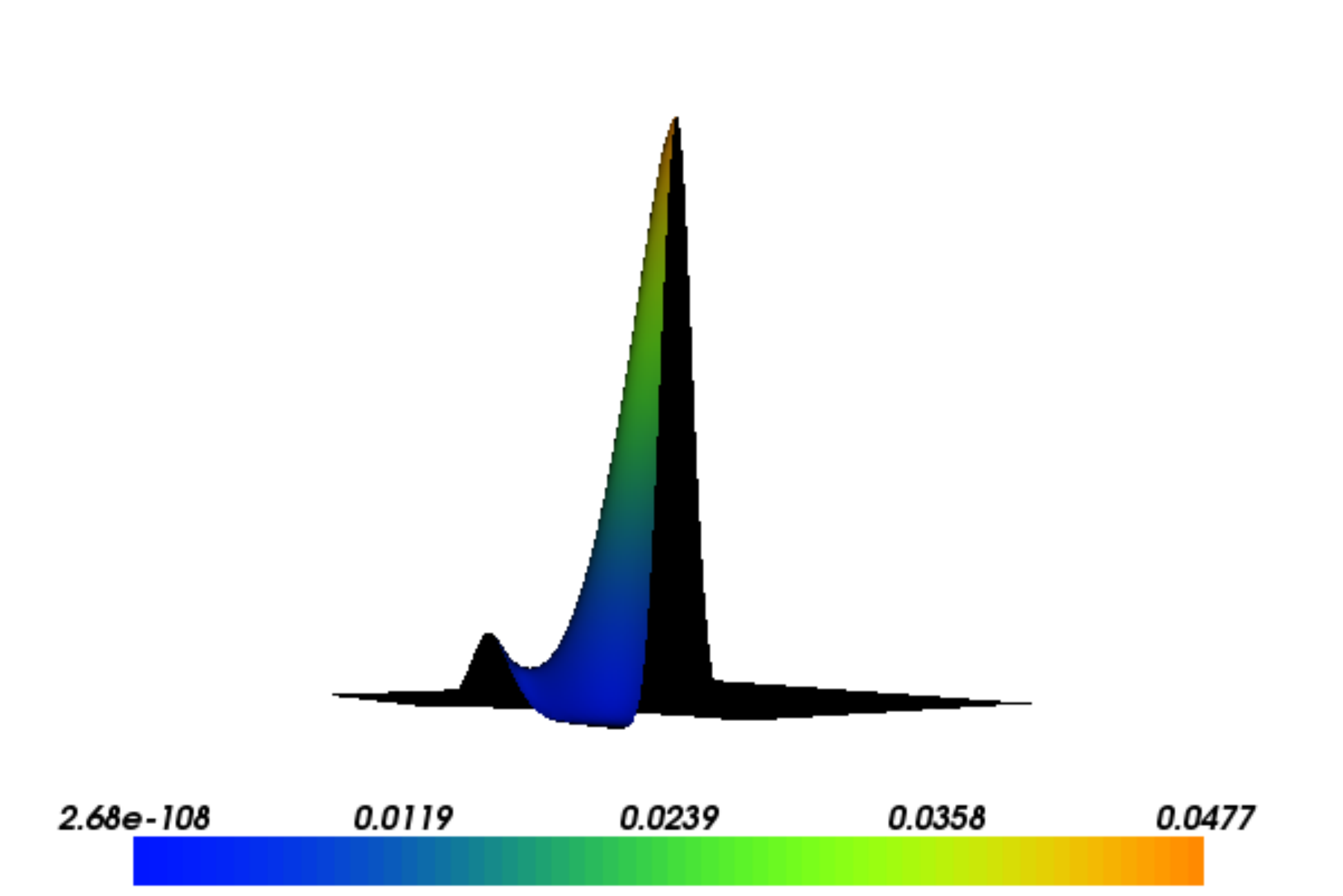}& \includegraphics[scale=.22]{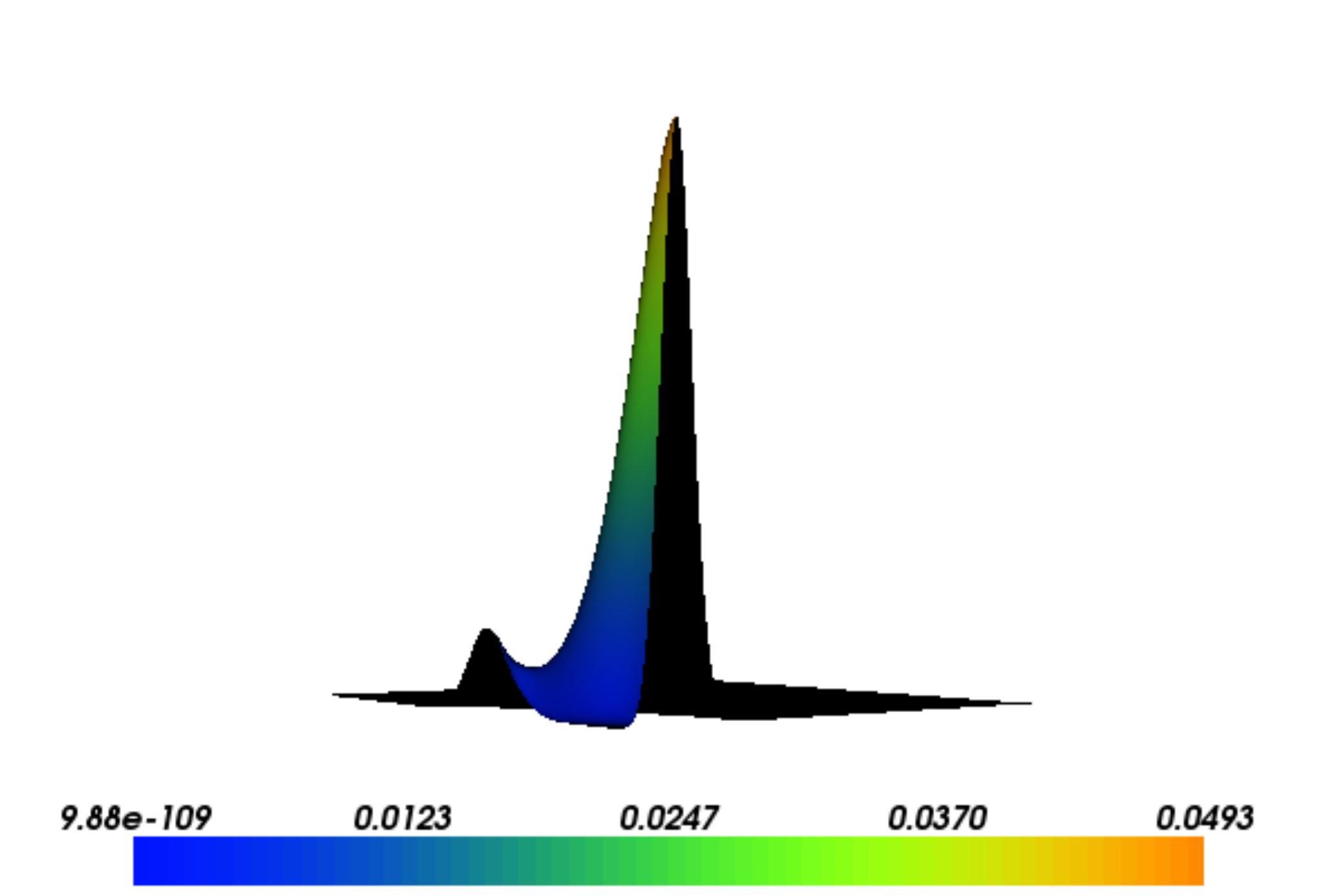}&\includegraphics[scale=.22]{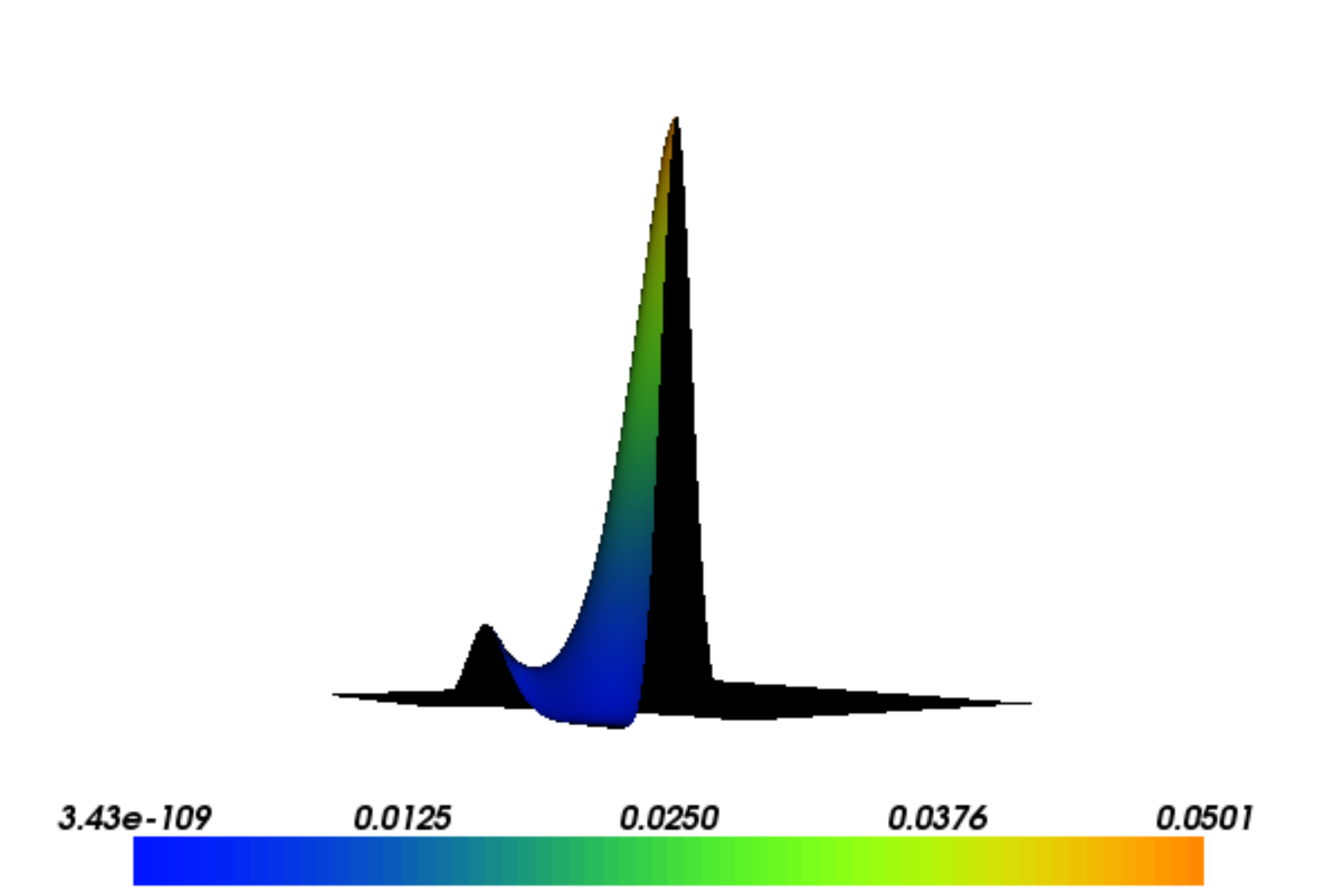}\\ \hline
          $t=901$ & \includegraphics[scale=.22]{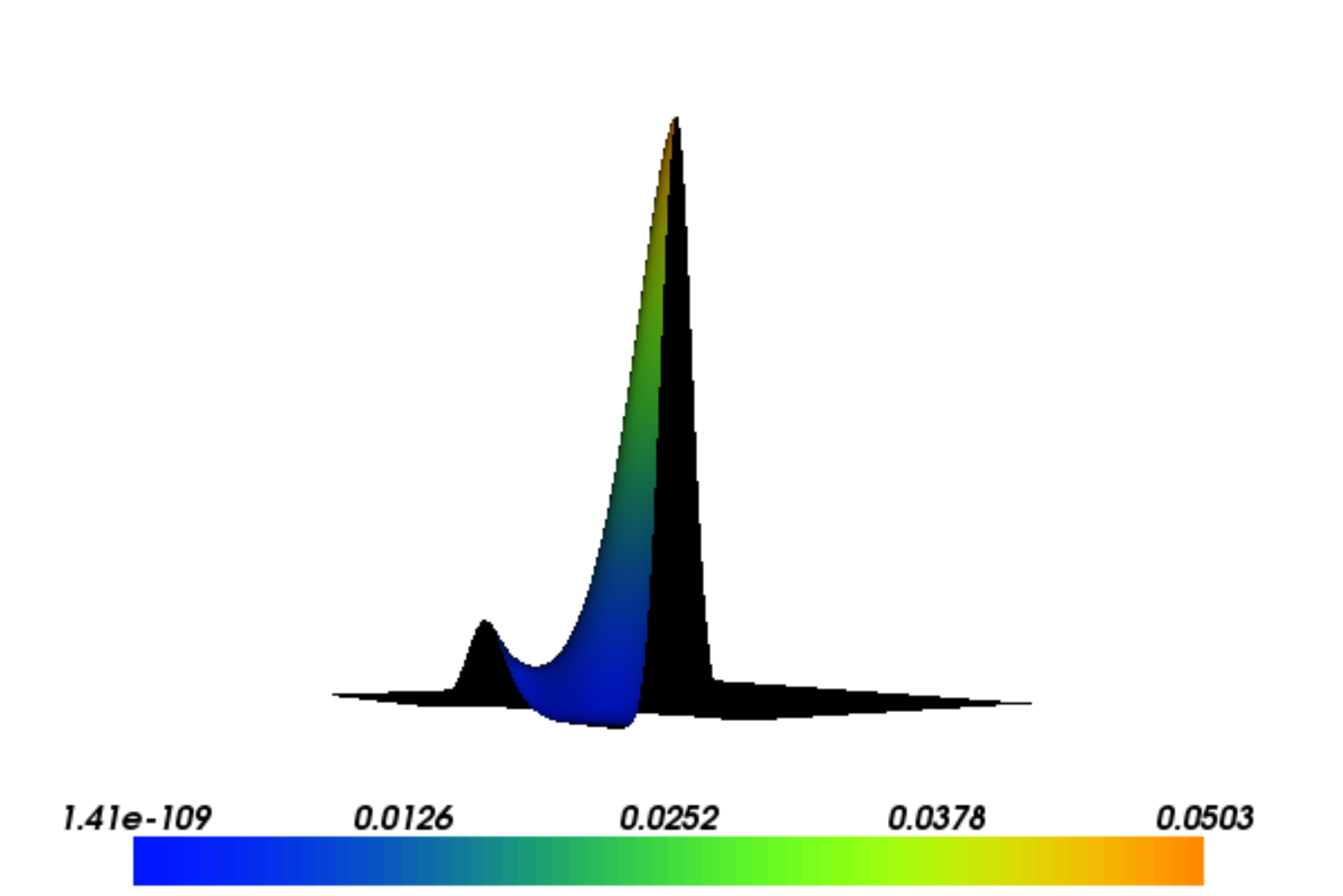}& \includegraphics[scale=.22]{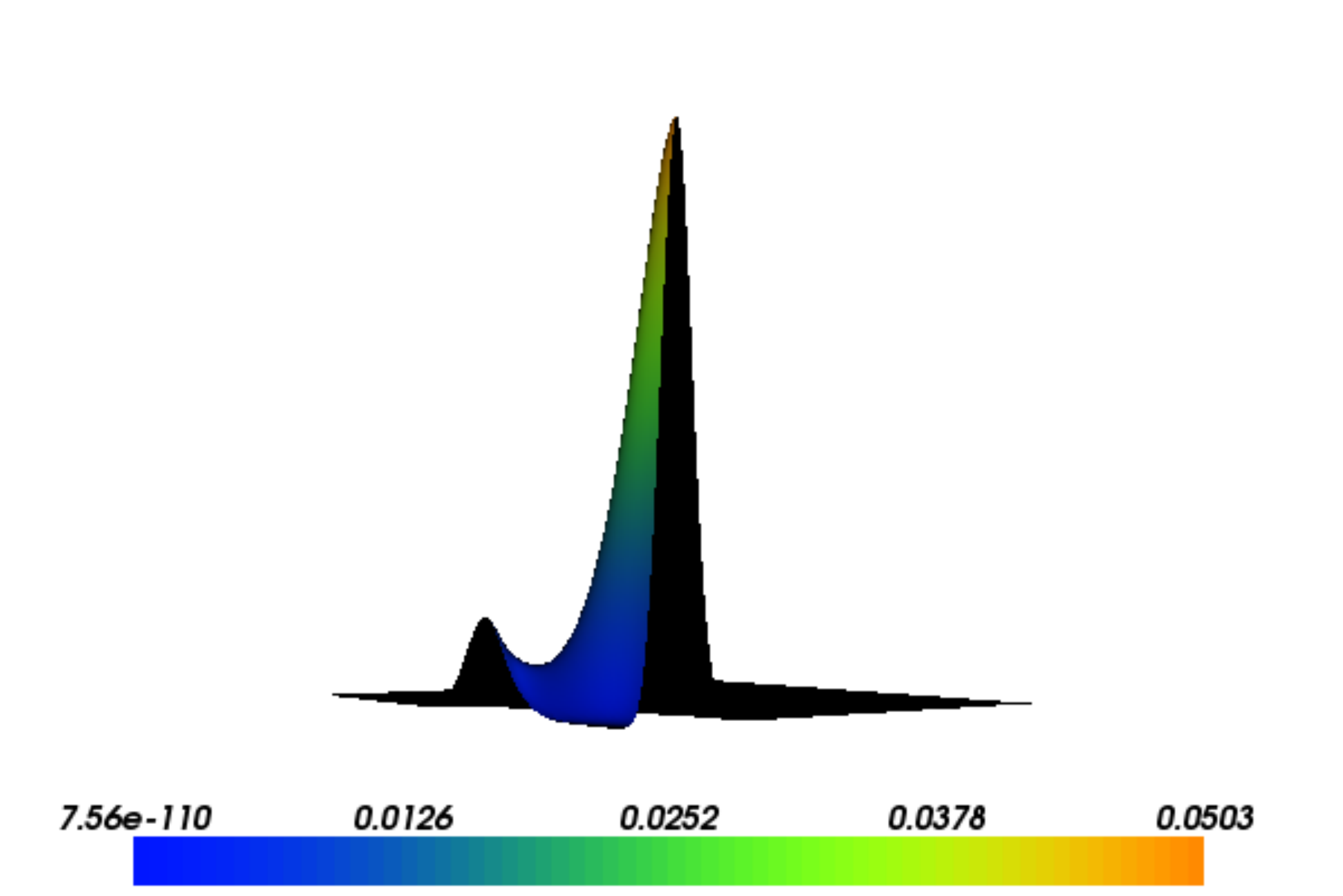}&\includegraphics[scale=.22]{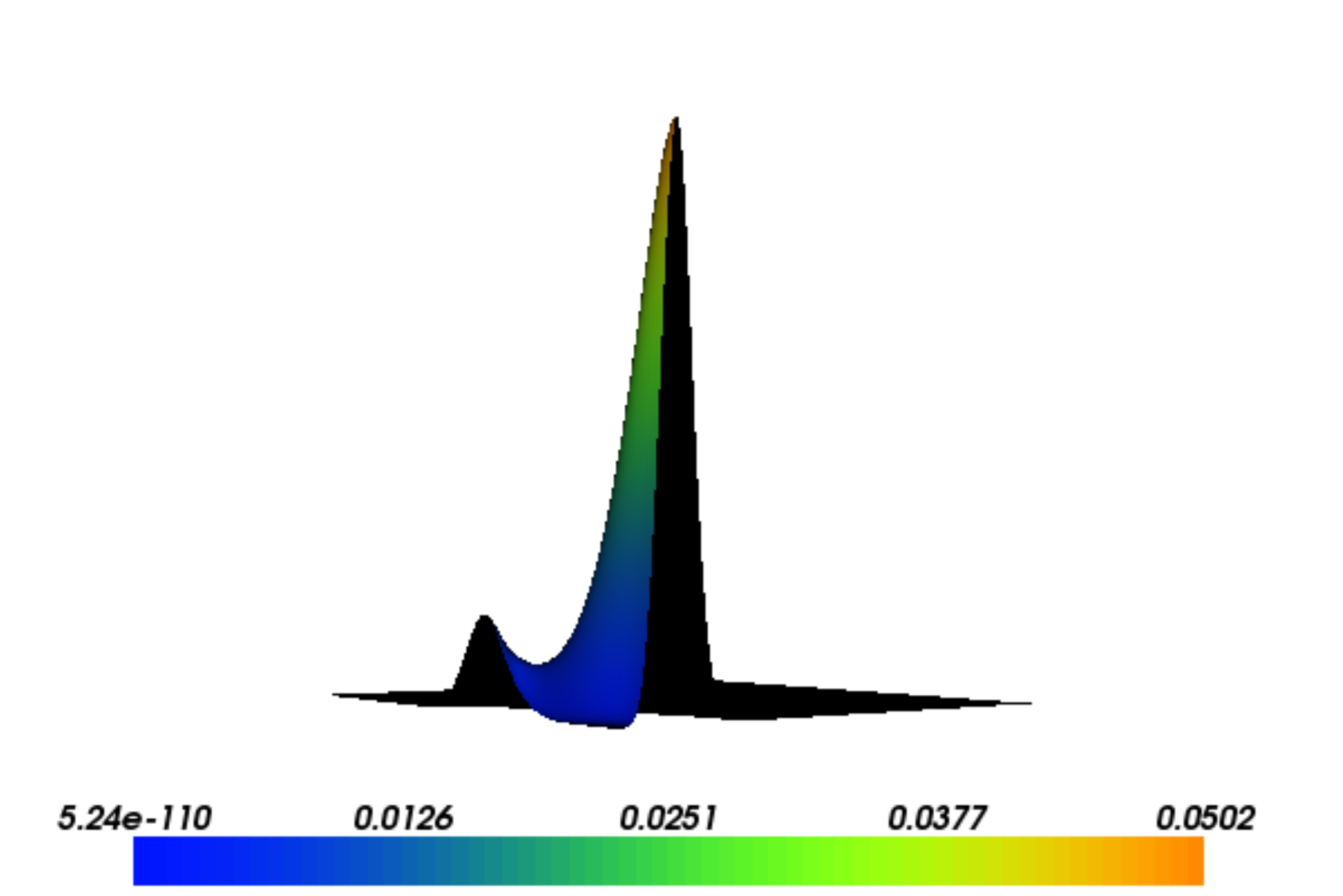}\\

          \hline
        \hline
        \end{tabular}
    \caption{\bf {The time evolution of a gaussian initial condition relaxing into the steady state for the 121 mutant.  The dominant peak is the lysogenic peak.}} 
        \label{tab:gt5}
    \end{table}%

\begin{table}[ht]

        \centering
        \begin{tabular}{|p{0.11\textwidth}|p{0.29\textwidth}|p{0.29\textwidth}|p{0.29\textwidth}|} 
          \hline
          \multicolumn{4}{|c|}{Time Evolution}
          \\ \hline \hline
          time&$+0$&$+100$&$+200$\\ \hline
           $t=1$ & \includegraphics[scale=.22]{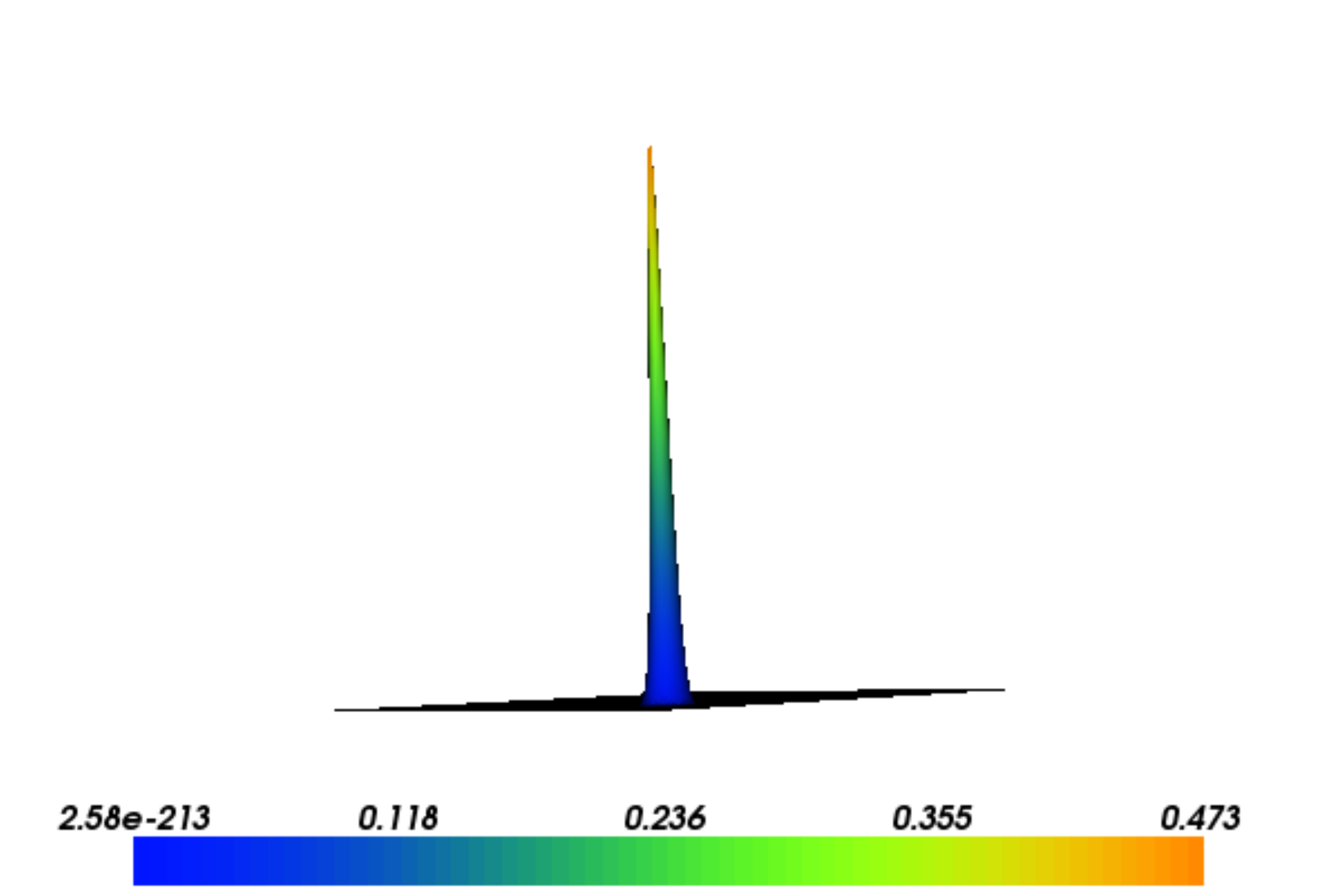}& \includegraphics[scale=.22]{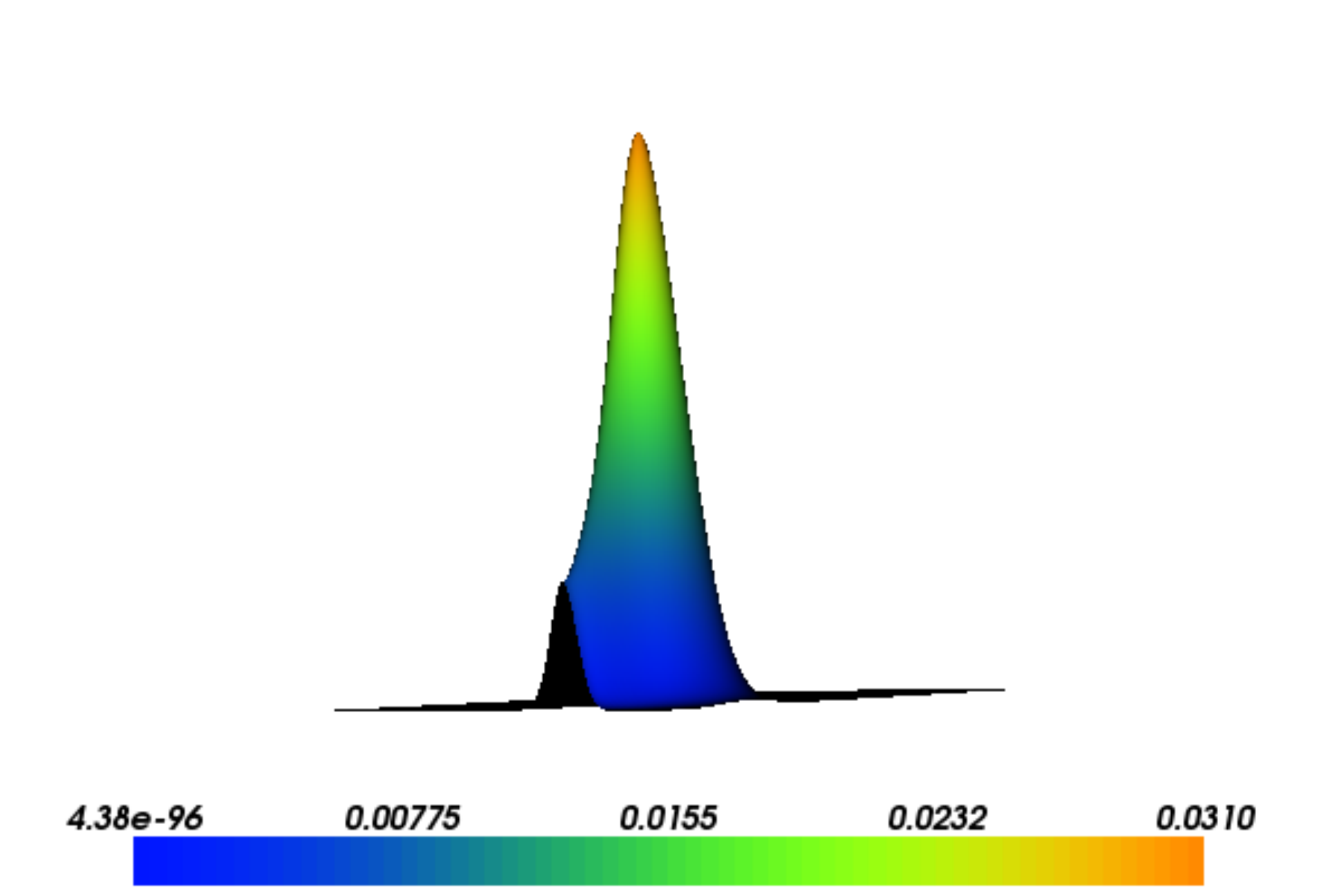}&\includegraphics[scale=.22]{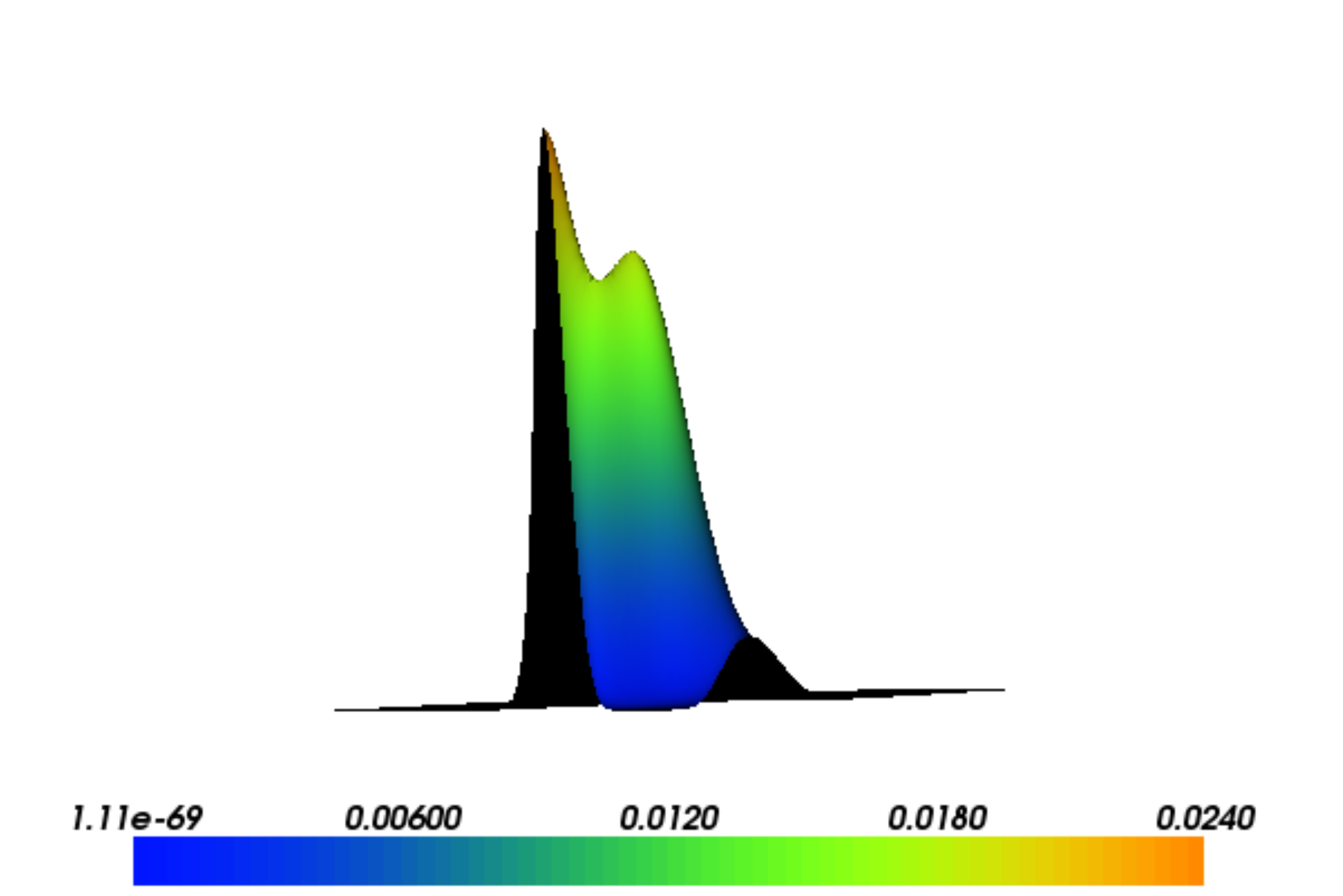}\\ \hline
          $t=301$ & \includegraphics[scale=.22]{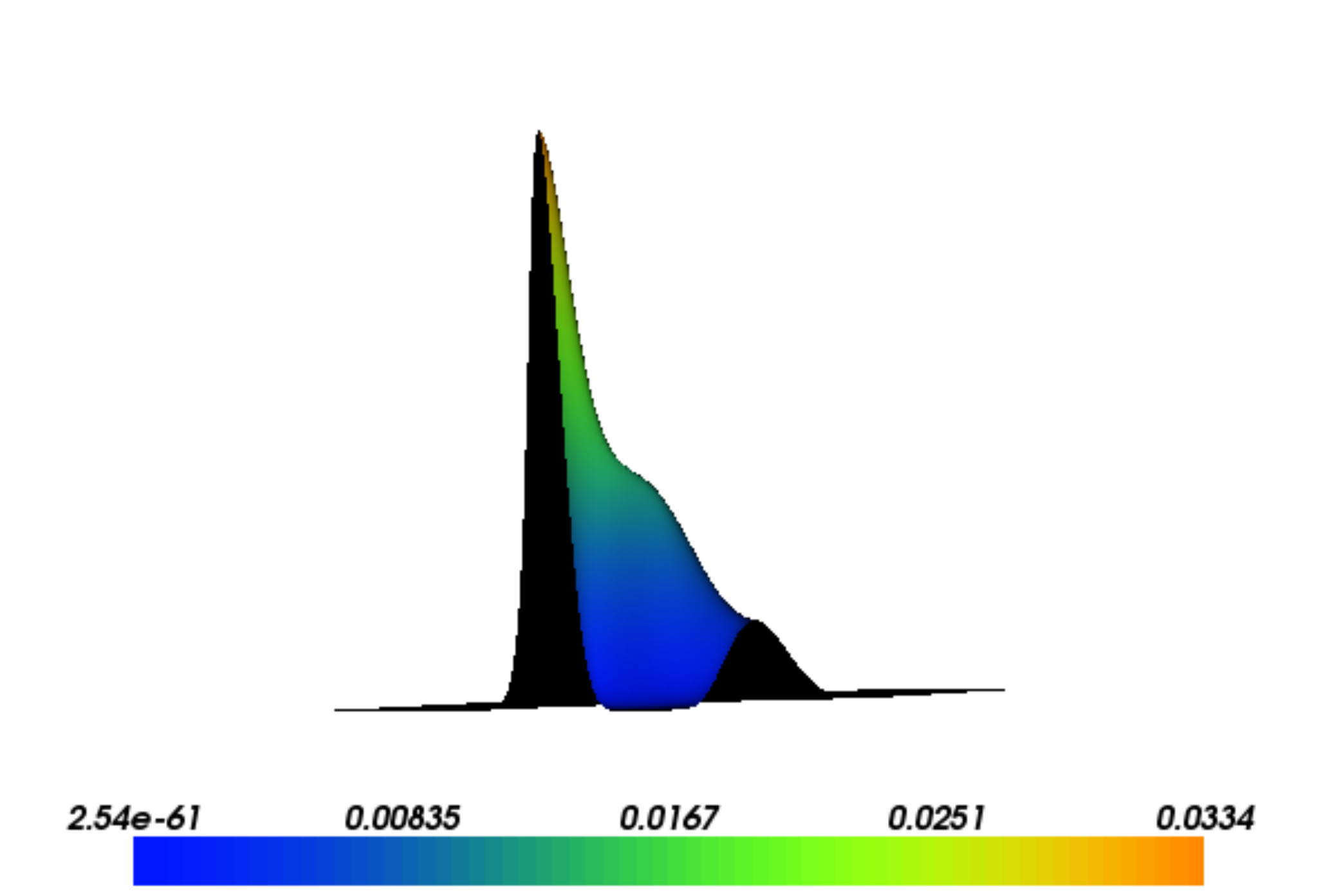}& \includegraphics[scale=.22]{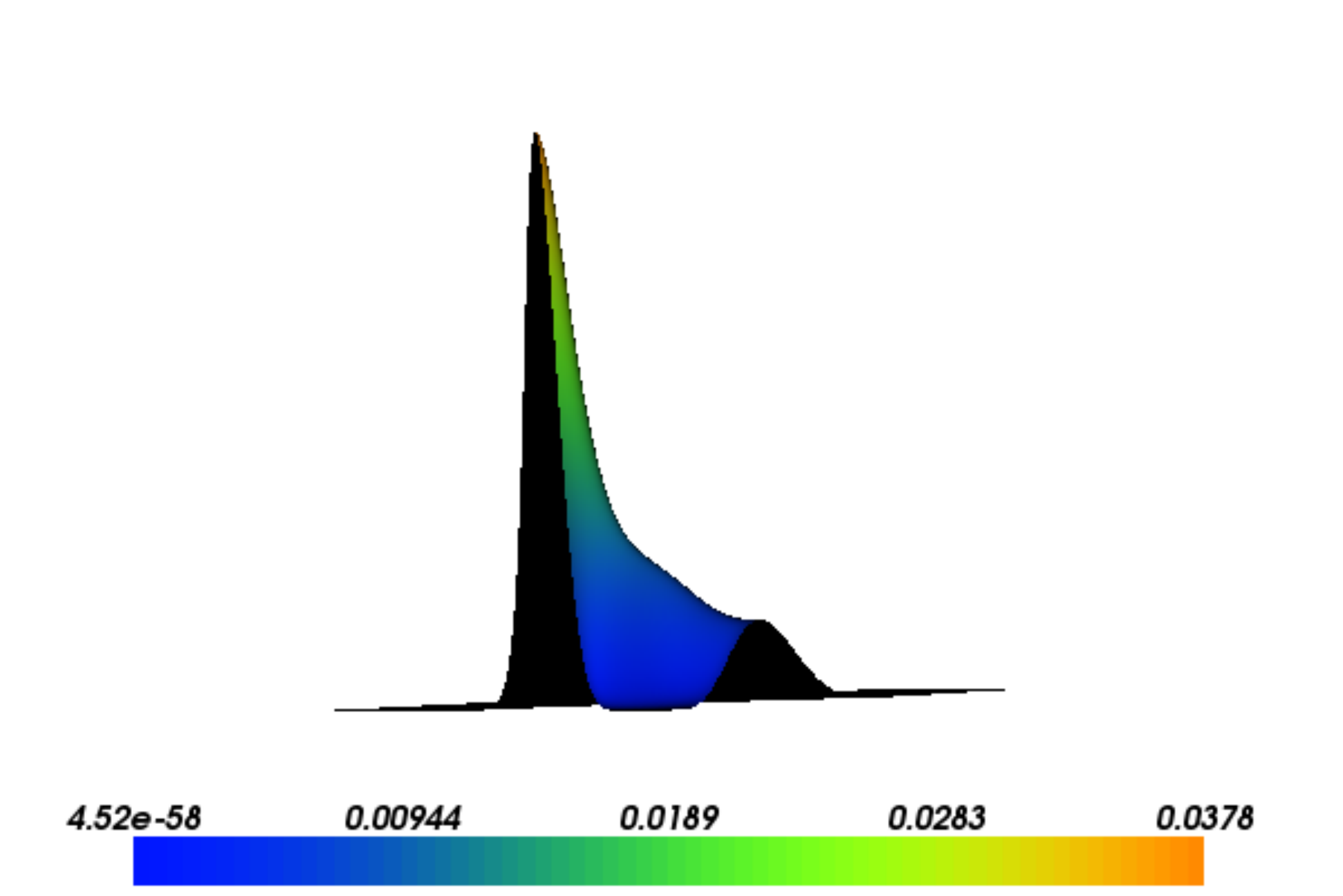}&\includegraphics[scale=.22]{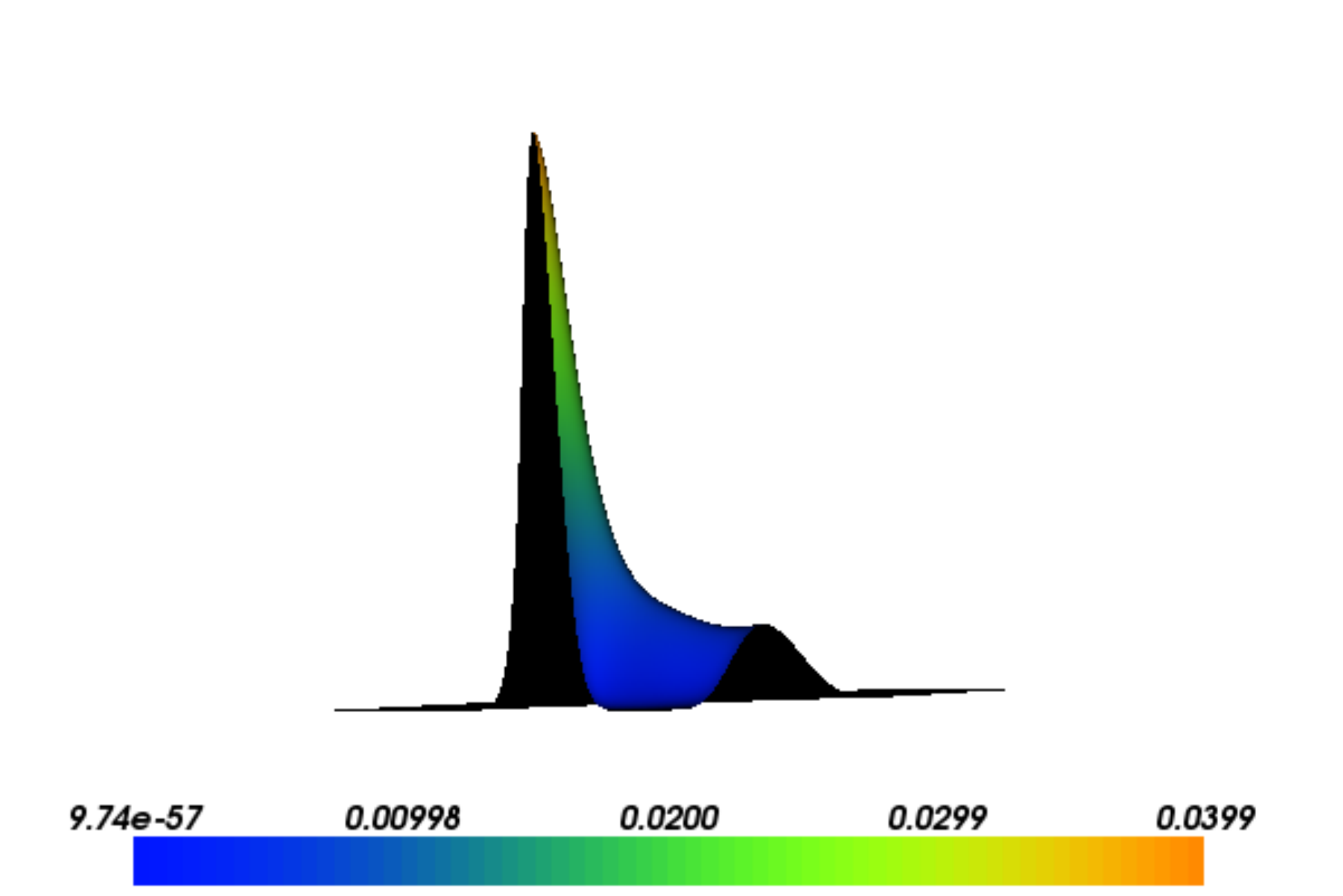}\\ \hline
          $t=601$ & \includegraphics[scale=.22]{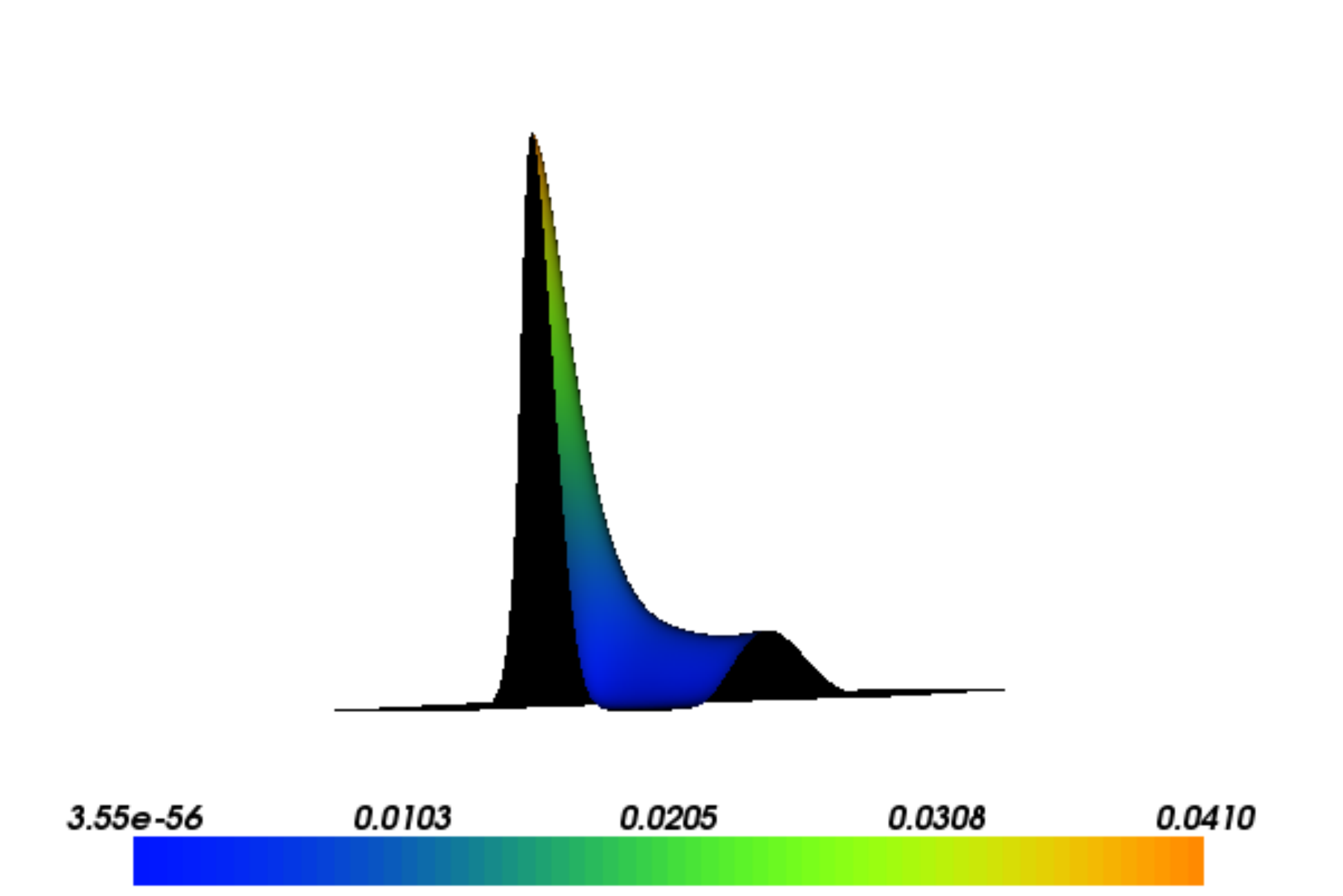}& \includegraphics[scale=.22]{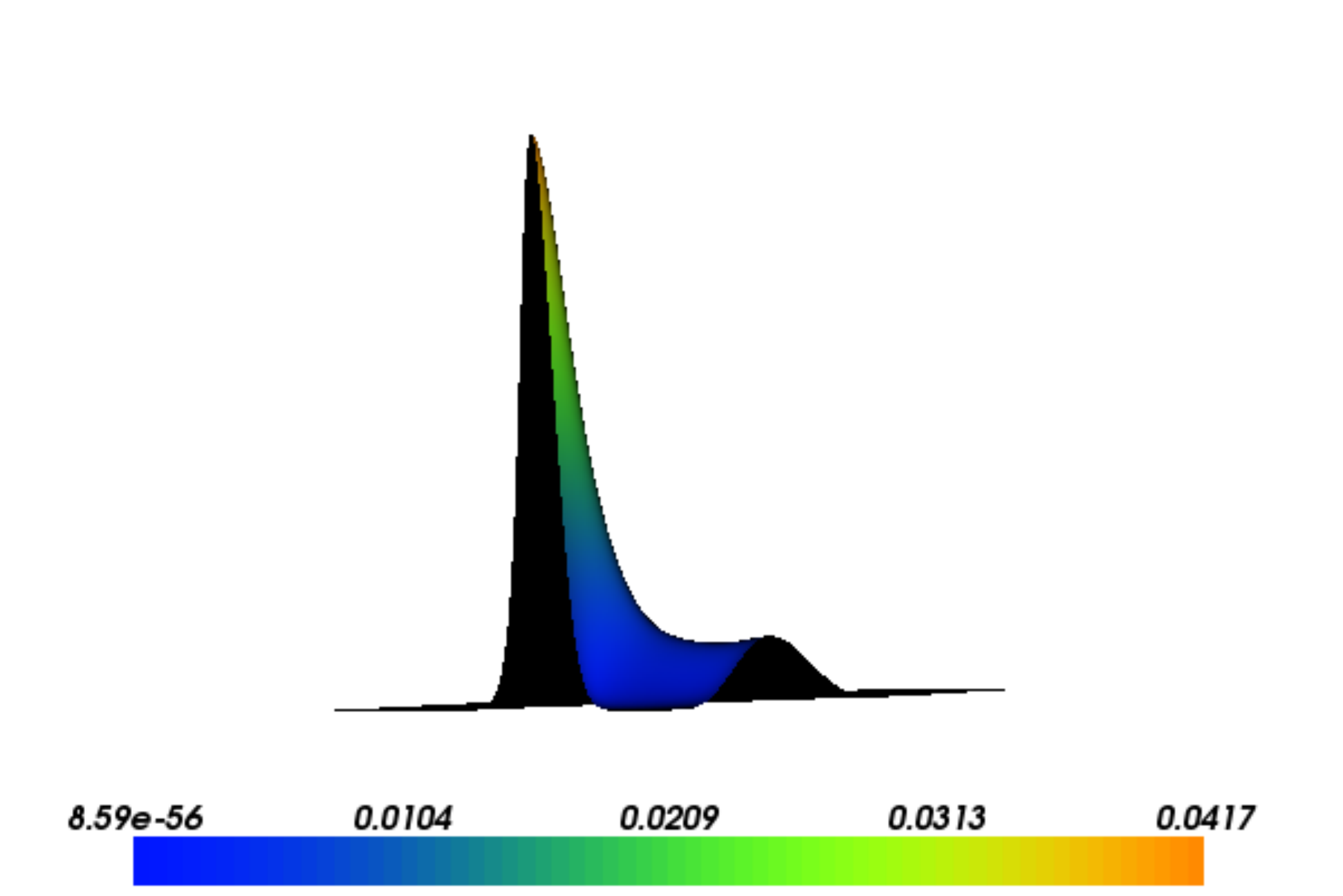}&\includegraphics[scale=.22]{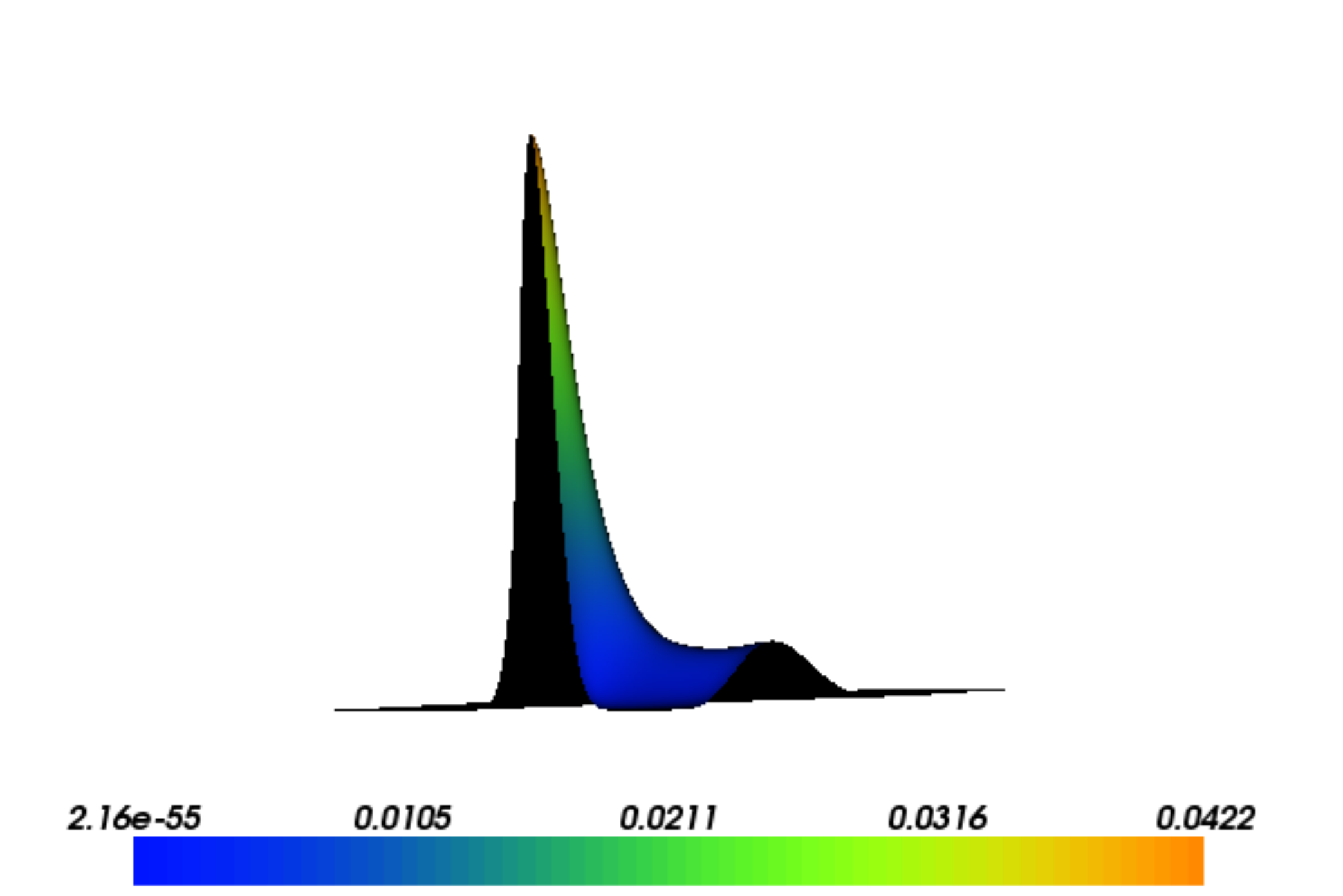}\\ \hline
          $t=901$ & \includegraphics[scale=.22]{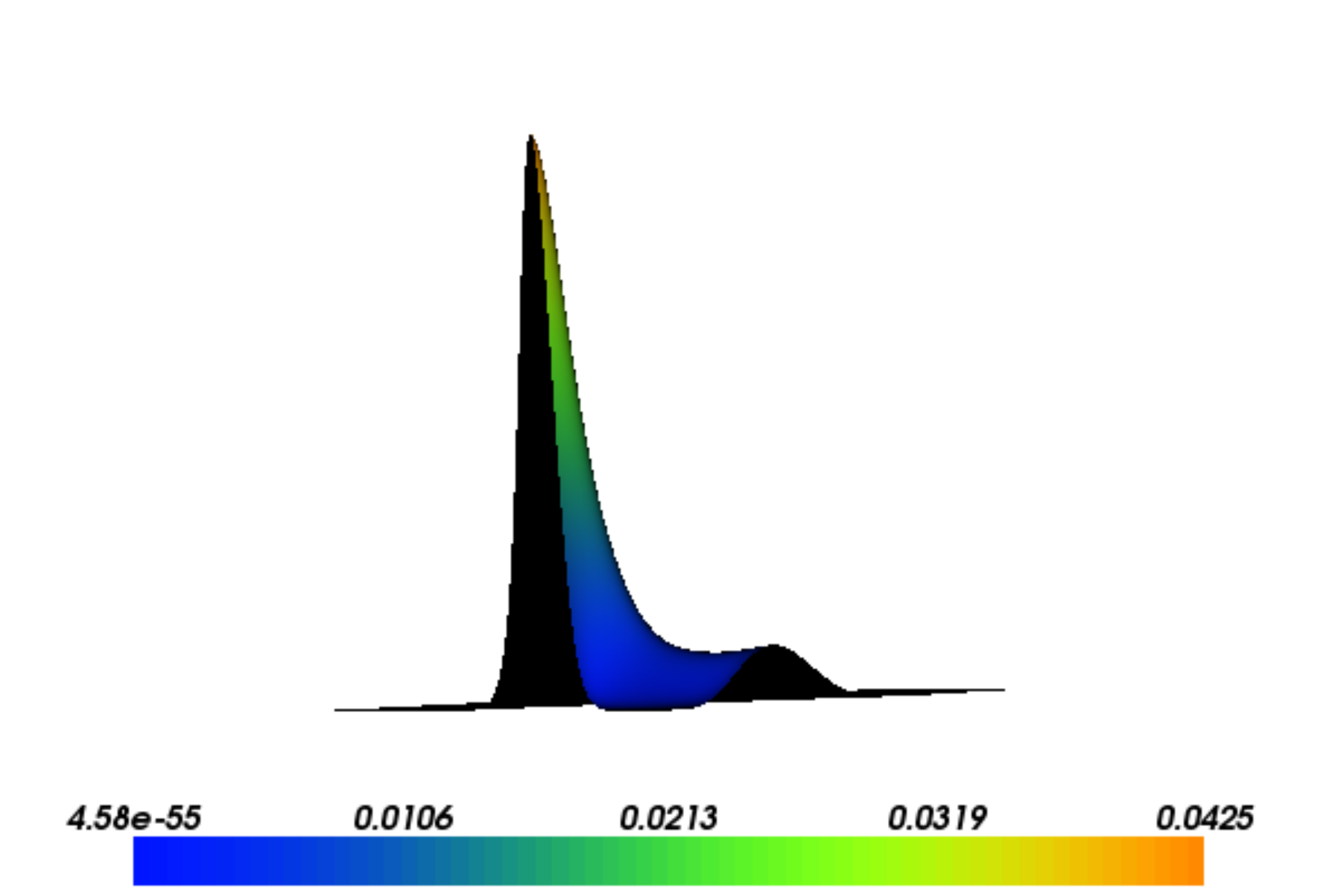}& \includegraphics[scale=.22]{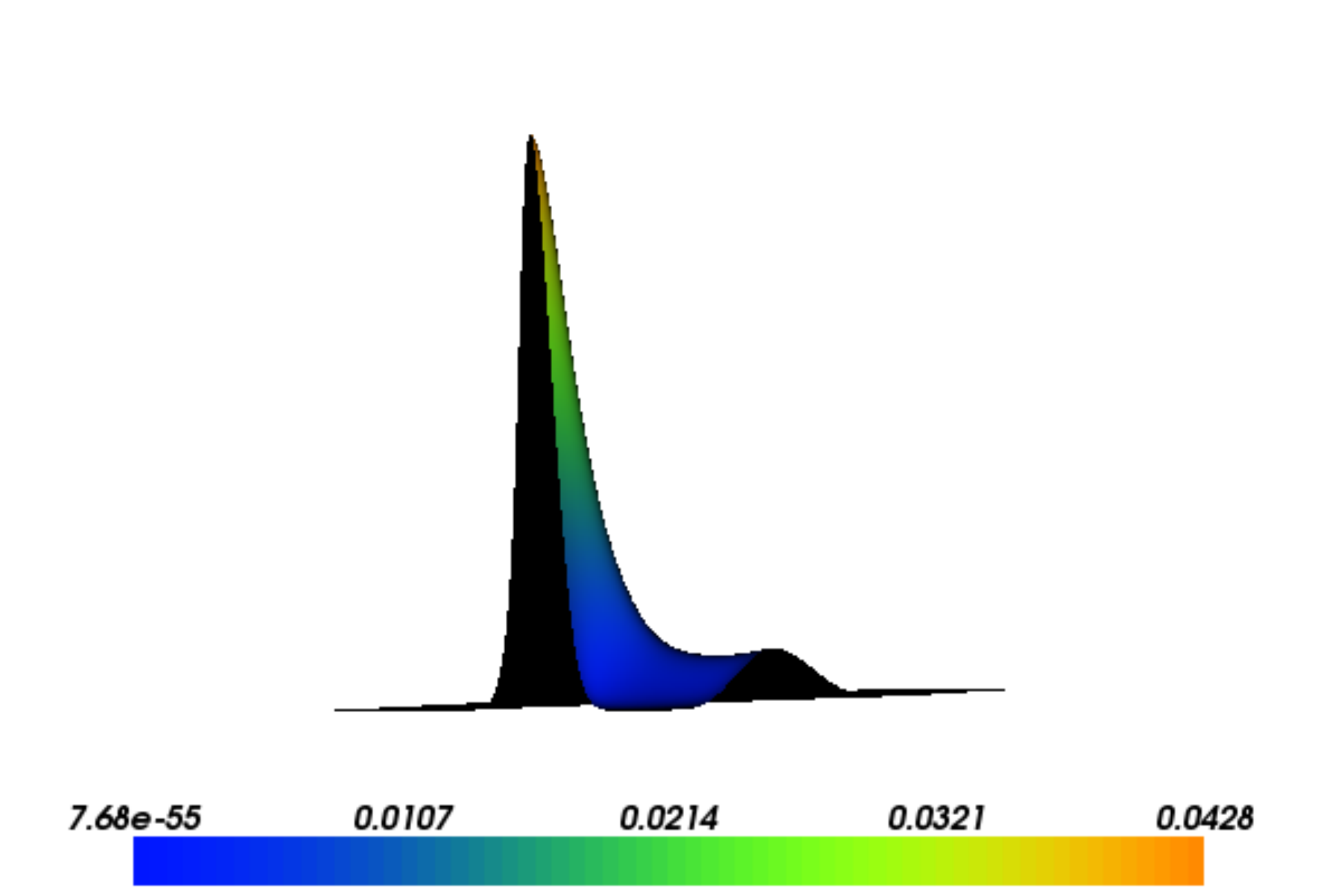}&\includegraphics[scale=.22]{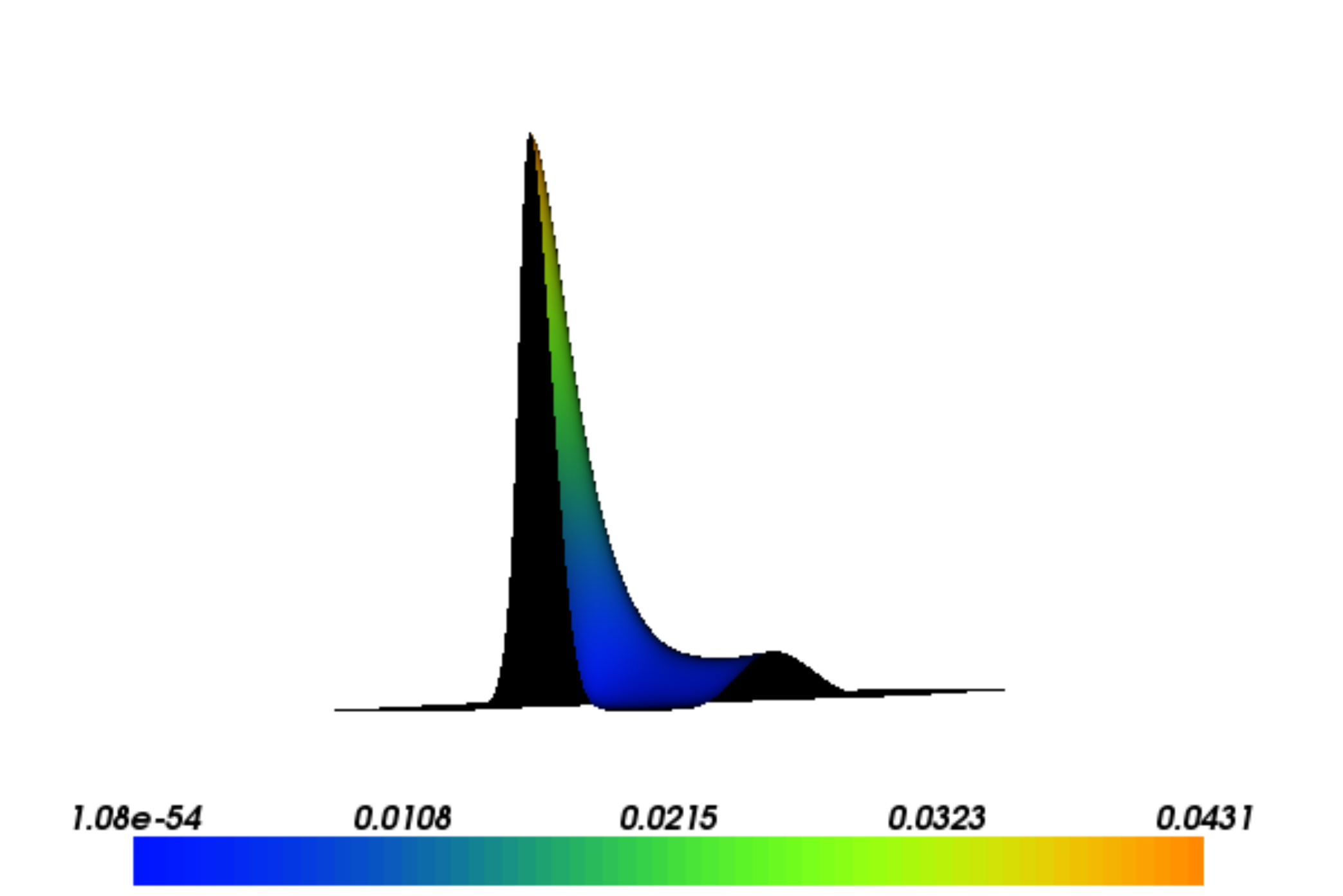}\\

          \hline
        \hline
        \end{tabular}
    \caption{\bf {The time evolution of a gaussian initial condition relaxing into the steady state for the 323 mutant.  The dominant peak is lytic.  The gap is reduce compared to wild type.}} 
        \label{tab:gt6}
    \end{table}%

 \begin{table}[ht]
     \centering
        \begin{tabular}{|p{0.11\textwidth}|p{0.29\textwidth}|p{0.29\textwidth}|p{0.29\textwidth}|} 
          \hline
          \multicolumn{4}{|c|}{Time Evolution}
          \\ \hline \hline
          time&$+0$&$+100$&$+200$\\ \hline
           $t=1$ & \includegraphics[scale=.22]{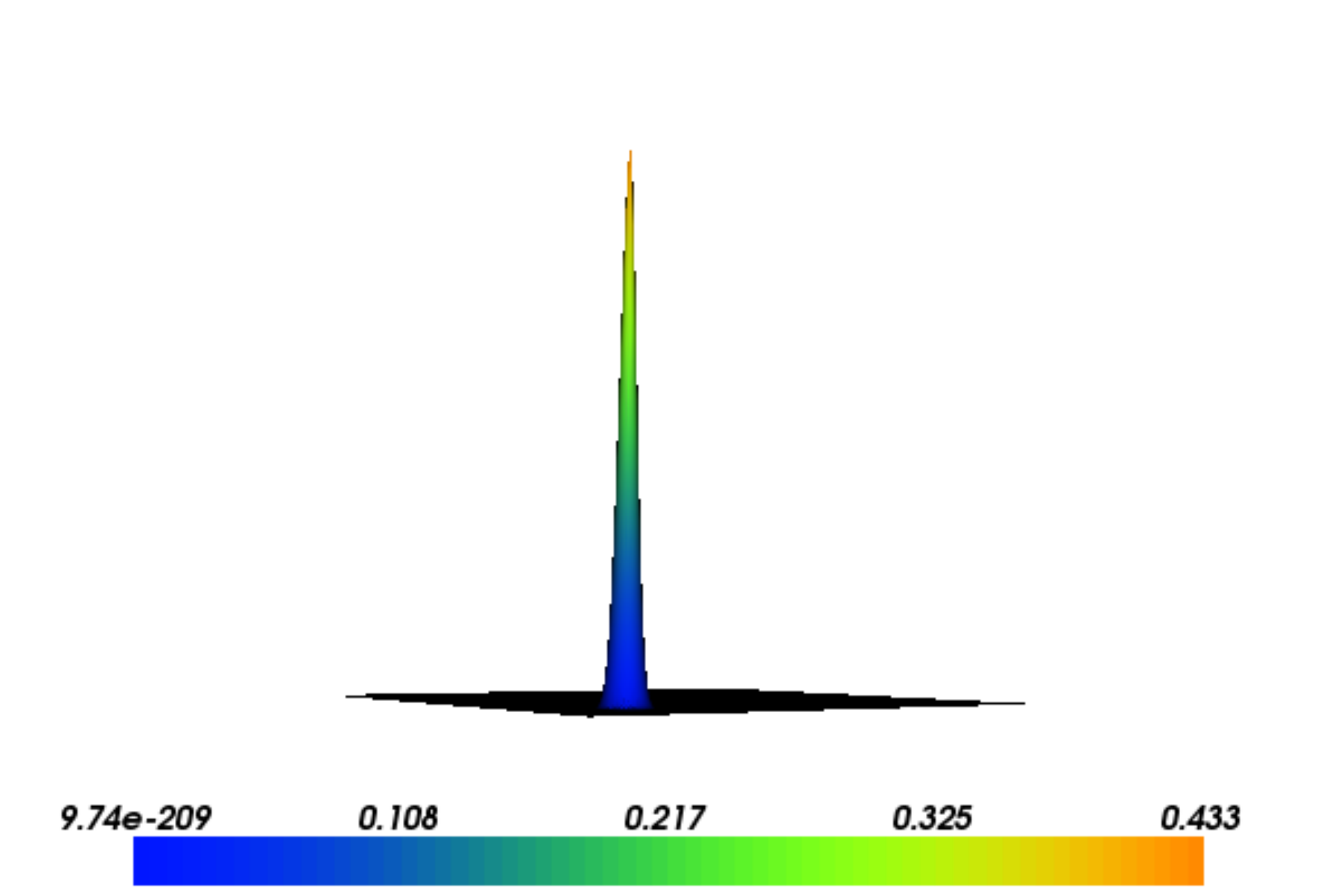}& \includegraphics[scale=.22]{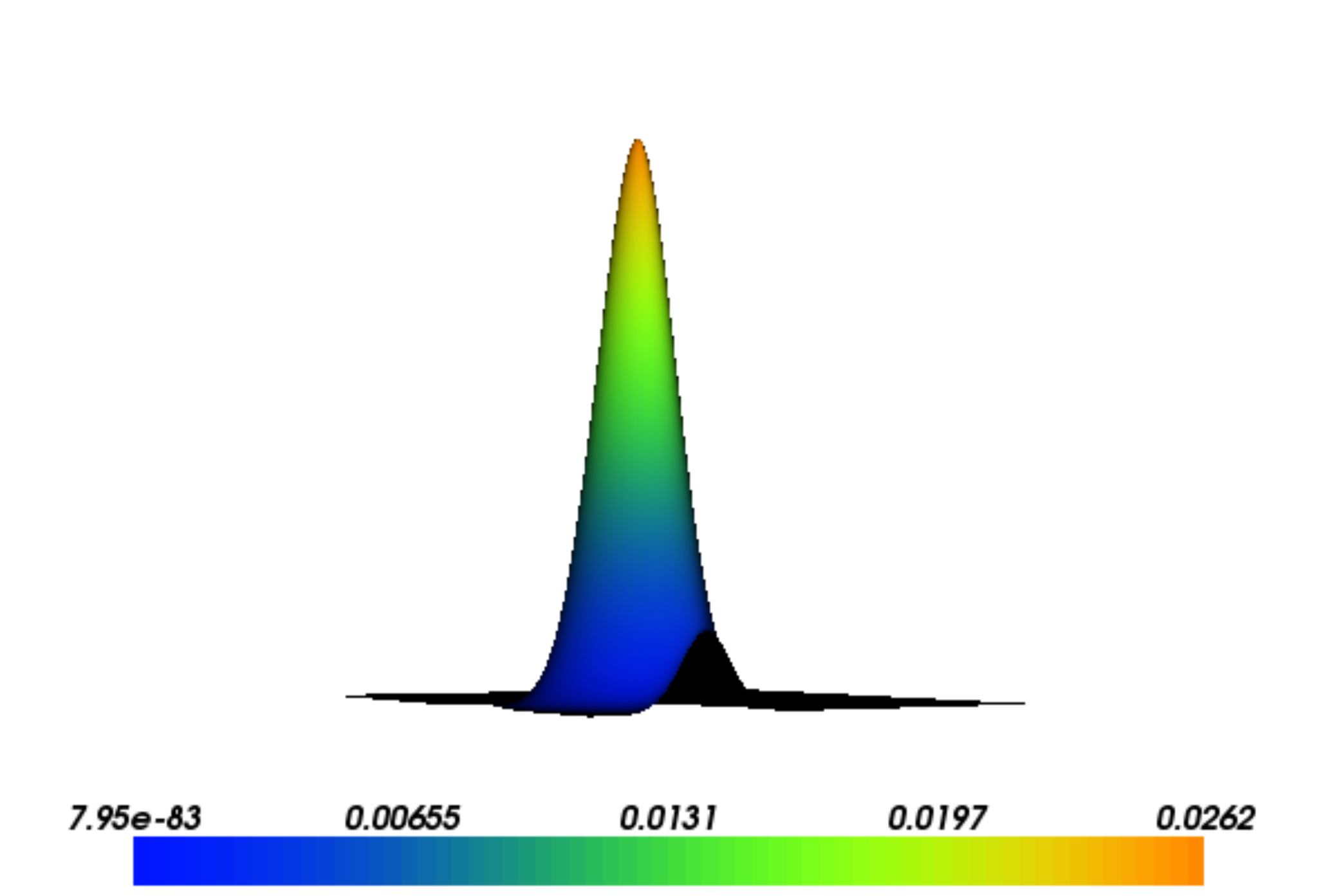}&\includegraphics[scale=.22]{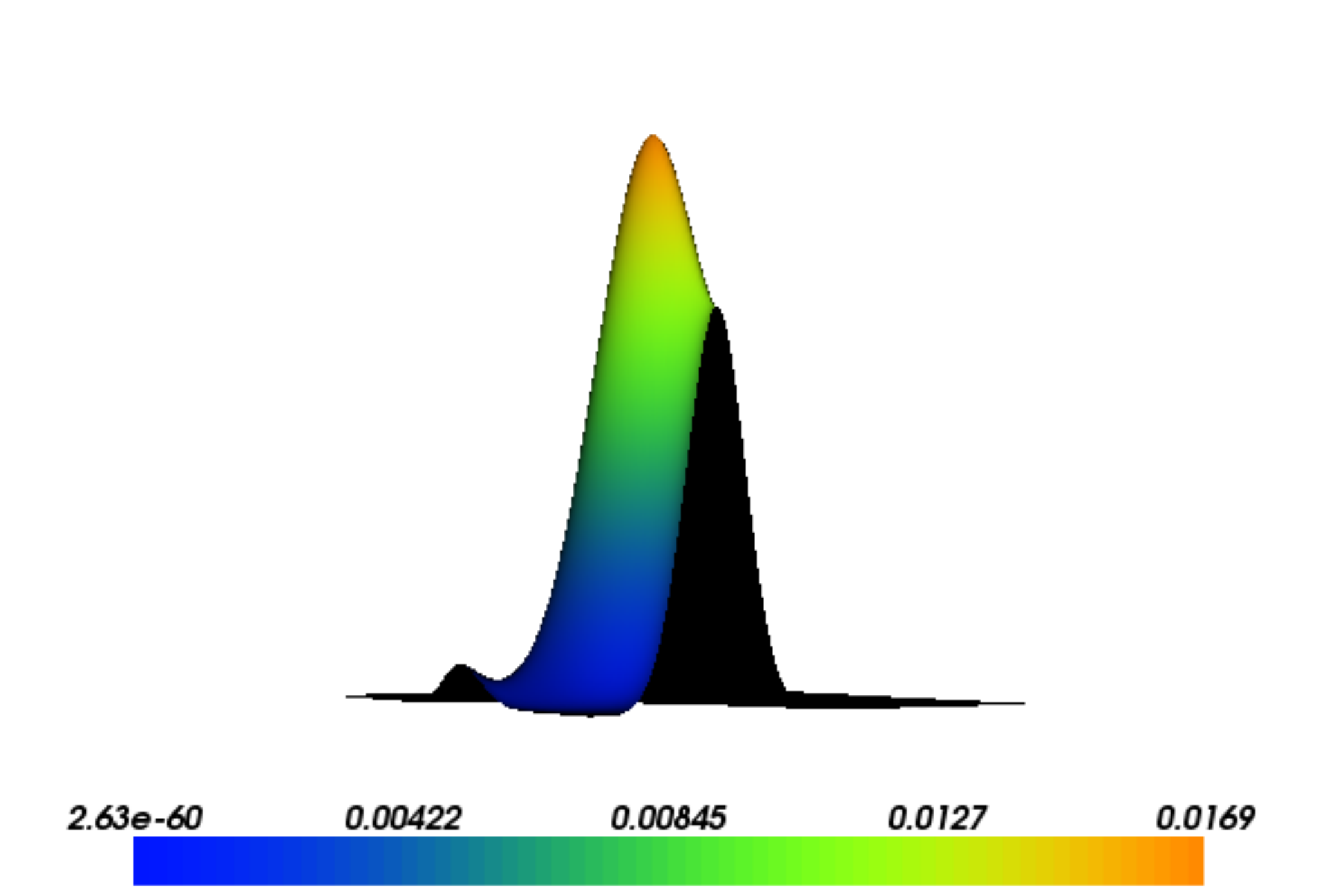}\\ \hline
          $t=301$ & \includegraphics[scale=.22]{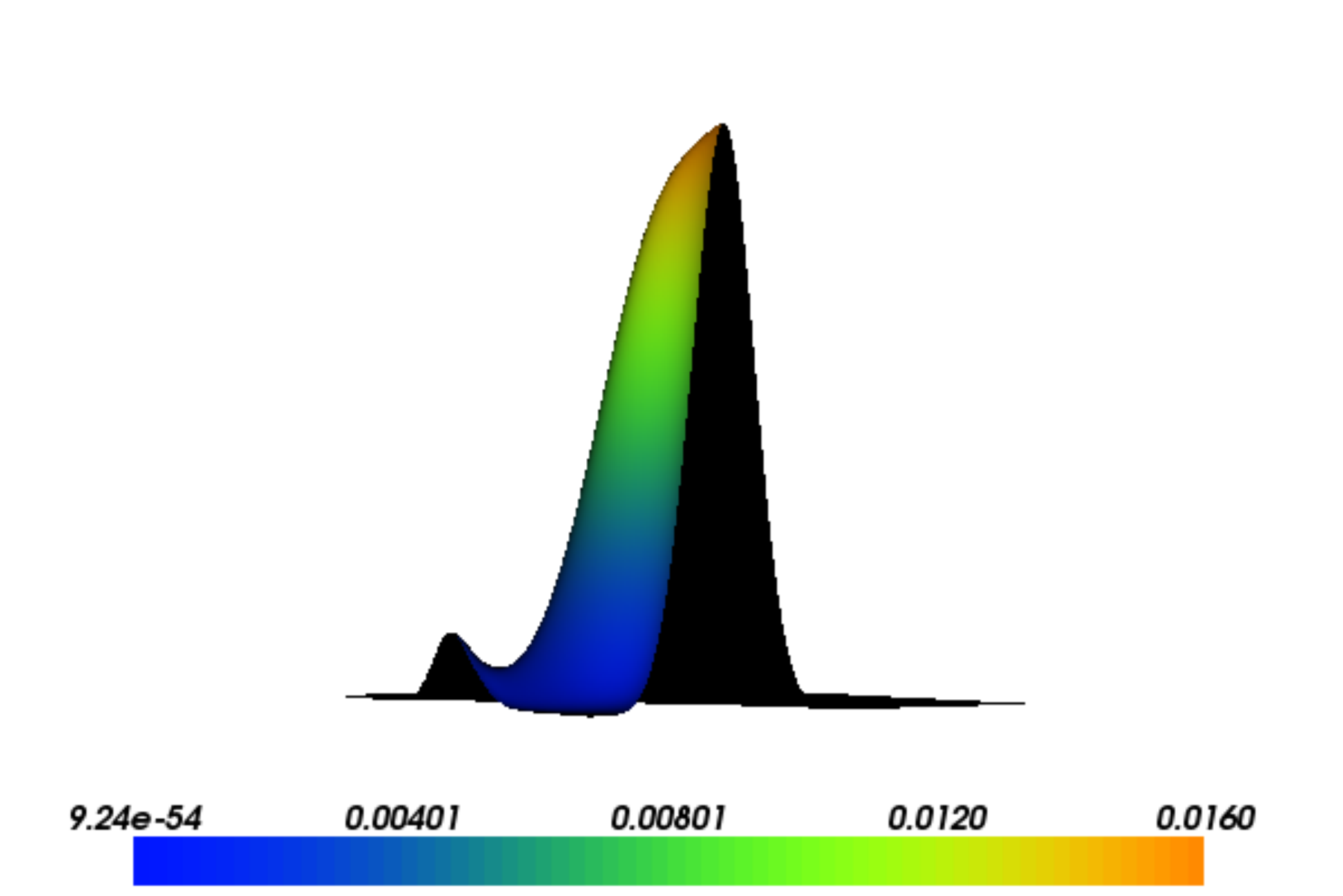}& \includegraphics[scale=.22]{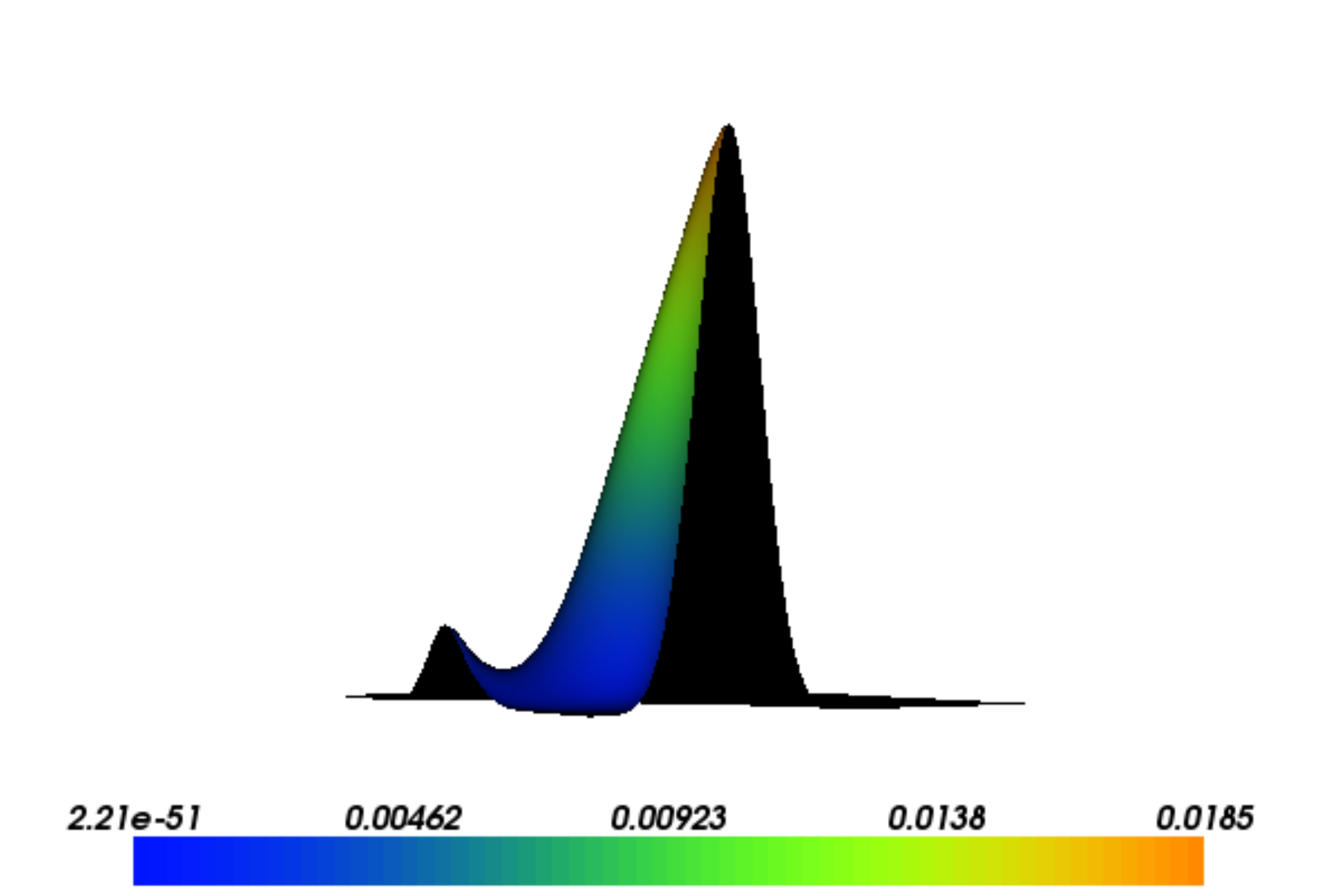}&\includegraphics[scale=.22]{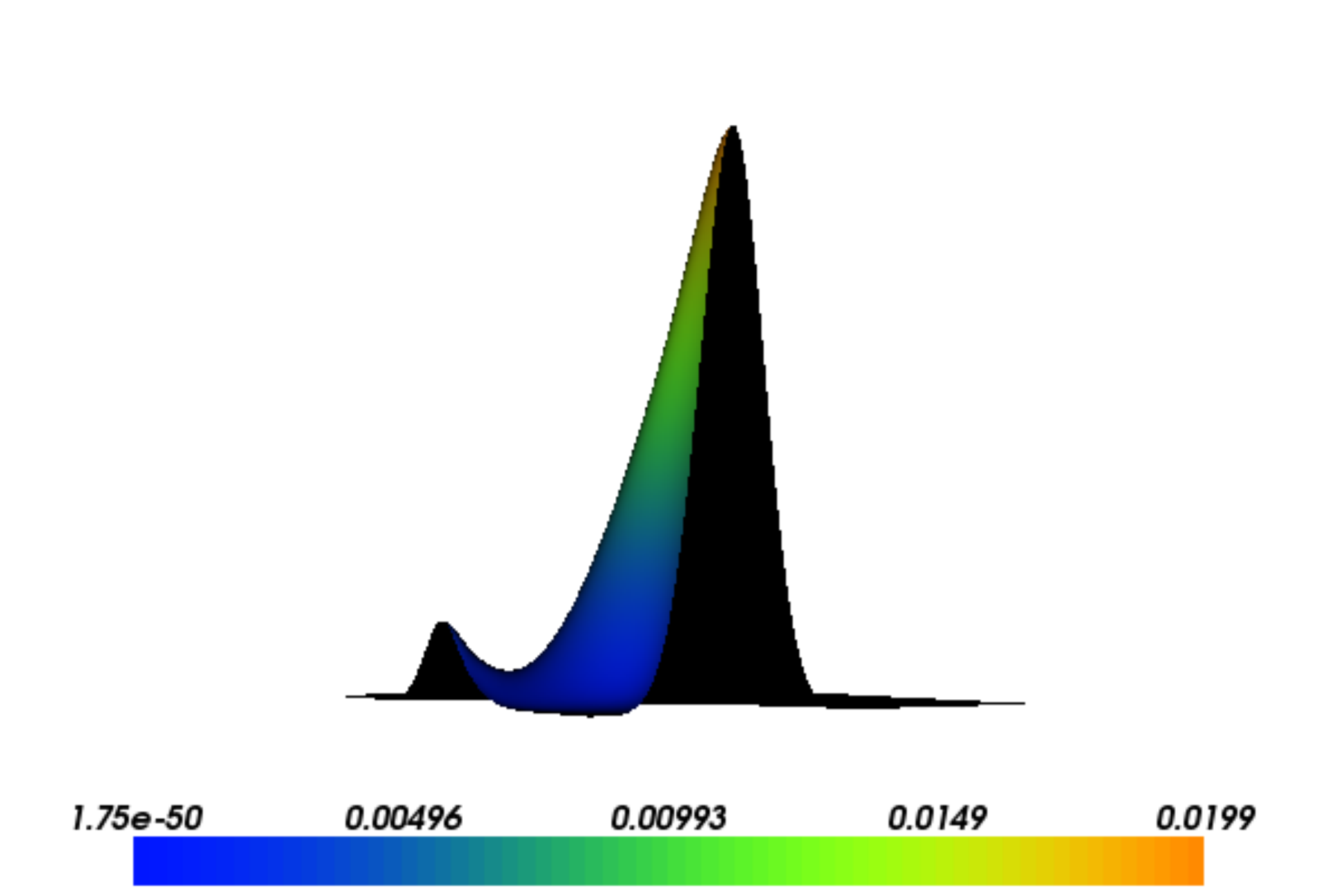}\\ \hline
          $t=601$ & \includegraphics[scale=.22]{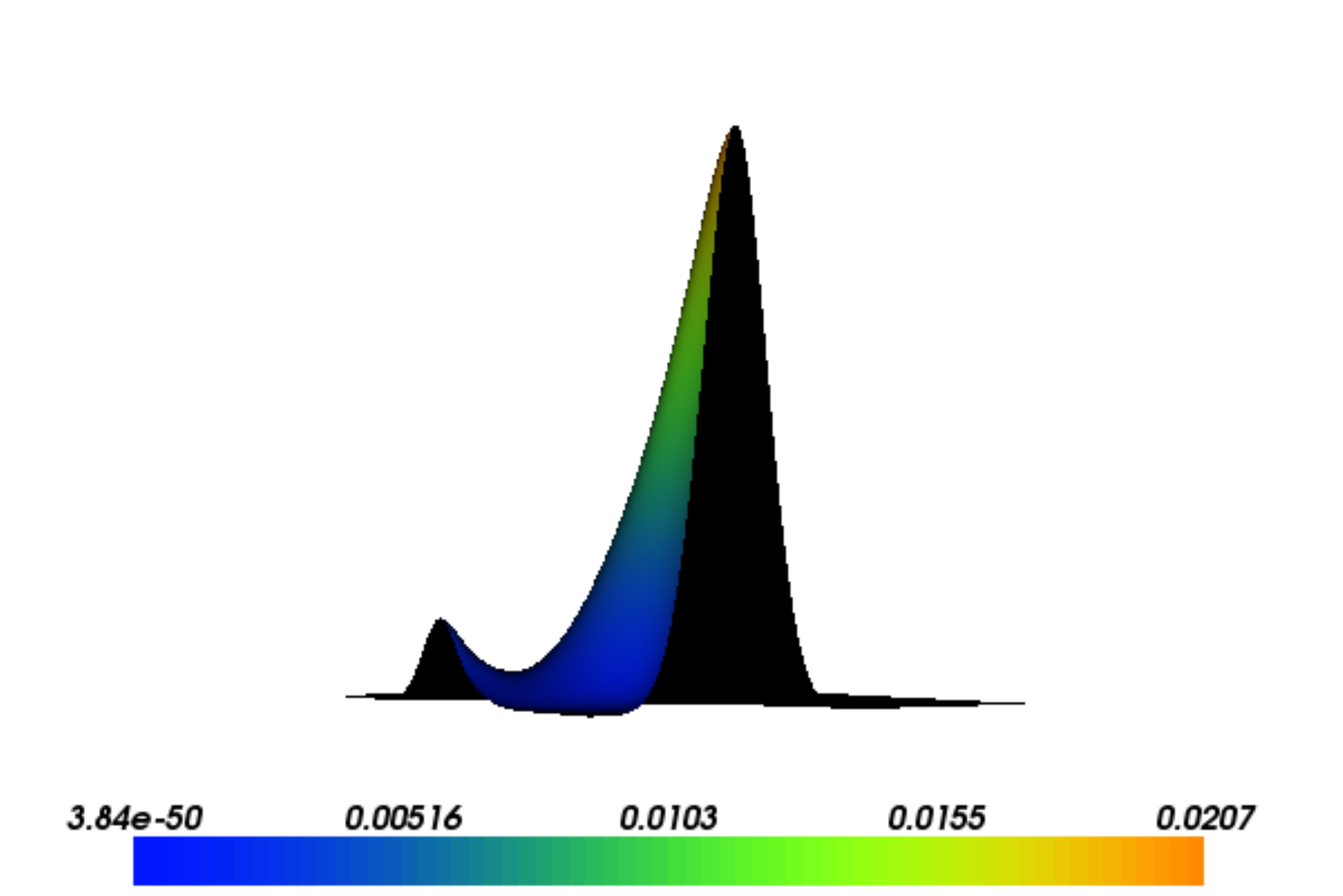}& \includegraphics[scale=.22]{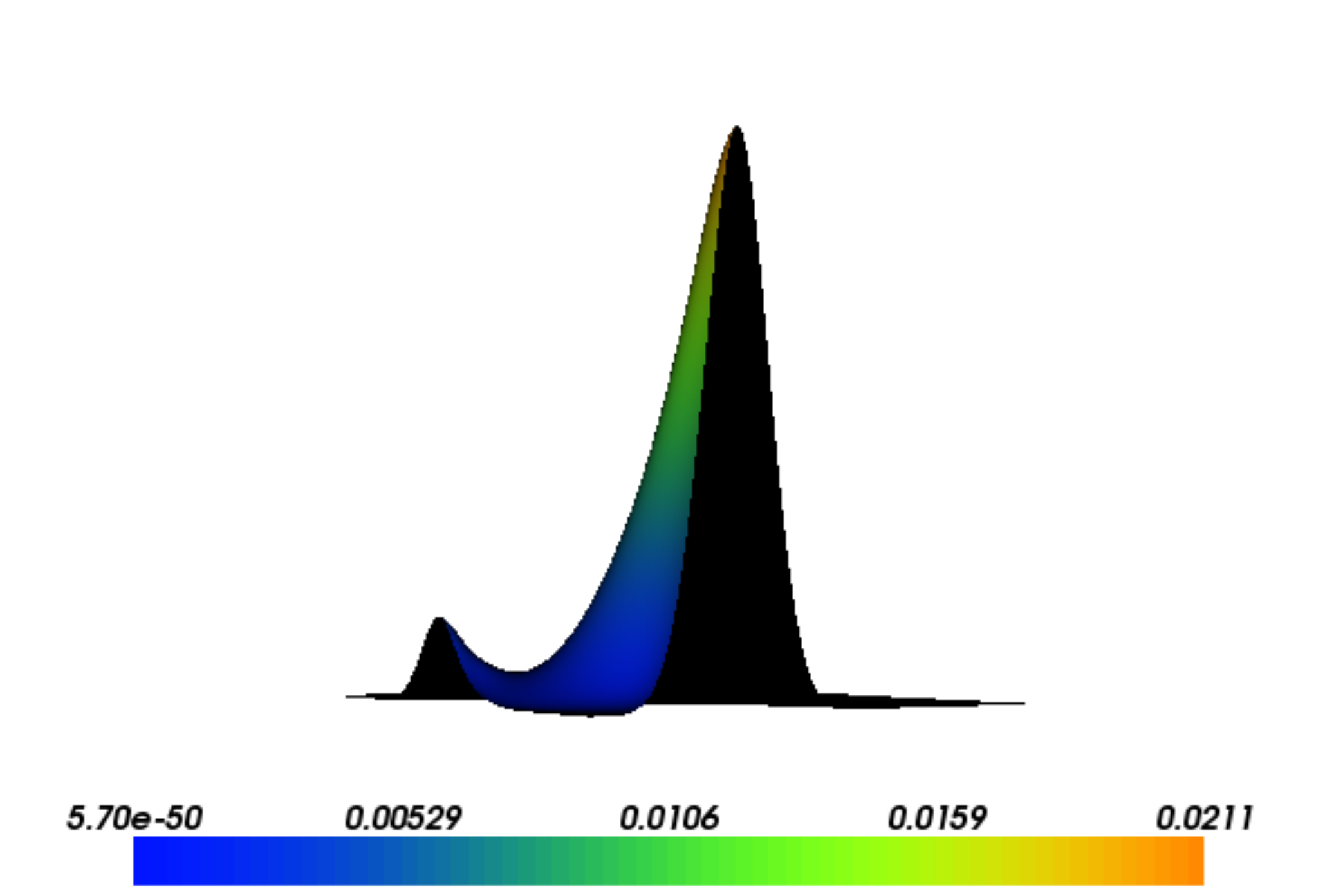}&\includegraphics[scale=.22]{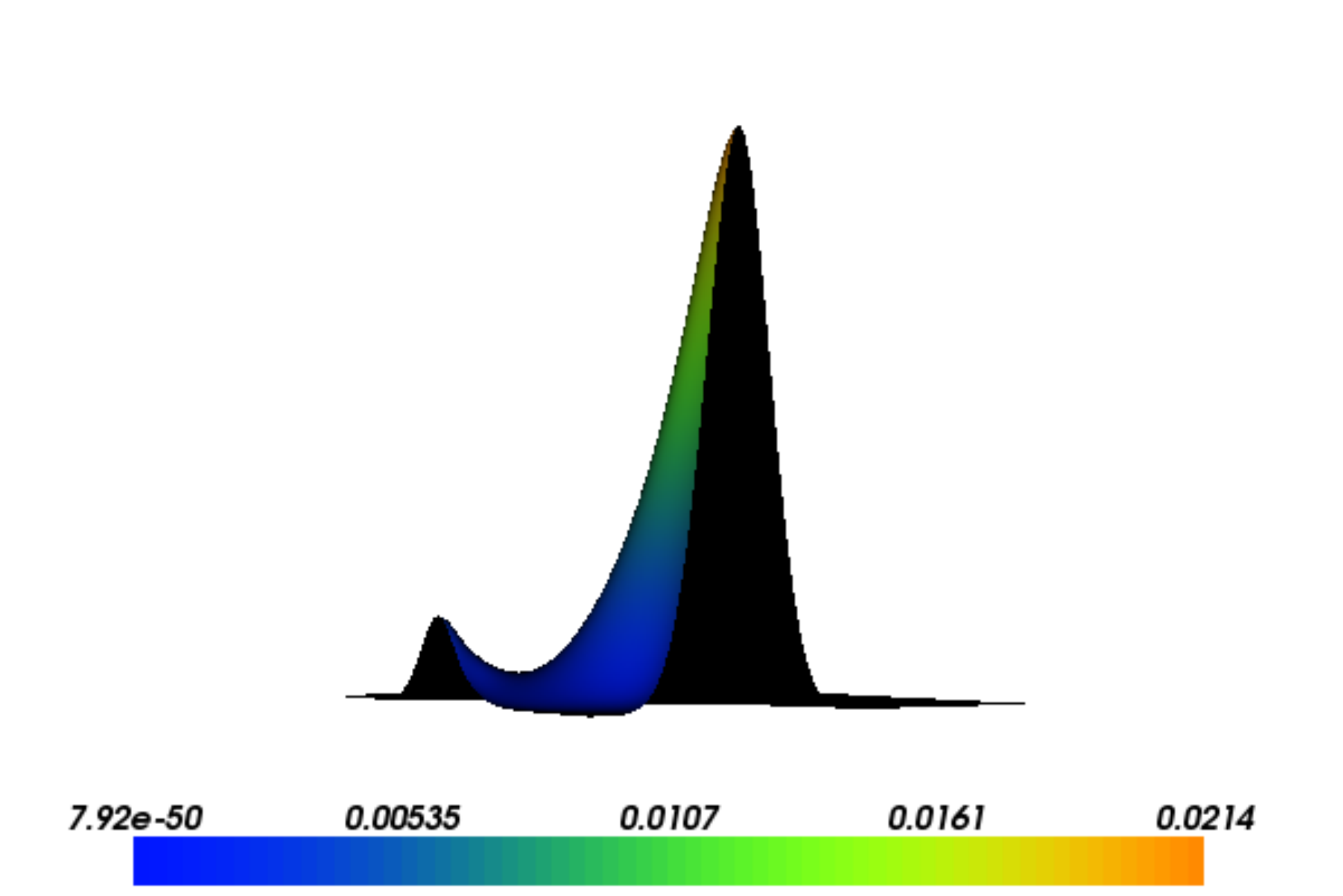}\\ \hline
          $t=901$ & \includegraphics[scale=.22]{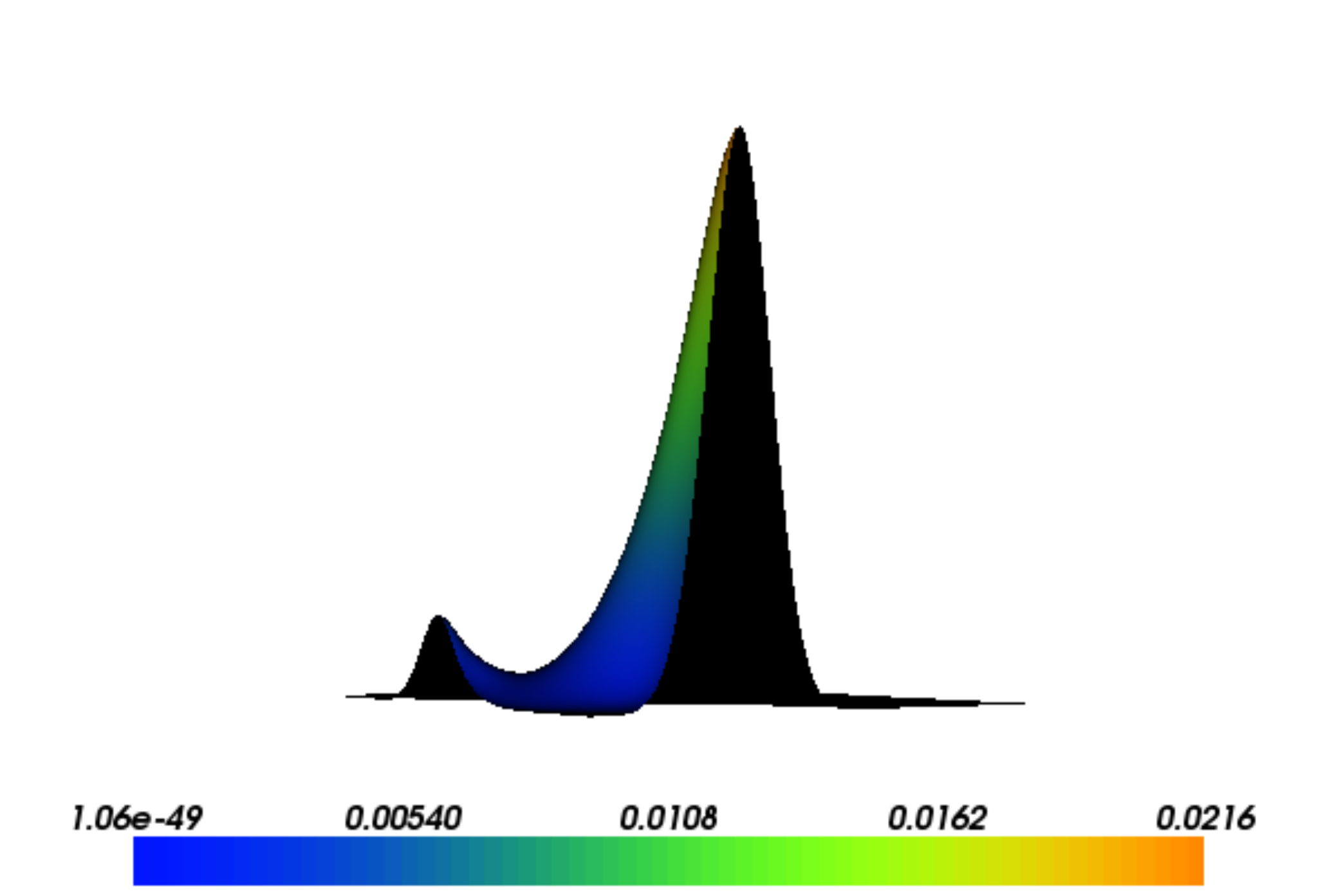}& \includegraphics[scale=.22]{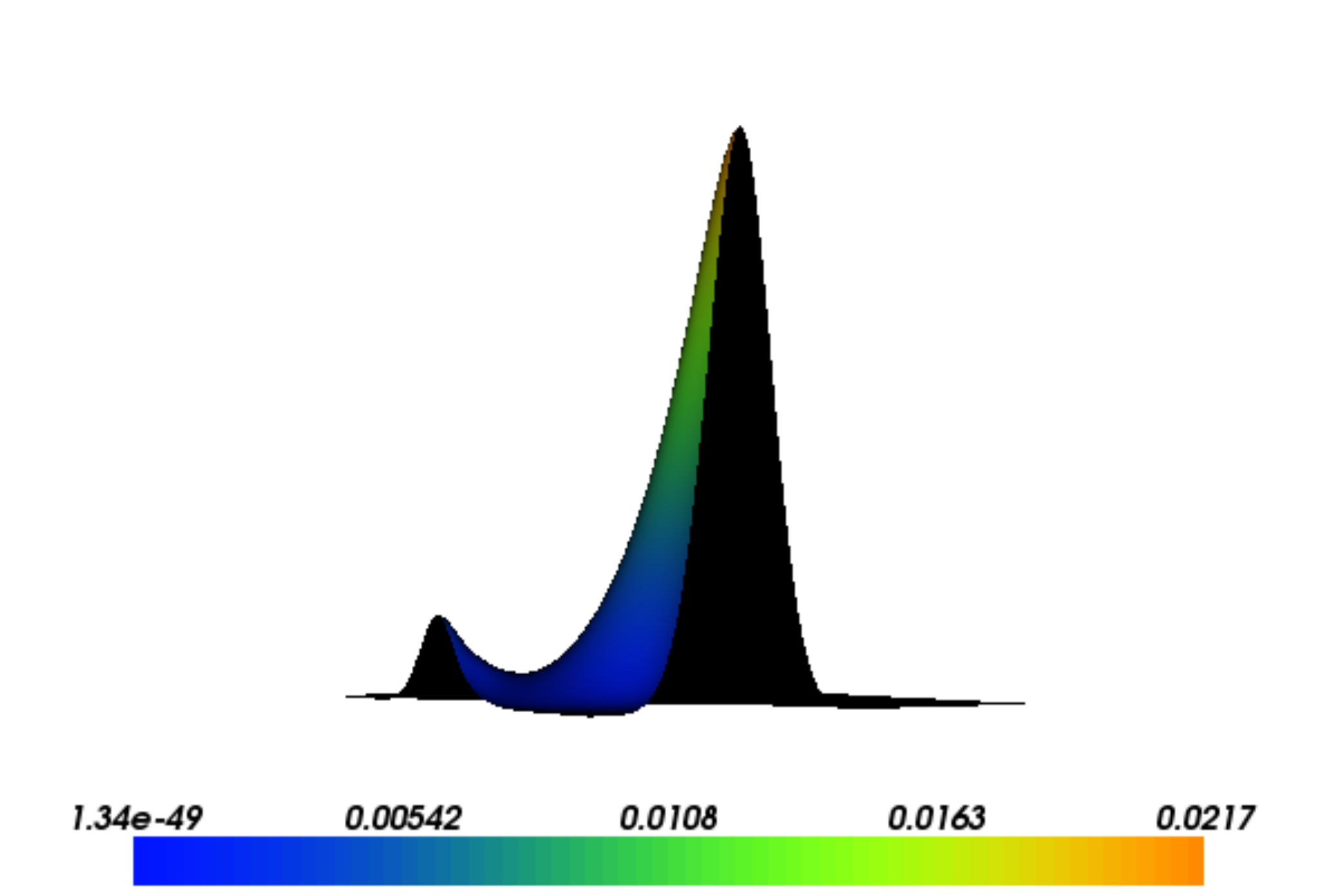}&\includegraphics[scale=.22]{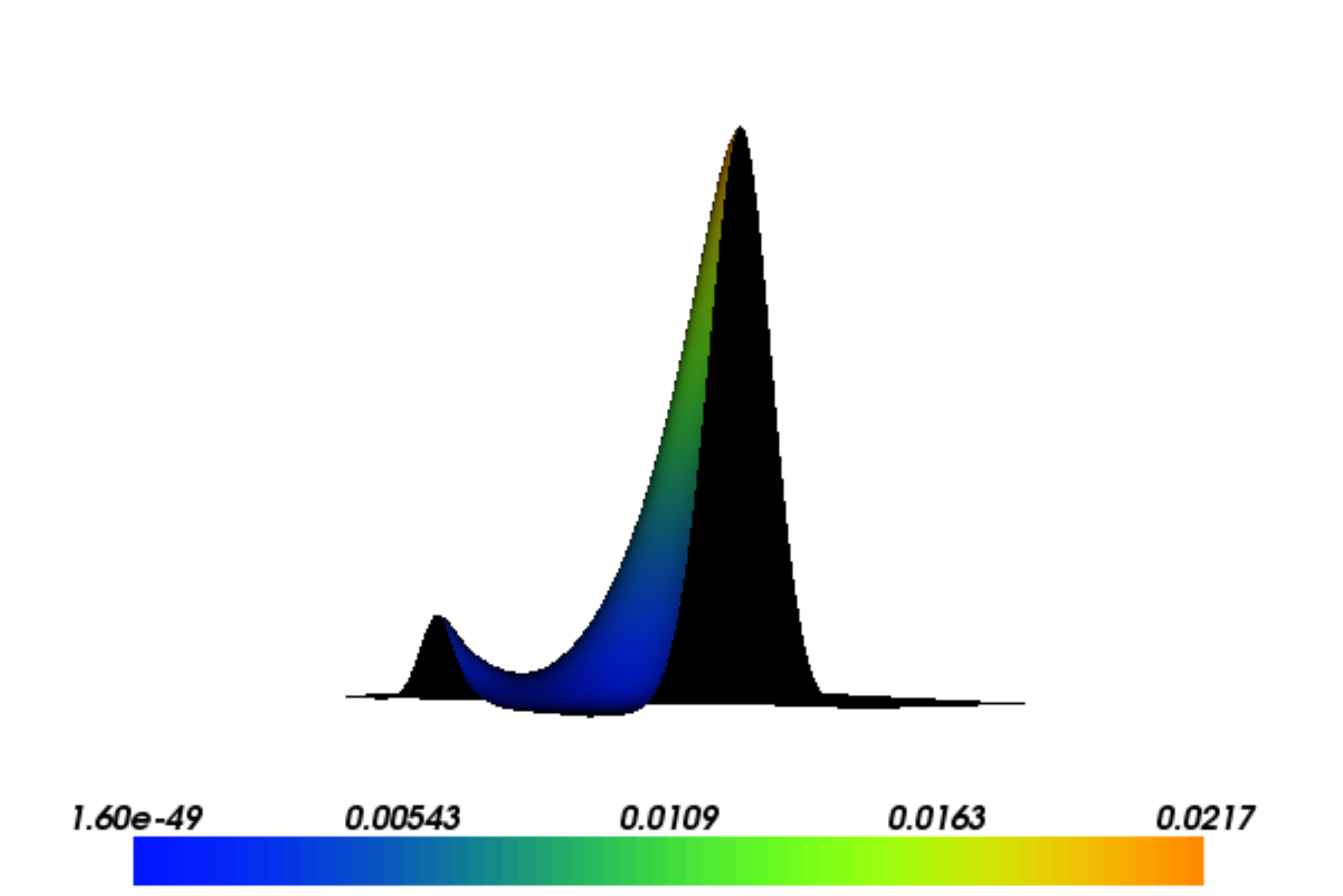}\\

          \hline
        \hline
        \end{tabular}
    \caption{\bf {The time evolution of a gaussian initial condition relaxing into the steady state for the 323p mutant.  The dominant peak is lysogenic.}}
        \label{tab:gt7}
    \end{table}%


\begin{table}[ht]
\noindent
     \resizebox{!}{9cm}{
        \begin{tabular}{|c|c|c|c|}
          \hline
          \multicolumn{4}{|c|}{$\lambda_{121}$ steady state and first passage time distributions}
          \\ \hline \hline
$O_{R1}$ &  \multicolumn{3}{|c|}{$TATCACCGCCAGAGGTA$}          \\ \hline
$O_{R2}$ &  \multicolumn{3}{|c|}{$TAACACCGTGCGTGTTG$}          \\ \hline        
$O_{R3}$ &  \multicolumn{3}{|c|}{$TATCACCGCCAGAGGTA$}          \\ \hline

             \includegraphics[scale=.50]{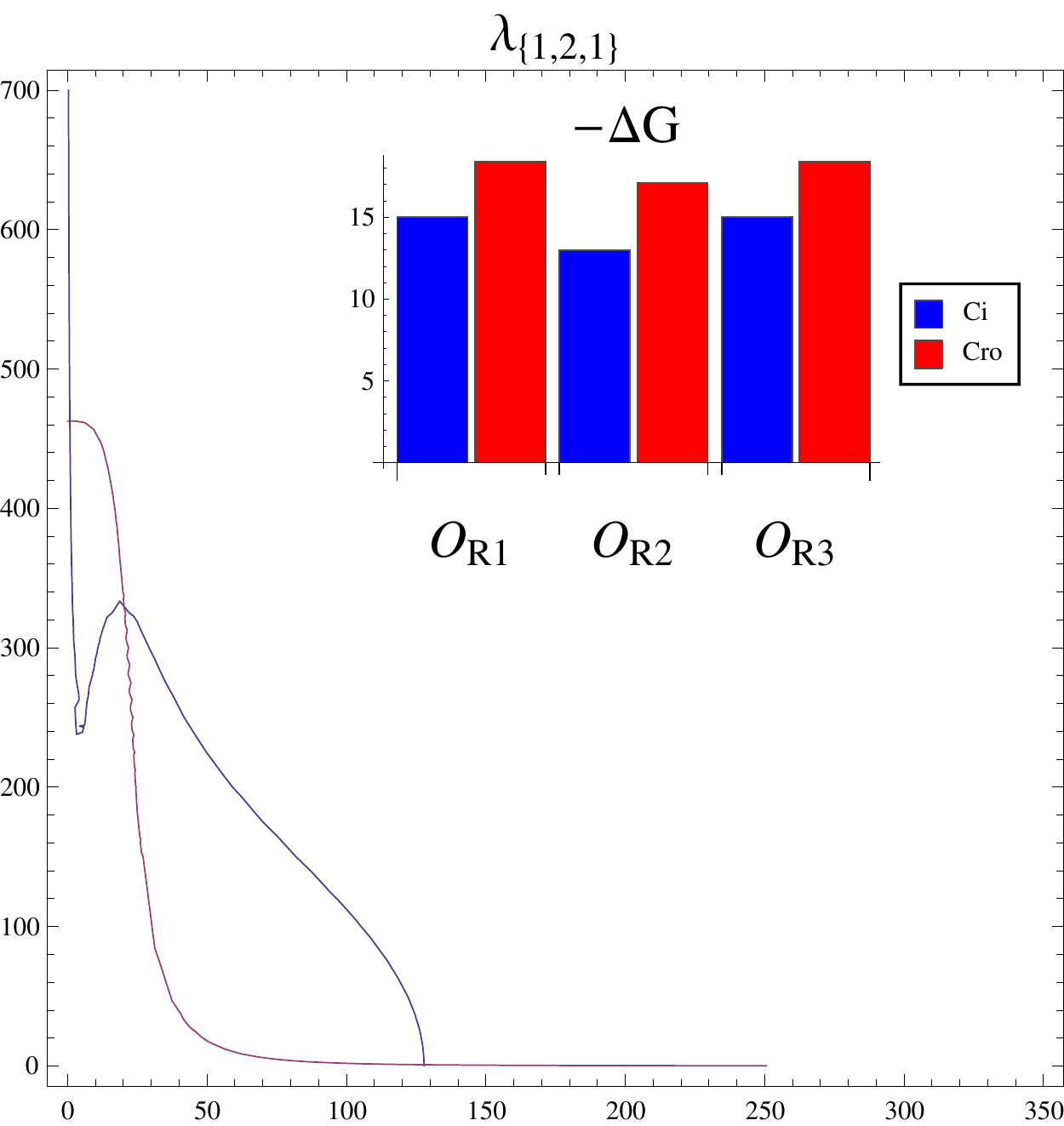}     
         &  \multicolumn{3}{|c|}{\includegraphics[scale=.5 ]{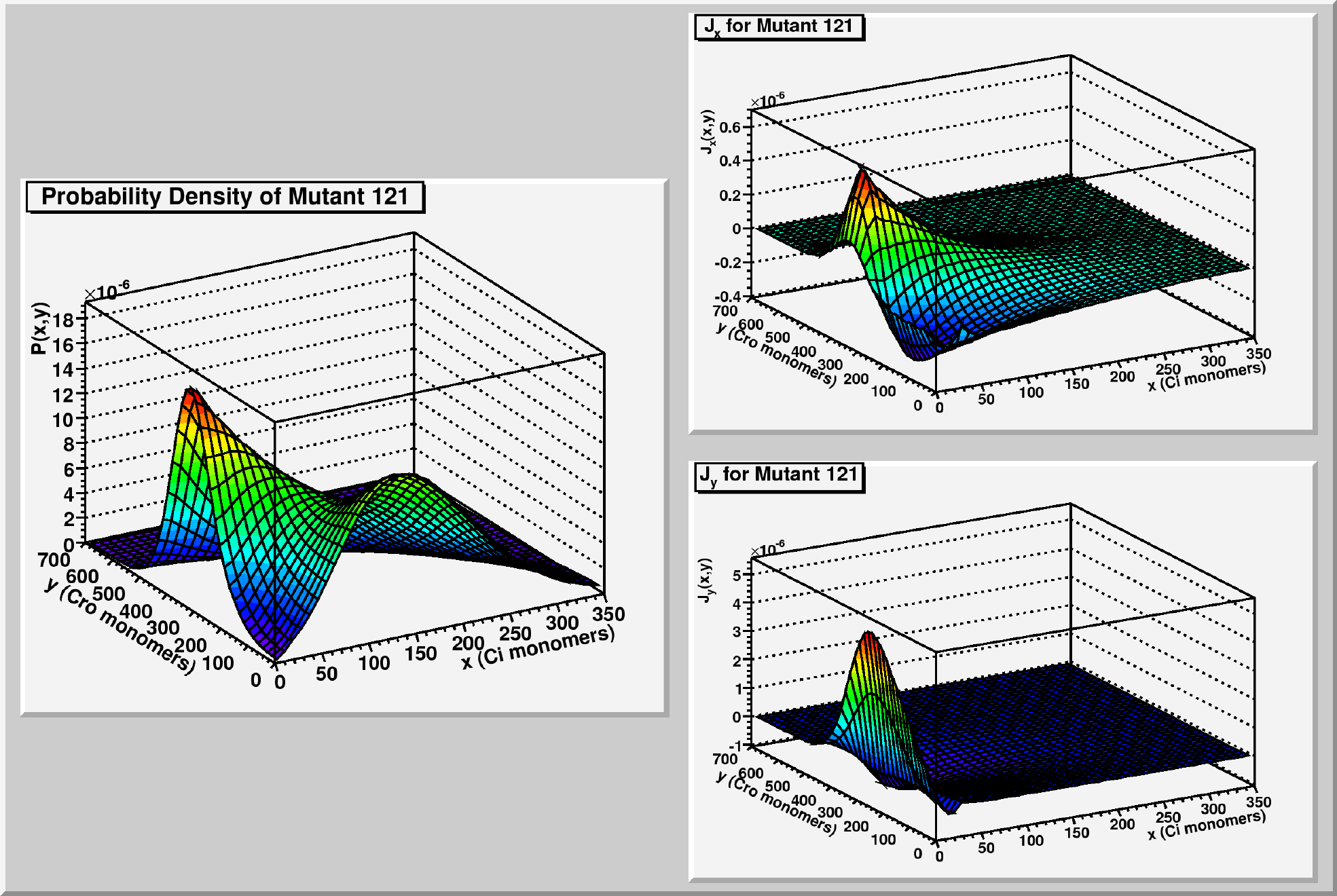}} \\ \hline    
          \multicolumn{4}{|c|}{\includegraphics[scale=.5 ]{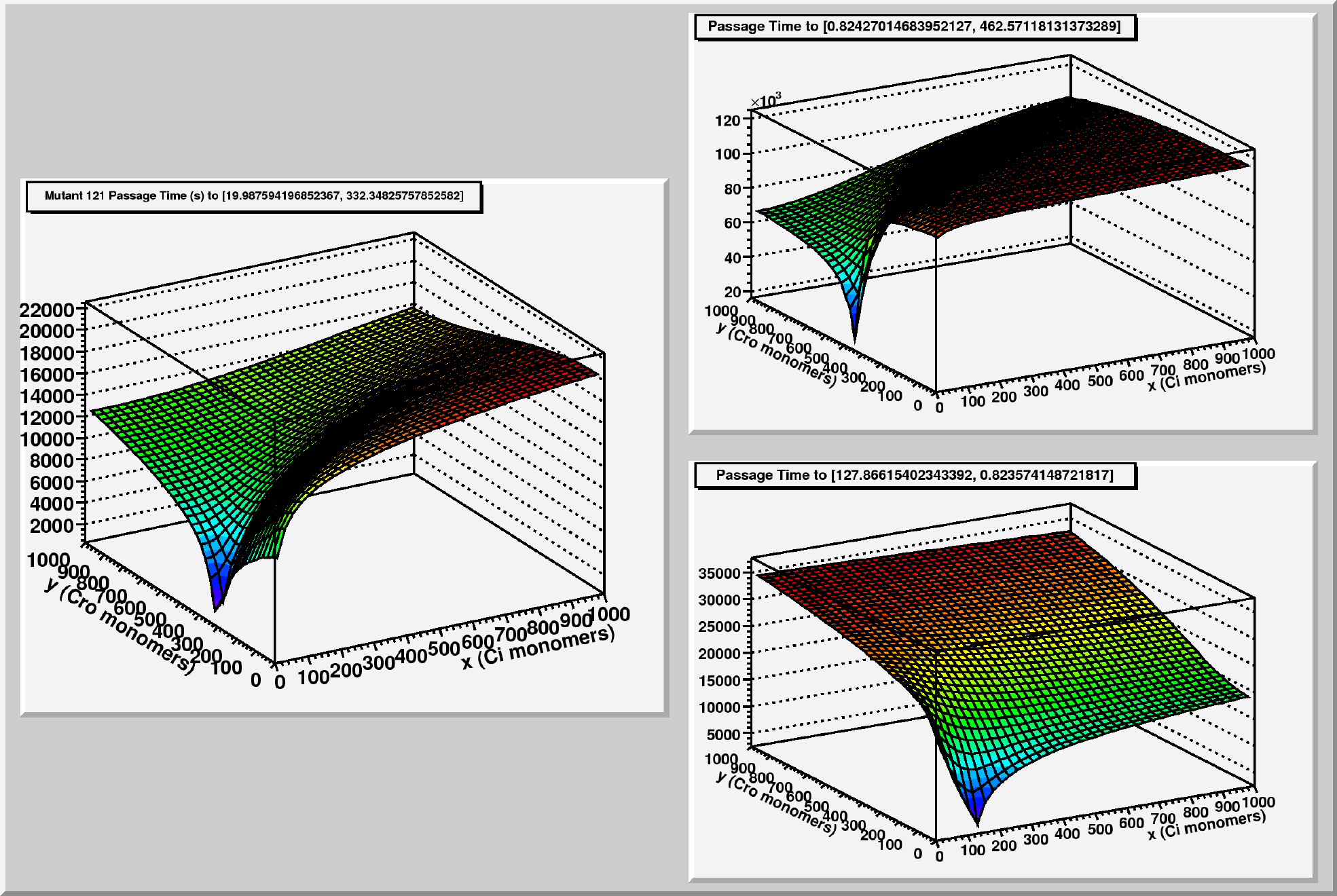}} \\ \hline     
           \multicolumn{3}{|c|}{$\lambda_{121}$ Assorted Expectation Values} &ratio to $\lambda_{123}$ \\ \hline
	 $<x>$ (Ci protein) &  $\frac{\int_\Omega x \rho(x,y) dxdy}{\int_\Omega \rho(x,y) dxdy}$ & 114.899 & 1.234 \\ \hline          
             $<y>$ (Cro protein) & $\frac{\int_\Omega y \rho(x,y) dxdy}{\int_\Omega \rho(x,y) dxdy}$ & 193.112 & 1.137  \\ \hline     
	 $<\tau_0>$ (seconds to ts) &  $\frac{\int_\Omega \tau_0(x,y) \rho(x,y) dxdy}{\int_\Omega \rho(x,y) dxdy}$ & 11177.264 & 0.527 \\ \hline
	 $<\tau_1>$ (seconds to lytic) &  $\frac{\int_\Omega \tau_1(x,y) \rho(x,y) dxdy}{\int_\Omega \rho(x,y) dxdy}$ & 103986.923 & 1.113   \\ \hline
	 $<\tau_2>$ (seconds to lysogen) &  $\frac{\int_\Omega \tau_2(x,y) \rho(x,y) dxdy}{\int_\Omega \rho(x,y) dxdy}$ & 22020.823 & 0.083  \\ \hline
         \hline
        \end{tabular}
        }
    \caption{\bf {Properties of mutant $\lambda_{121}$}}
        \label{tab:l121}
    \end{table}%

\begin{table}[ht]
\noindent
     \resizebox{!}{8.5cm}{
        \begin{tabular}{|c|c|c|}
          \hline
          \multicolumn{3}{|c|}{$\lambda_{123}$ steady state and first passage time distributions}          \\ \hline \hline
$O_{R1}$ &  \multicolumn{2}{|c|}{$TATCACCGCCAGAGGTA$}          \\ \hline
$O_{R2}$ &  \multicolumn{2}{|c|}{$TAACACCGTGCGTGTTG$}          \\ \hline
$O_{R3}$ &  \multicolumn{2}{|c|}{$TATCACCGCAAGGGATA$}          \\ \hline             \includegraphics[scale=.50]{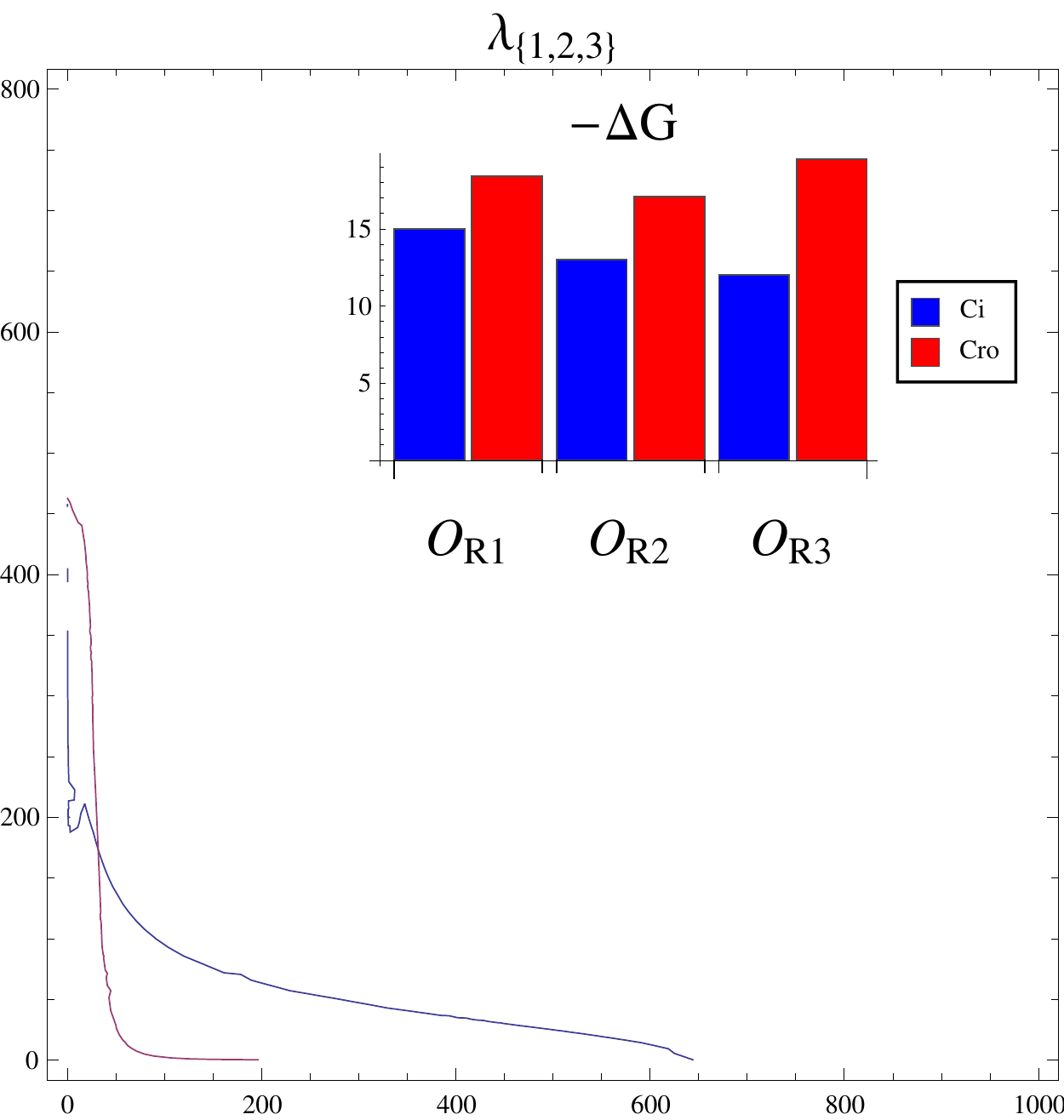}     
         &  \multicolumn{2}{|c|}{\includegraphics[scale=.50 ]{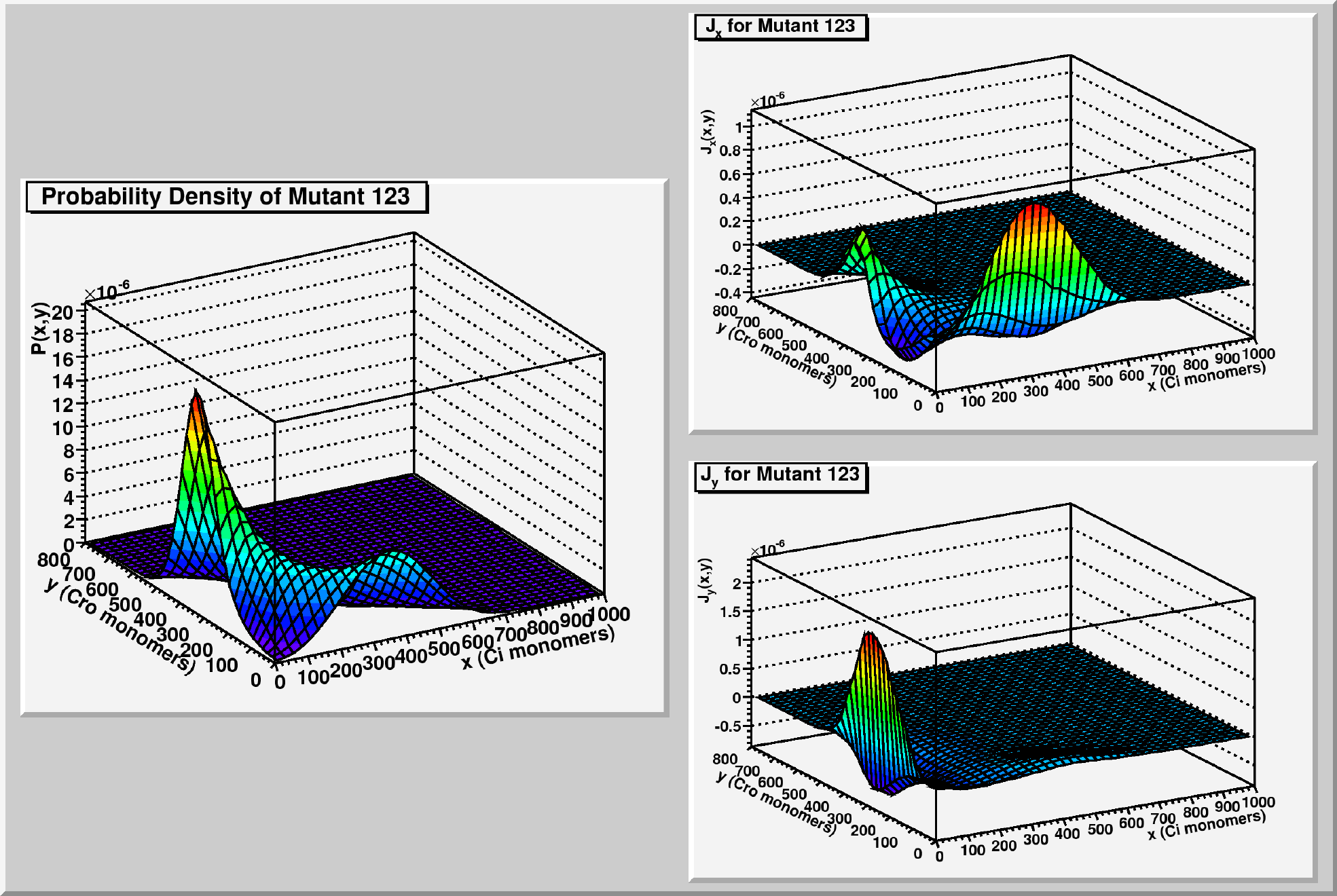}} \\ \hline    
          \multicolumn{3}{|c|}{\includegraphics[scale=.50 ]{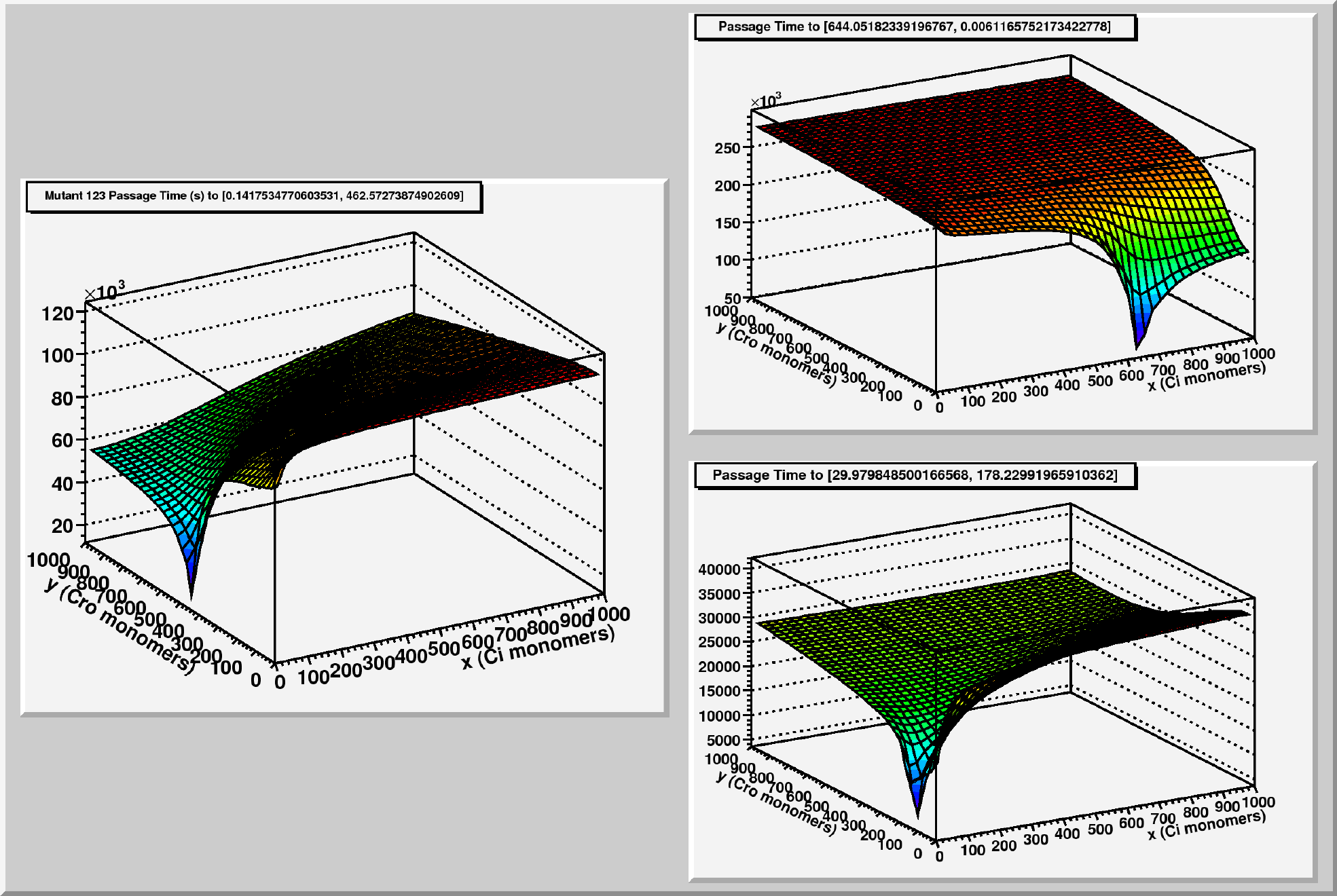}} \\ \hline     
           \multicolumn{3}{|c|}{$\lambda_{123}$ Assorted Expectation Values}    \\ \hline
	 $<x>$ (Ci protein) &  $\int_\Omega x \rho(x,y) dxdy/\int_\Omega \rho(x,y) dxdy$ & 93.076  \\ \hline          
             $<y>$ (Cro protein) & $\int_\Omega y \rho(x,y) dxdy/\int_\Omega \rho(x,y) dxdy$ & 169.777 \\ \hline     
	 $<\tau_0>$ (seconds to lytic) &  $\int_\Omega \tau_0(x,y) \rho(x,y) dxdy/\int_\Omega \rho(x,y) dxdy$ & 93387.22  \\ \hline
	 $<\tau_1>$ (seconds to lysogenic) &  $\int_\Omega \tau_1(x,y) \rho(x,y) dxdy/\int_\Omega \rho(x,y) dxdy$ & 263858.749 \\ \hline
	 $<\tau_2>$ (seconds to transition state) &  $\int_\Omega \tau_2(x,y) \rho(x,y) dxdy/\int_\Omega \rho(x,y) dxdy$ & 21214.802   \\ \hline
         \hline
        \end{tabular}
}
    \caption{\bf {Properties of wild-type $\lambda_{123}$}}
        \label{tab:l123}
    \end{table}%

\begin{table}[ht]
\noindent
     \resizebox{!}{9cm}{
        \begin{tabular}{|c|c|c|c|}
          \hline
          \multicolumn{4}{|c|}{$\lambda_{131}$ steady state and first passage time distributions}
          \\ \hline \hline
$O_{R1}$ &  \multicolumn{3}{|c|}{$TATCACCGCCAGAGGTA$}          \\ \hline
$O_{R2}$ &  \multicolumn{3}{|c|}{$TATCACCGCAAGGGATA$}          \\ \hline
$O_{R3}$ &  \multicolumn{3}{|c|}{$TATCACCGCCAGAGGTA$}          \\ \hline
             \includegraphics[scale=.30]{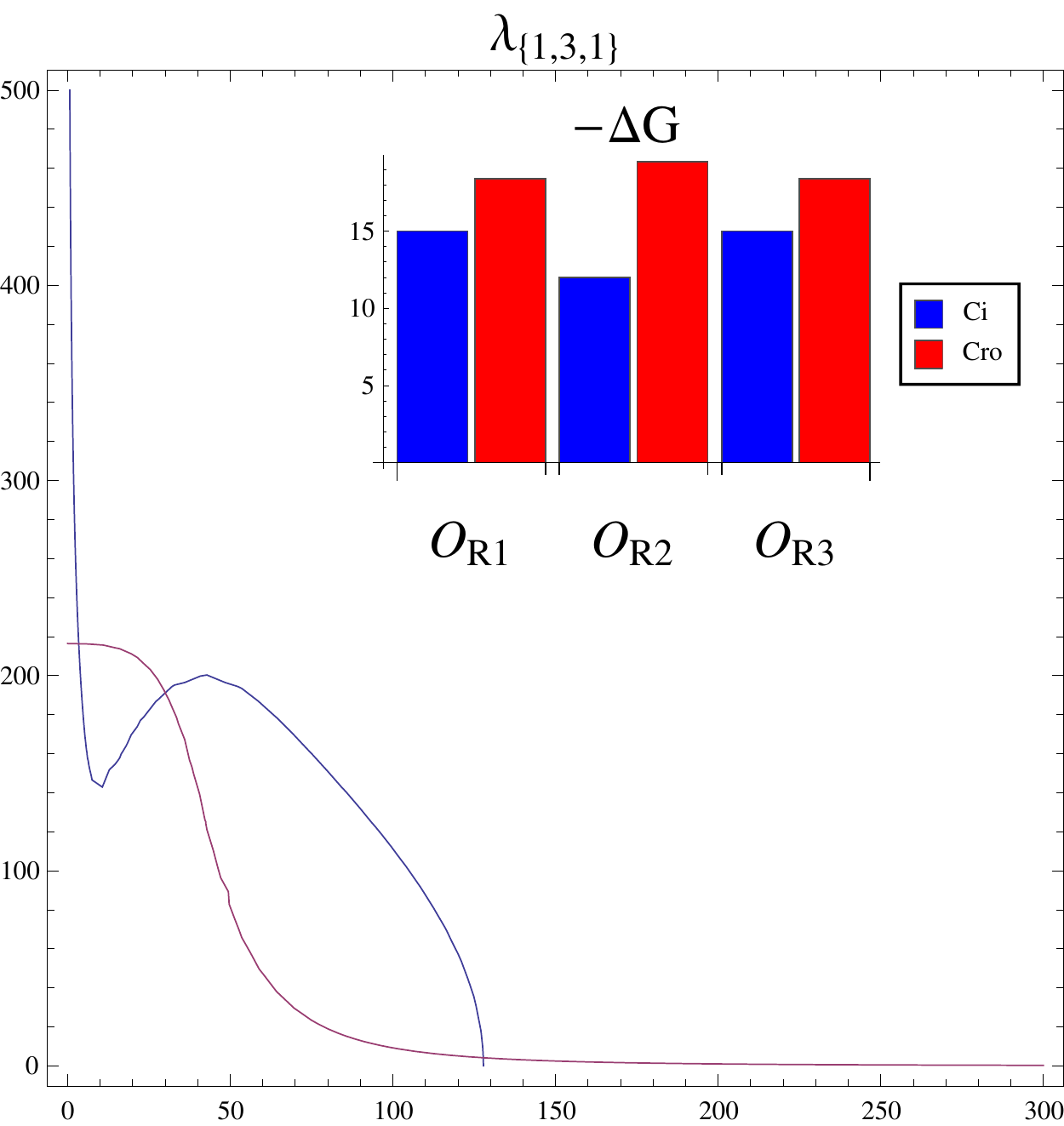}     
         &  \multicolumn{3}{|c|}{\includegraphics[scale=.50 ]{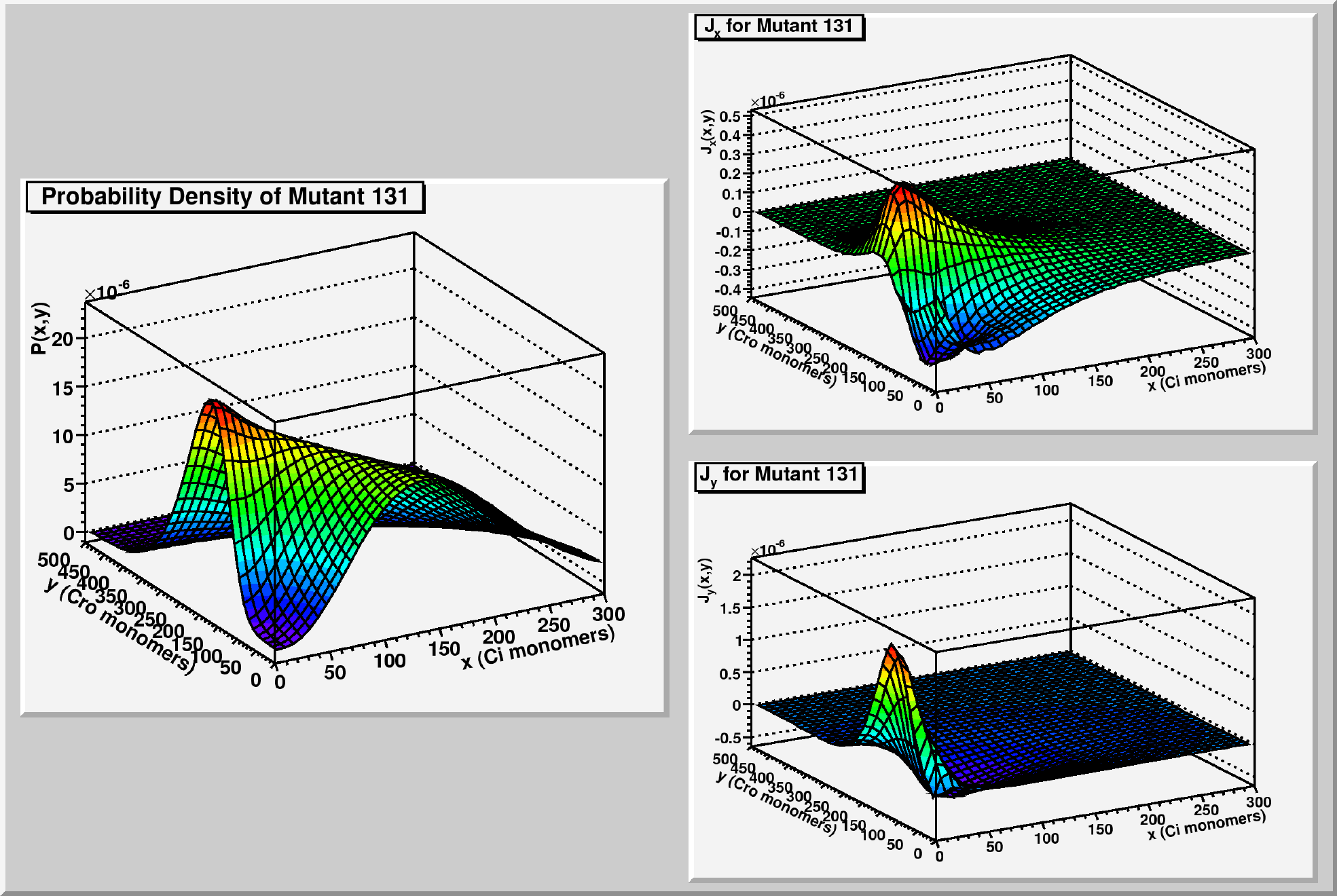}} \\ \hline    
          \multicolumn{4}{|c|}{\includegraphics[scale=.50 ]{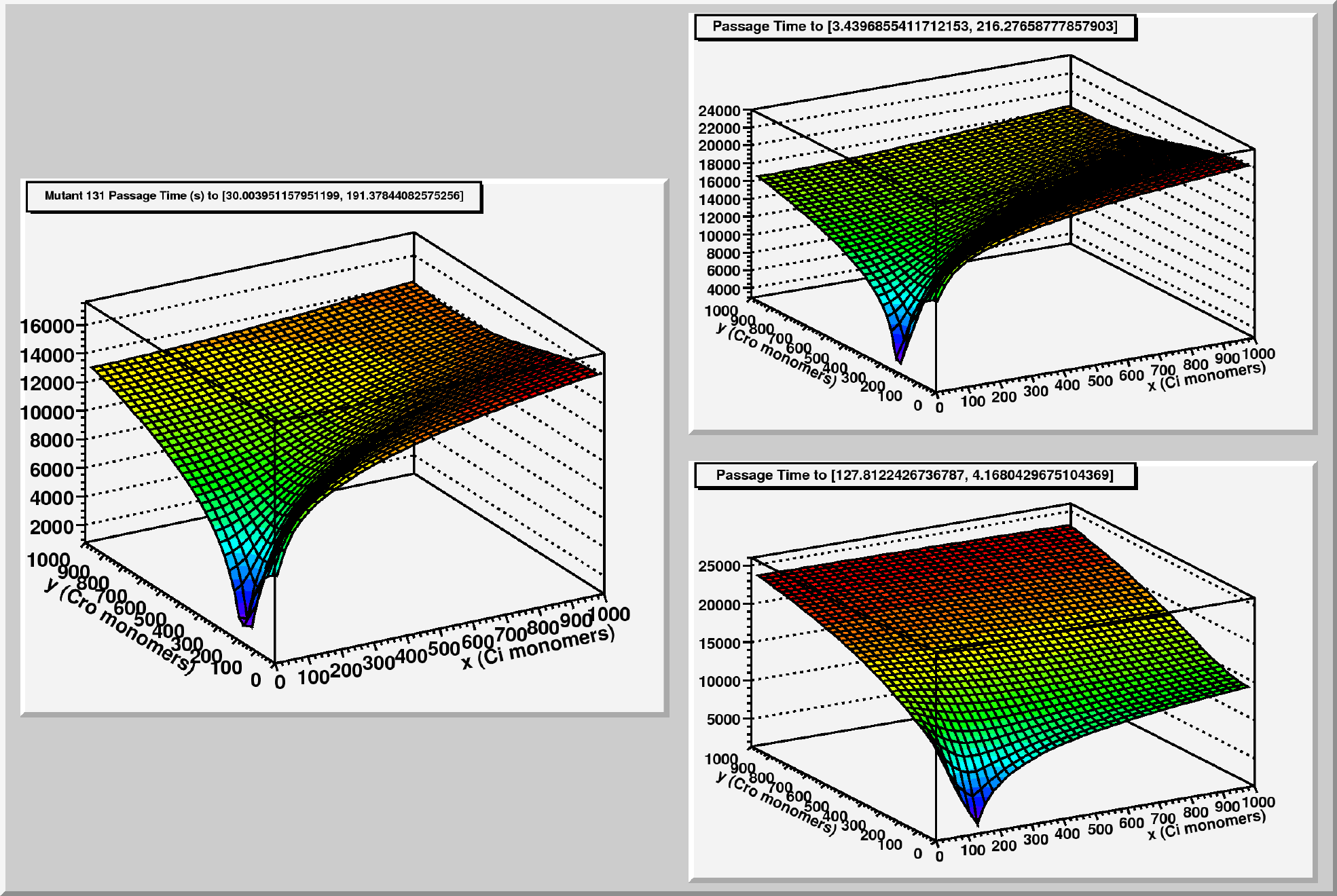}} \\ \hline     
           \multicolumn{3}{|c|}{$\lambda_{131}$ Assorted Expectation Values} & ratio to $\lambda_{123}$    \\ \hline
	 $<x>$ (Ci protein) &  $\frac{\int_\Omega x \rho(x,y) dxdy}{\int_\Omega \rho(x,y) dxdy}$ & 117.454 & 1.262  \\ \hline          
             $<y>$ (Cro protein) & $\frac{\int_\Omega y \rho(x,y) dxdy}{\int_\Omega \rho(x,y) dxdy}$ & 143.183 & 0.843\\ \hline     
	 $<\tau_0>$ (seconds to ts) &  $\frac{\int_\Omega \tau_0(x,y) \rho(x,y) dxdy}{\int_\Omega \rho(x,y) dxdy}$ & 8084.495 & 0.381   \\ \hline
	 $<\tau_1>$ (seconds to lysogenic) &  $\frac{\int_\Omega \tau_1(x,y) \rho(x,y) dxdy}{\int_\Omega \rho(x,y) dxdy}$ & 13499.860 & 0.051 \\ \hline
	 $<\tau_2>$ (seconds to lytic) &  $\frac{\int_\Omega \tau_2(x,y) \rho(x,y) dxdy}{\int_\Omega \rho(x,y) dxdy}$ & 12908.285 & 0.138  \\ \hline
         \hline
        \end{tabular}
}

    \caption{\bf {Properties of mutant $\lambda_{131}$}}
        \label{tab:l131}
    \end{table}%

\begin{table}[ht]
\noindent
     \resizebox{!}{9cm}{
        \begin{tabular}{|c|c|c|c|}
          \hline
          \multicolumn{4}{|c|}{$\lambda_{133}$ steady state and first passage time distributions}
          \\ \hline \hline
$O_{R1}$ &  \multicolumn{3}{|c|}{$TATCACCGCCAGAGGTA$}          \\ \hline
$O_{R2}$ &  \multicolumn{3}{|c|}{$TATCACCGCAAGGGATA$}          \\ \hline
$O_{R3}$ &  \multicolumn{3}{|c|}{$TATCACCGCAAGGGATA$}          \\ \hline

             \includegraphics[scale=.50]{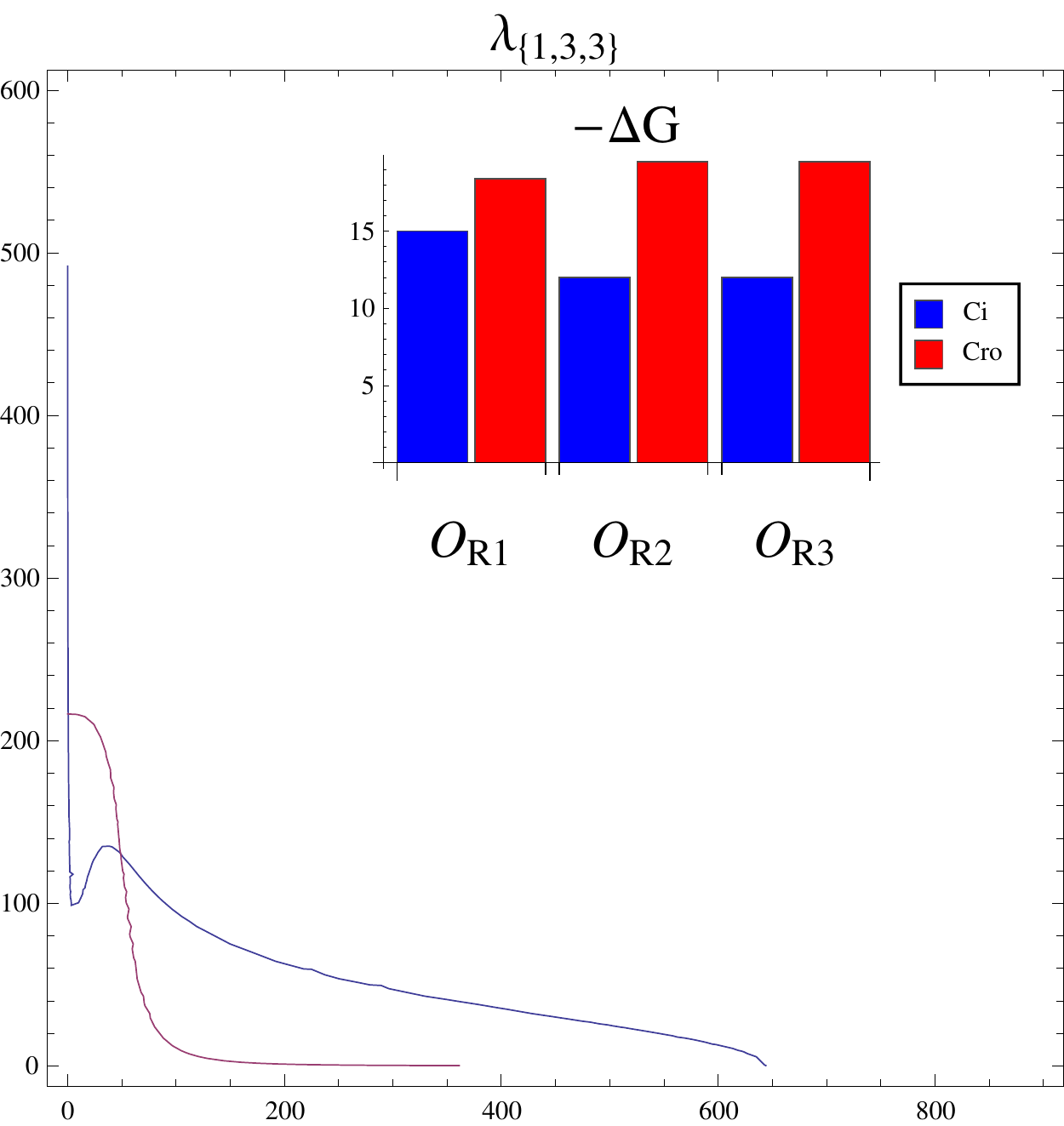}     
         &  \multicolumn{3}{|c|}{\includegraphics[scale=.50 ]{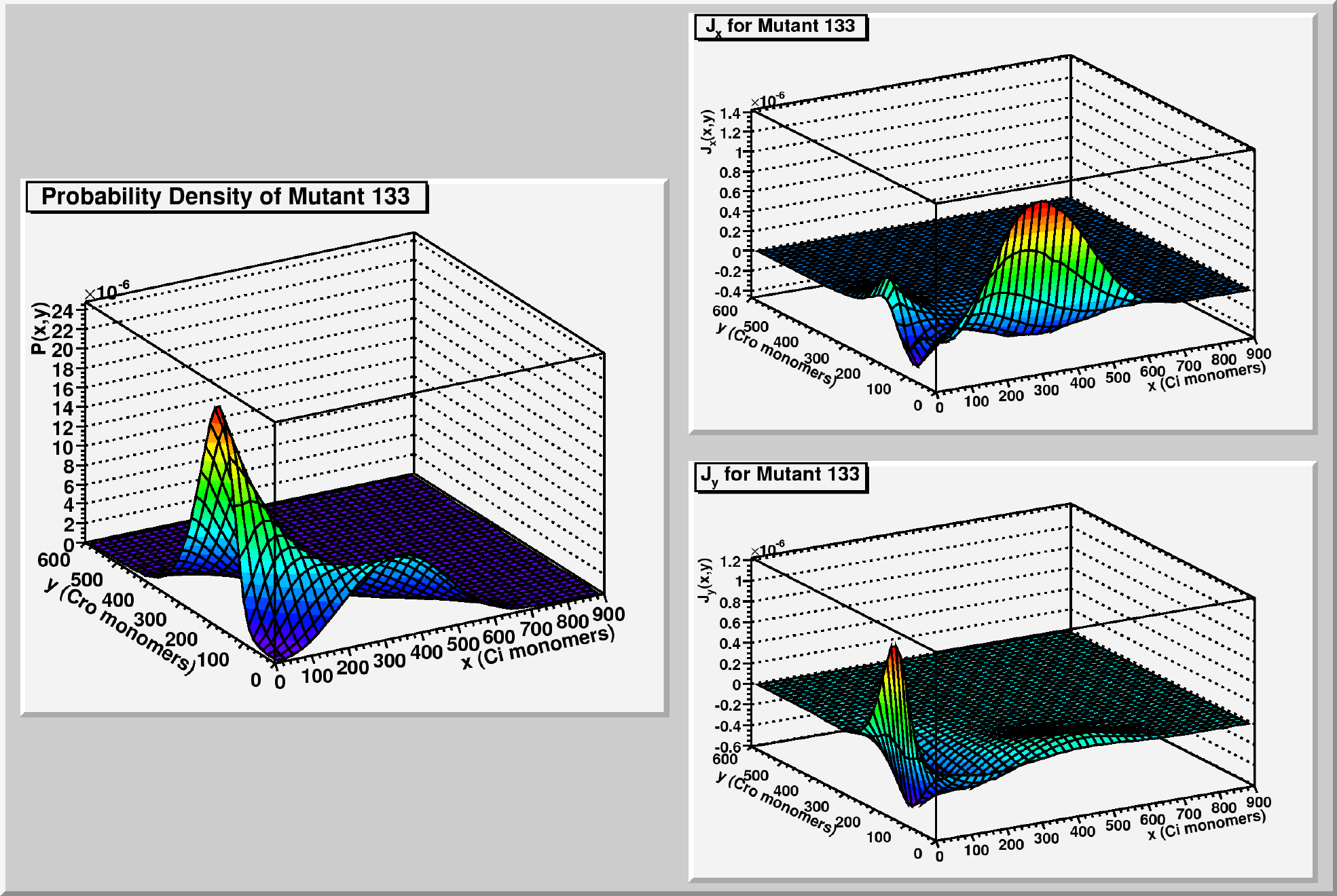}} \\ \hline    
          \multicolumn{4}{|c|}{\includegraphics[scale=.50 ]{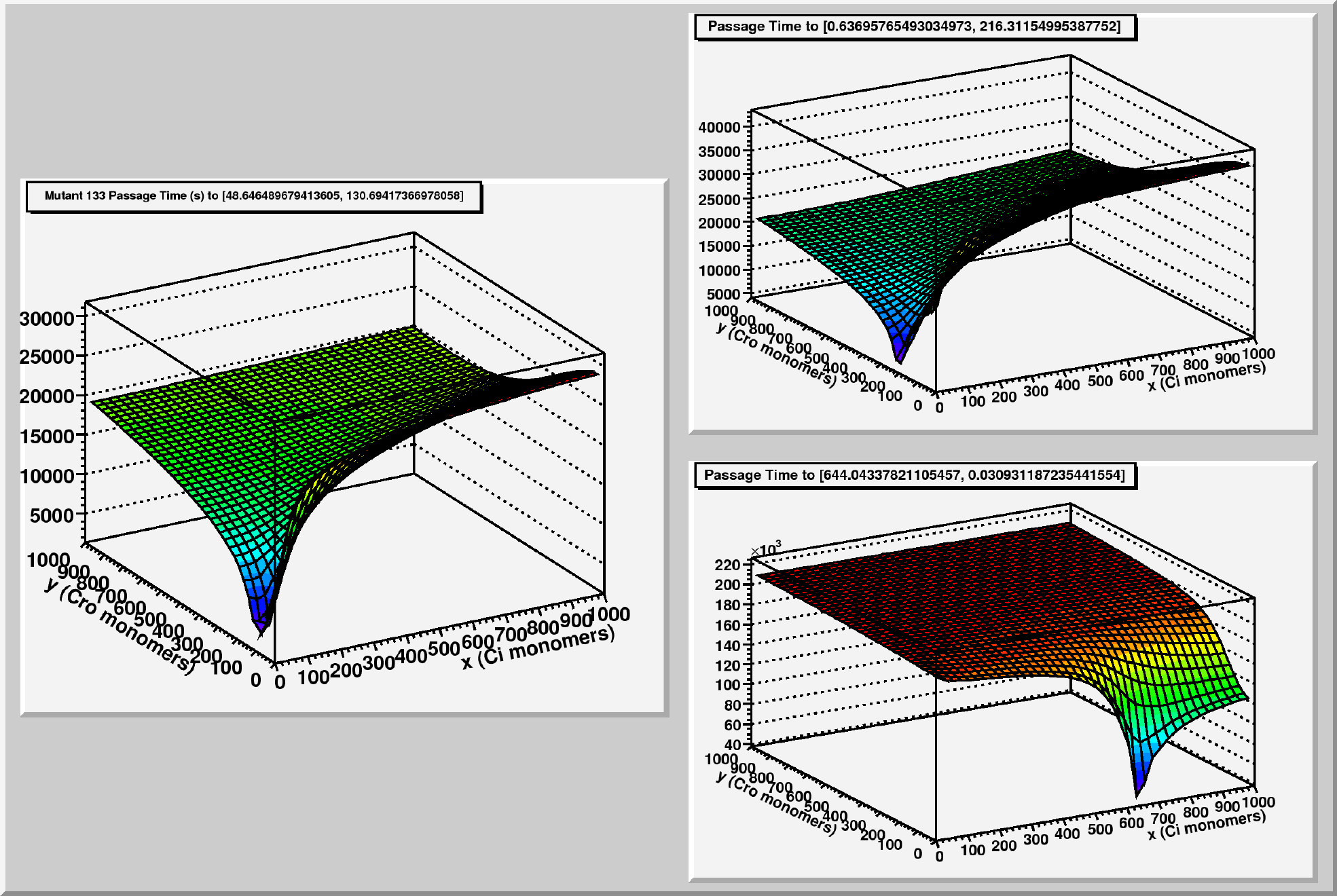}} \\ \hline     
           \multicolumn{3}{|c|}{$\lambda_{133}$ Assorted Expectation Values} & ratio to $\lambda_{123}$   \\ \hline
	 $<x>$ (Ci protein) &  $\frac{\int_\Omega x \rho(x,y) dxdy}{\int_\Omega \rho(x,y) dxdy}$ & 93.227 & 1.002 \\ \hline          
             $<y>$ (Cro protein) & $\frac{\int_\Omega y \rho(x,y) dxdy}{\int_\Omega \rho(x,y) dxdy}$ & 137.624 & 0.8106 \\ \hline     
	 $<\tau_0>$ (seconds to ts) &  $\frac{\int_\Omega \tau_0(x,y) \rho(x,y) dxdy}{\int_\Omega \rho(x,y) dxdy}$ & 12264.638 &0.578  \\ \hline
	 $<\tau_1>$ (seconds to lytic) &  $\frac{\int_\Omega \tau_1(x,y) \rho(x,y) dxdy}{\int_\Omega \rho(x,y) dxdy}$ & 21628.286 & 0.232 \\ \hline
	 $<\tau_2>$ (seconds to lysogenic) &  $\frac{\int_\Omega \tau_2(x,y) \rho(x,y) dxdy}{\int_\Omega \rho(x,y) dxdy}$ & 198518.768 &0.752  \\ \hline
         \hline
        \end{tabular}
}
    \caption{\bf {Properties of mutant $\lambda_{133}$}}
        \label{tab:l133}
    \end{table}%

\begin{table}[ht]
\noindent
     \resizebox{!}{9cm}{
        \begin{tabular}{|c|c|c|c|}
          \hline
          \multicolumn{4}{|c|}{$\lambda_{211}$ steady state and first passage time distributions}
          \\ \hline \hline
$O_{R1}$ &  \multicolumn{3}{|c|}{$TAACACCGTGCGTGTTG$}          \\ \hline
$O_{R2}$ &  \multicolumn{3}{|c|}{$TATCACCGCCAGAGGTA$}          \\ \hline
$O_{R3}$ &  \multicolumn{3}{|c|}{$TATCACCGCCAGAGGTA$}          \\ \hline
             \includegraphics[scale=.50]{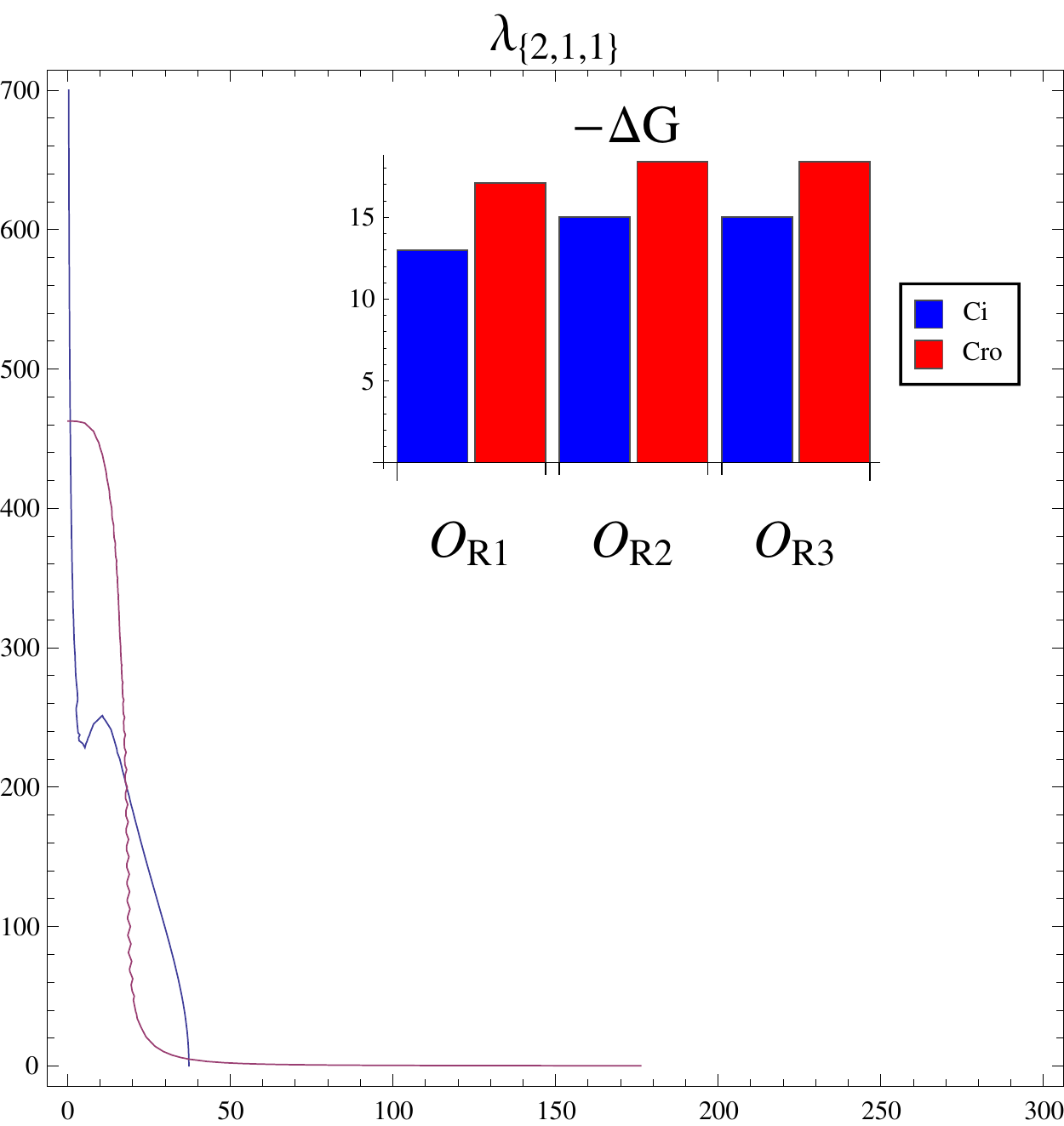}     
         &  \multicolumn{3}{|c|}{\includegraphics[scale=.50 ]{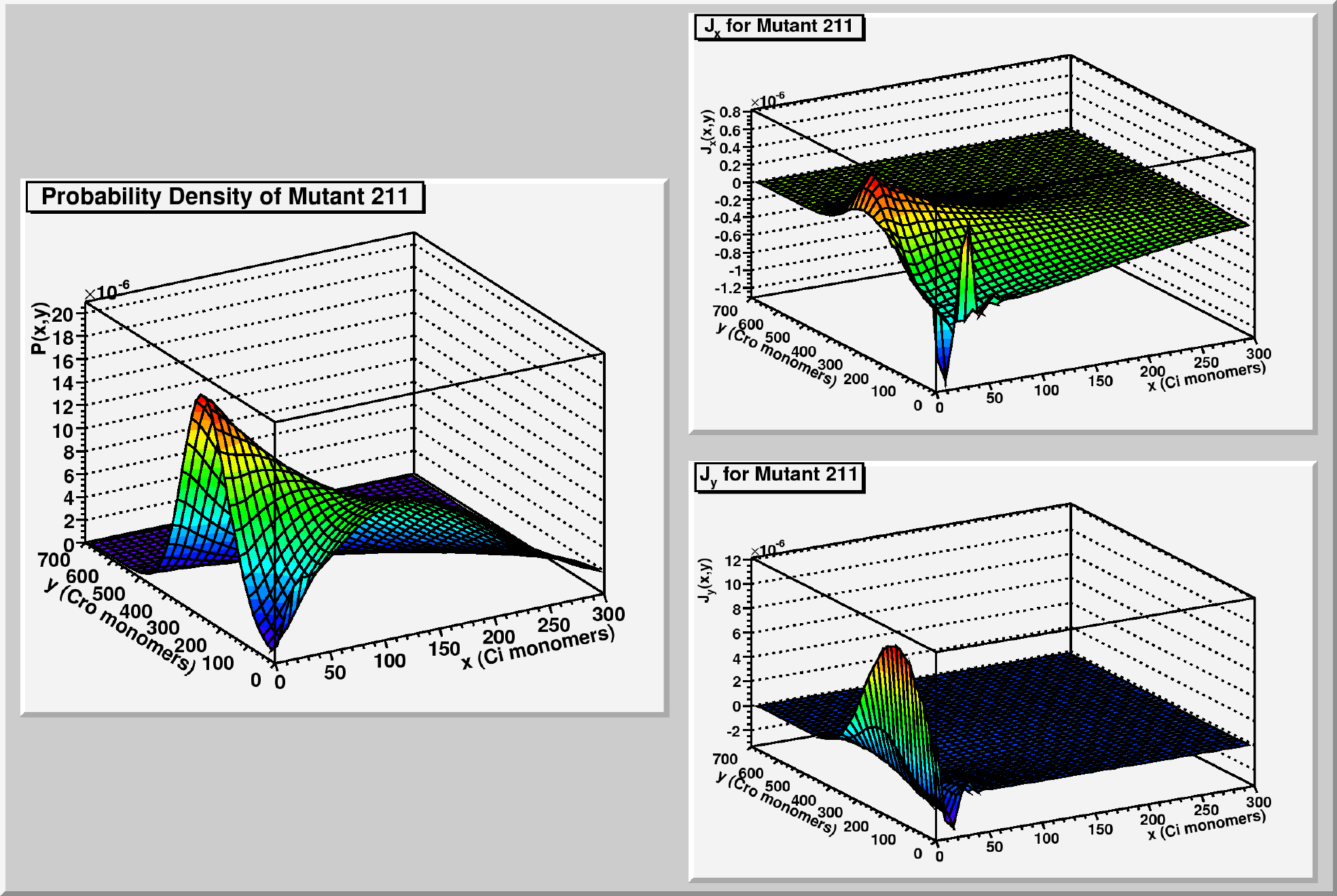}} \\ \hline    
          \multicolumn{4}{|c|}{\includegraphics[scale=.50 ]{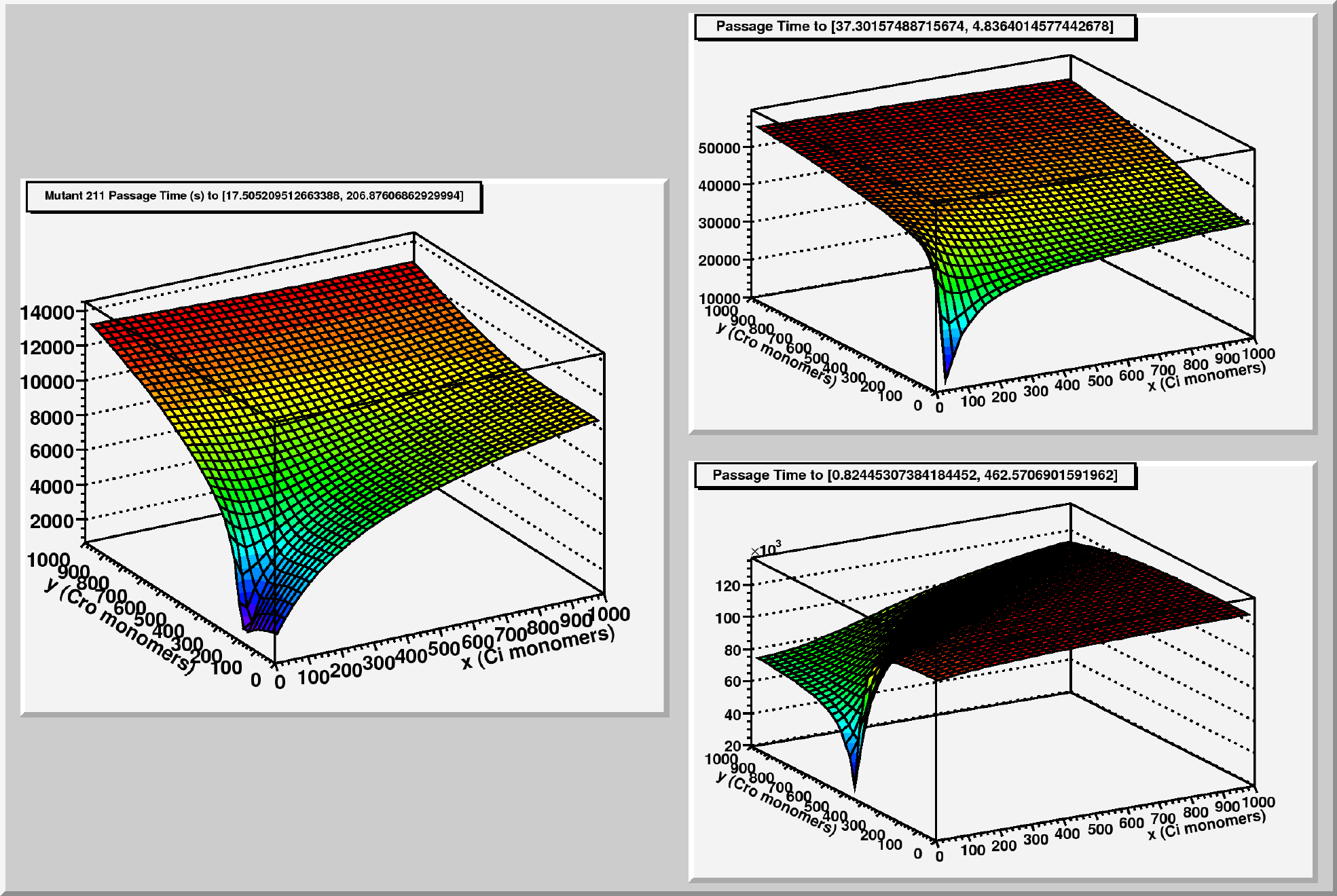}} \\ \hline     
           \multicolumn{3}{|c|}{$\lambda_{211}$ Assorted Expectation Values} & ratio to $\lambda_{123}$   \\ \hline
	 $<x>$ (Ci protein) &  $\frac{\int_\Omega x \rho(x,y) dxdy}{\int_\Omega \rho(x,y) dxdy}$ & 329.760 & 3.543 \\ \hline          
             $<y>$ (Cro protein) & $\frac{\int_\Omega y \rho(x,y) dxdy}{\int_\Omega \rho(x,y) dxdy}$ & 120.570 & 0.710 \\ \hline     
	 $<\tau_0>$ (seconds to lytic) &  $\frac{\int_\Omega \tau_0(x,y) \rho(x,y) dxdy}{\int_\Omega \rho(x,y) dxdy}$ & 7388.126 & 0.079 \\ \hline
	 $<\tau_1>$ (seconds to ts) &  $\frac{\int_\Omega \tau_1(x,y) \rho(x,y) dxdy}{\int_\Omega \rho(x,y) dxdy}$ & 39165.111 & 1.846 \\ \hline
	 $<\tau_2>$ (seconds to lysogenic) &  $\frac{\int_\Omega \tau_2(x,y) \rho(x,y) dxdy}{\int_\Omega \rho(x,y) dxdy}$ & 122448.847 & 0.464   \\ \hline
         \hline
        \end{tabular}
}
     \caption{\bf {Properties of mutant $\lambda_{211}$}}
         \label{tab:l211}
     \end{table}%

\begin{table}[ht]
\noindent
     \resizebox{!}{9cm}{
        \begin{tabular}{|c|c|c|c|}
          \hline
          \multicolumn{4}{|c|}{$\lambda_{213}$ steady state and first passage time distributions}
          \\ \hline \hline
$O_{R1}$ &  \multicolumn{3}{|c|}{$TAACACCGTGCGTGTTG$}          \\ \hline
$O_{R2}$ &  \multicolumn{3}{|c|}{$TATCACCGCCAGAGGTA$}          \\ \hline
$O_{R3}$ &  \multicolumn{3}{|c|}{$TATCACCGCAAGGGATA$}          \\ \hline
             \includegraphics[scale=.50]{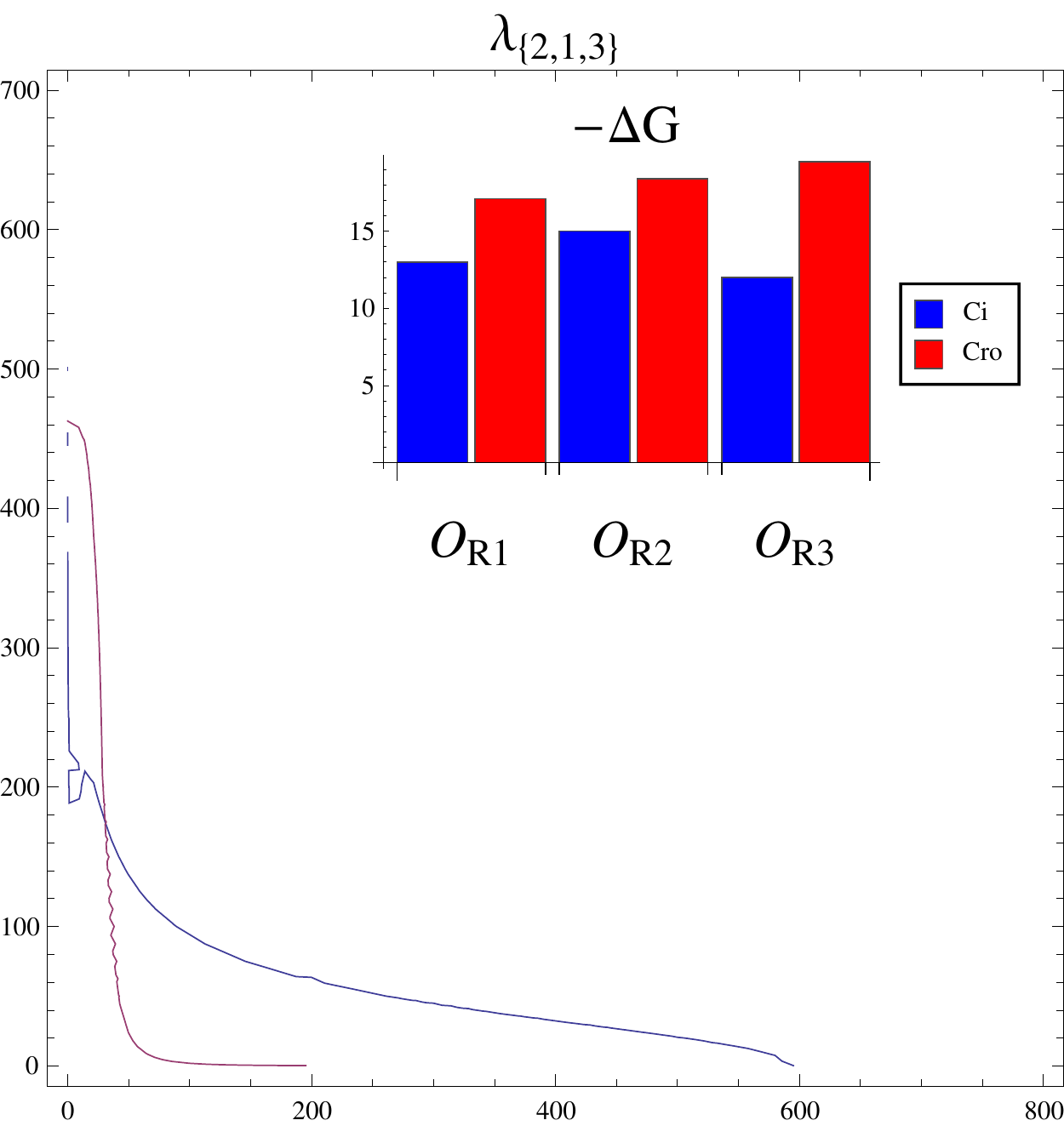}     
         &  \multicolumn{3}{|c|}{\includegraphics[scale=.50 ]{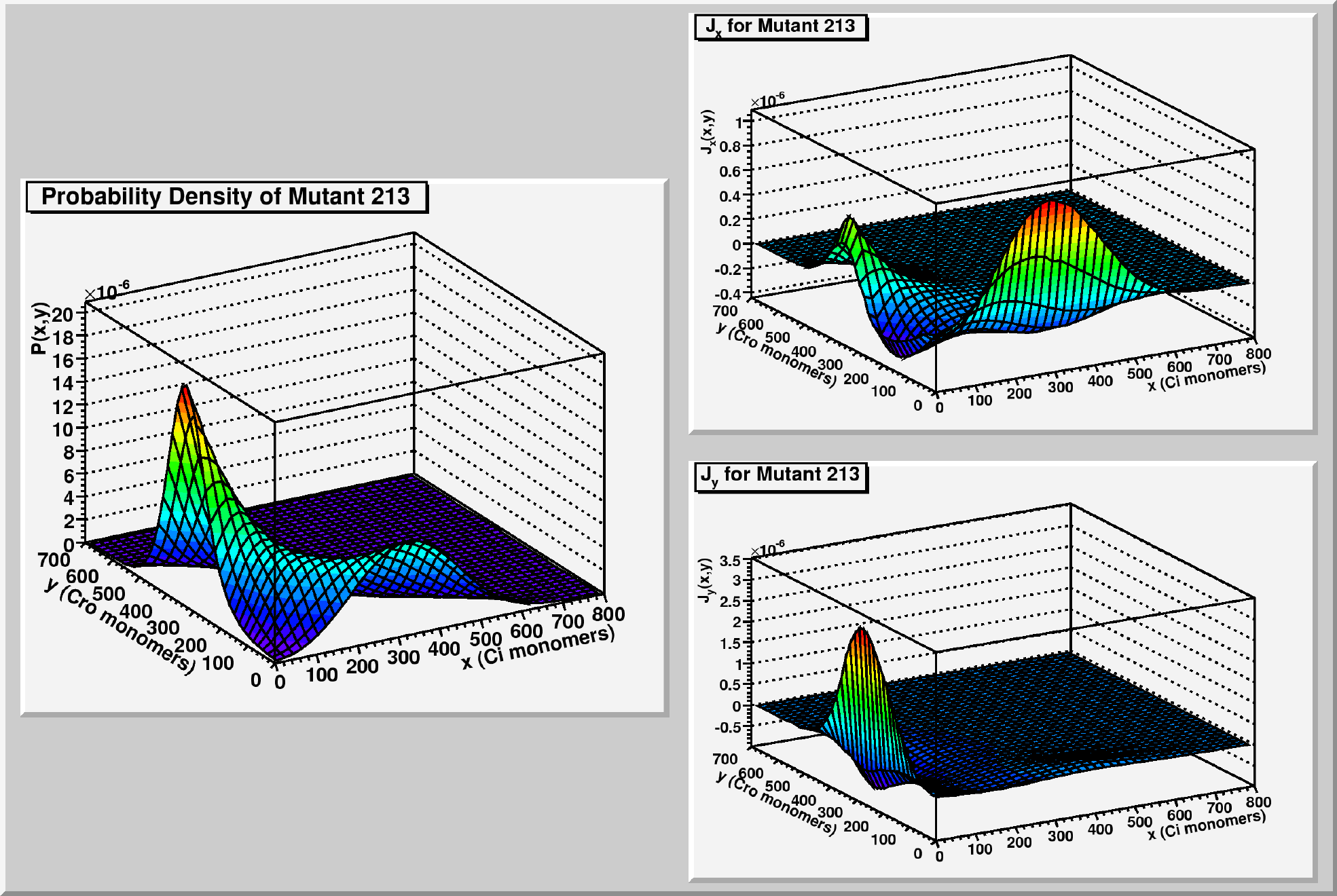}} \\ \hline    
          \multicolumn{4}{|c|}{\includegraphics[scale=.50 ]{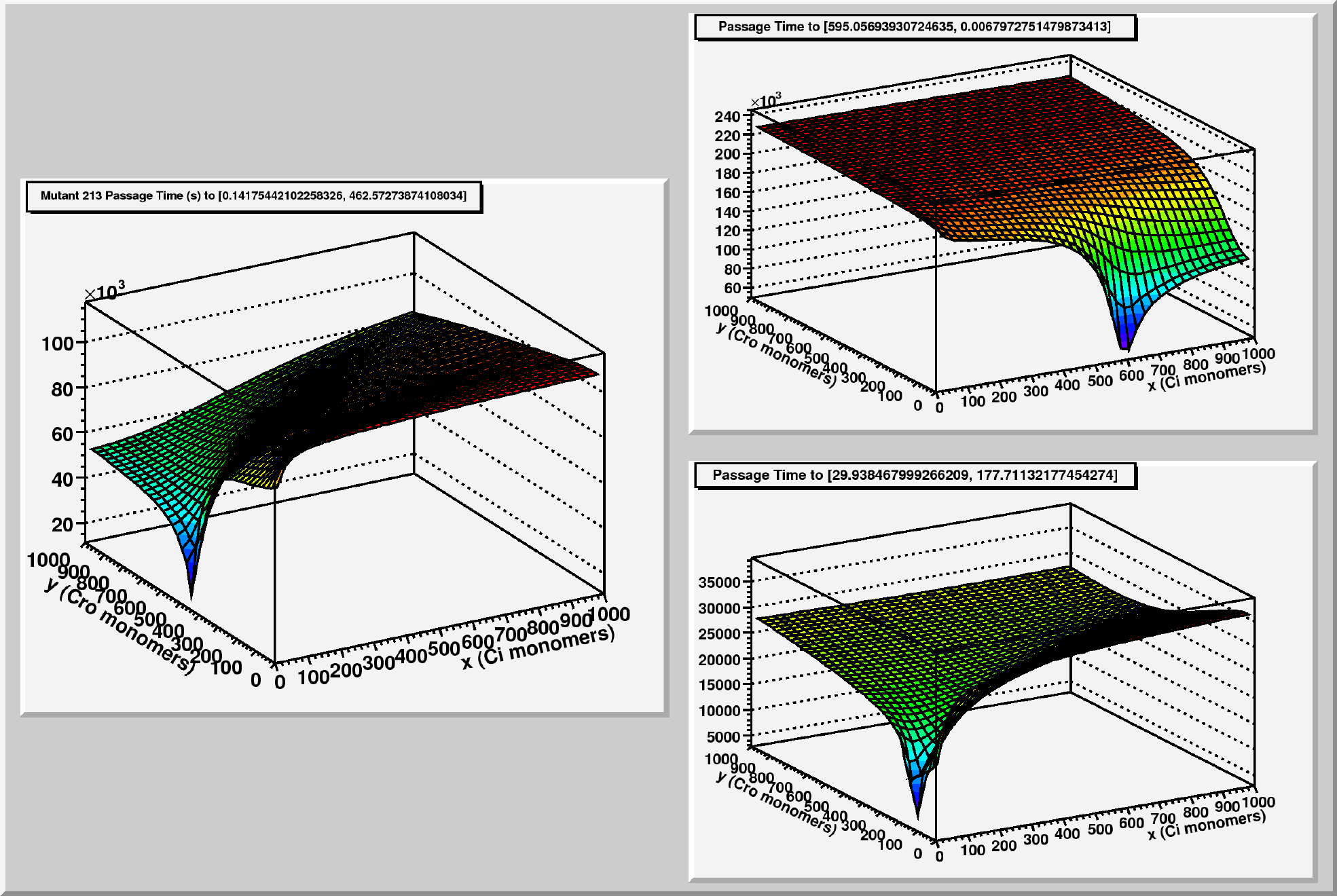}} \\ \hline     
           \multicolumn{3}{|c|}{$\lambda_{213}$ Assorted Expectation Values}& ratio to $\lambda_{123}$       \\ \hline
	 $<x>$ (Ci protein) &  $\frac{\int_\Omega x \rho(x,y) dxdy}{\int_\Omega \rho(x,y) dxdy}$ & 105.183 & 1.130 \\ \hline          
             $<y>$ (Cro protein) & $\frac{\int_\Omega y \rho(x,y) dxdy}{\int_\Omega \rho(x,y) dxdy}$ & 141.037 & 0.831 \\ \hline     
	 $<\tau_0>$ (seconds to lytic) &  $\frac{\int_\Omega \tau_0(x,y) \rho(x,y) dxdy}{\int_\Omega \rho(x,y) dxdy}$ & 90915.013 & 0.974  \\ \hline
	 $<\tau_1>$ (seconds to lysogenic) &  $\frac{\int_\Omega \tau_1(x,y) \rho(x,y) dxdy}{\int_\Omega \rho(x,y) dxdy}$ & 212601.812 & 0.806 \\ \hline
	 $<\tau_2>$ (seconds to ts) &  $\frac{\int_\Omega \tau_2(x,y) \rho(x,y) dxdy}{\int_\Omega \rho(x,y) dxdy}$ & 20741.420 & 0.978 \\ \hline
         \hline
        \end{tabular}
}
    \caption{\bf {Properties of mutant $\lambda_{213}$}}
        \label{tab:l213}
    \end{table}%

\begin{table}[h]
\noindent
     \resizebox{!}{8cm}{
        \begin{tabular}{|c|c|c|}
          \hline
          \multicolumn{3}{|c|}{$\lambda_{222}$ steady state and first passage time distributions}
          \\ \hline \hline
             \includegraphics[scale=.50]{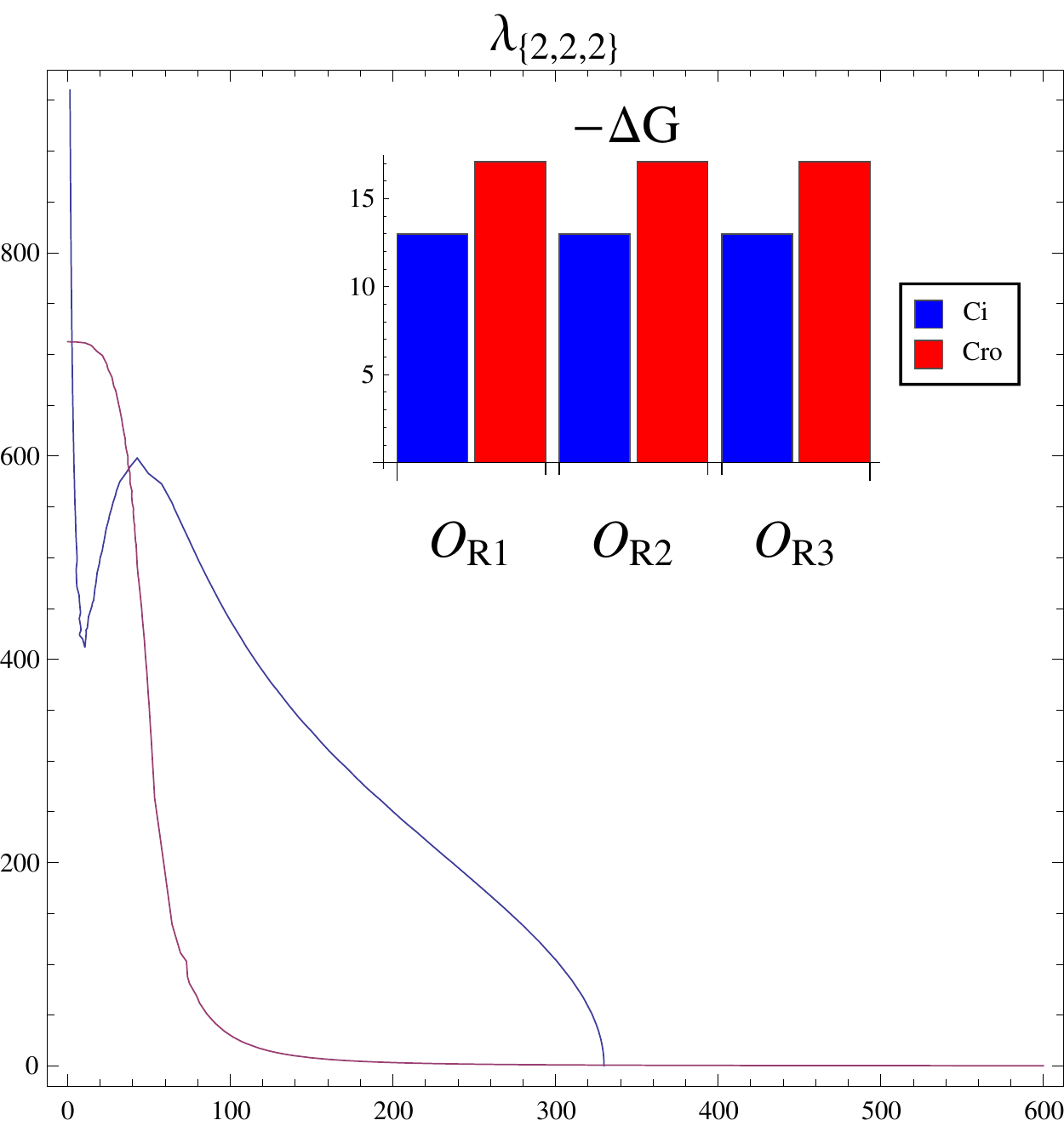}     
         &  \multicolumn{2}{|c|}{\includegraphics[scale=.50 ]{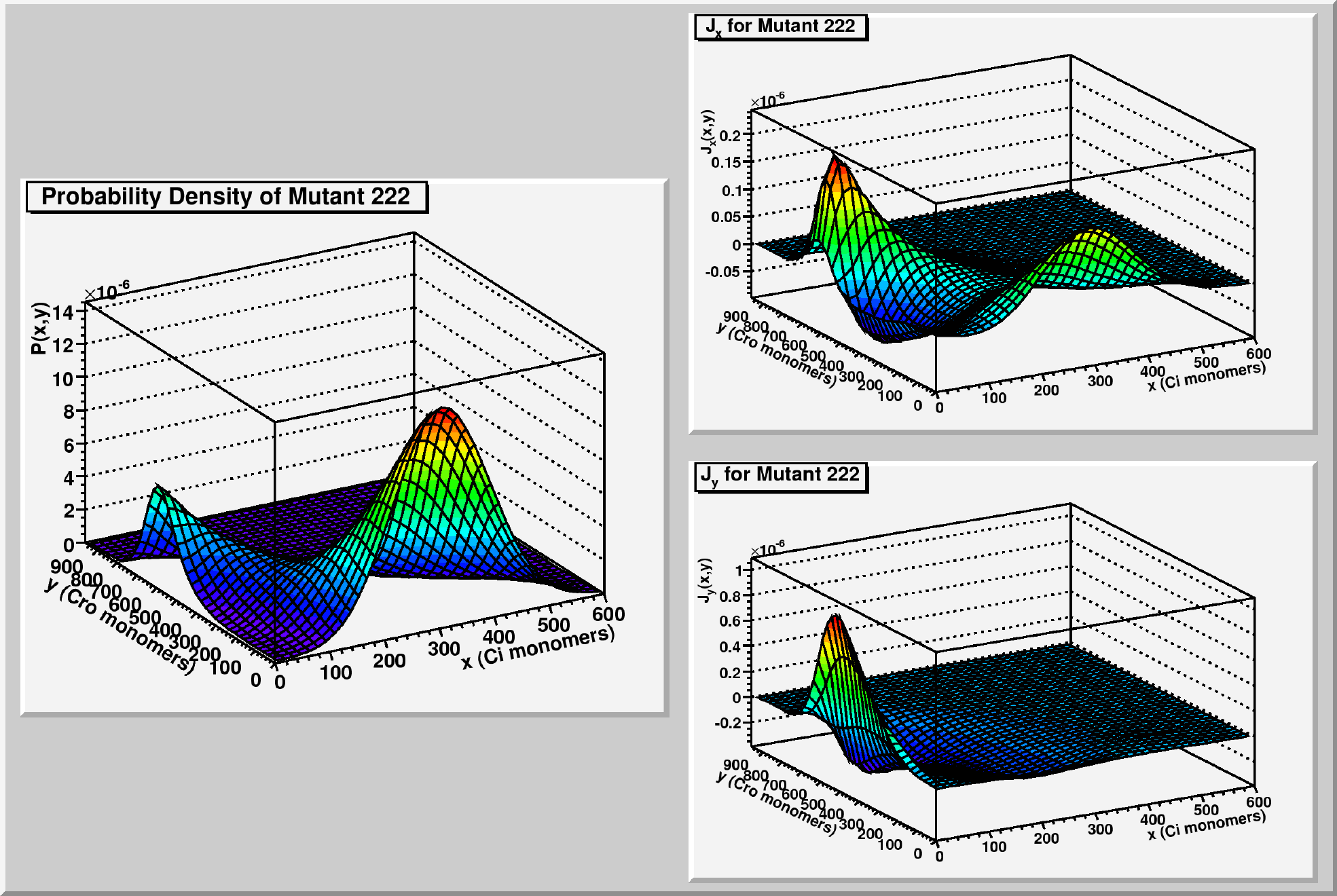}} \\ \hline    
          \multicolumn{3}{|c|}{\includegraphics[scale=.50 ]{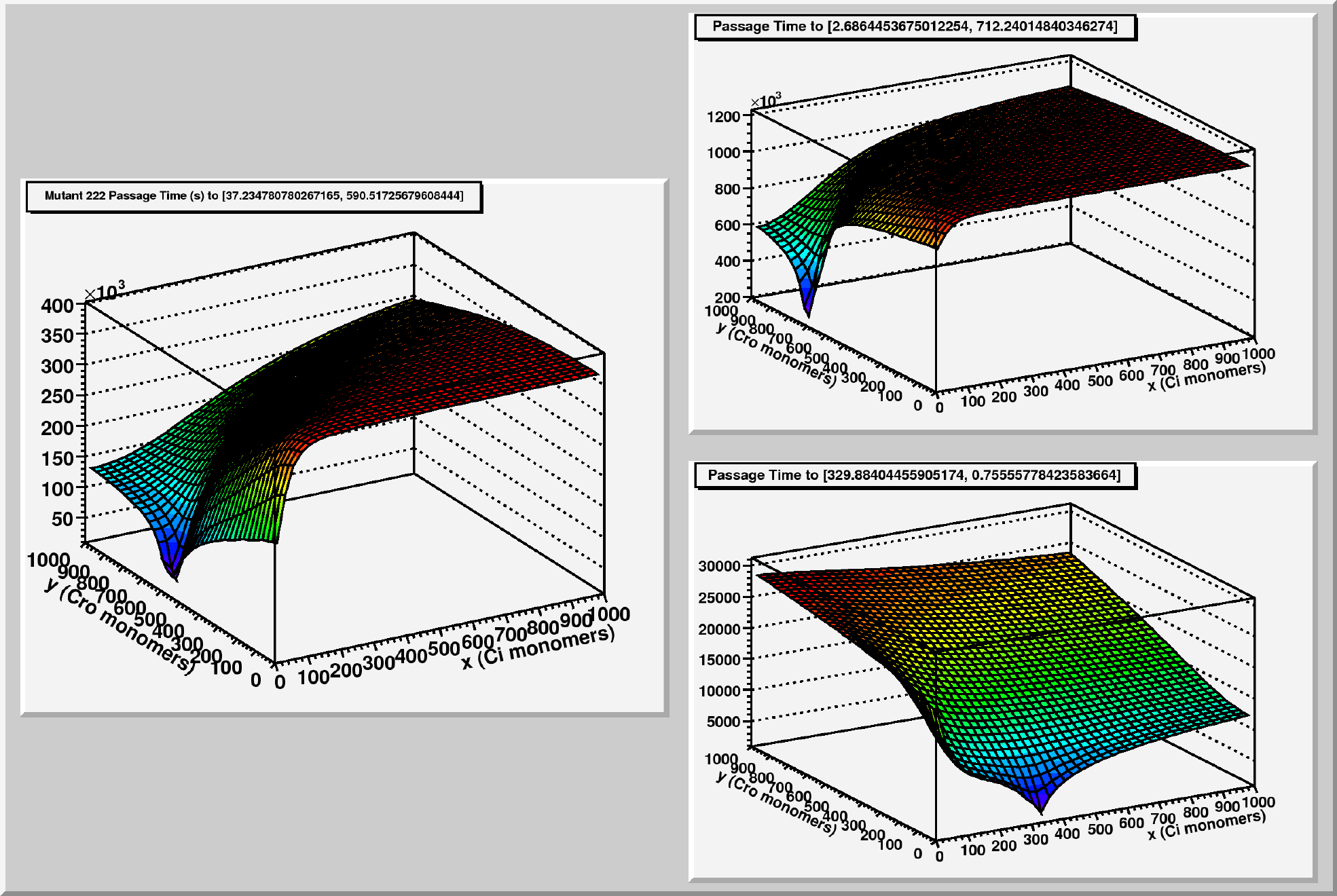}} \\ \hline     
           \multicolumn{3}{|c|}{$\lambda_{222}$ Assorted Expectation Values}    \\ \hline
	 $<x>$ (Ci protein) &  $\int_\Omega x \rho(x,y) dxdy/\int_\Omega \rho(x,y) dxdy$ & 230.600  \\ \hline          
             $<y>$ (Cro protein) & $\int_\Omega y \rho(x,y) dxdy/\int_\Omega \rho(x,y) dxdy$ & 216.398 \\ \hline     
	 $<\tau_0>$ (seconds) &  $\int_\Omega \tau_0(x,y) \rho(x,y) dxdy/\int_\Omega \rho(x,y) dxdy$ & 305052.530  \\ \hline
	 $<\tau_1>$ (seconds) &  $\int_\Omega \tau_1(x,y) \rho(x,y) dxdy/\int_\Omega \rho(x,y) dxdy$ & 1065951 \\ \hline
	 $<\tau_2>$ (seconds) &  $\int_\Omega \tau_2(x,y) \rho(x,y) dxdy/\int_\Omega \rho(x,y) dxdy$ & 13739.885   \\ \hline
         \hline
        \end{tabular}
}

    \caption{\bf {Properties of mutant $\lambda_{222}$}}
        \label{tab:l222}
    \end{table}%

\begin{table}[h]
\noindent
     \resizebox{!}{8cm}{
        \begin{tabular}{|c|c|c|}
          \hline
          \multicolumn{3}{|c|}{$\lambda_{223}$ steady state and first passage time distributions}
          \\ \hline \hline
             \includegraphics[scale=.50]{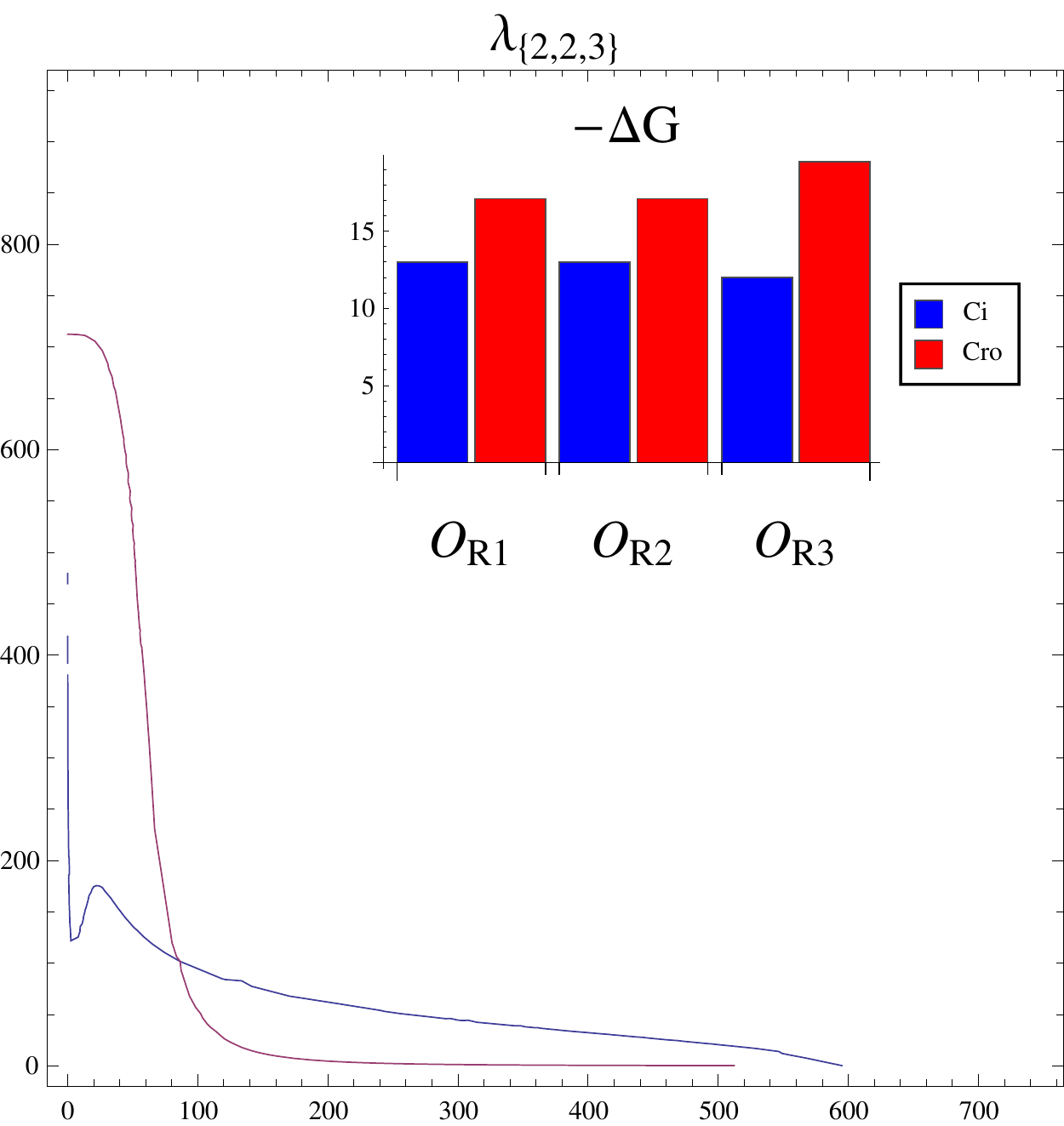}     
         &  \multicolumn{2}{|c|}{\includegraphics[scale=.50 ]{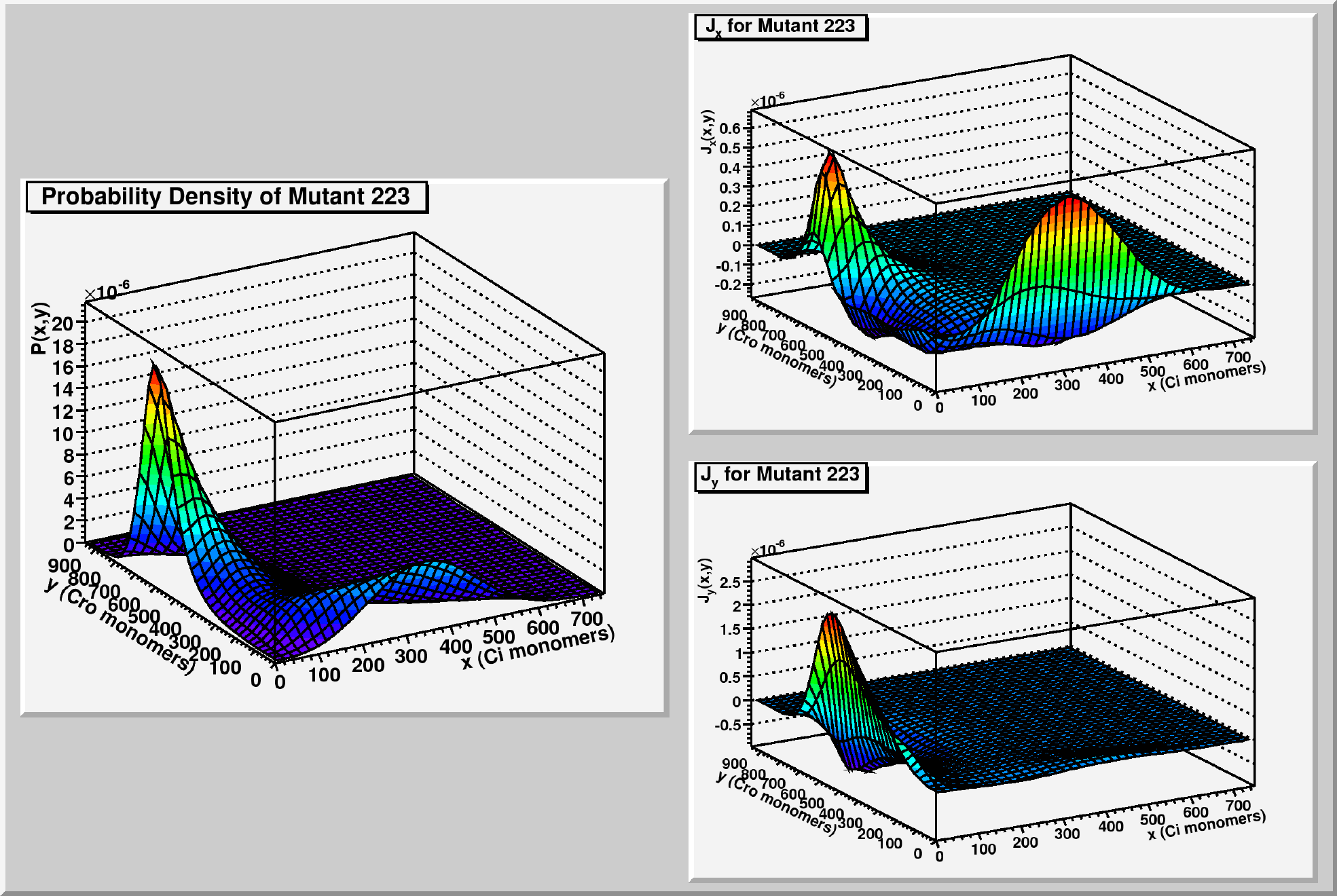}} \\ \hline    
          \multicolumn{3}{|c|}{\includegraphics[scale=.50 ]{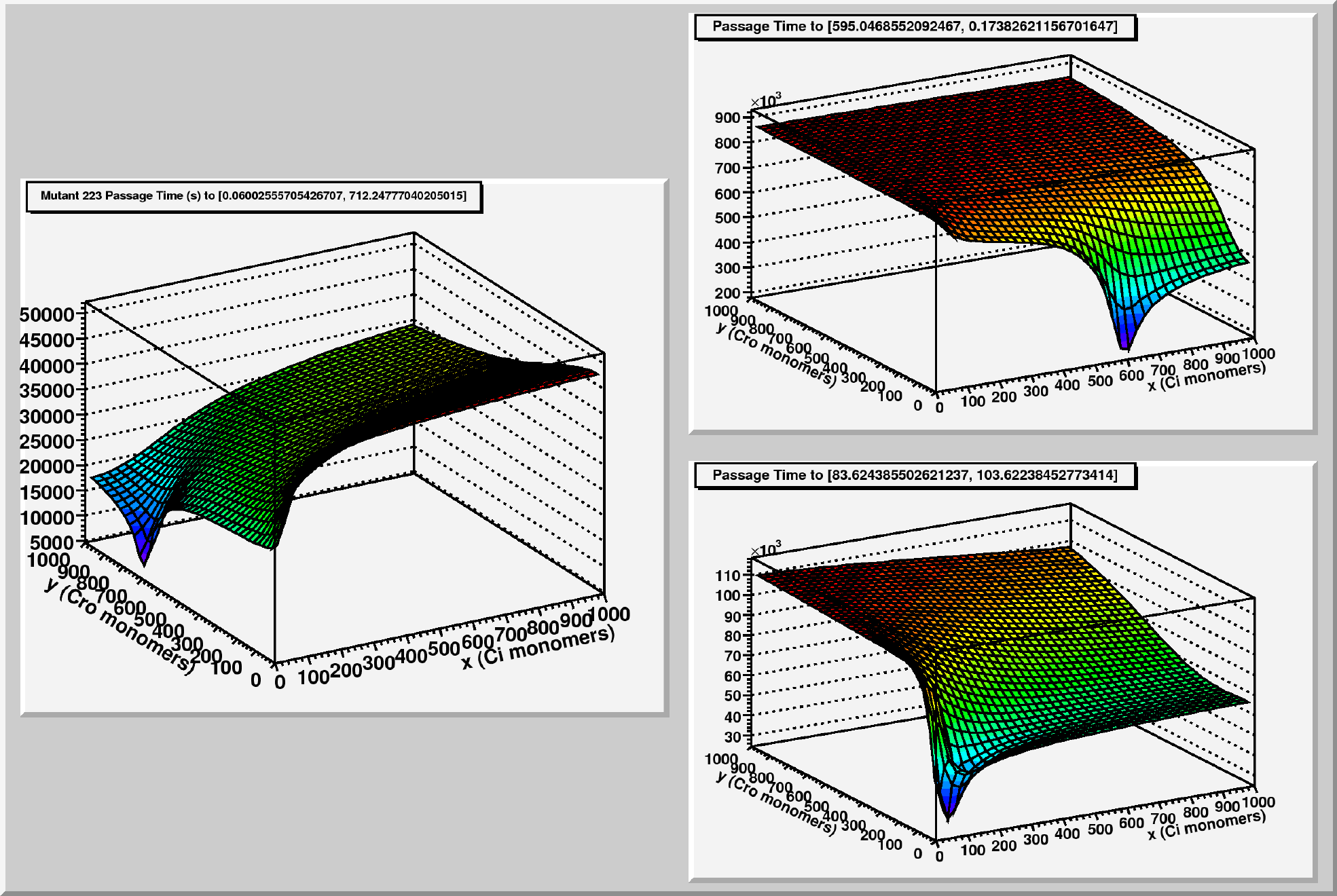}} \\ \hline     
           \multicolumn{3}{|c|}{$\lambda_{223}$ Assorted Expectation Values}    \\ \hline
	 $<x>$ (Ci protein) &  $\int_\Omega x \rho(x,y) dxdy/\int_\Omega \rho(x,y) dxdy$ & 102.346  \\ \hline          
             $<y>$ (Cro protein) & $\int_\Omega y \rho(x,y) dxdy/\int_\Omega \rho(x,y) dxdy$ & 202.666 \\ \hline     
	 $<\tau_0>$ (seconds) &  $\int_\Omega \tau_0(x,y) \rho(x,y) dxdy/\int_\Omega \rho(x,y) dxdy$ & 30623.759  \\ \hline
	 $<\tau_1>$ (seconds) &  $\int_\Omega \tau_1(x,y) \rho(x,y) dxdy/\int_\Omega \rho(x,y) dxdy$ & 826935.205 \\ \hline
	 $<\tau_2>$ (seconds) &  $\int_\Omega \tau_2(x,y) \rho(x,y) dxdy/\int_\Omega \rho(x,y) dxdy$ & 815868.173  \\ \hline
         \hline
        \end{tabular}
}
    \caption{\bf {Properties of mutant $\lambda_{223}$}}
        \label{tab:l223}
    \end{table}%

\begin{table}[ht]
\noindent
     \resizebox{!}{8cm}{
        \begin{tabular}{|c|c|c|}
          \hline
          \multicolumn{3}{|c|}{$\lambda_{232}$ steady state and first passage time distributions}
          \\ \hline \hline
             \includegraphics[scale=.50]{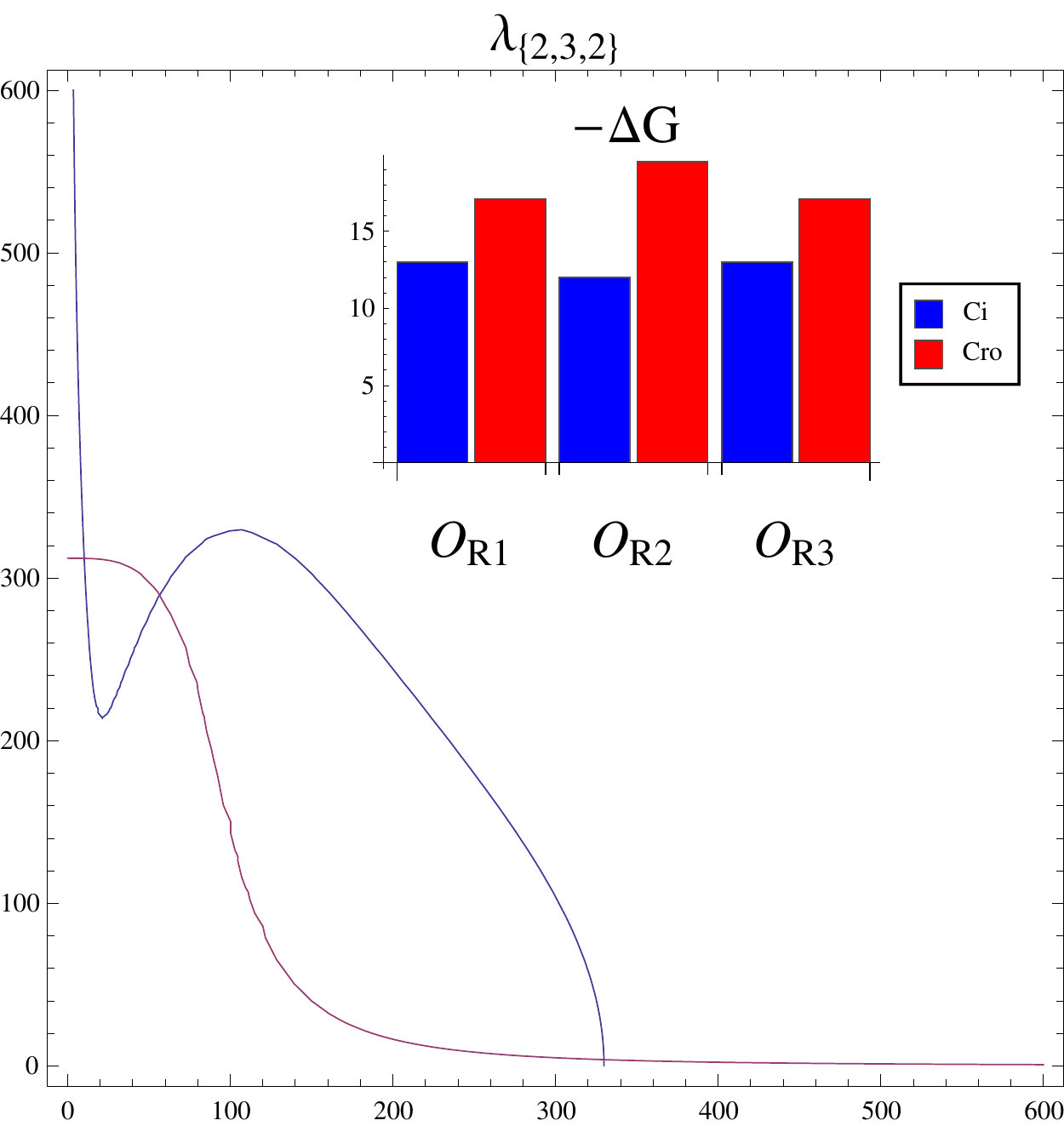}     
         &  \multicolumn{2}{|c|}{\includegraphics[scale=.50 ]{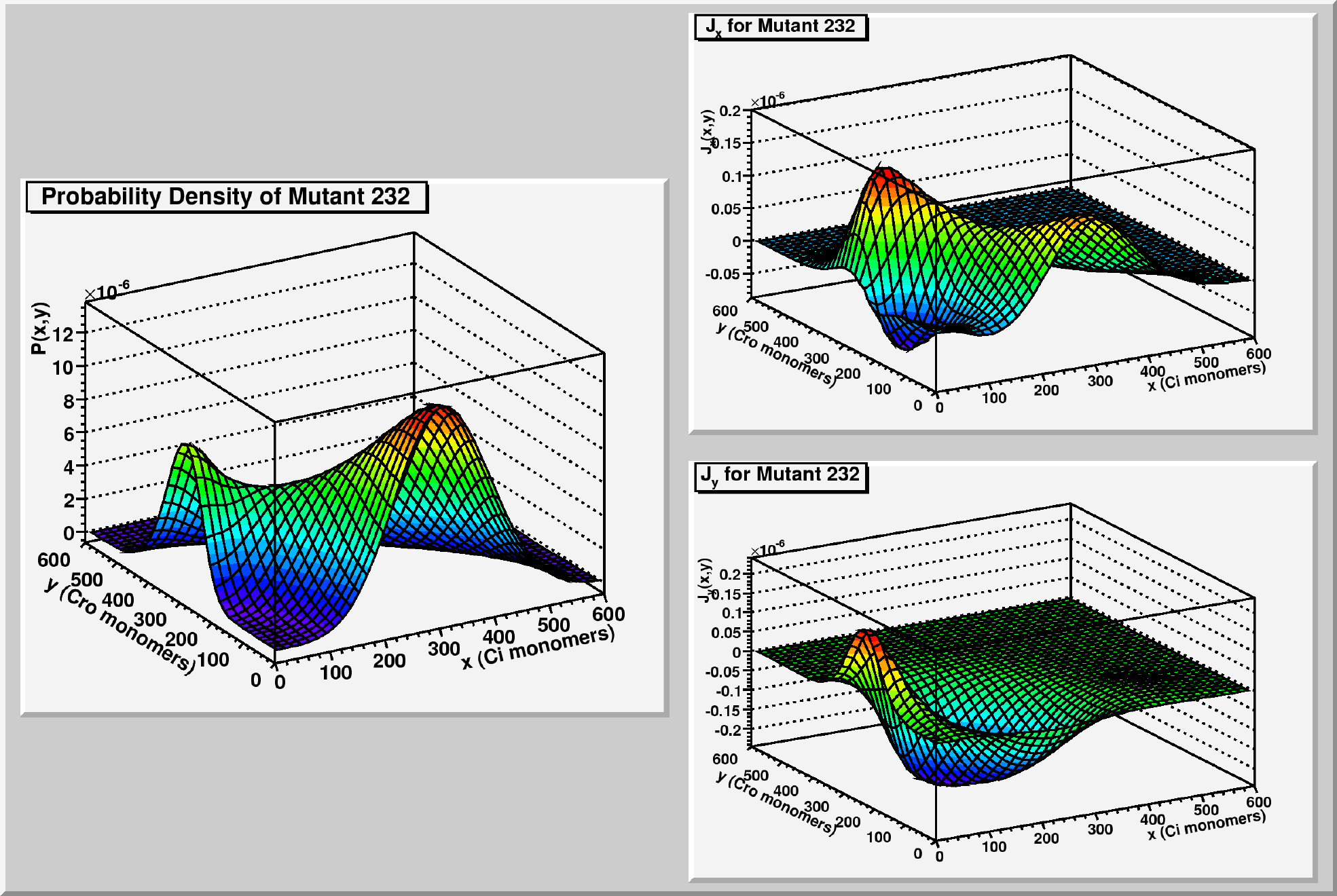}} \\ \hline    
          \multicolumn{3}{|c|}{\includegraphics[scale=.50 ]{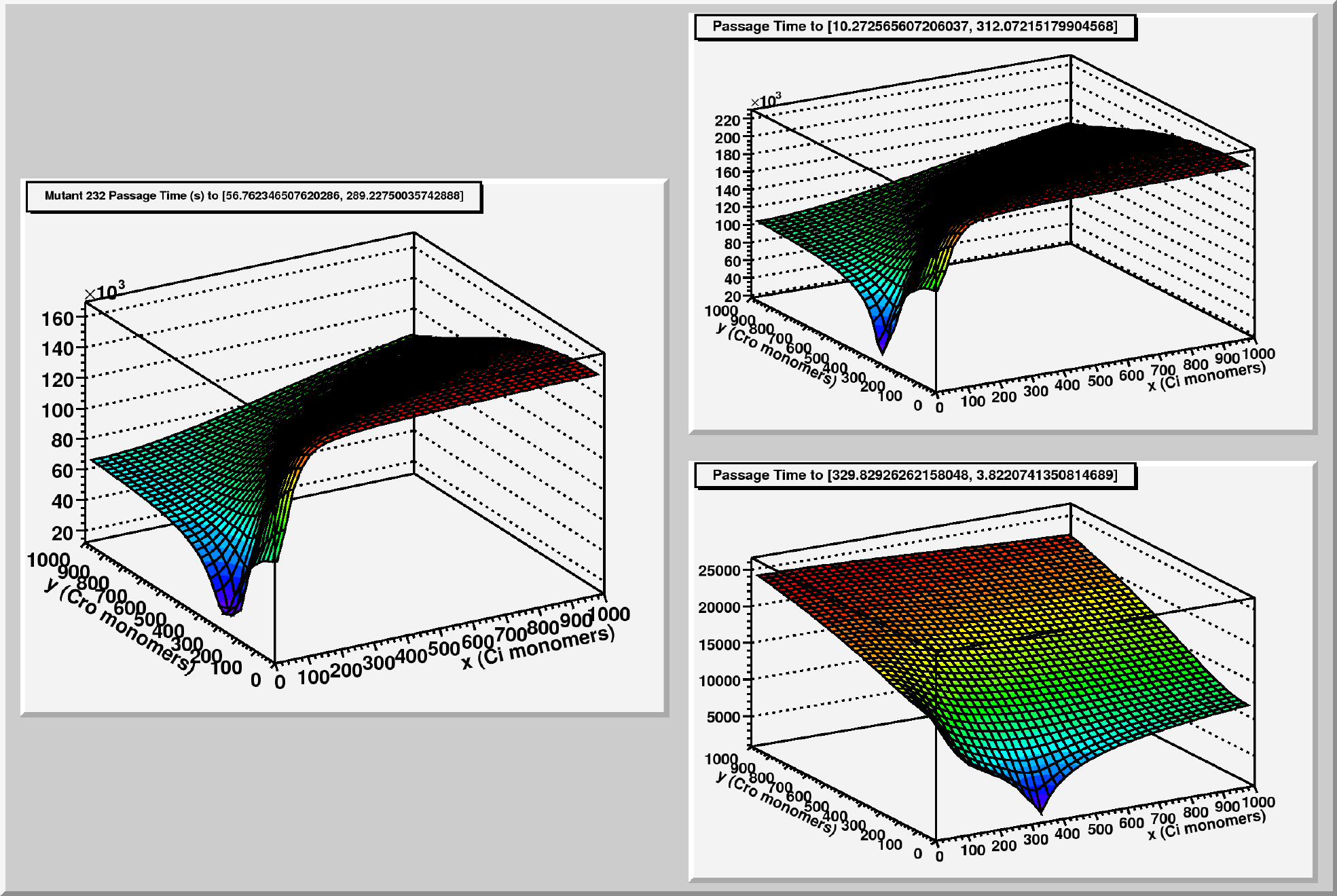}} \\ \hline     
           \multicolumn{3}{|c|}{$\lambda_{232}$ Assorted Expectation Values}    \\ \hline
	 $<x>$ (Ci protein) &  $\int_\Omega x \rho(x,y) dxdy/\int_\Omega \rho(x,y) dxdy$ & 231.352  \\ \hline          
             $<y>$ (Cro protein) & $\int_\Omega y \rho(x,y) dxdy/\int_\Omega \rho(x,y) dxdy$ & 160.462 \\ \hline     
	 $<\tau_0>$ (seconds) &  $\int_\Omega \tau_0(x,y) \rho(x,y) dxdy/\int_\Omega \rho(x,y) dxdy$ & 122528.722  \\ \hline
	 $<\tau_1>$ (seconds) &  $\int_\Omega \tau_1(x,y) \rho(x,y) dxdy/\int_\Omega \rho(x,y) dxdy$ & 175675.148 \\ \hline
	 $<\tau_2>$ (seconds) &  $\int_\Omega \tau_2(x,y) \rho(x,y) dxdy/\int_\Omega \rho(x,y) dxdy$ & 11888.417  \\ \hline
         \hline
        \end{tabular}
}

    \caption{\bf {Properties of mutant $\lambda_{232}$}}
        \label{tab:l232}
    \end{table}%

\begin{table}[ht]
\noindent
     \resizebox{!}{8cm}{
        \begin{tabular}{|c|c|c|}
          \hline
          \multicolumn{3}{|c|}{$\lambda_{233}$ steady state and first passage time distributions}
          \\ \hline \hline
             \includegraphics[scale=.50]{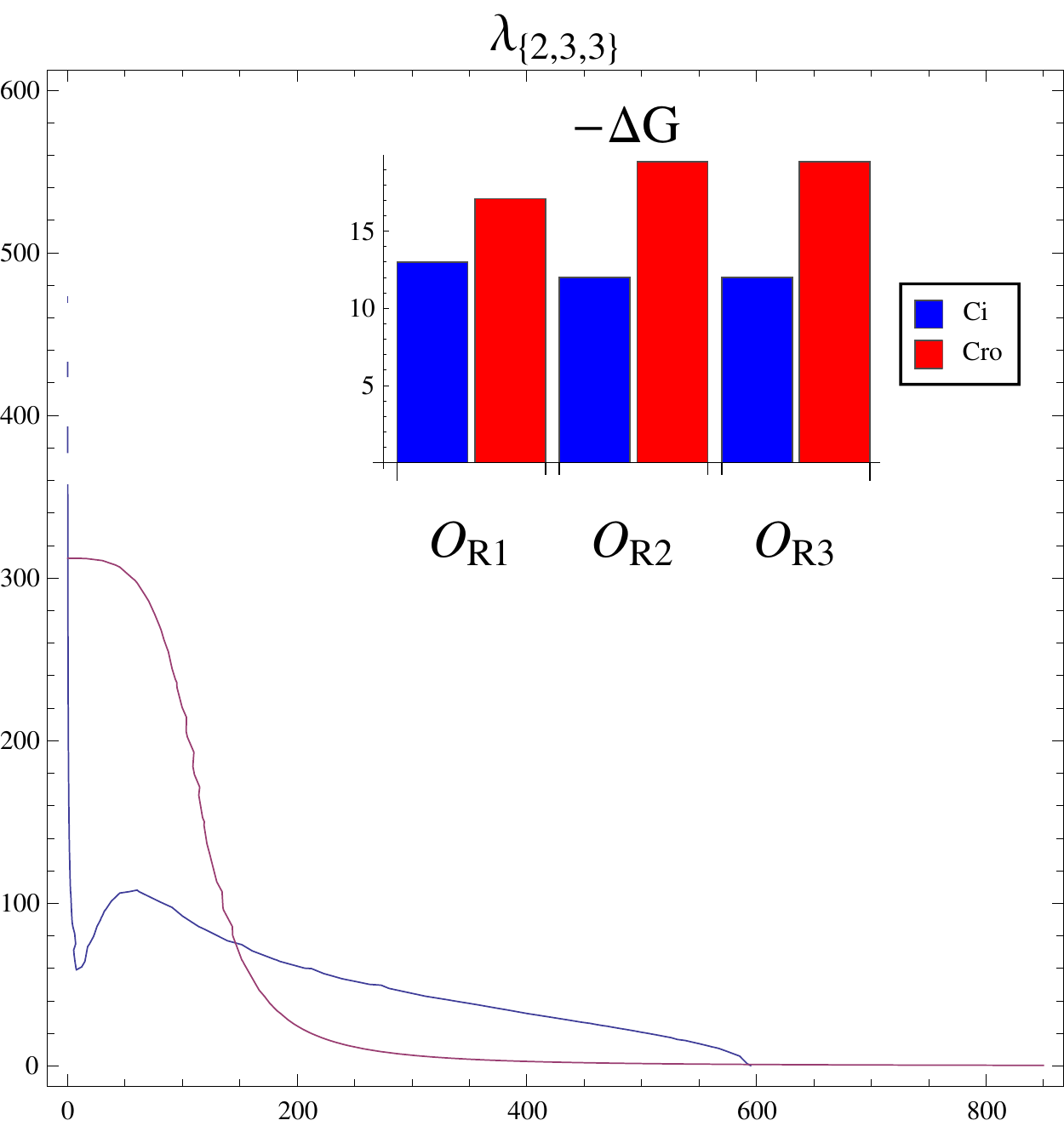}     
         &  \multicolumn{2}{|c|}{\includegraphics[scale=.50 ]{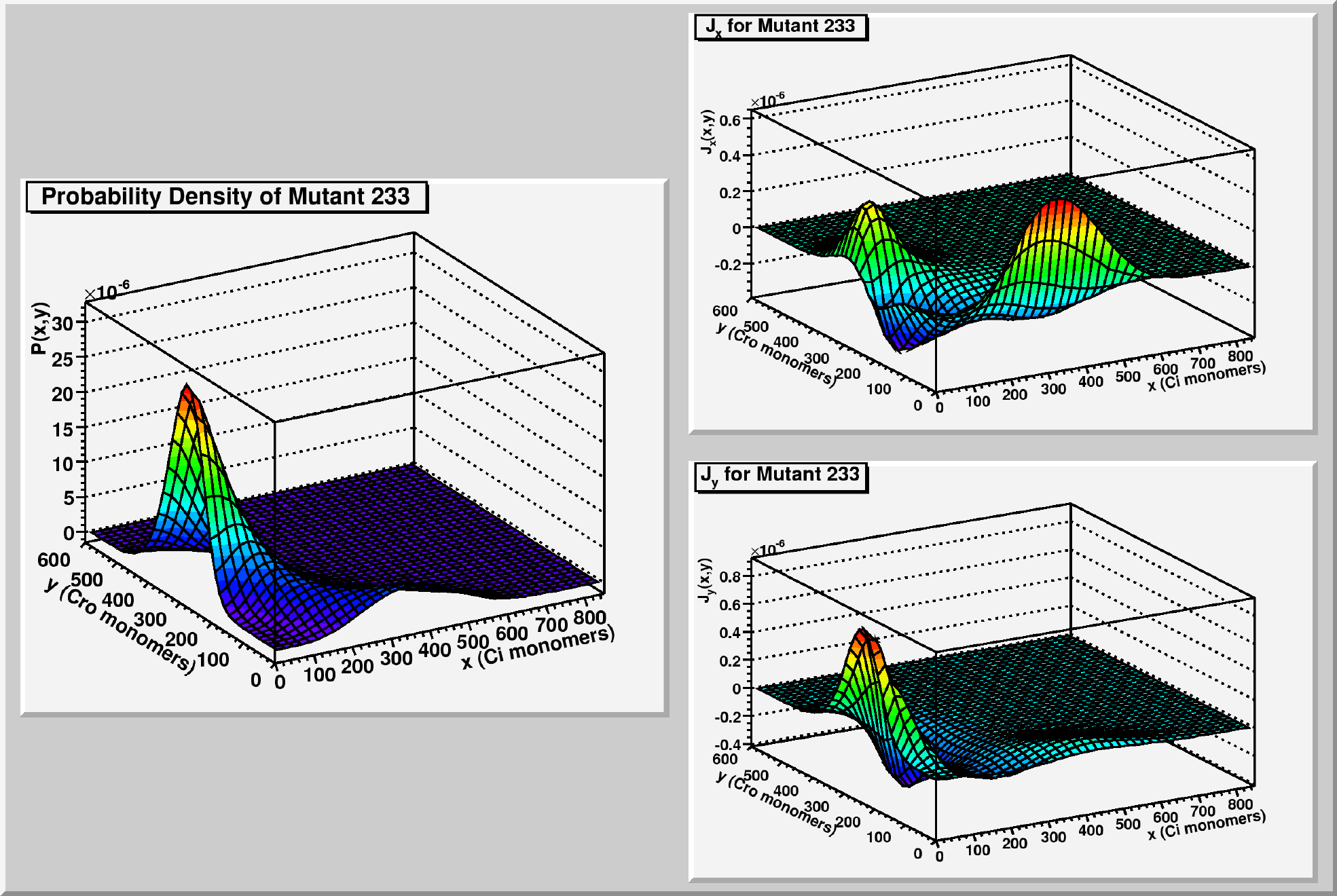}} \\ \hline    
          \multicolumn{3}{|c|}{\includegraphics[scale=.50 ]{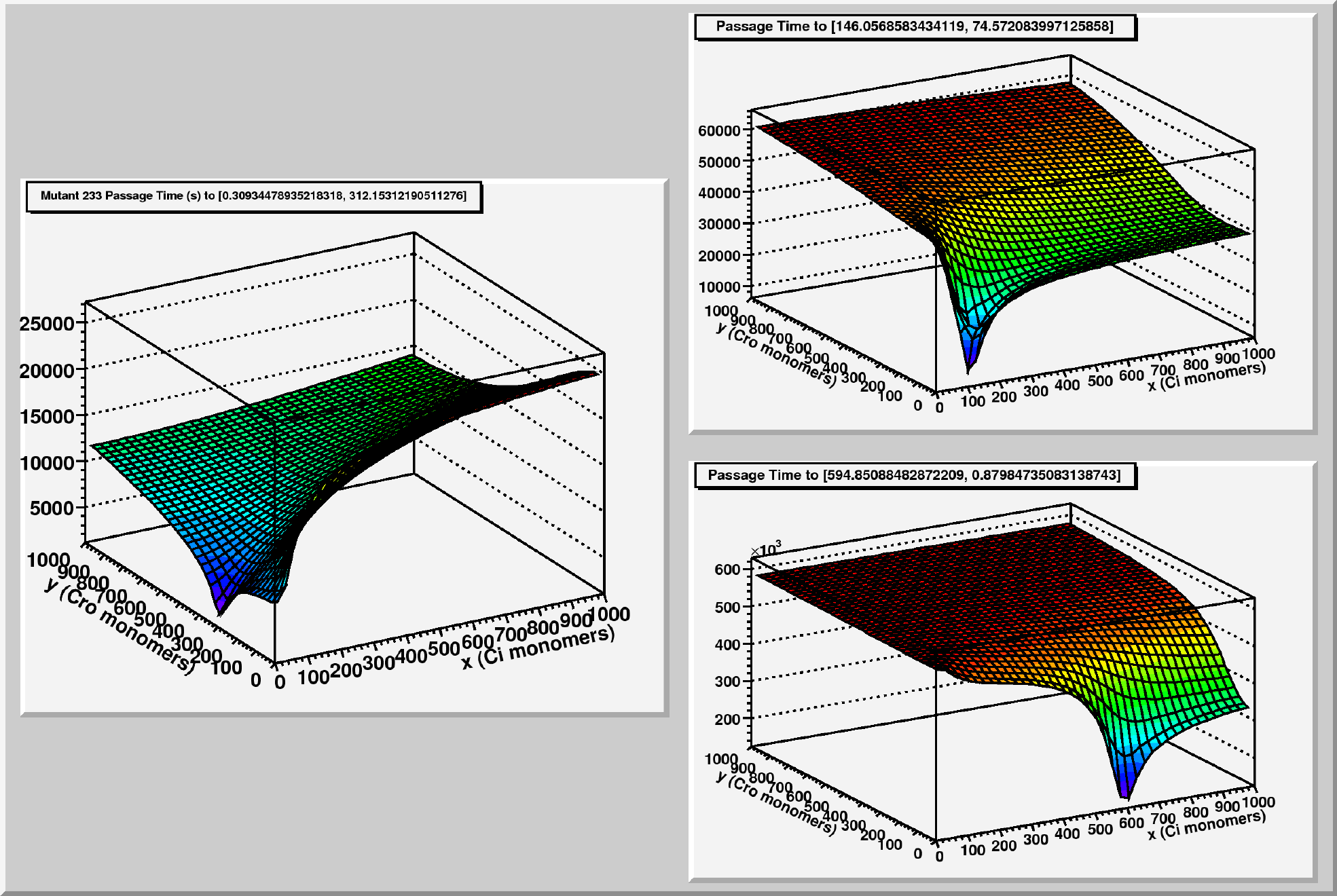}} \\ \hline     
           \multicolumn{3}{|c|}{$\lambda_{233}$ Assorted Expectation Values}    \\ \hline
	 $<x>$ (Ci protein) &  $\int_\Omega x \rho(x,y) dxdy/\int_\Omega \rho(x,y) dxdy$ & 103.439  \\ \hline          
             $<y>$ (Cro protein) & $\int_\Omega y \rho(x,y) dxdy/\int_\Omega \rho(x,y) dxdy$ & 149.505 \\ \hline     
	 $<\tau_0>$ (seconds) &  $\int_\Omega \tau_0(x,y) \rho(x,y) dxdy/\int_\Omega \rho(x,y) dxdy$ & 10320.871  \\ \hline
	 $<\tau_1>$ (seconds) &  $\int_\Omega \tau_1(x,y) \rho(x,y) dxdy/\int_\Omega \rho(x,y) dxdy$ & 43522.861 \\ \hline
	 $<\tau_2>$ (seconds) &  $\int_\Omega \tau_2(x,y) \rho(x,y) dxdy/\int_\Omega \rho(x,y) dxdy$ & 564417.963  \\ \hline
         \hline
        \end{tabular}
}
    \caption{\bf {Properties of mutant $\lambda_{233}$}}
        \label{tab:l233}
    \end{table}%

\begin{table}[ht]
\noindent
     \resizebox{!}{8cm}{
        \begin{tabular}{|c|c|c|}
          \hline
          \multicolumn{3}{|c|}{$\lambda_{312}$ steady state and first passage time distributions}
          \\ \hline \hline
             \includegraphics[scale=.50]{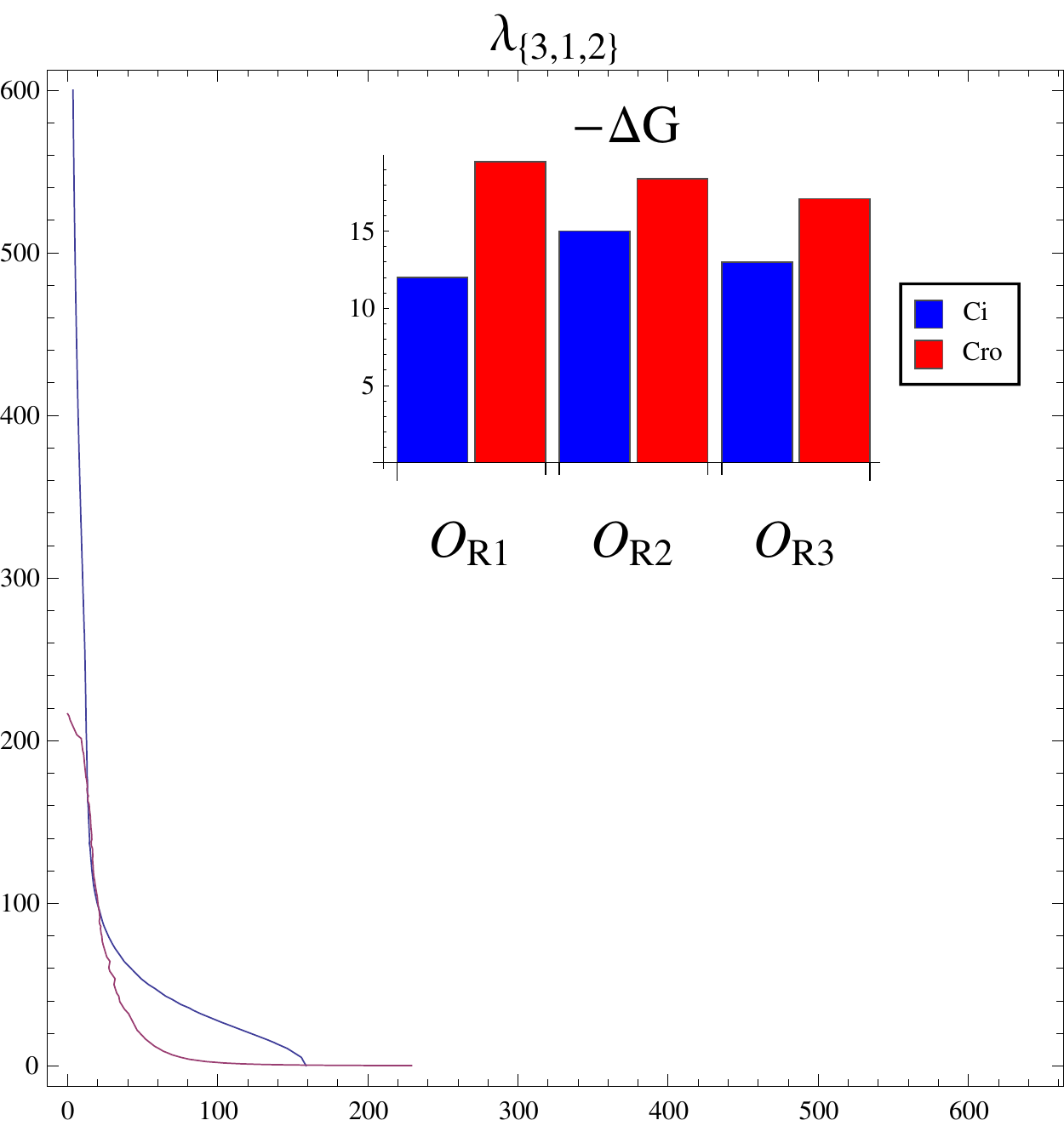}     
         &  \multicolumn{2}{|c|}{\includegraphics[scale=.50 ]{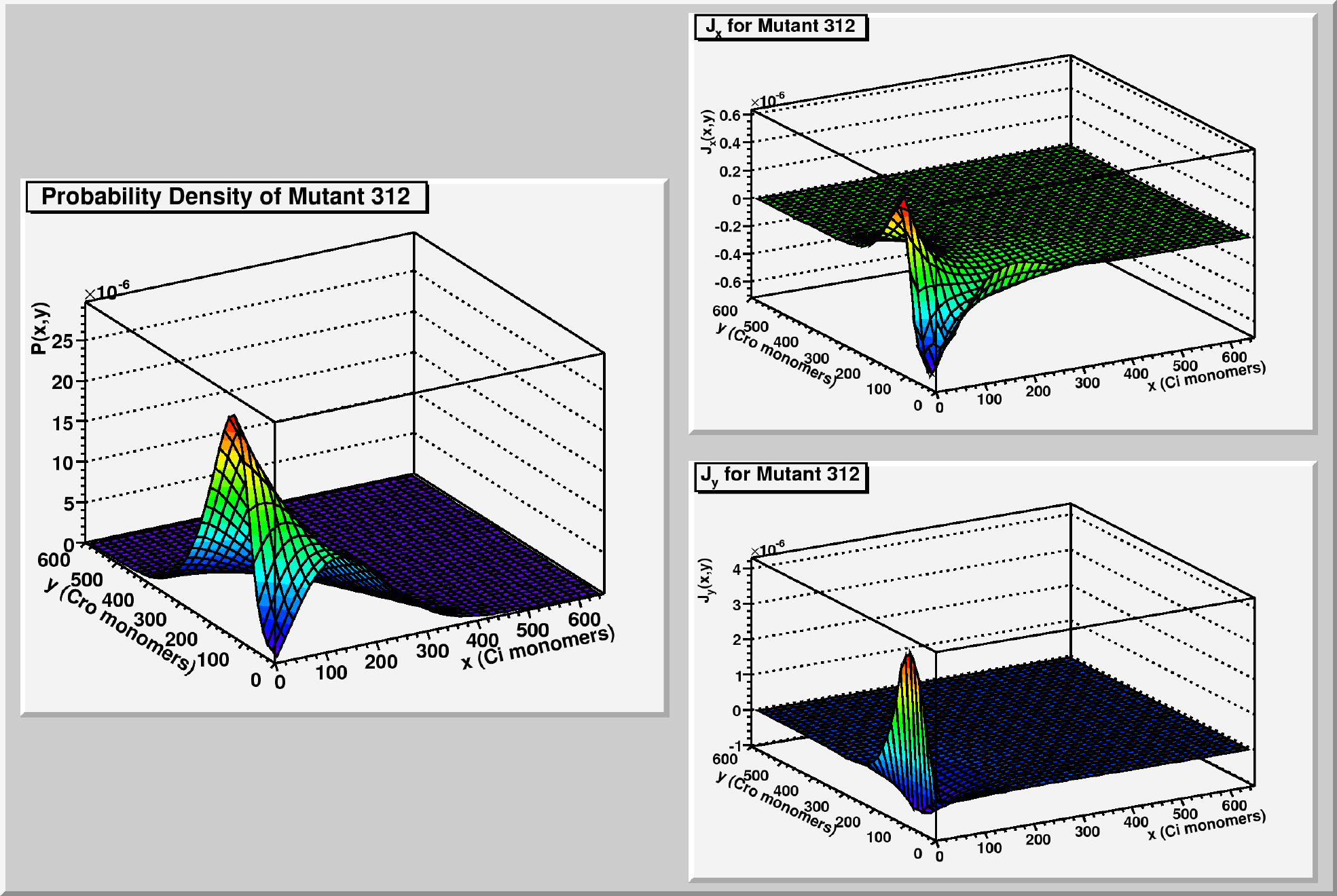}} \\ \hline    
          \multicolumn{3}{|c|}{\includegraphics[scale=.50 ]{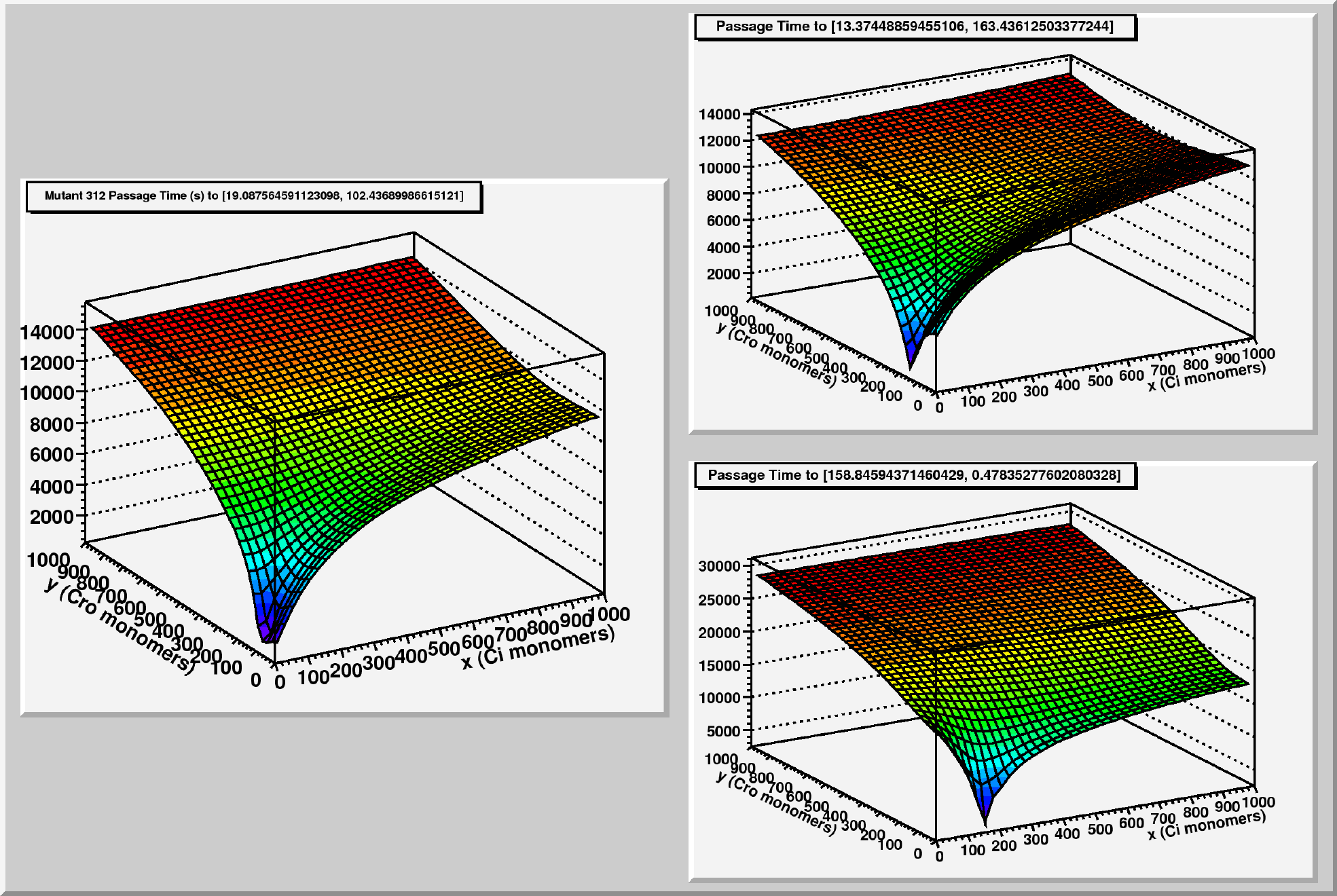}} \\ \hline     
           \multicolumn{3}{|c|}{$\lambda_{312}$ Assorted Expectation Values}    \\ \hline
	 $<x>$ (Ci protein) &  $\int_\Omega x \rho(x,y) dxdy/\int_\Omega \rho(x,y) dxdy$ & 159.2588  \\ \hline          
             $<y>$ (Cro protein) & $\int_\Omega y \rho(x,y) dxdy/\int_\Omega \rho(x,y) dxdy$ & 199.500 \\ \hline     
	 $<\tau_0>$ (seconds) &  $\int_\Omega \tau_0(x,y) \rho(x,y) dxdy/\int_\Omega \rho(x,y) dxdy$ & 7637.354  \\ \hline
	 $<\tau_1>$ (seconds) &  $\int_\Omega \tau_1(x,y) \rho(x,y) dxdy/\int_\Omega \rho(x,y) dxdy$ & 7311.862 \\ \hline
	 $<\tau_2>$ (seconds) &  $\int_\Omega \tau_2(x,y) \rho(x,y) dxdy/\int_\Omega \rho(x,y) dxdy$ & 19007.010  \\ \hline
         \hline
        \end{tabular}
}
    \caption{\bf {Properties of mutant $\lambda_{312}$}}
        \label{tab:l312}
    \end{table}%


\begin{table}[ht]
\noindent
     \resizebox{!}{8cm}{
        \begin{tabular}{|c|c|c|}
          \hline
          \multicolumn{3}{|c|}{$\lambda_{313}$ steady state and first passage time distributions}
          \\ \hline \hline
             \includegraphics[scale=.50]{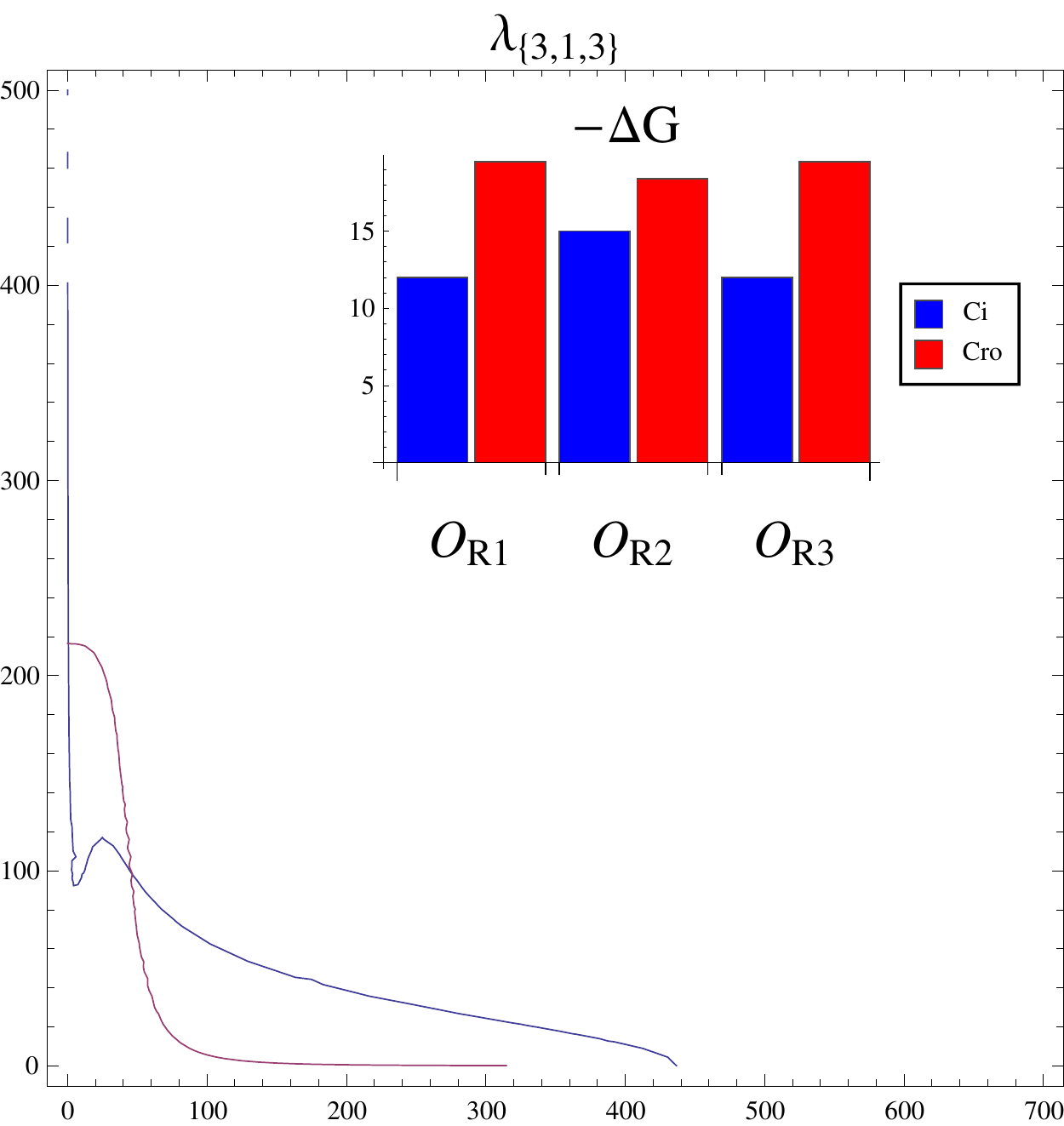}     
         &  \multicolumn{2}{|c|}{\includegraphics[scale=.50 ]{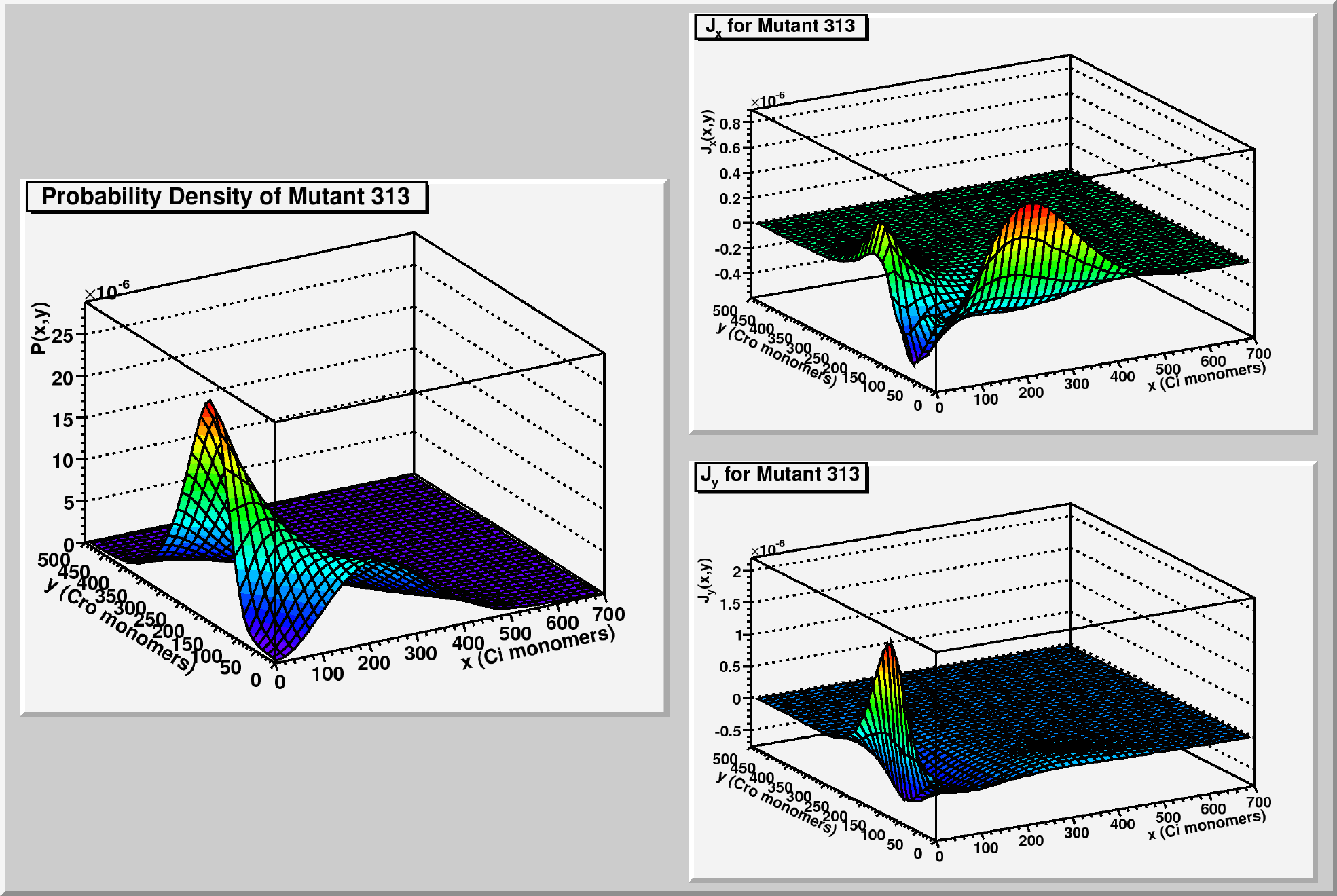}} \\ \hline    
          \multicolumn{3}{|c|}{\includegraphics[scale=.50 ]{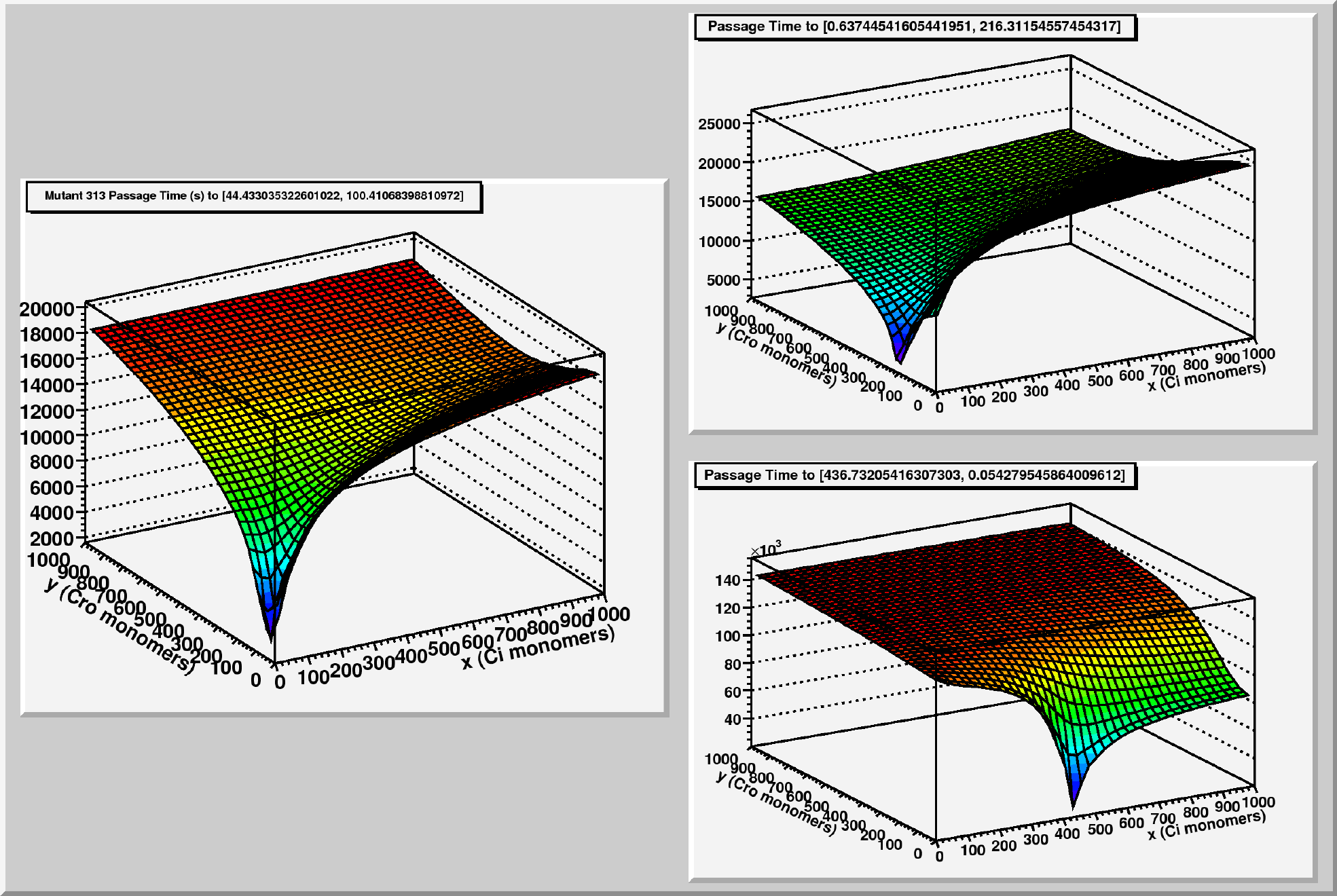}} \\ \hline     
           \multicolumn{3}{|c|}{$\lambda_{313}$ Assorted Expectation Values}    \\ \hline
	 $<x>$ (Ci protein) &  $\int_\Omega x \rho(x,y) dxdy/\int_\Omega \rho(x,y) dxdy$ & 130.454  \\ \hline          
             $<y>$ (Cro protein) & $\int_\Omega y \rho(x,y) dxdy/\int_\Omega \rho(x,y) dxdy$ & 146.169 \\ \hline     
	 $<\tau_0>$ (seconds) &  $\int_\Omega \tau_0(x,y) \rho(x,y) dxdy/\int_\Omega \rho(x,y) dxdy$ & 10679.975  \\ \hline
	 $<\tau_1>$ (seconds) &  $\int_\Omega \tau_1(x,y) \rho(x,y) dxdy/\int_\Omega \rho(x,y) dxdy$ & 13379.790 \\ \hline
	 $<\tau_2>$ (seconds) &  $\int_\Omega \tau_2(x,y) \rho(x,y) dxdy/\int_\Omega \rho(x,y) dxdy$ & 129138.611  \\ \hline
         \hline
        \end{tabular}
}
    \caption{\bf {Properties of mutant $\lambda_{313}$}}
        \label{tab:l313}
    \end{table}%

\begin{table}[ht]
\noindent
     \resizebox{!}{8cm}{
        \begin{tabular}{|c|c|c|}
          \hline
          \multicolumn{3}{|c|}{$\lambda_{323}$ steady state and first passage time distributions}
          \\ \hline \hline
             \includegraphics[scale=.50]{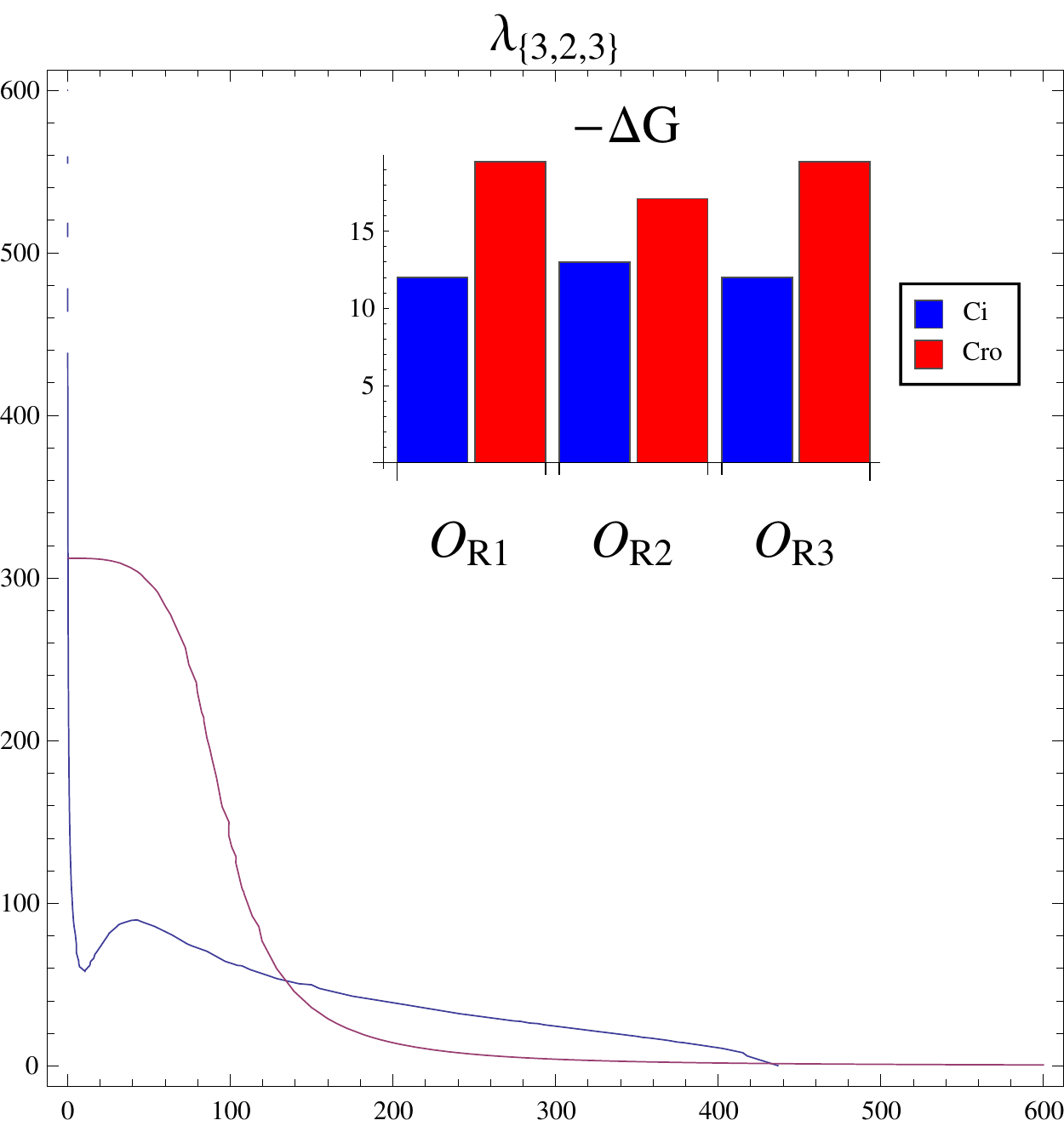}     
         &  \multicolumn{2}{|c|}{\includegraphics[scale=.50 ]{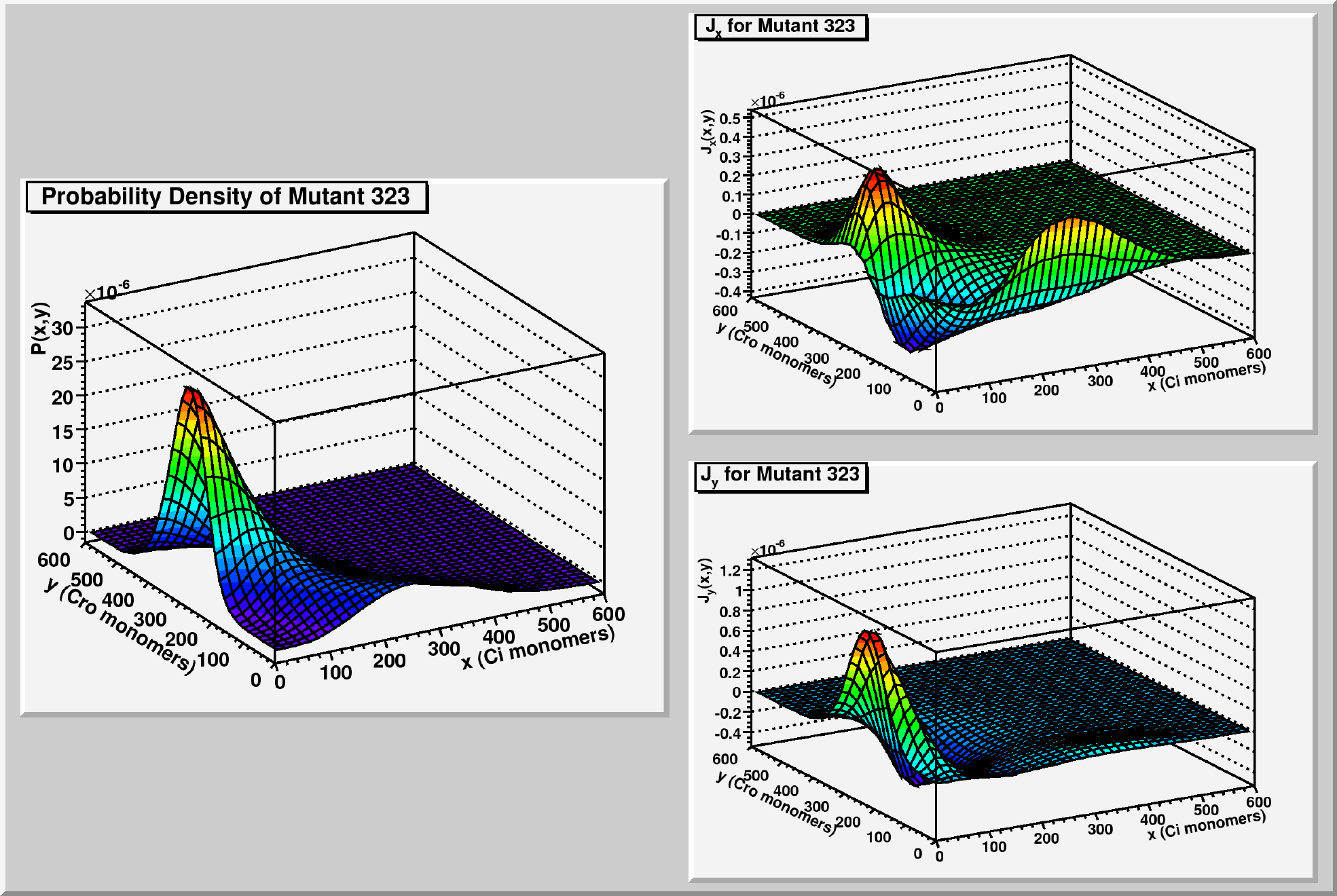}} \\ \hline    
          \multicolumn{3}{|c|}{\includegraphics[scale=.50 ]{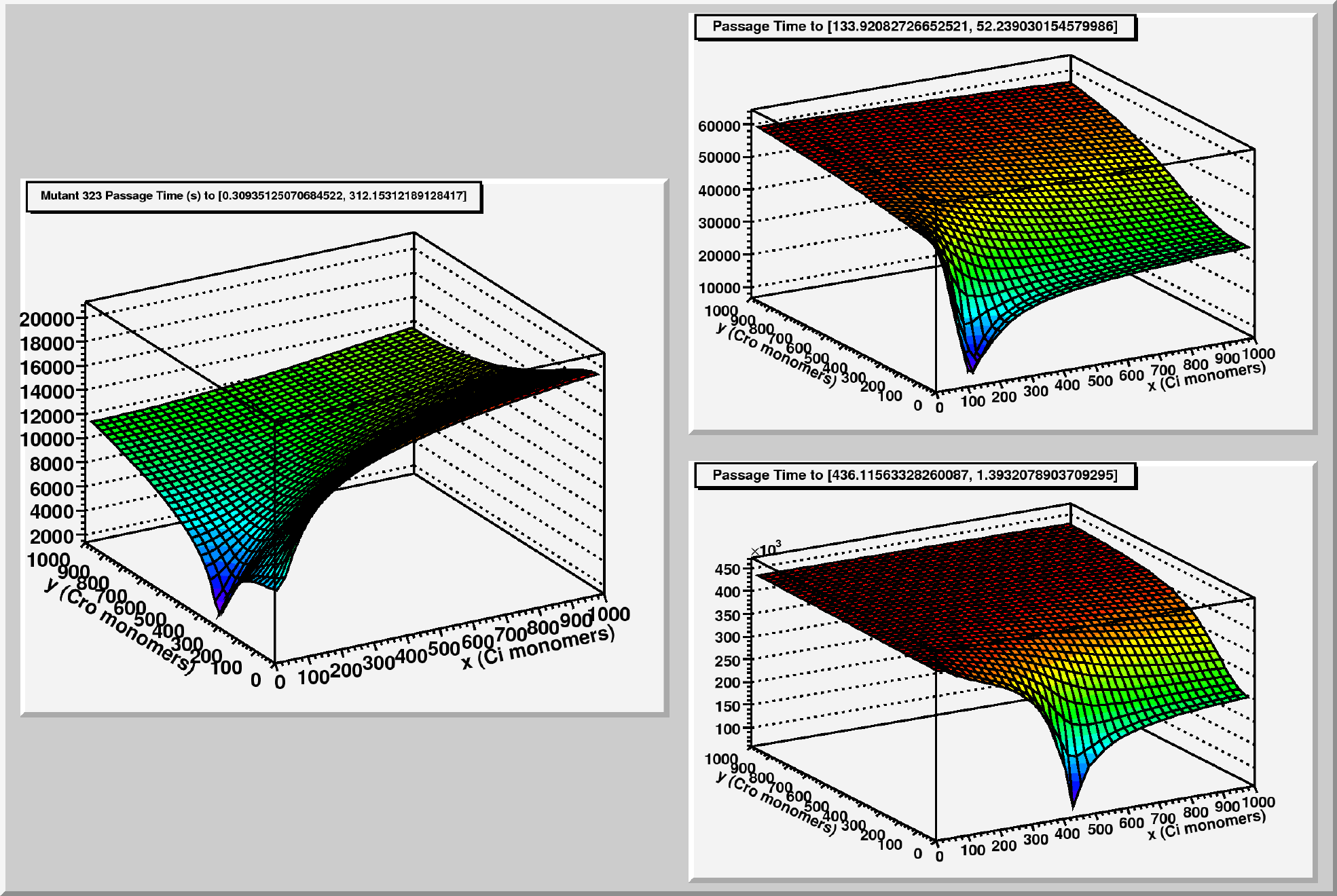}} \\ \hline     
           \multicolumn{3}{|c|}{$\lambda_{323}$ Assorted Expectation Values}    \\ \hline
	 $<x>$ (Ci protein) &  $\int_\Omega x \rho(x,y) dxdy/\int_\Omega \rho(x,y) dxdy$ & 117.832  \\ \hline          
             $<y>$ (Cro protein) & $\int_\Omega y \rho(x,y) dxdy/\int_\Omega \rho(x,y) dxdy$ & 223.495 \\ \hline     
	 $<\tau_0>$ (seconds) &  $\int_\Omega \tau_0(x,y) \rho(x,y) dxdy/\int_\Omega \rho(x,y) dxdy$ & 8119.700  \\ \hline
	 $<\tau_1>$ (seconds) &  $\int_\Omega \tau_1(x,y) \rho(x,y) dxdy/\int_\Omega \rho(x,y) dxdy$ & 45614.933 \\ \hline
	 $<\tau_2>$ (seconds) &  $\int_\Omega \tau_2(x,y) \rho(x,y) dxdy/\int_\Omega \rho(x,y) dxdy$ & 413661.916  \\ \hline
         \hline
        \end{tabular}
}
    \caption{\bf {Properties of mutant $\lambda_{323}$}}
        \label{tab:l323}
    \end{table}%

\clearpage

\begin{table}[ht]
        \centering
        \begin{tabular}{|p{0.12\textwidth}|p{0.29\textwidth}|p{0.29\textwidth}|p{0.29\textwidth}|}
          \hline
          \multicolumn{4}{|c|}{Time Evolution}
          \\ \hline \hline
          time&$+0$&$+1200$&$+2400$ \\ \hline
          $t=0$ & \includegraphics[scale=.22]{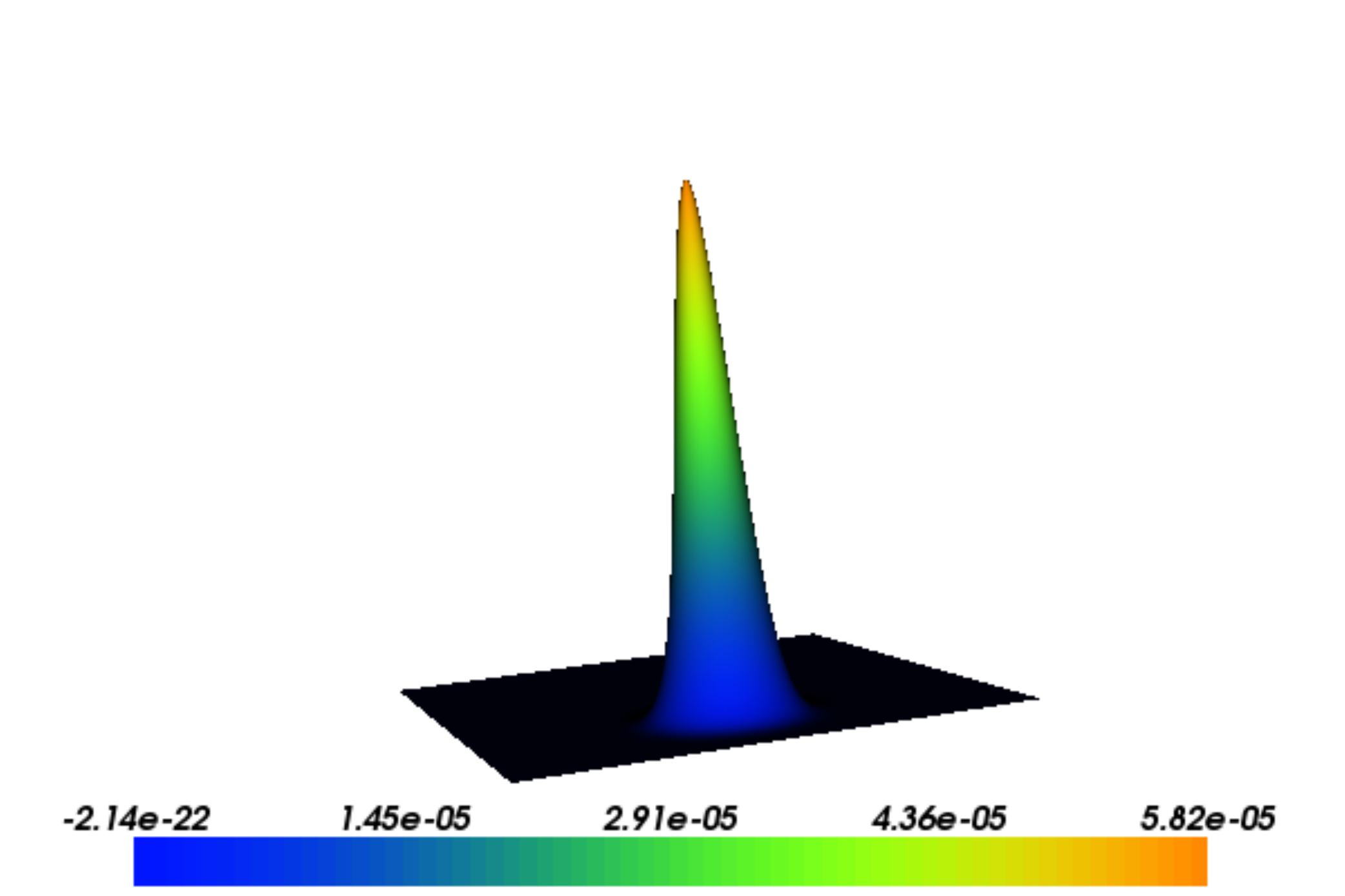}& \includegraphics[scale=.22]{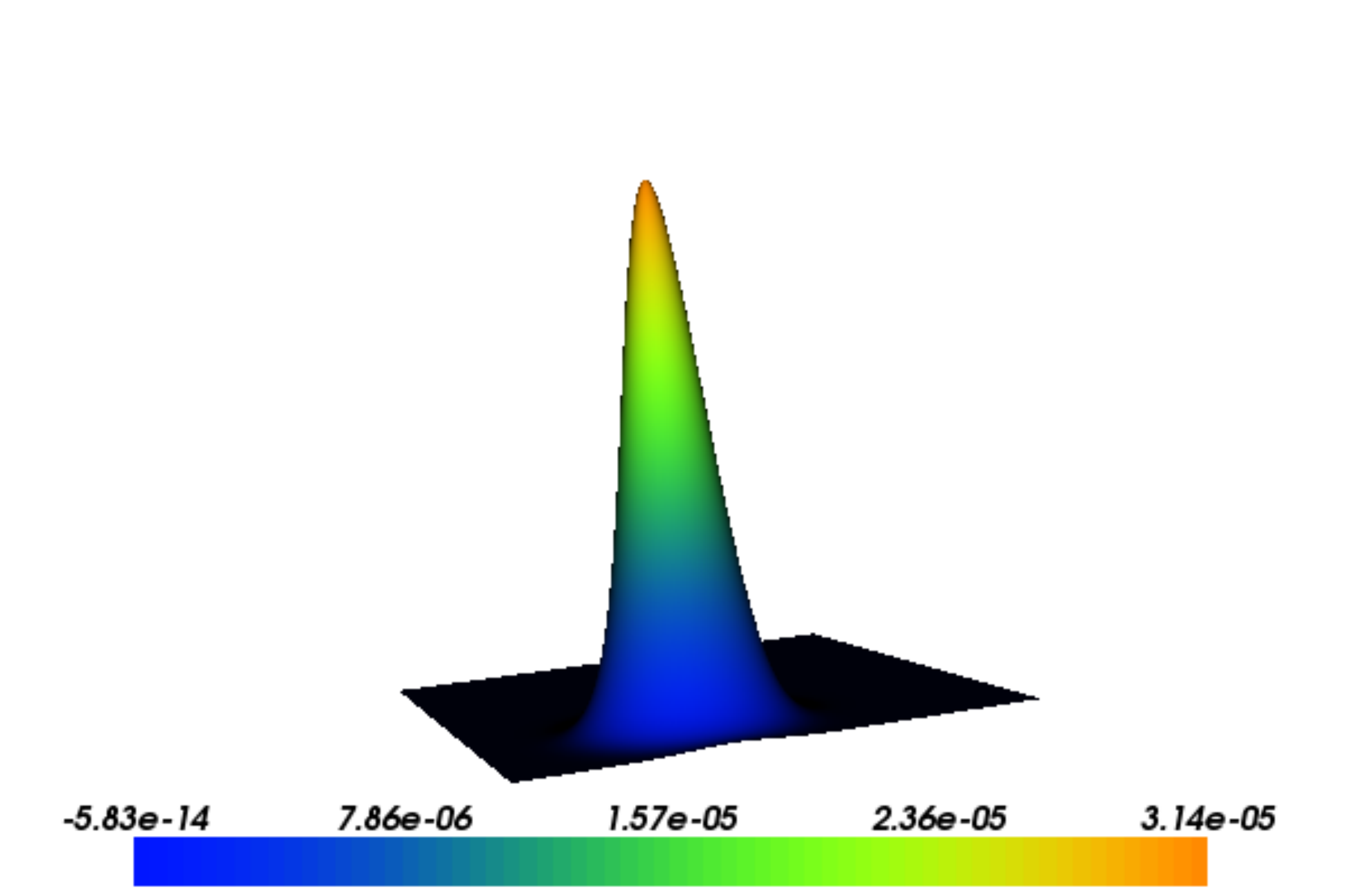}&\includegraphics[scale=.22]{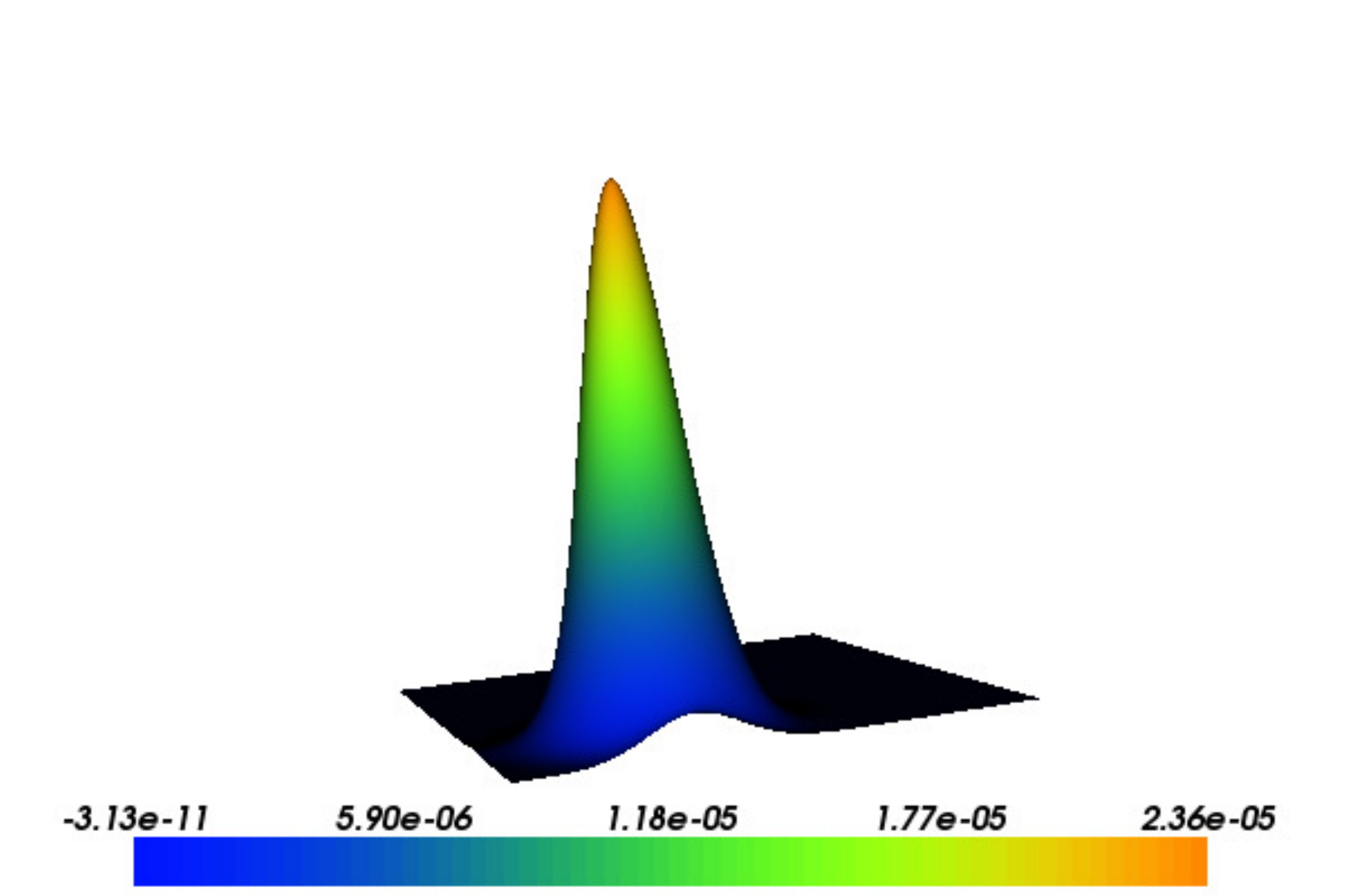}\\ \hline
          $t=3600$ & \includegraphics[scale=.22]{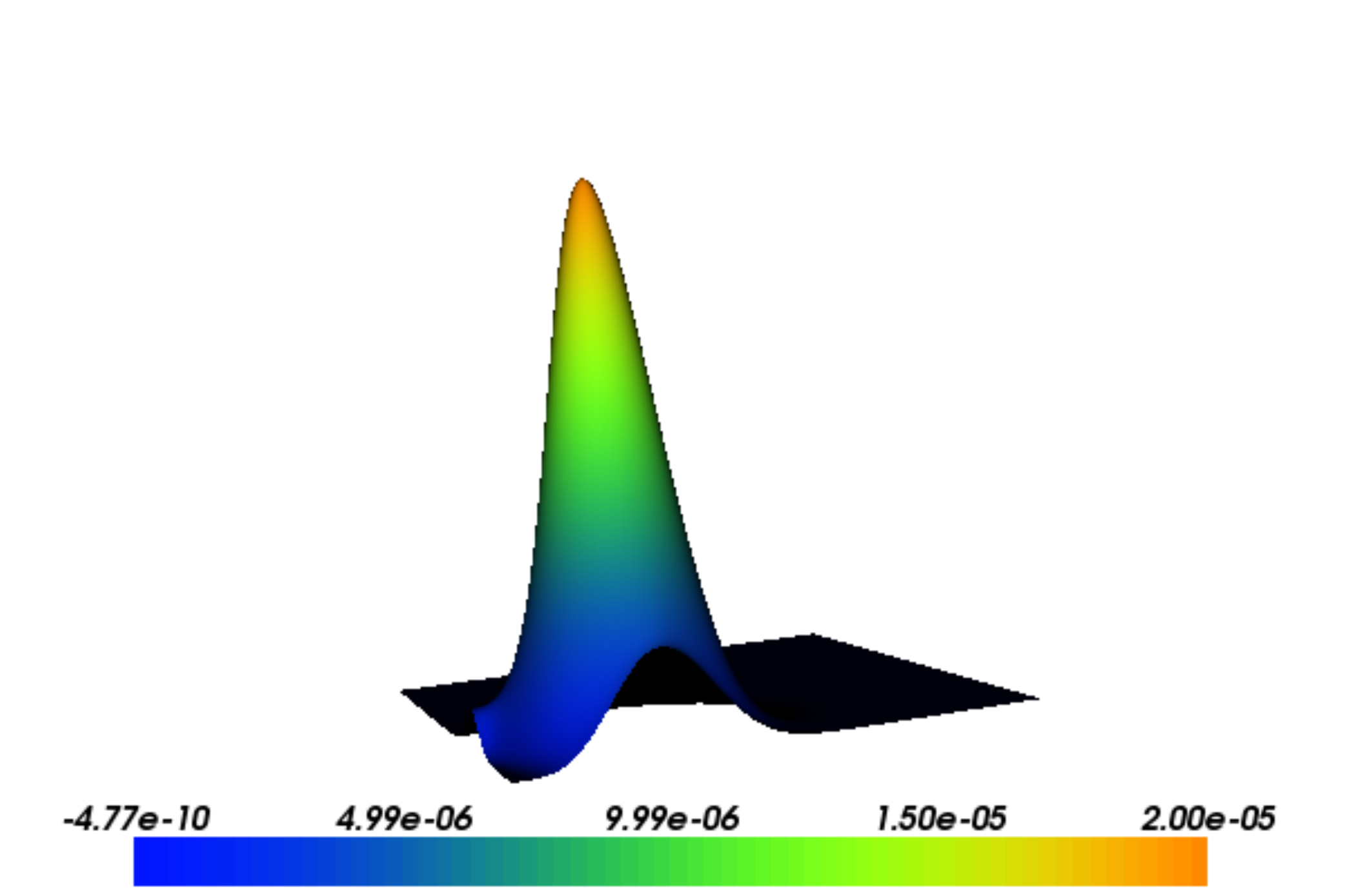}& \includegraphics[scale=.22]{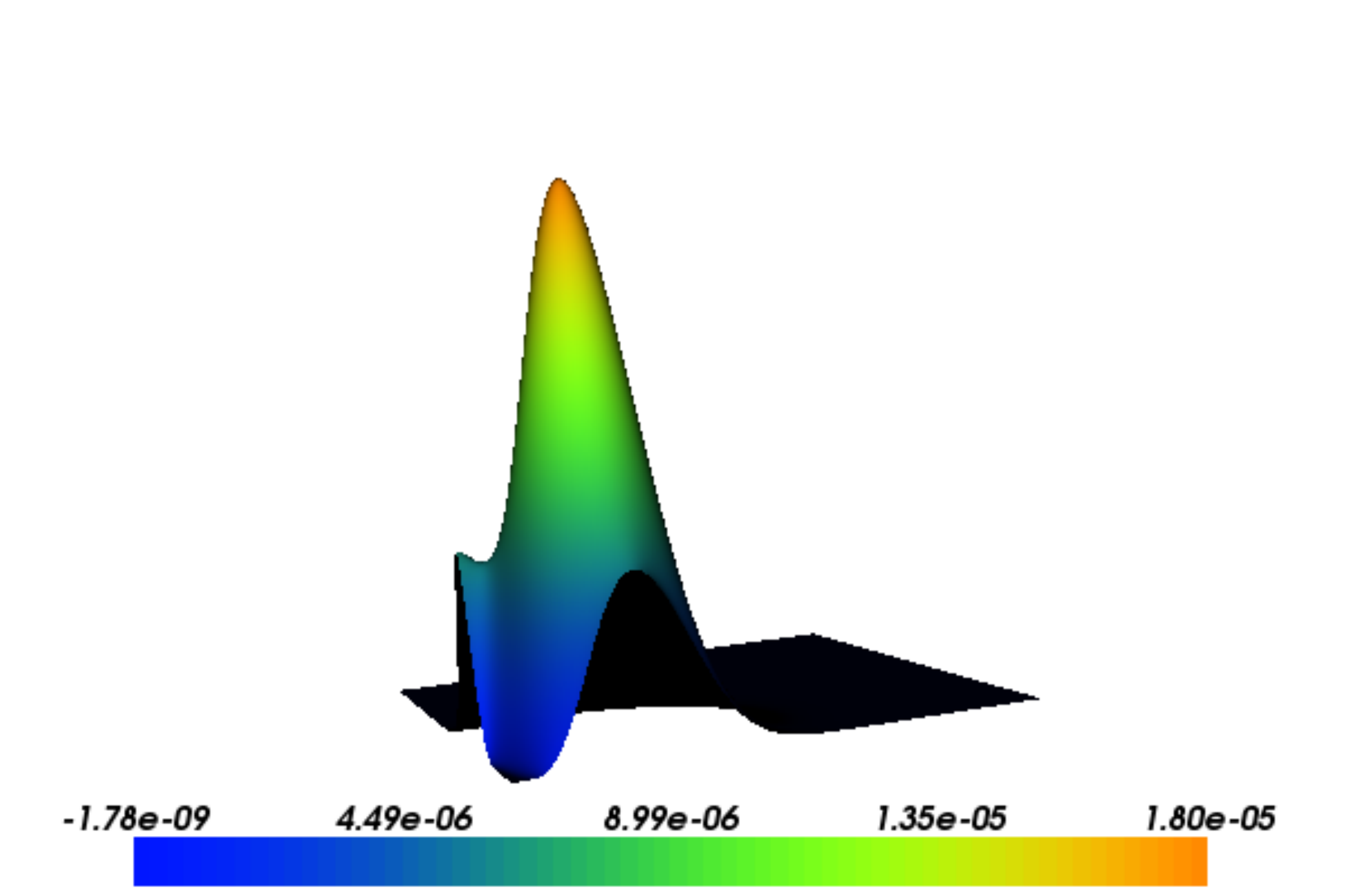}&\includegraphics[scale=.22]{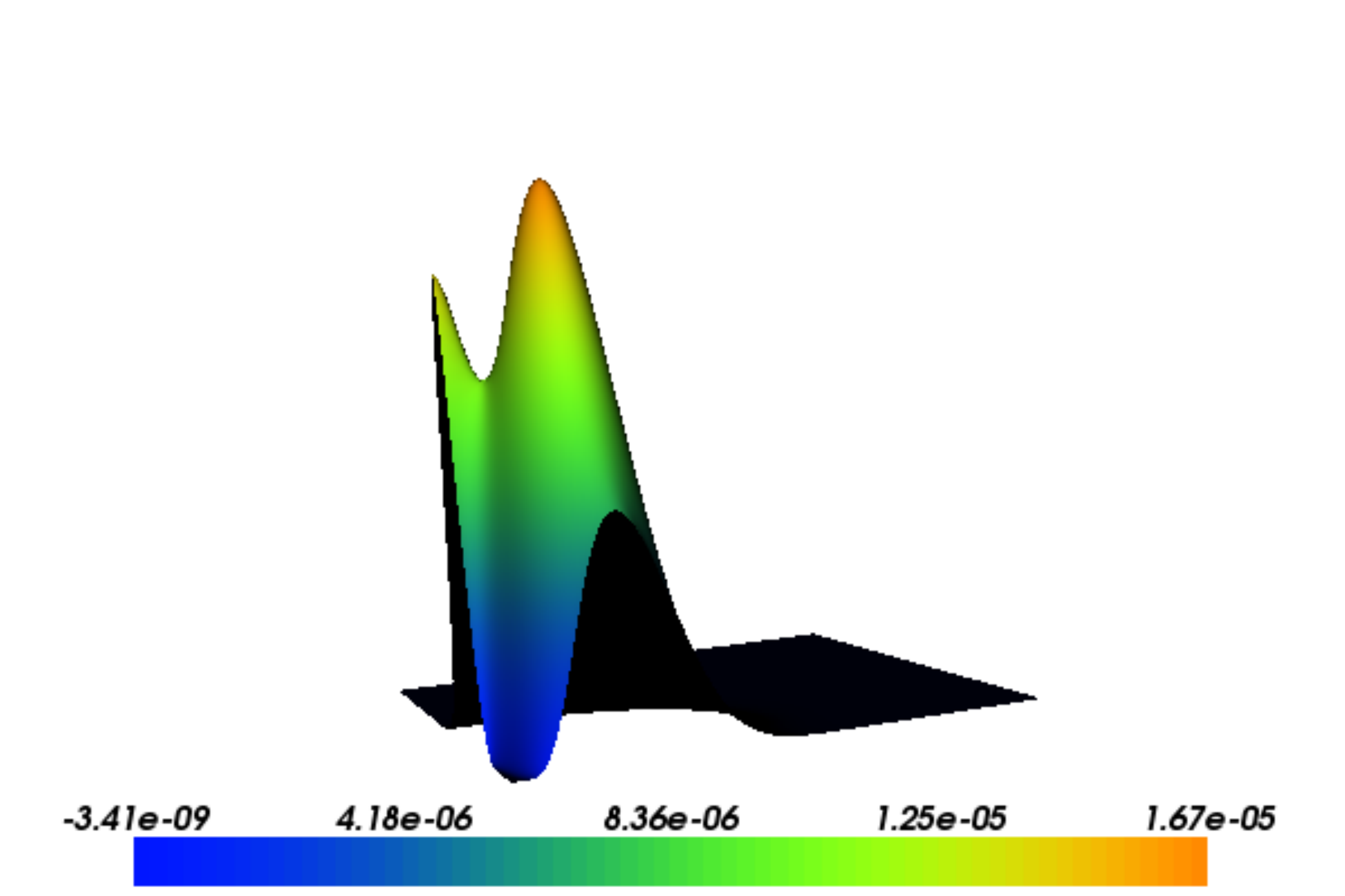}\\ \hline
          $t=7200$ & \includegraphics[scale=.22]{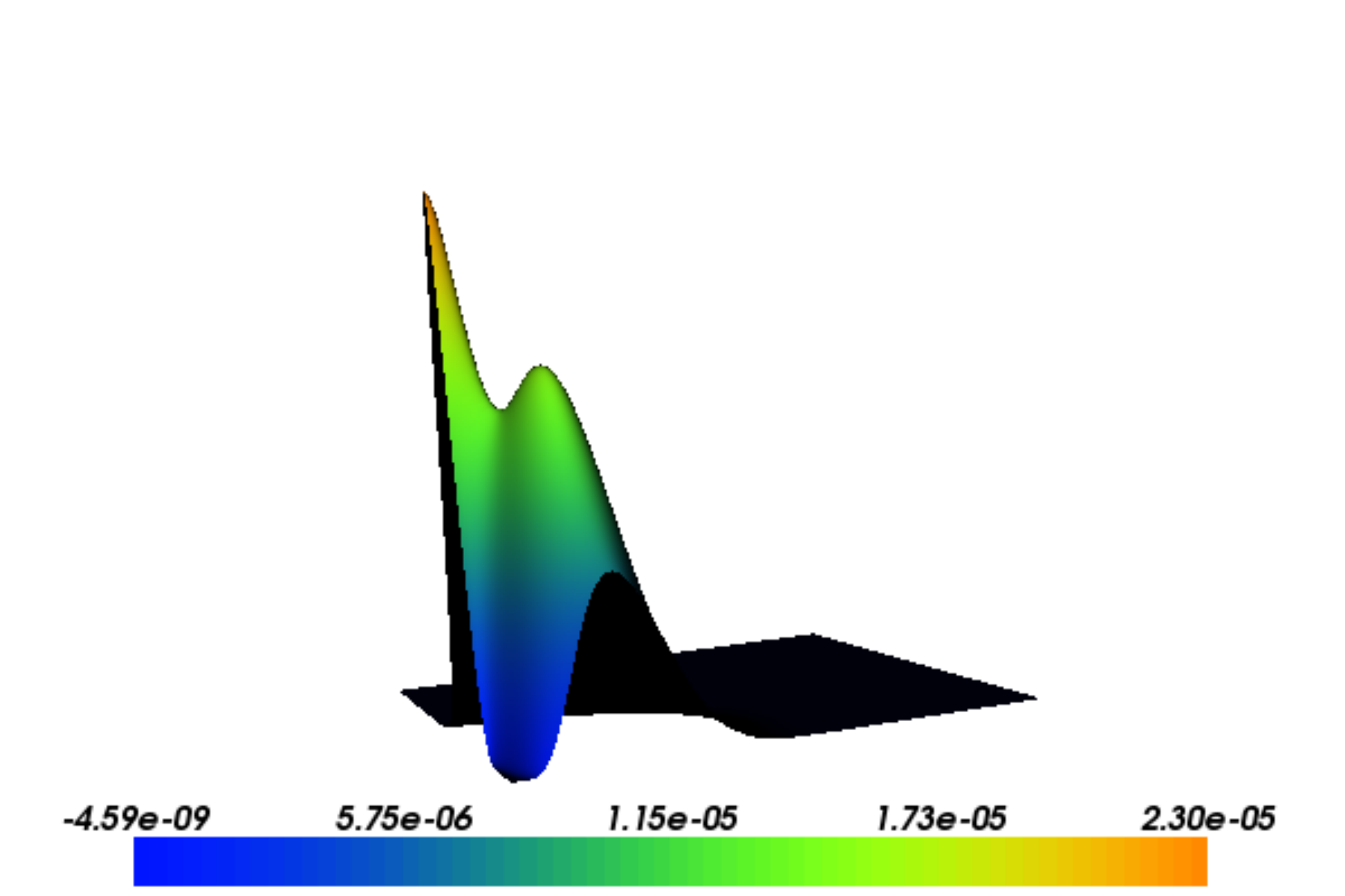}& \includegraphics[scale=.22]{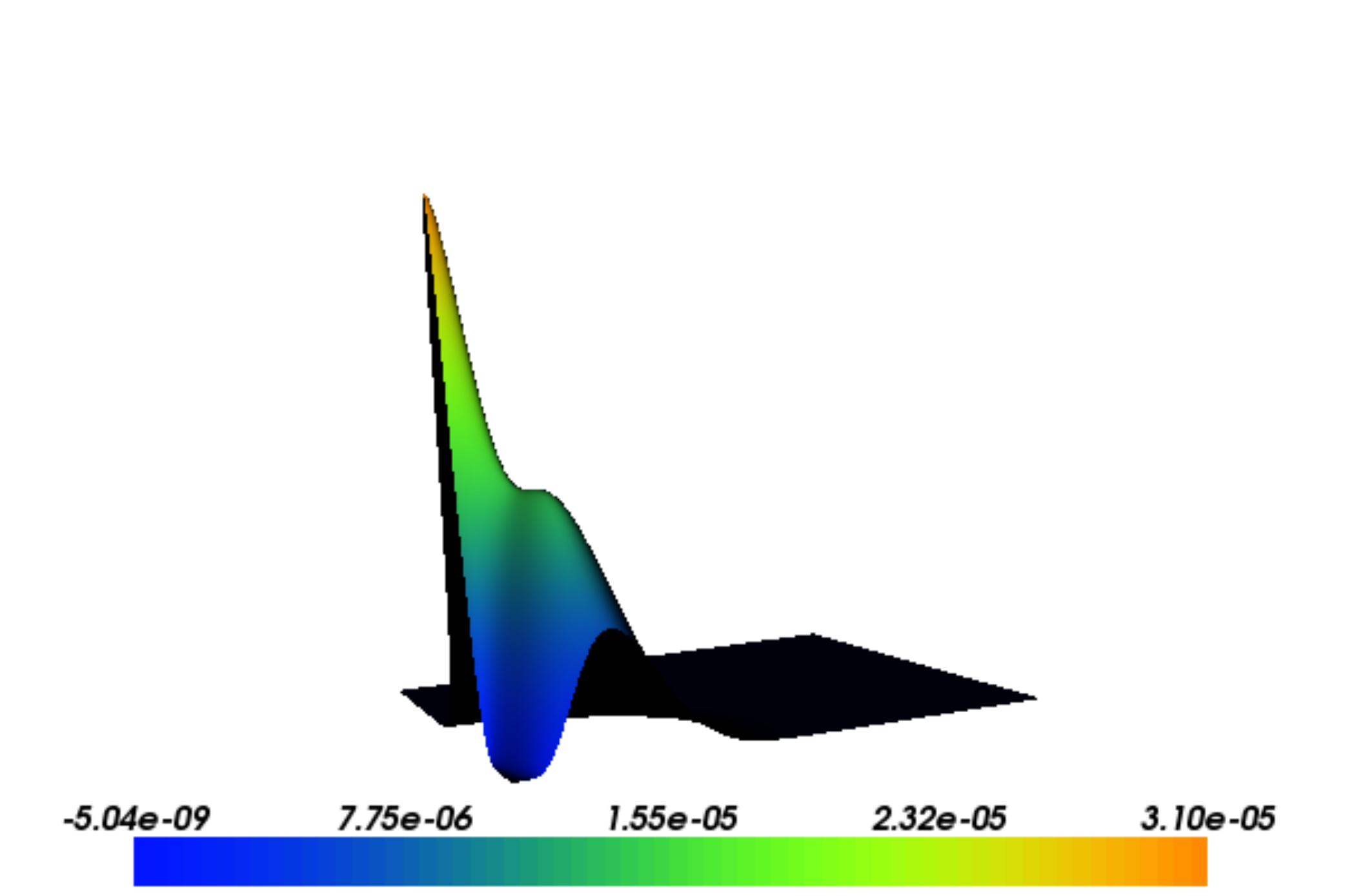}&\includegraphics[scale=.22]{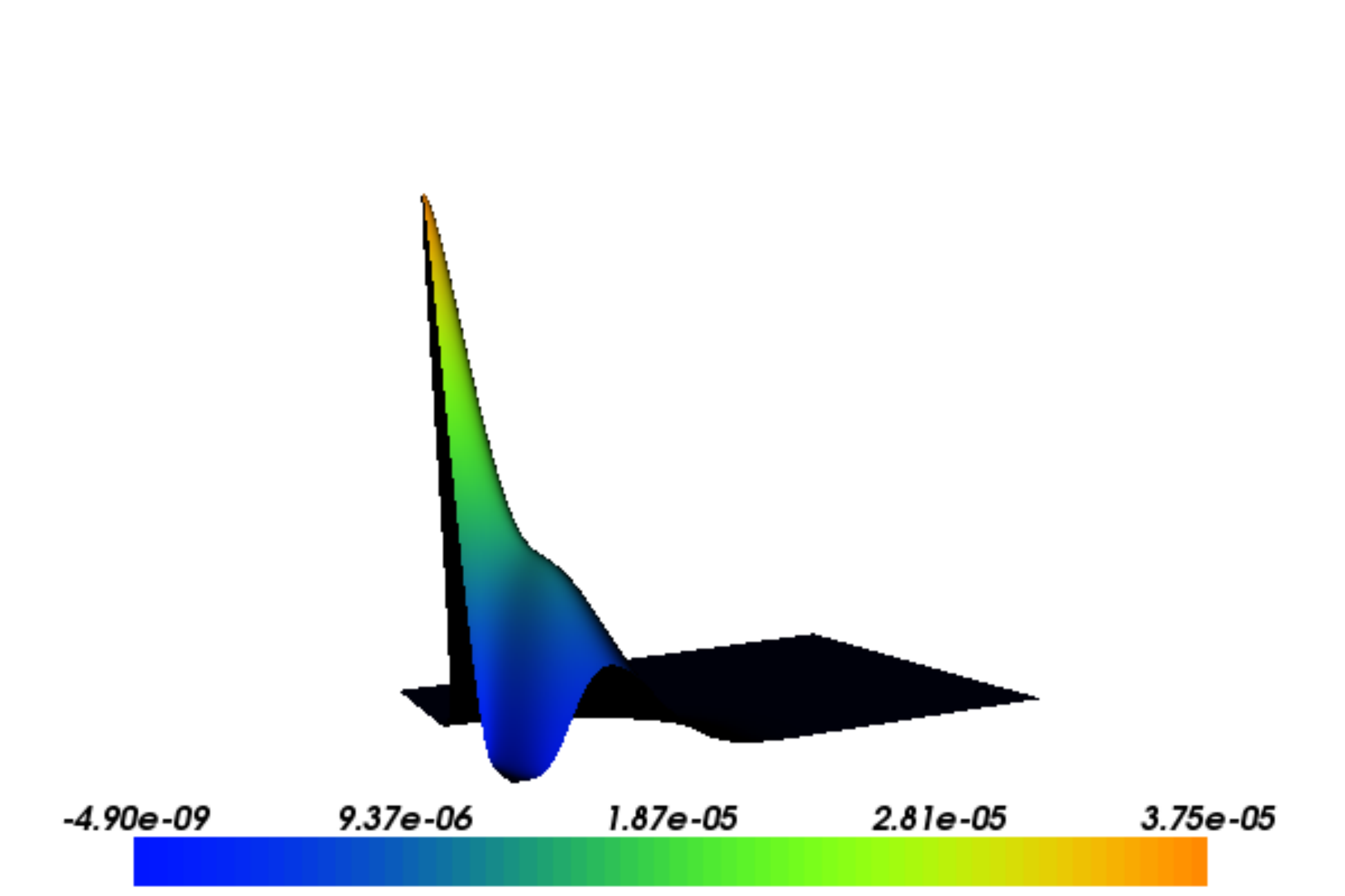}\\ \hline
          $t=10800$ & \includegraphics[scale=.22]{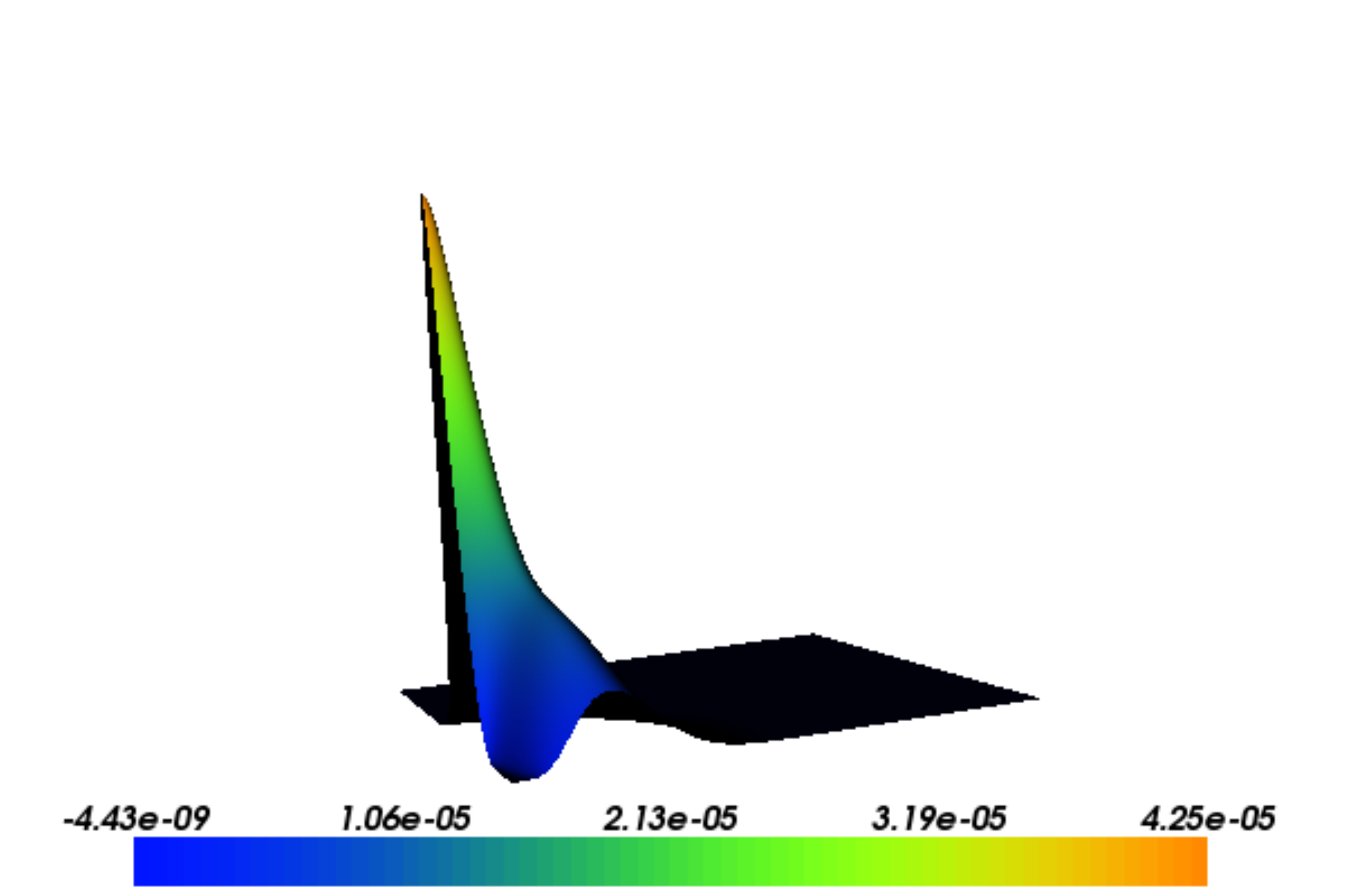}& \includegraphics[scale=.22]{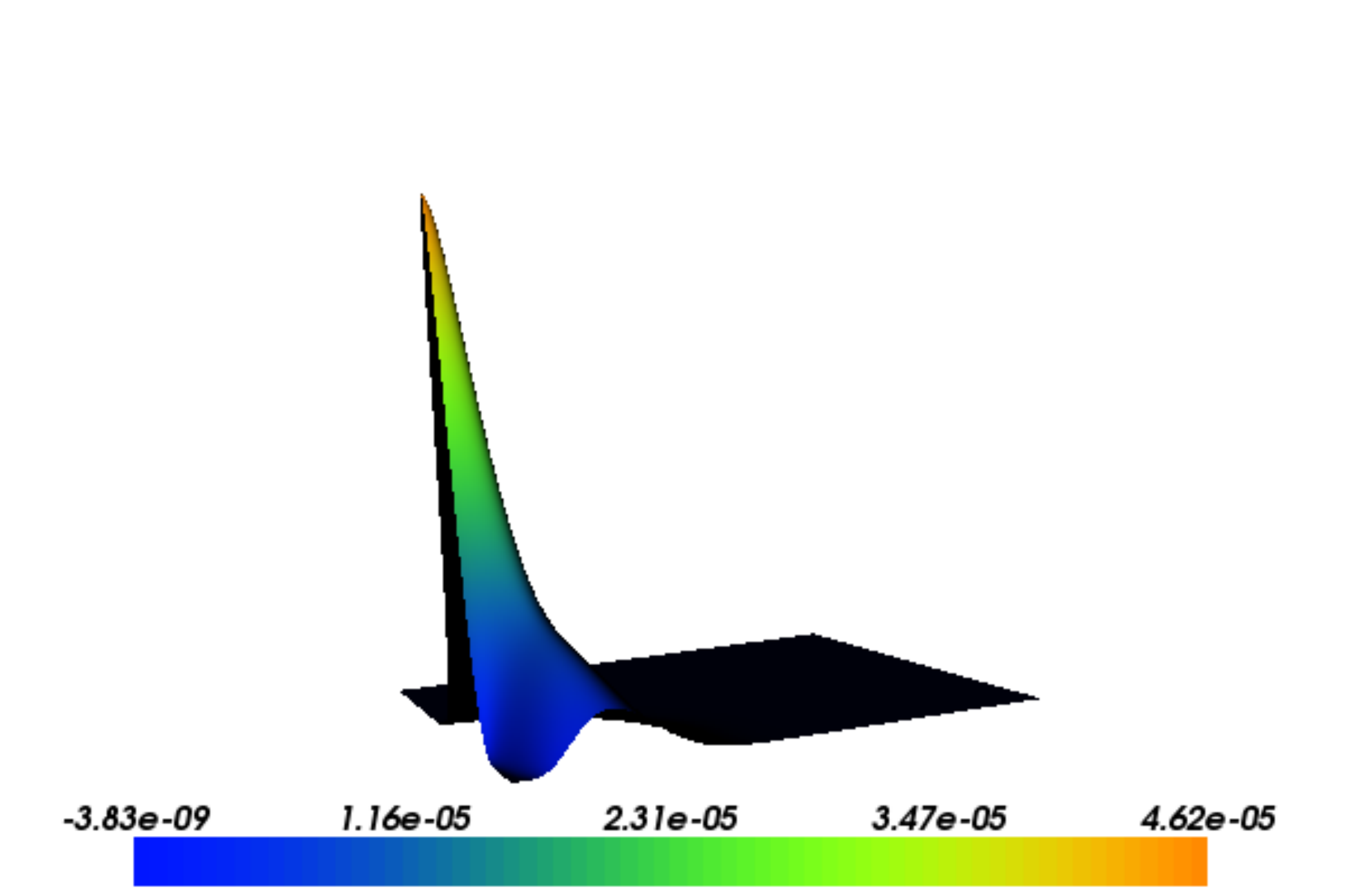}&\includegraphics[scale=.22]{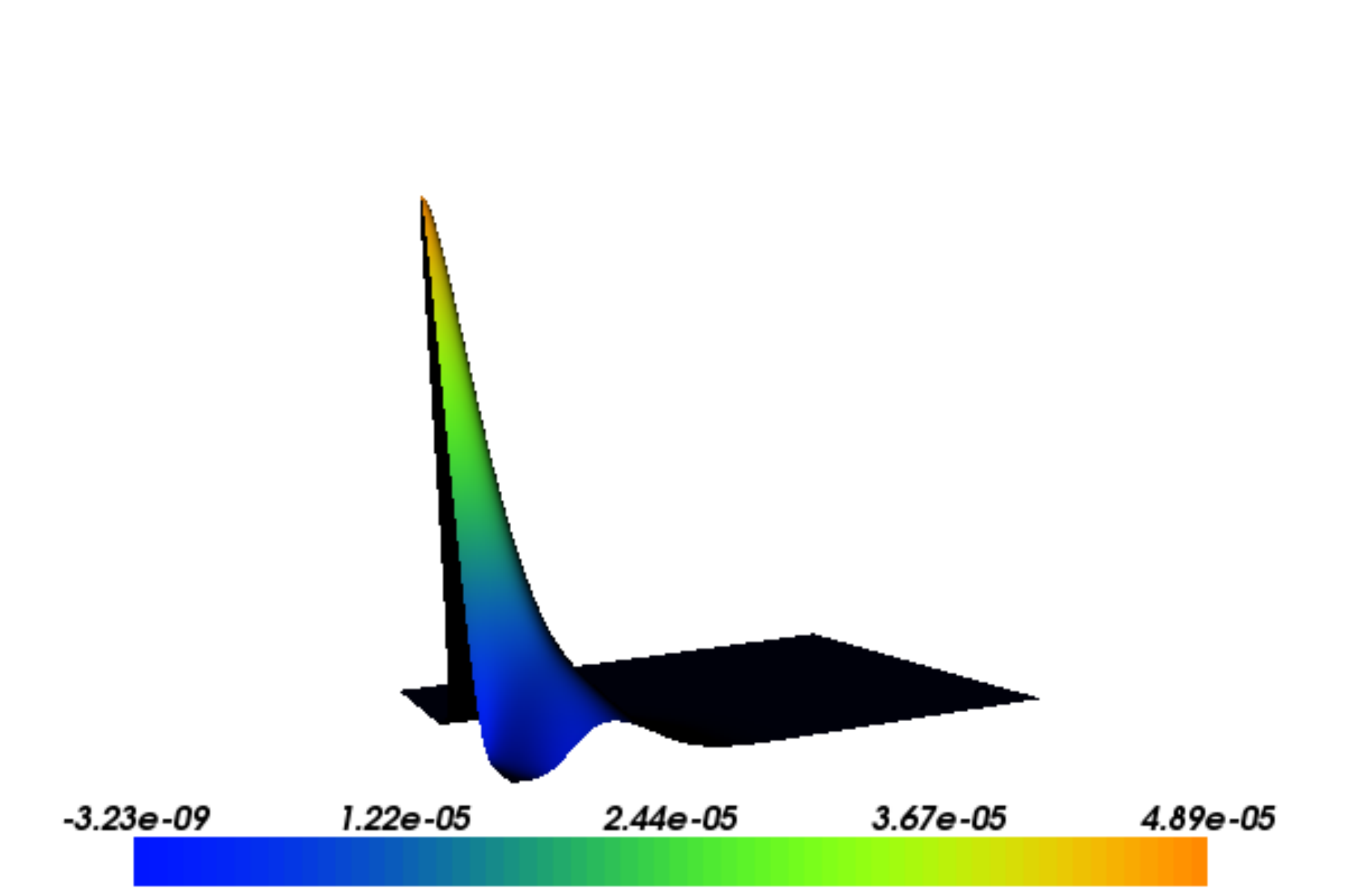}\\ \hline
        \hline
        \end{tabular}
    \caption{\bf {The time evolution of a gaussian initial condition relaxing into the steady state for the wild type with RNA polymerase in the partition function and constant diffusion tensor.  Notice that the two states, lytic and lysogenic, are gapped only for the window of time from 4800 to 8400 seconds.  Although the emphasis in the literature is with steady states, it is an interesting possibility that the two state phenomena is a strictly out of equilibrium phenomena, existing only for a window of time in the cell.  A bacteria cell cycle is around 1800 seconds, suggesting that for many distributions the steady state is not reachable in the necessary time scale.}}    
        \label{rnapdiff}
    \end{table}%

\begin{table}[ht]
        \centering
        \begin{tabular}{|p{0.12\textwidth}|p{0.29\textwidth}|p{0.29\textwidth}|p{0.29\textwidth}|}
          \hline
          \multicolumn{4}{|c|}{Time Evolution}
          \\ \hline \hline
          time&$+0$&$+1200$&$+2400$ \\ \hline
          $t=1200$ & \includegraphics[scale=.22]{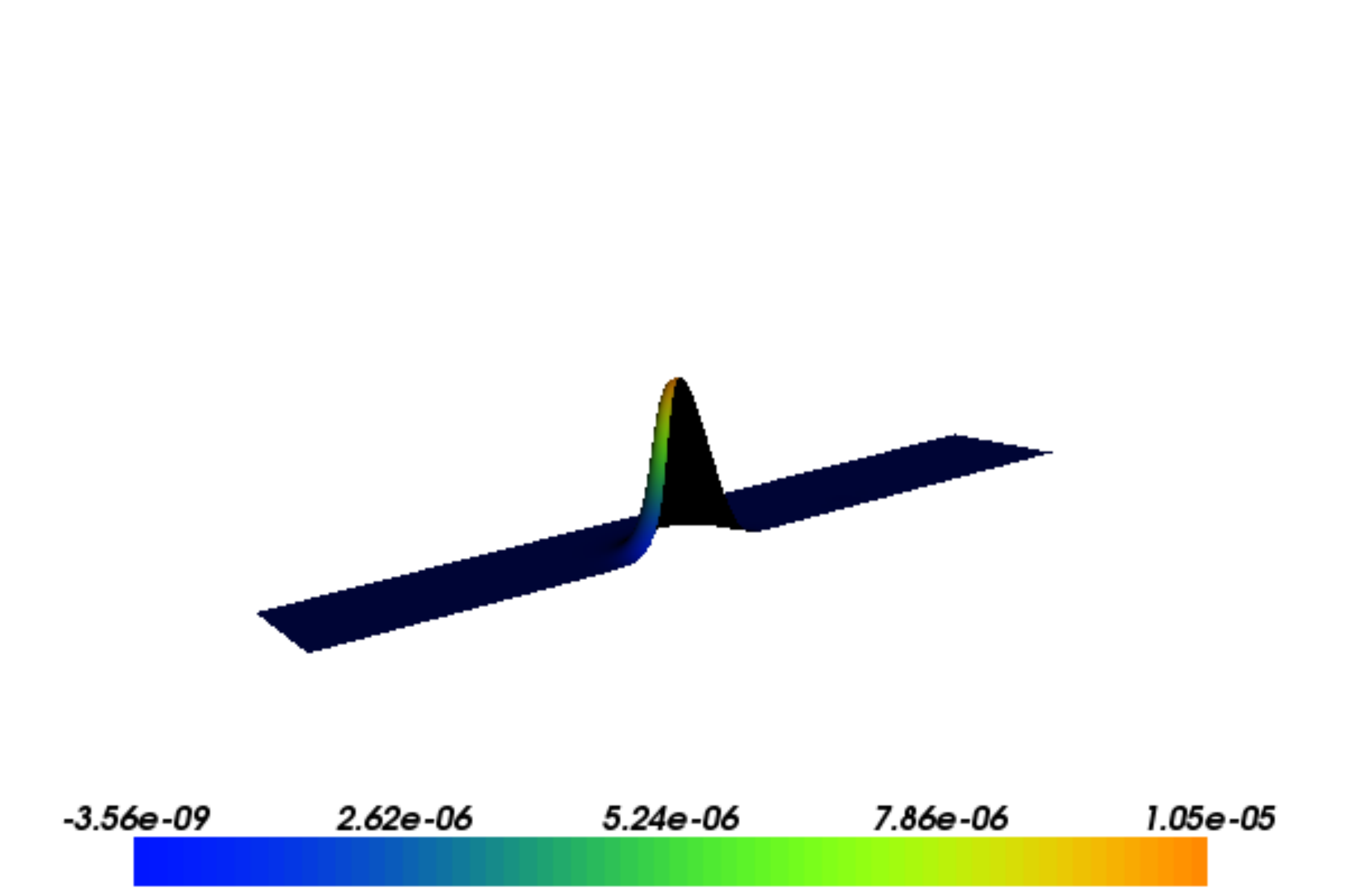}& \includegraphics[scale=.22]{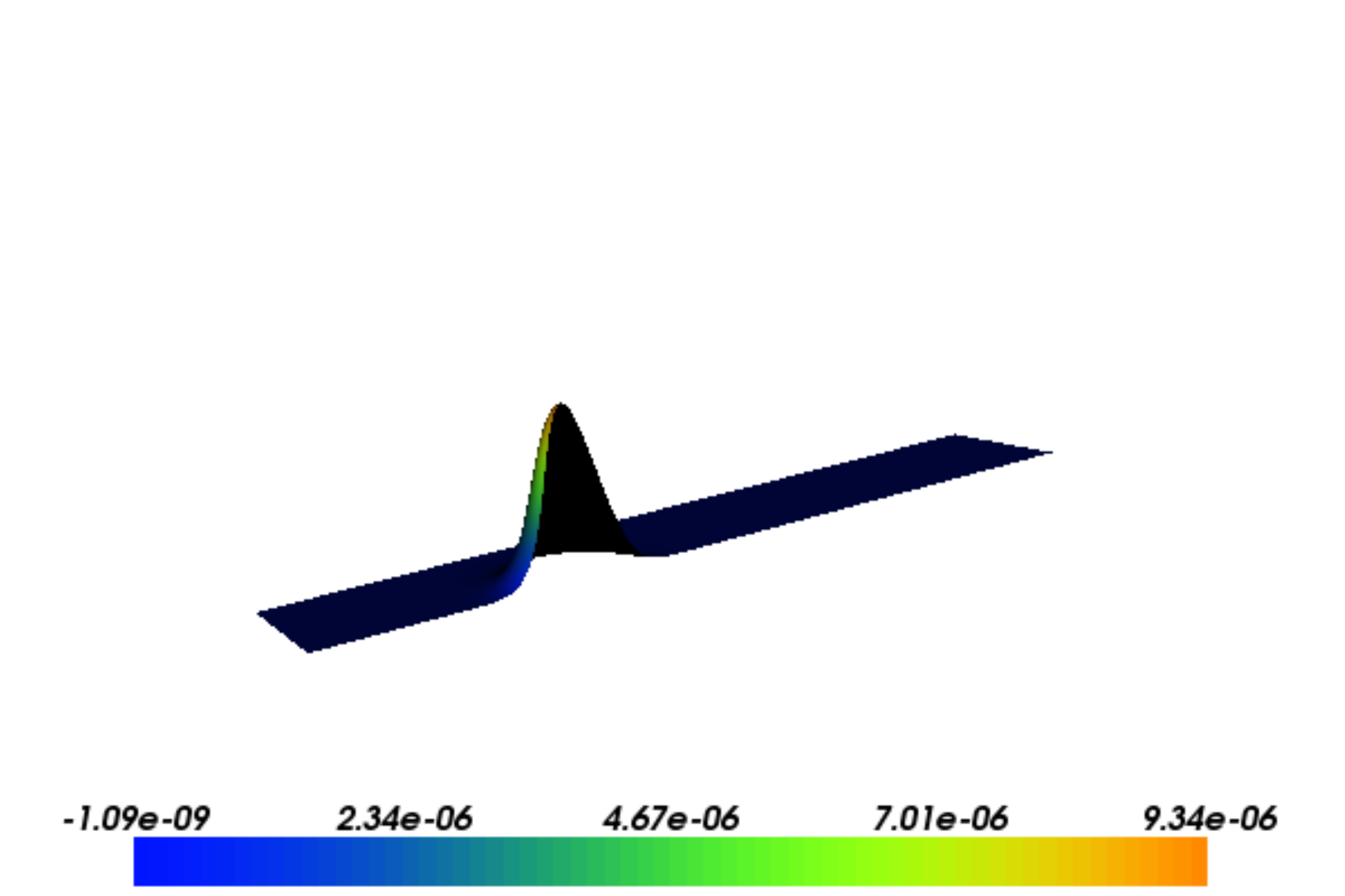}&\includegraphics[scale=.22]{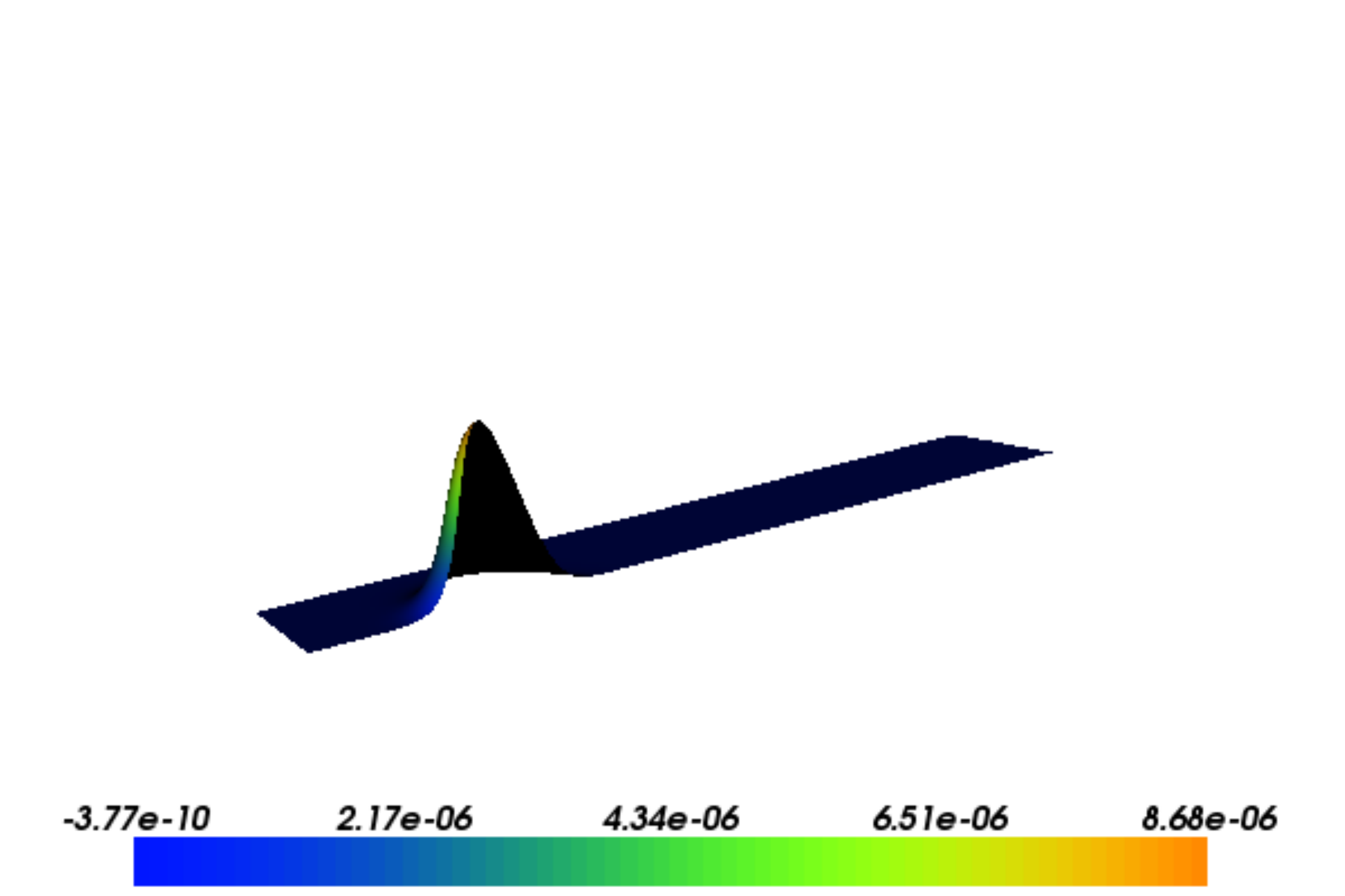}\\ \hline
          $t=4800$ & \includegraphics[scale=.22]{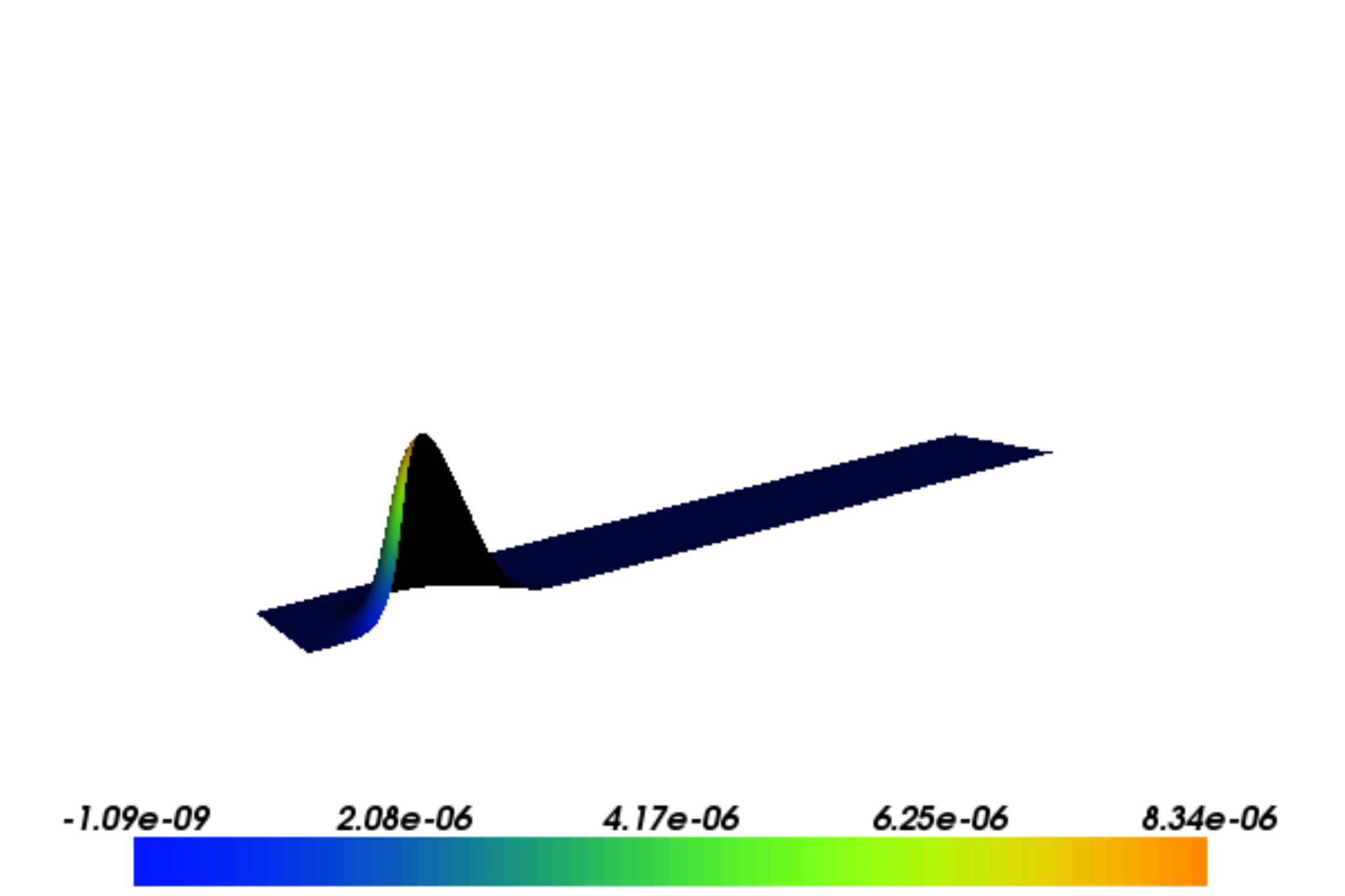}& \includegraphics[scale=.22]{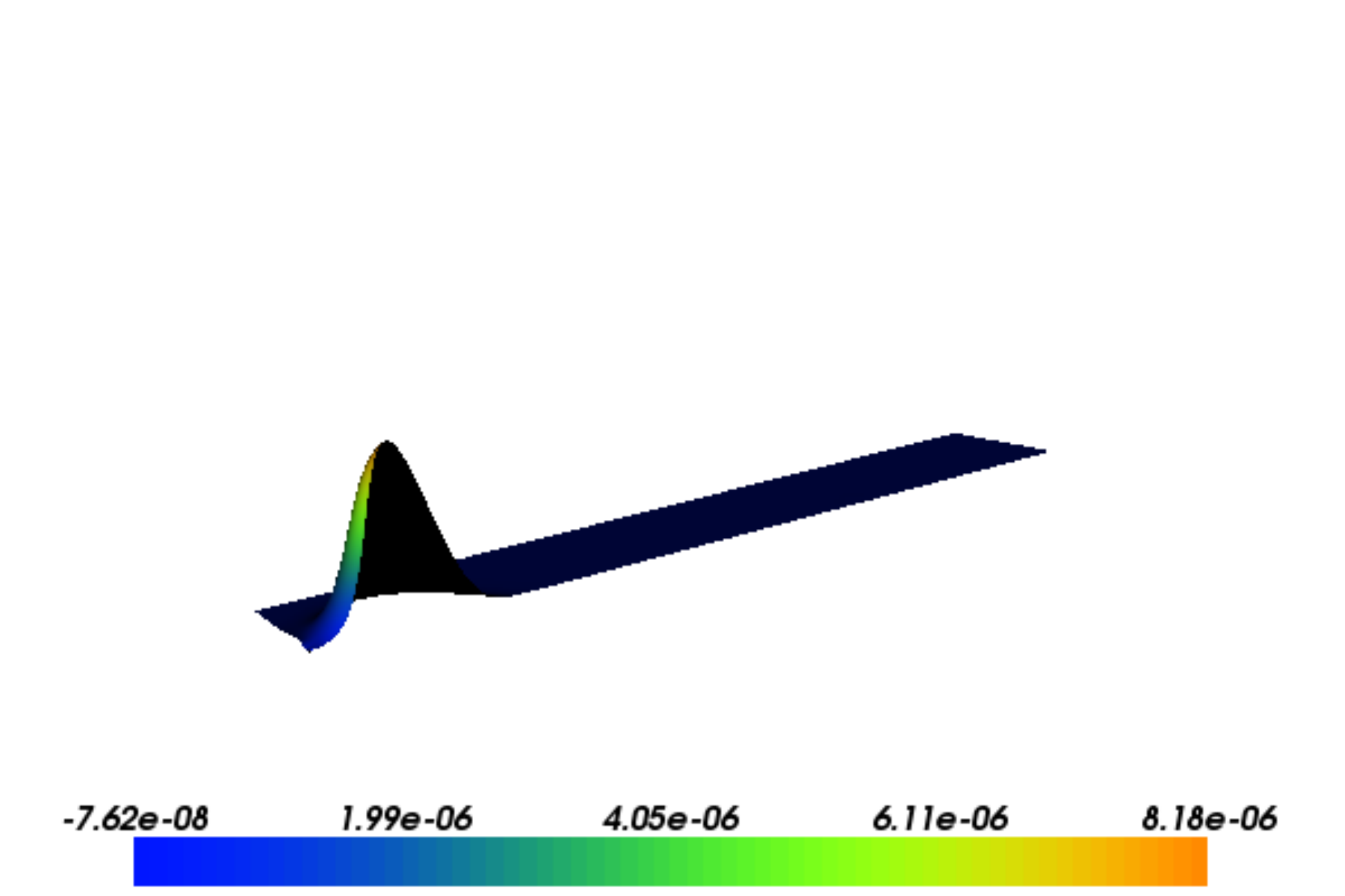}&\includegraphics[scale=.22]{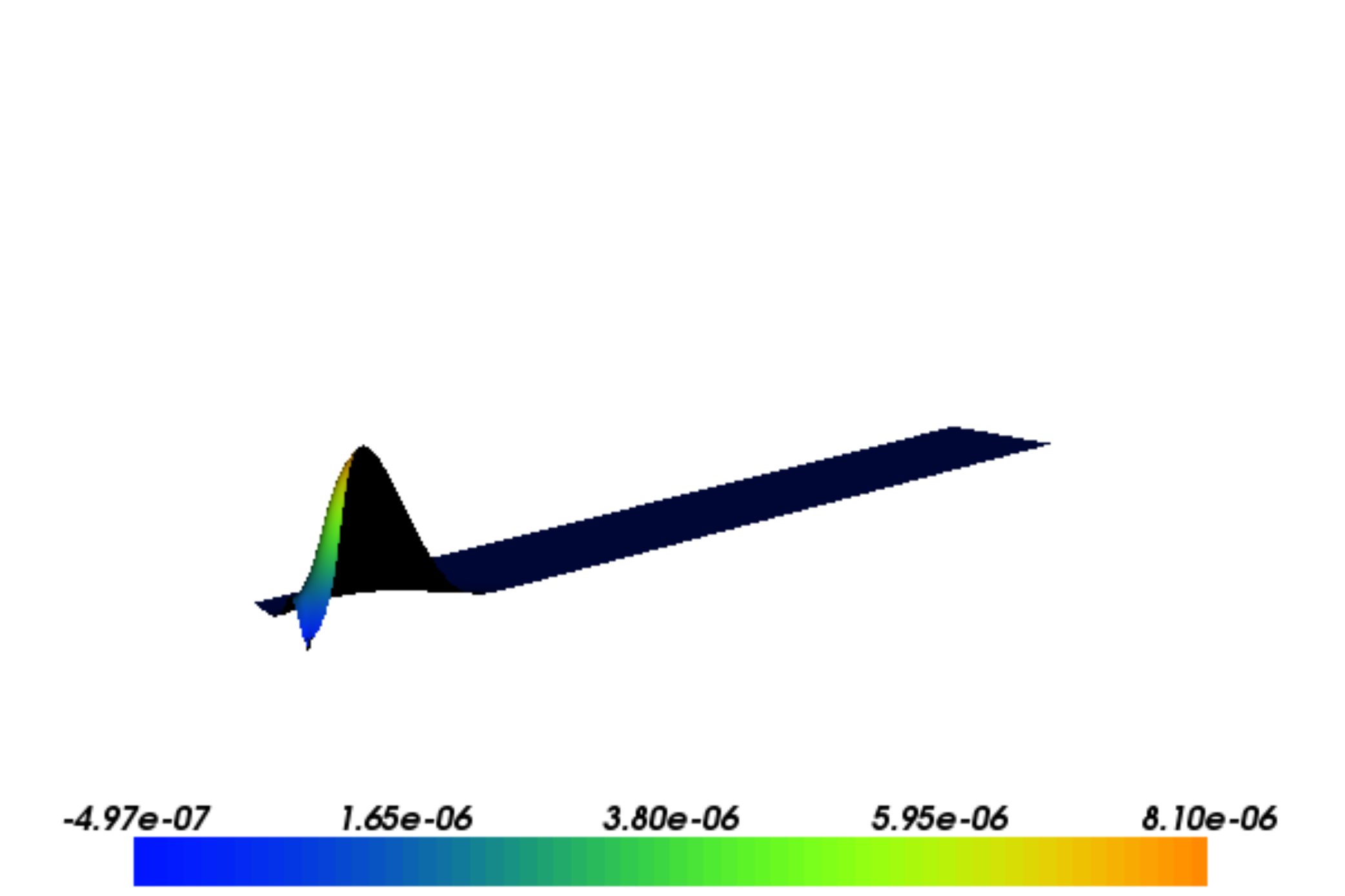}\\ \hline
          $t=8400$ & \includegraphics[scale=.22]{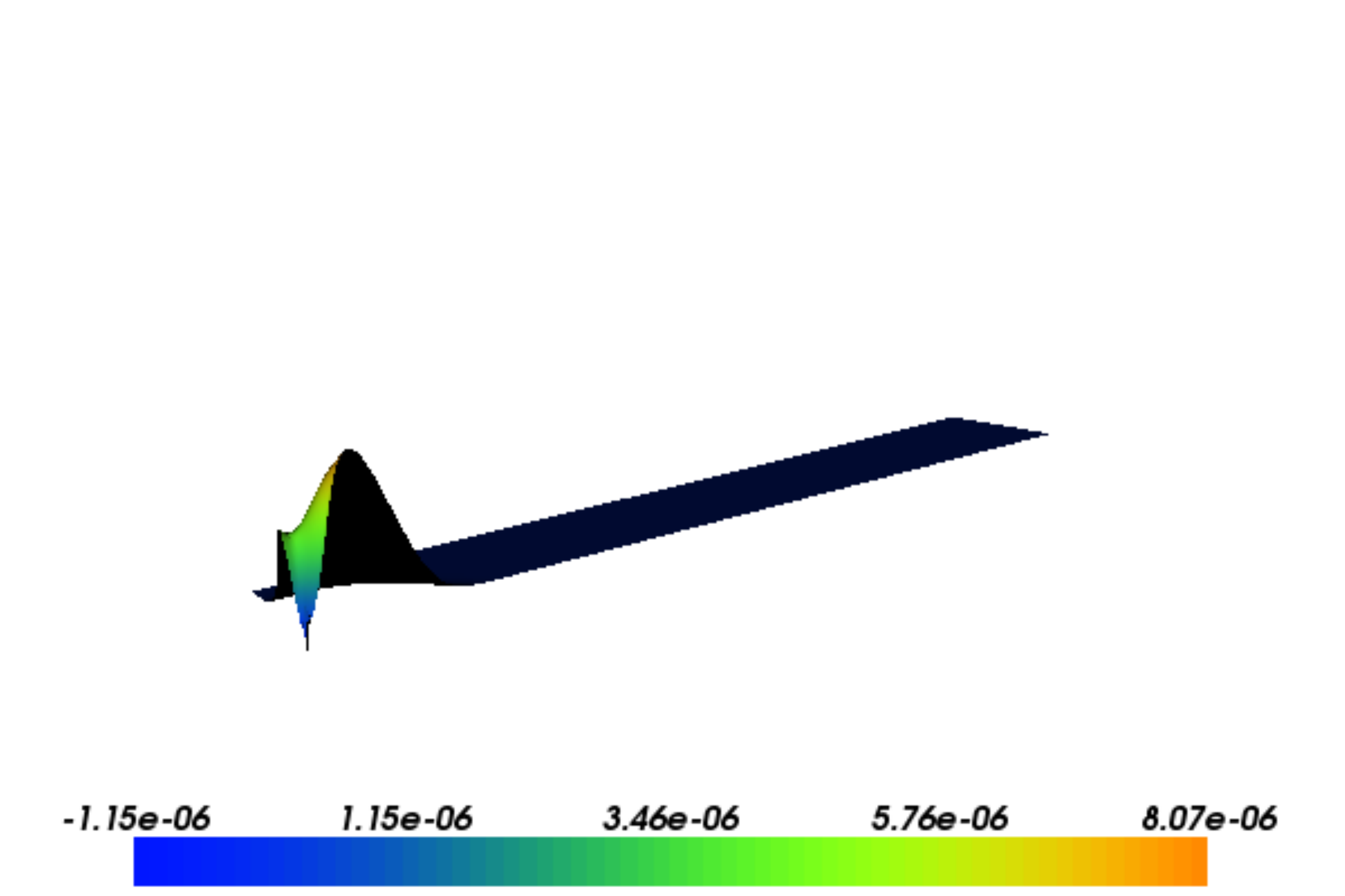}& \includegraphics[scale=.22]{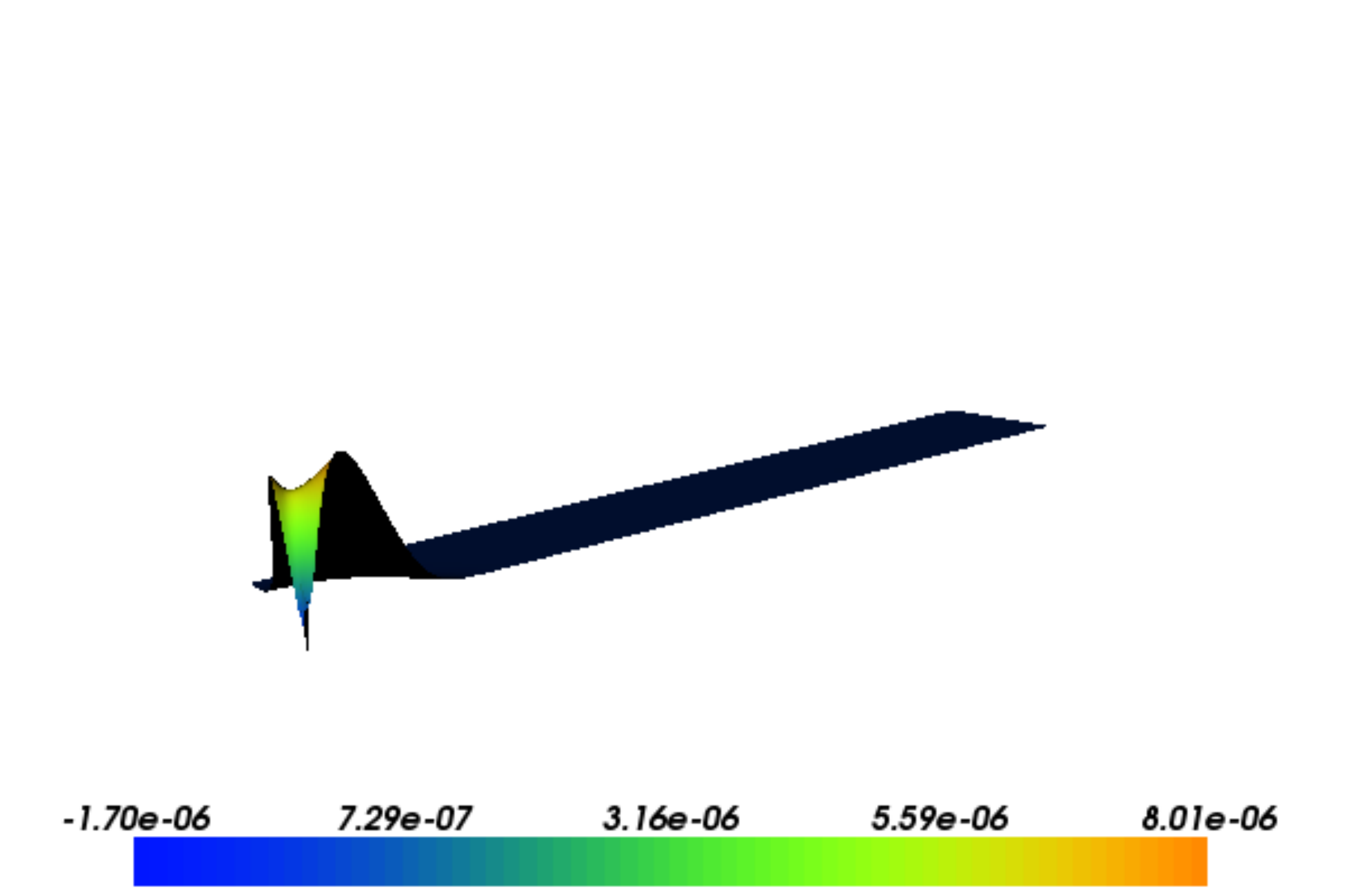}&\includegraphics[scale=.22]{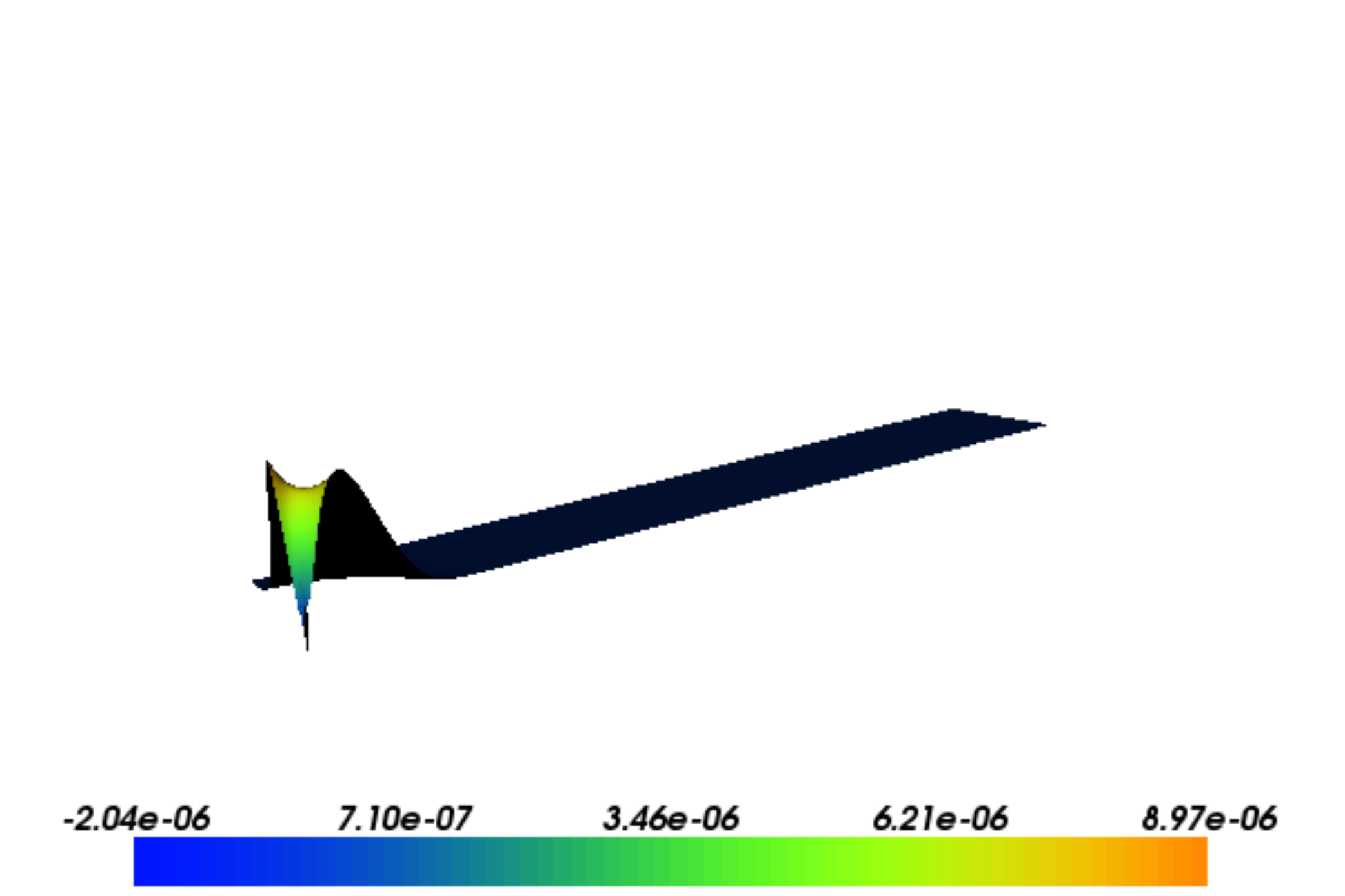}\\ \hline
          $t=12000$ & \includegraphics[scale=.22]{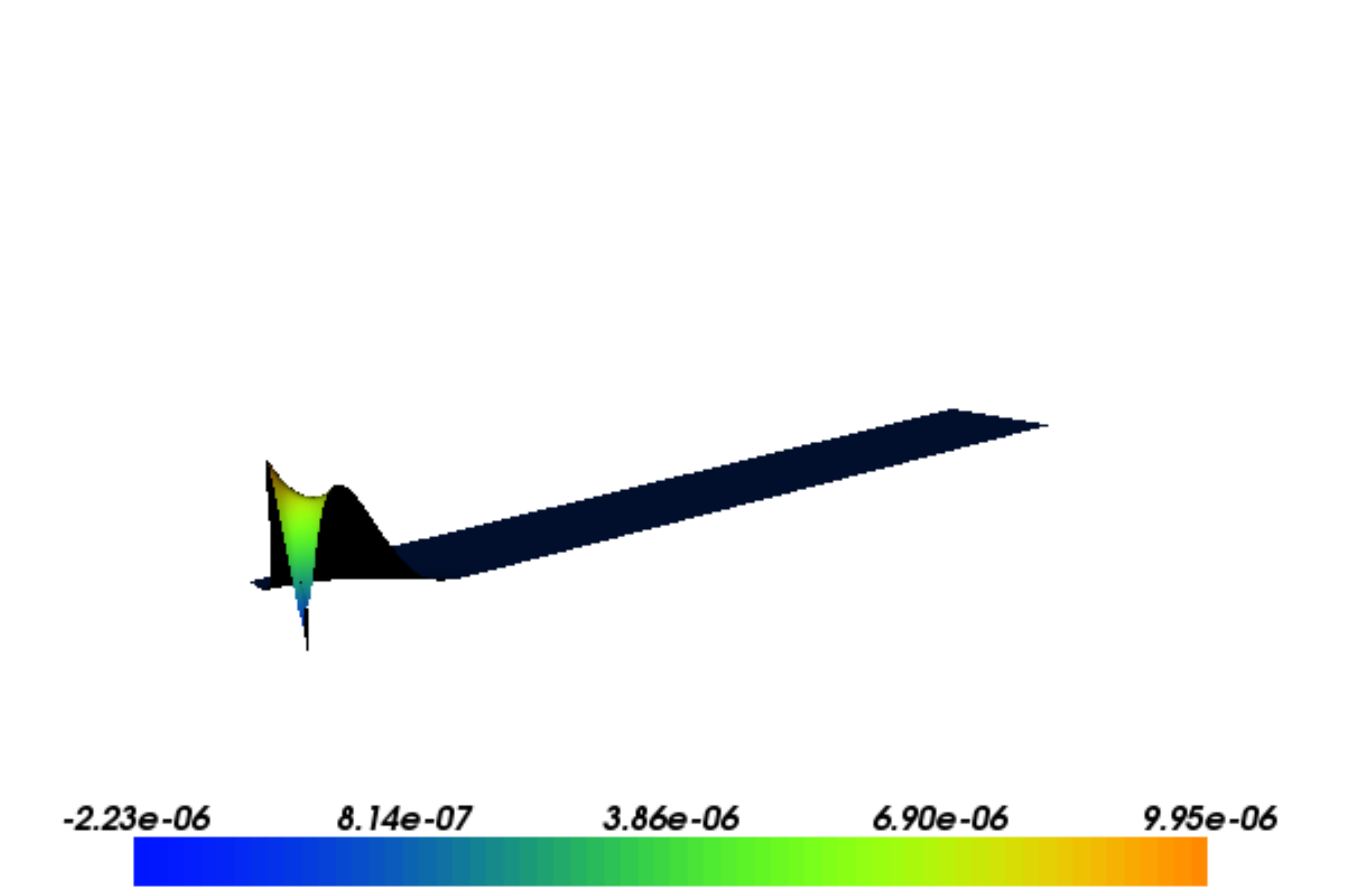}& \includegraphics[scale=.22]{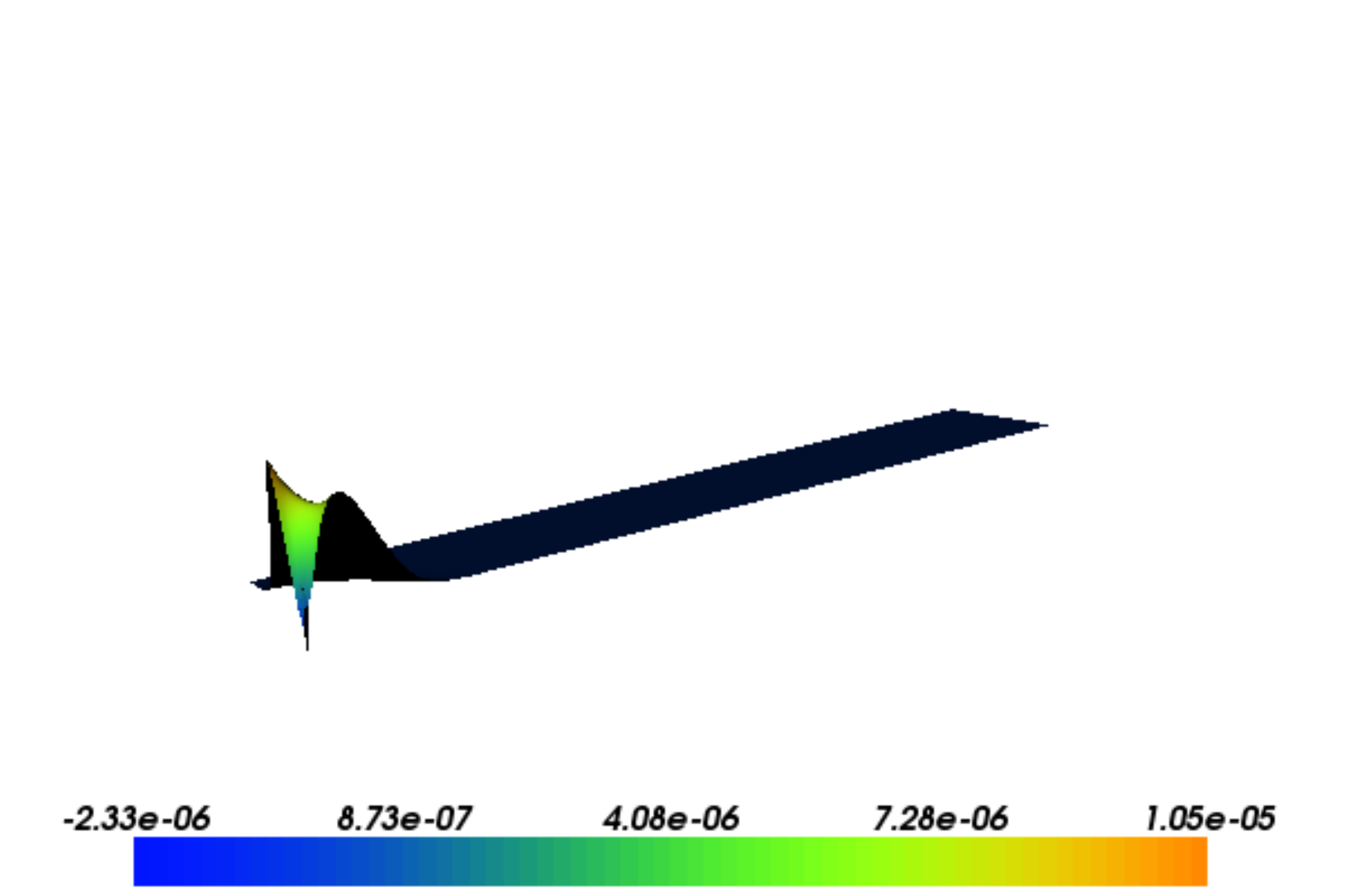}&\includegraphics[scale=.22]{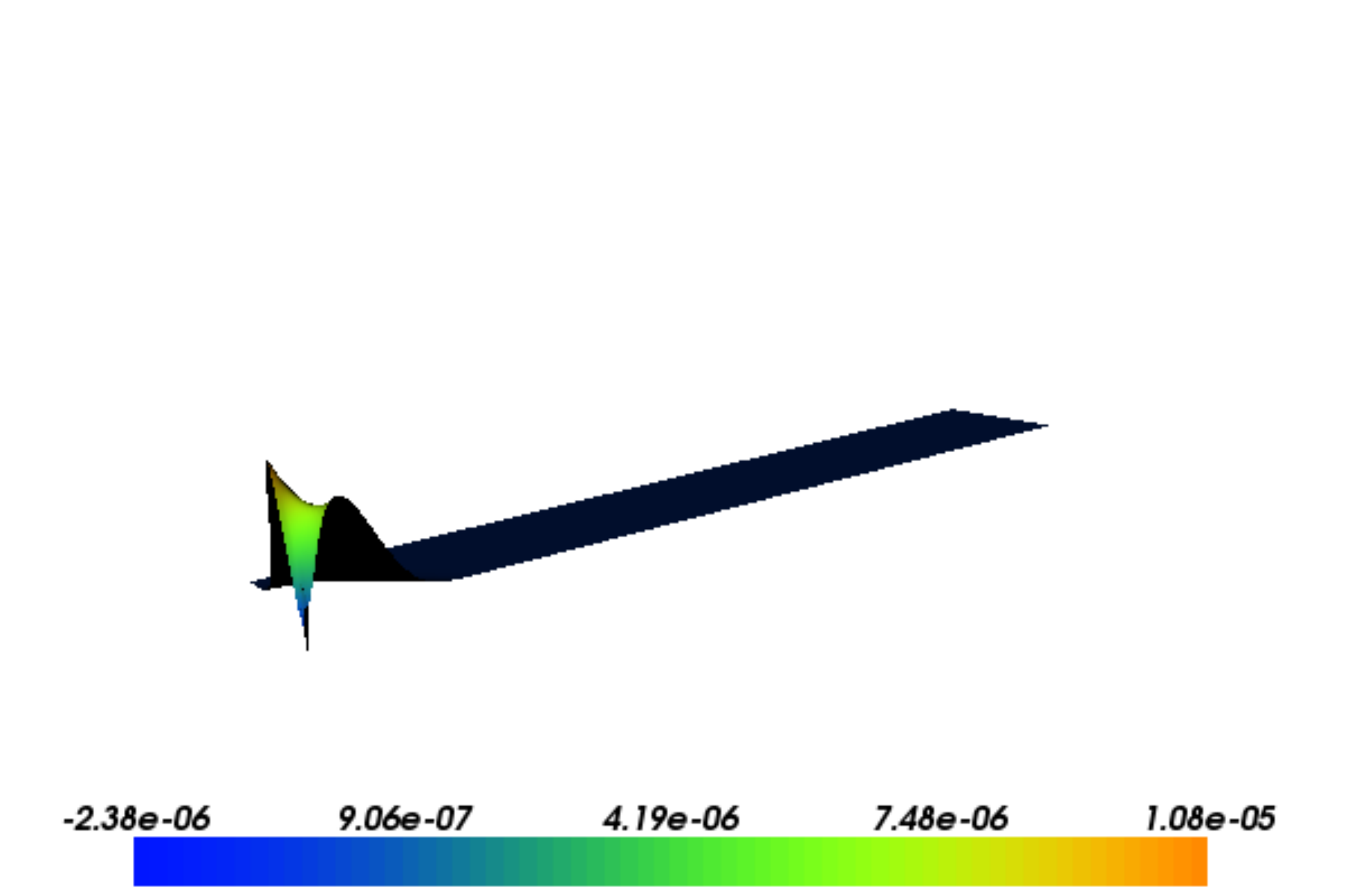}\\ \hline
        \hline
        \end{tabular}
    \caption{\bf {The time evolution of a gaussian initial condition relaxing into the steady state for the wild type with RNA polymerase in the partition function and constant diffusion tensor with different parameters from \ref{rnapdiff} .  The lysogenic peak in the classical dynamics is near 1500 cI proteins.  To find a distribution with a second peak required this enlarged domain and was facilitated by an initial conditition with many cI (3500).  The short axis is 0-1000 Cro proteins, and the long axis is 0-5000 cI proteins.  One can see the global effects through the large displacement of the lysogenic peak from its stationary point.}}    
        \label{rnapdiff2}
    \end{table}%

\begin{table}[ht]
        \centering
        \begin{tabular}{|p{0.11\textwidth}|p{0.29\textwidth}|p{0.29\textwidth}|p{0.29\textwidth}|}
          \hline
          \multicolumn{4}{|c|}{Time Evolution}
          \\ \hline \hline
          time&$+0$&$+600$&$+1200$ \\ \hline
          $t=0$ & \includegraphics[scale=.22]{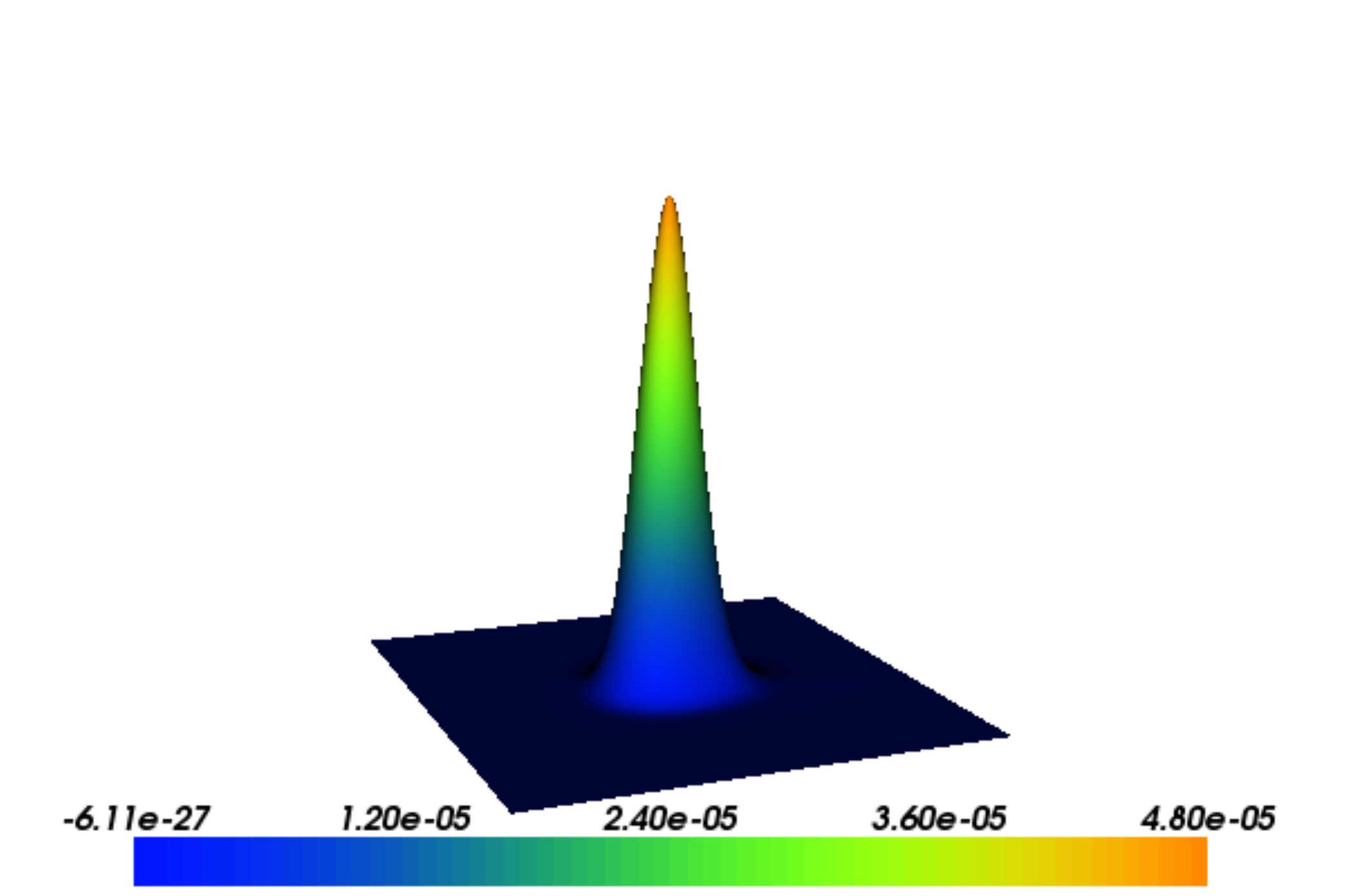}& \includegraphics[scale=.22]{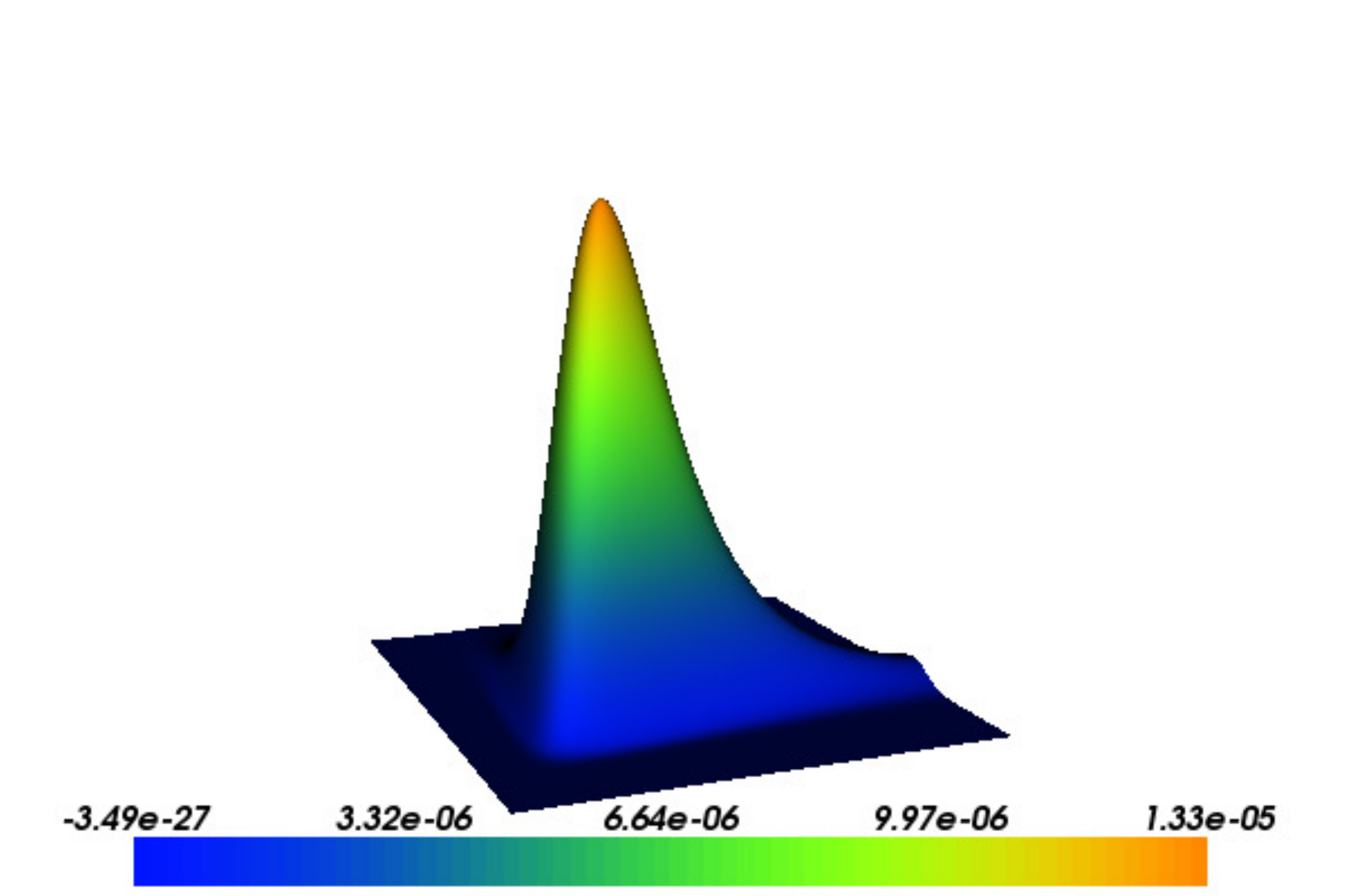}&\includegraphics[scale=.22]{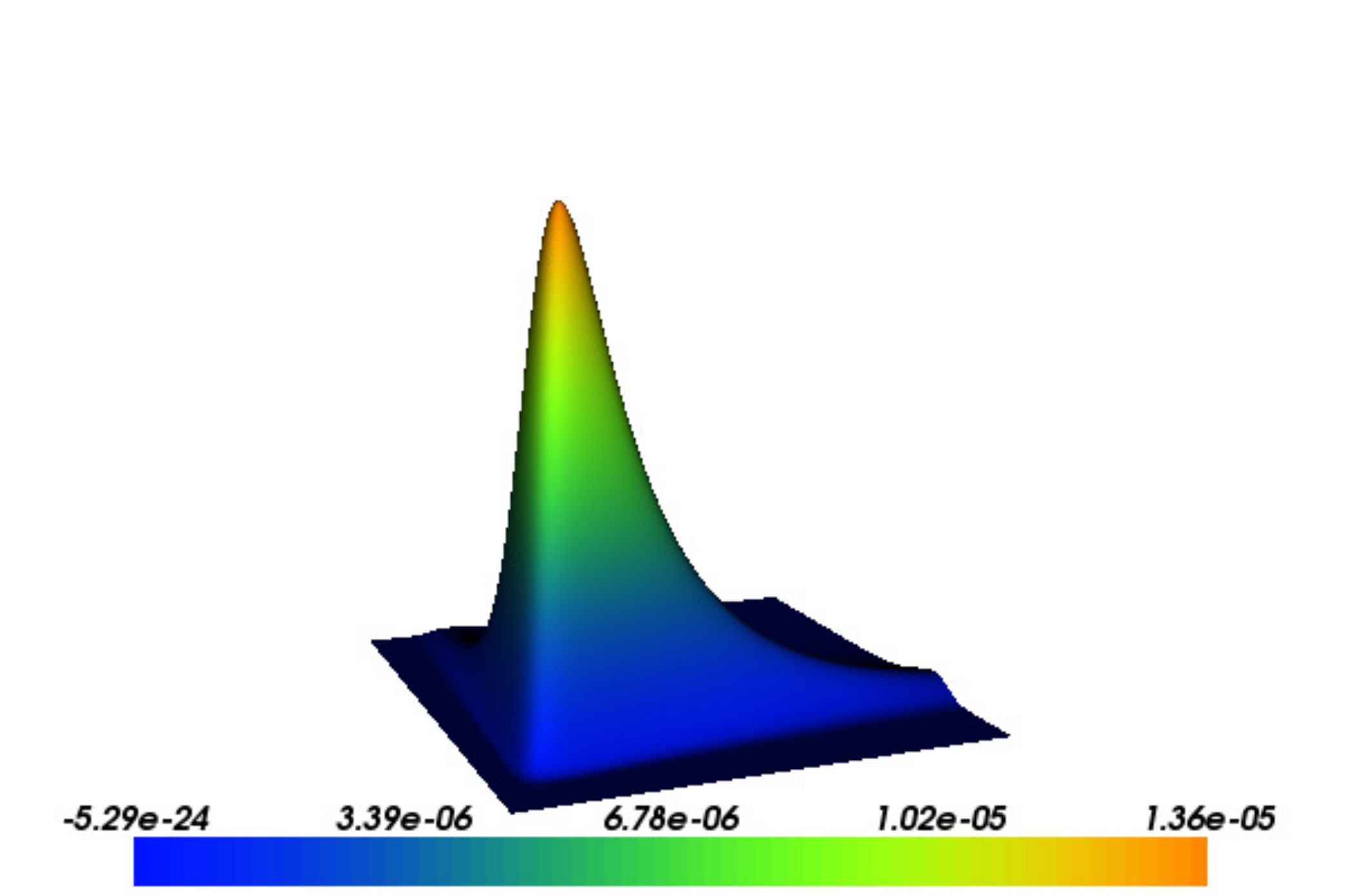}\\ \hline
          $t=301$ & \includegraphics[scale=.22]{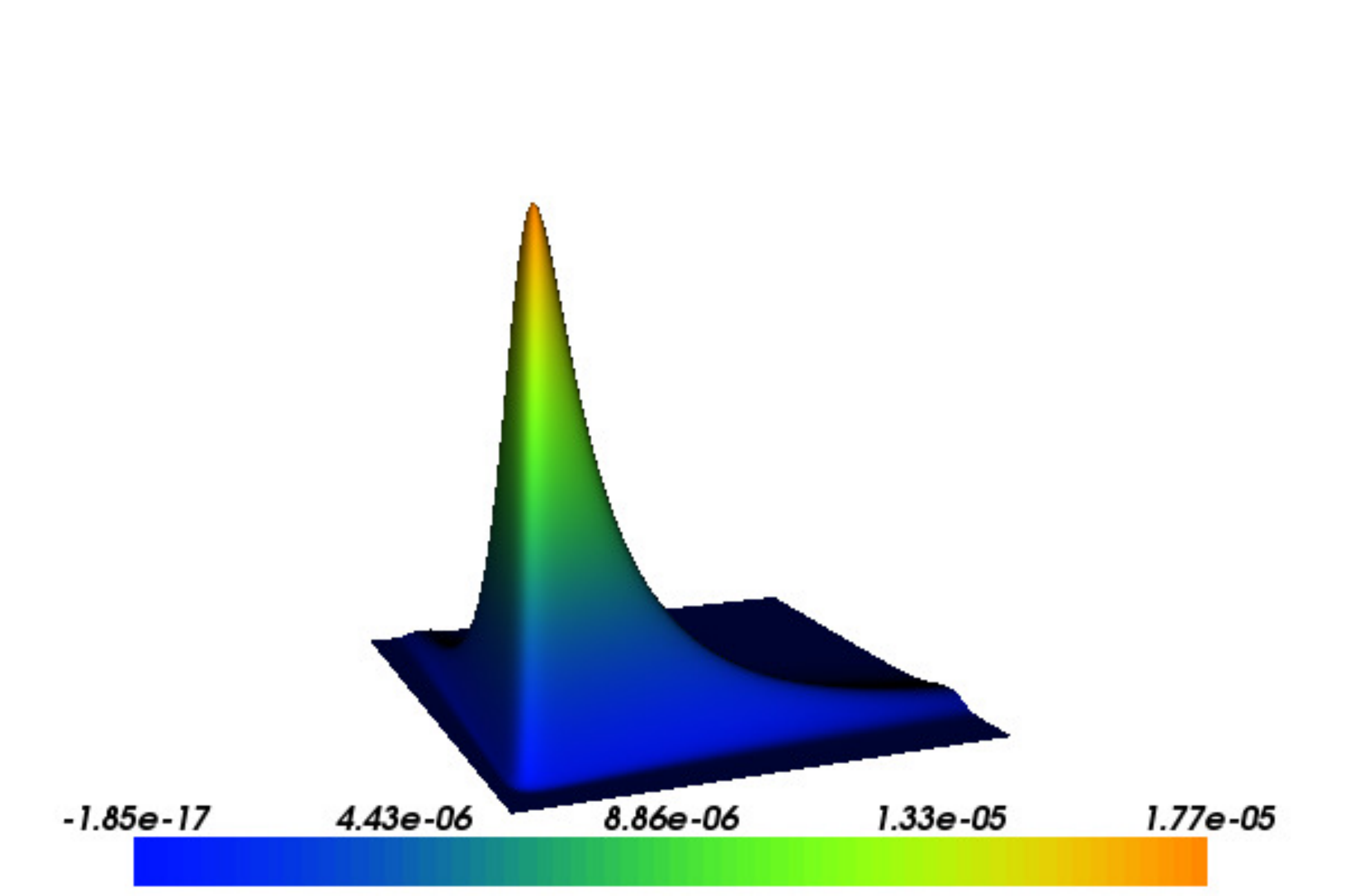}& \includegraphics[scale=.22]{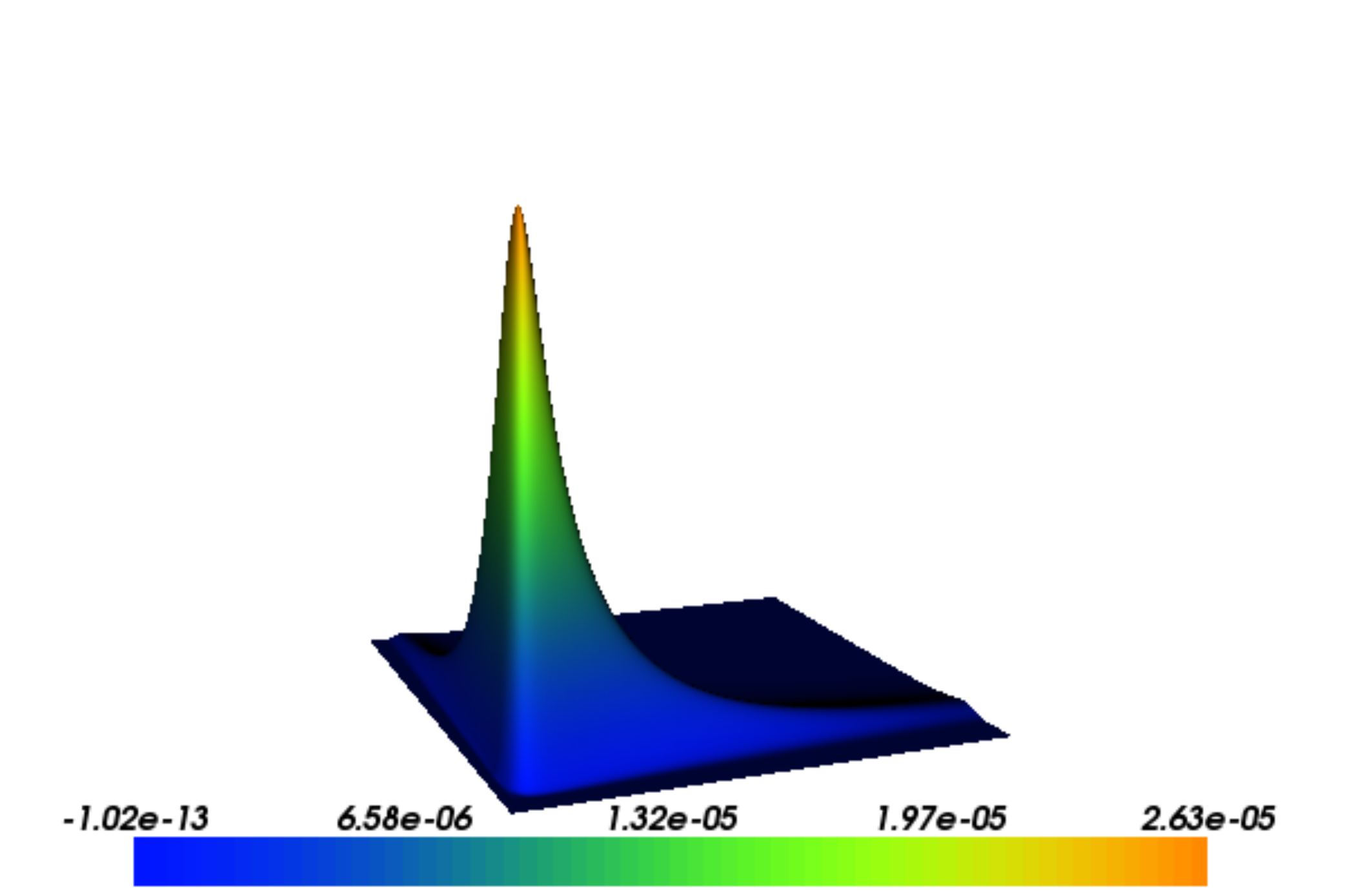}&\includegraphics[scale=.22]{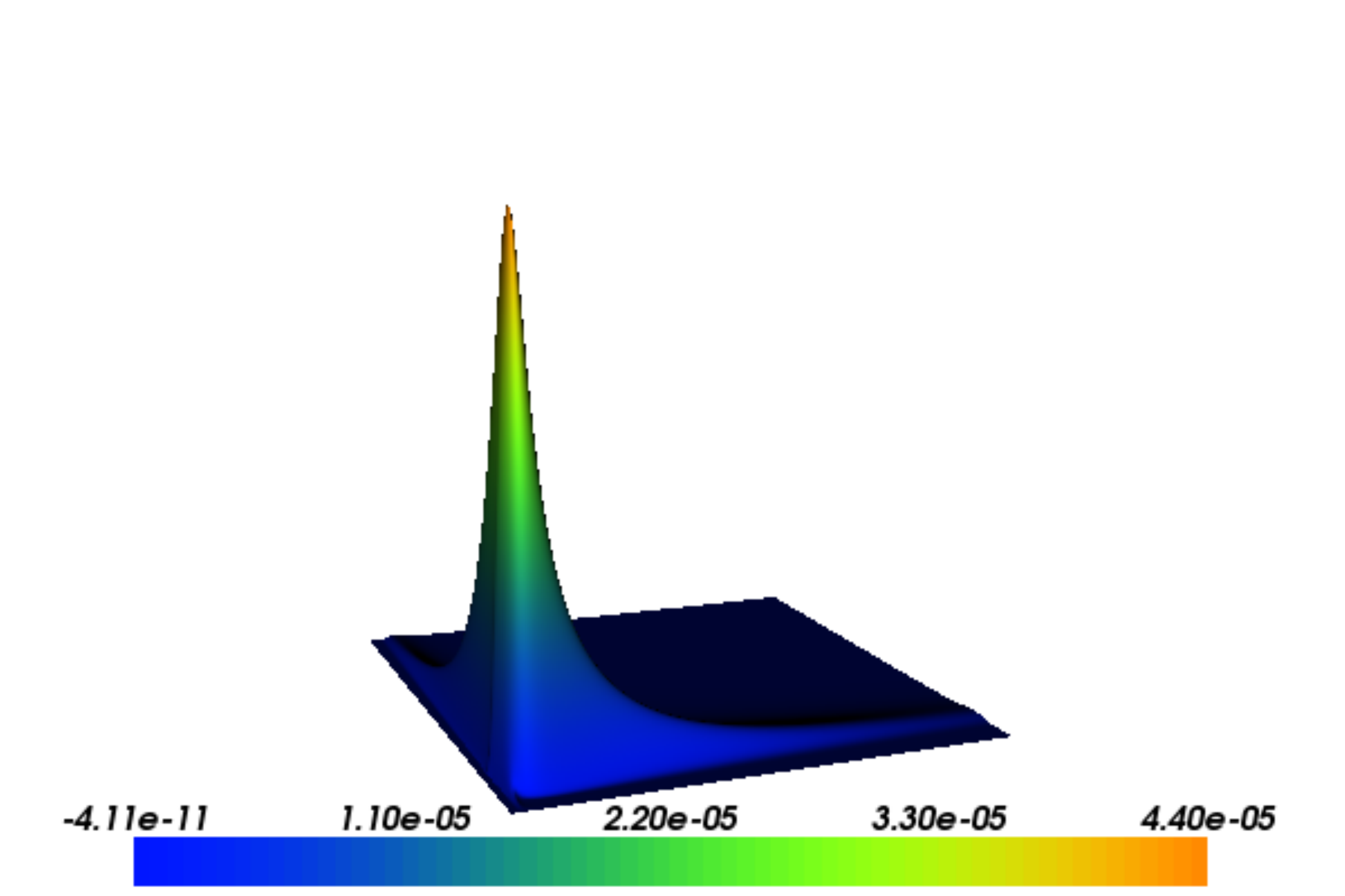}\\ \hline
          $t=601$ & \includegraphics[scale=.22]{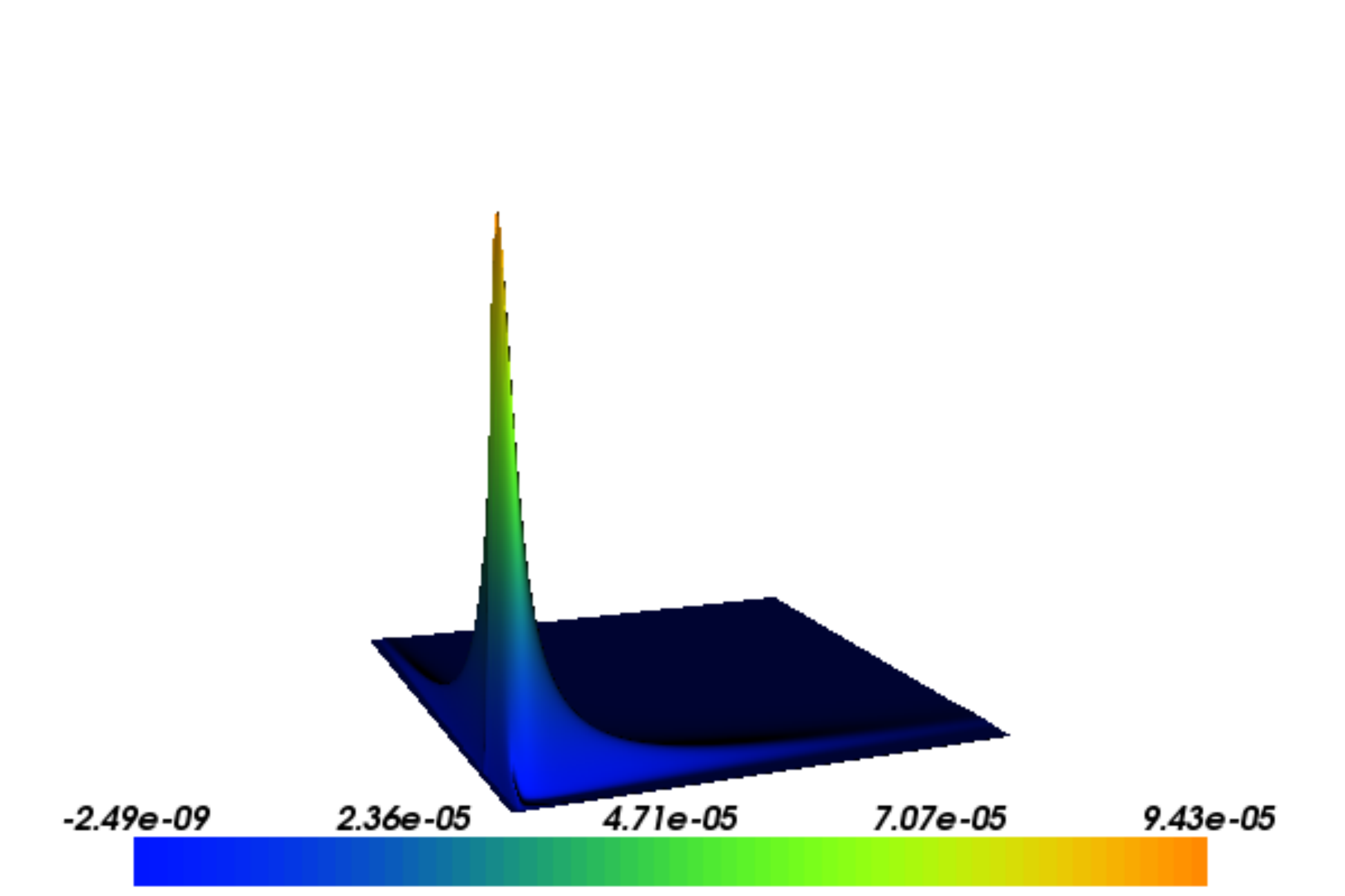}& \includegraphics[scale=.22]{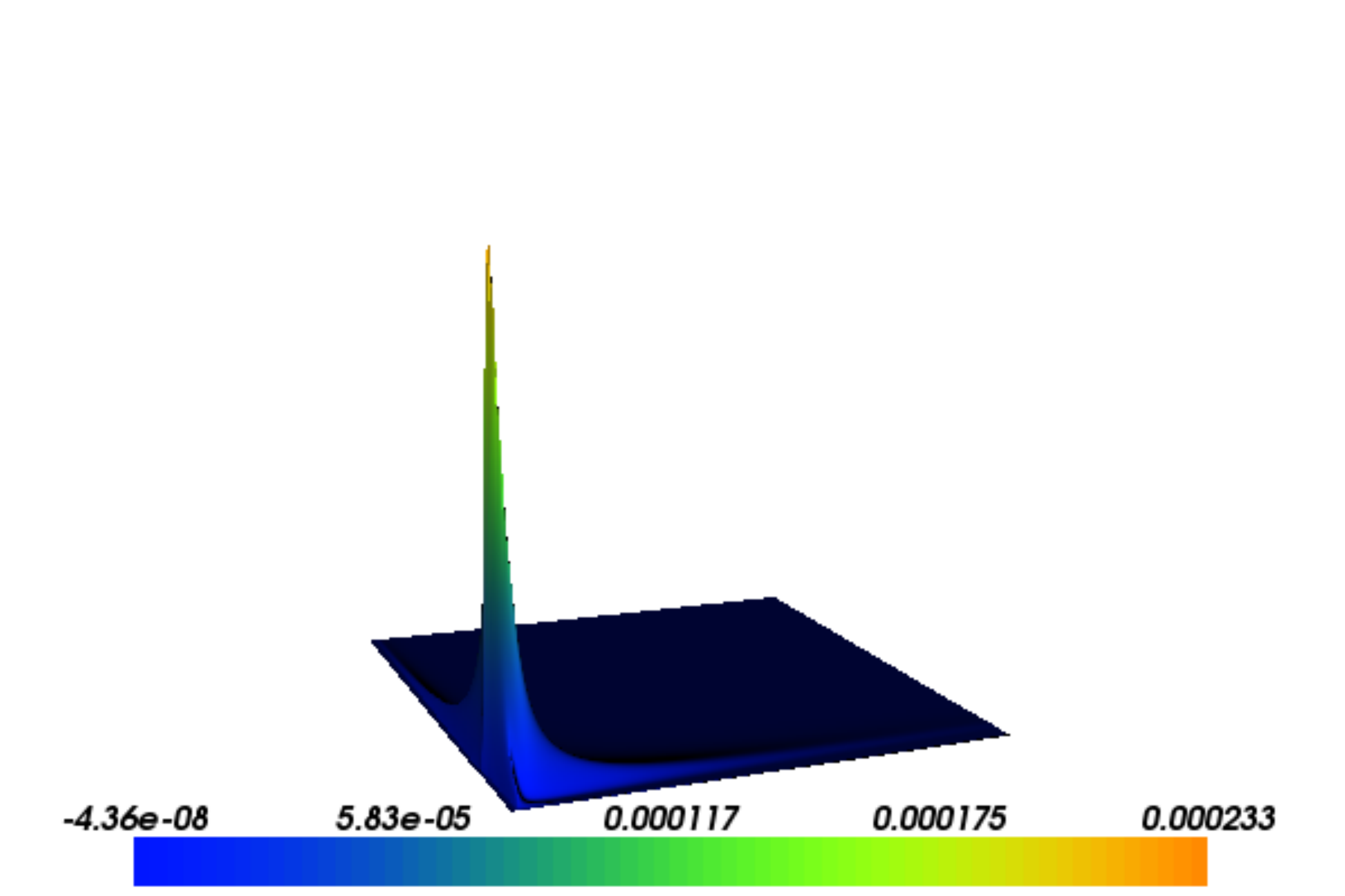}&\includegraphics[scale=.22]{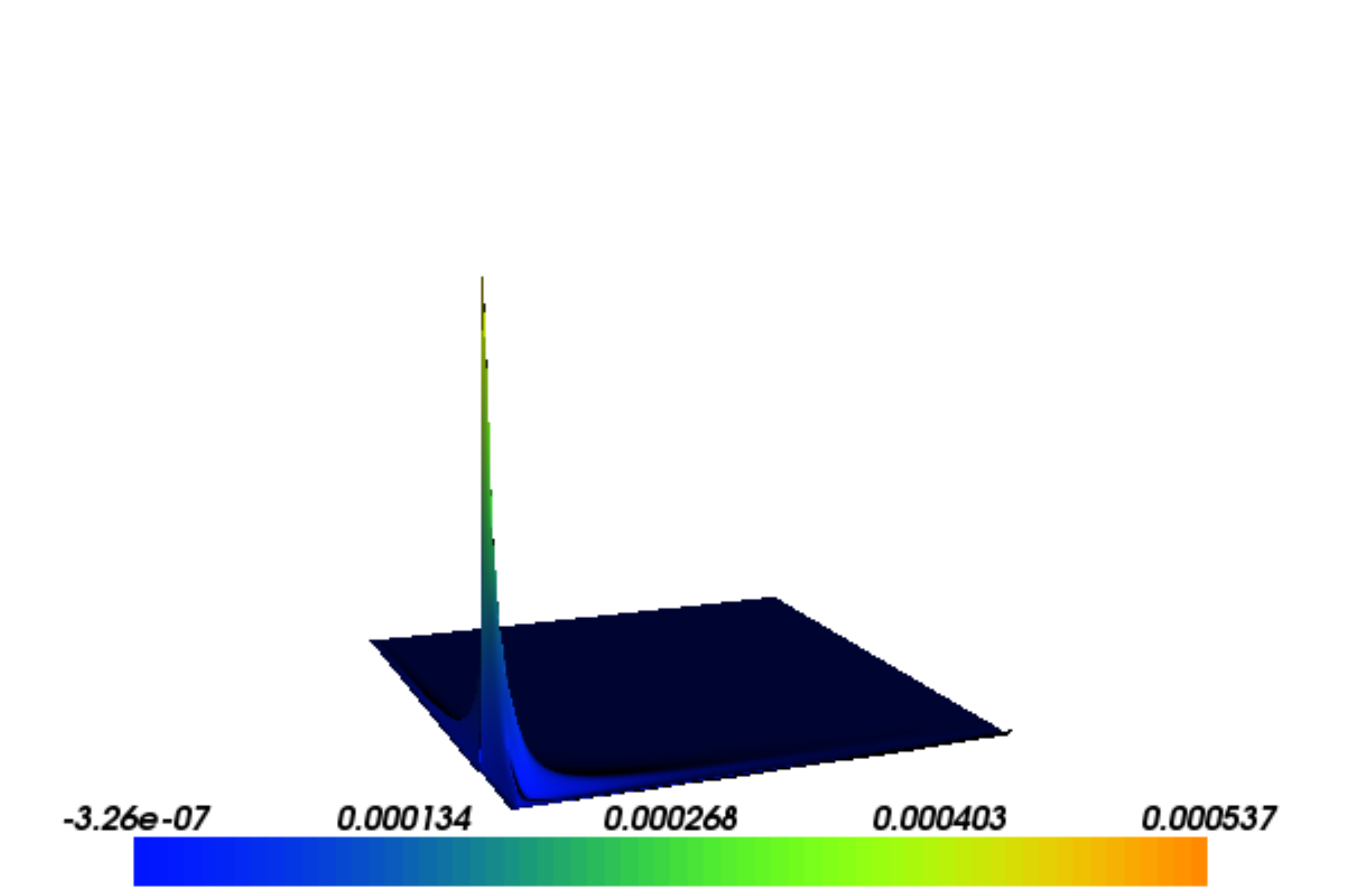}\\ \hline
        \hline
        \end{tabular}
    \caption{\bf {The time evolution of a gaussian initial condition relaxing into the steady state for the wild type with concentration-space dependant diffusion. The peak corresponds to the lytic state.  Two strong tails develop, one extending towards potentially another peak.}}    
        \label{rnapspacediff}
    \end{table}%

\begin{table}[ht]
\noindent 
\resizebox{!}{5.5cm}{     
        \begin{tabular}{|c|c|c|}
          \hline
          \multicolumn{3}{|c|}{$\lambda_{1**}$ Path Entropy and Path Energy Densities}
          \\ \hline \hline
             \includegraphics[scale=.27]{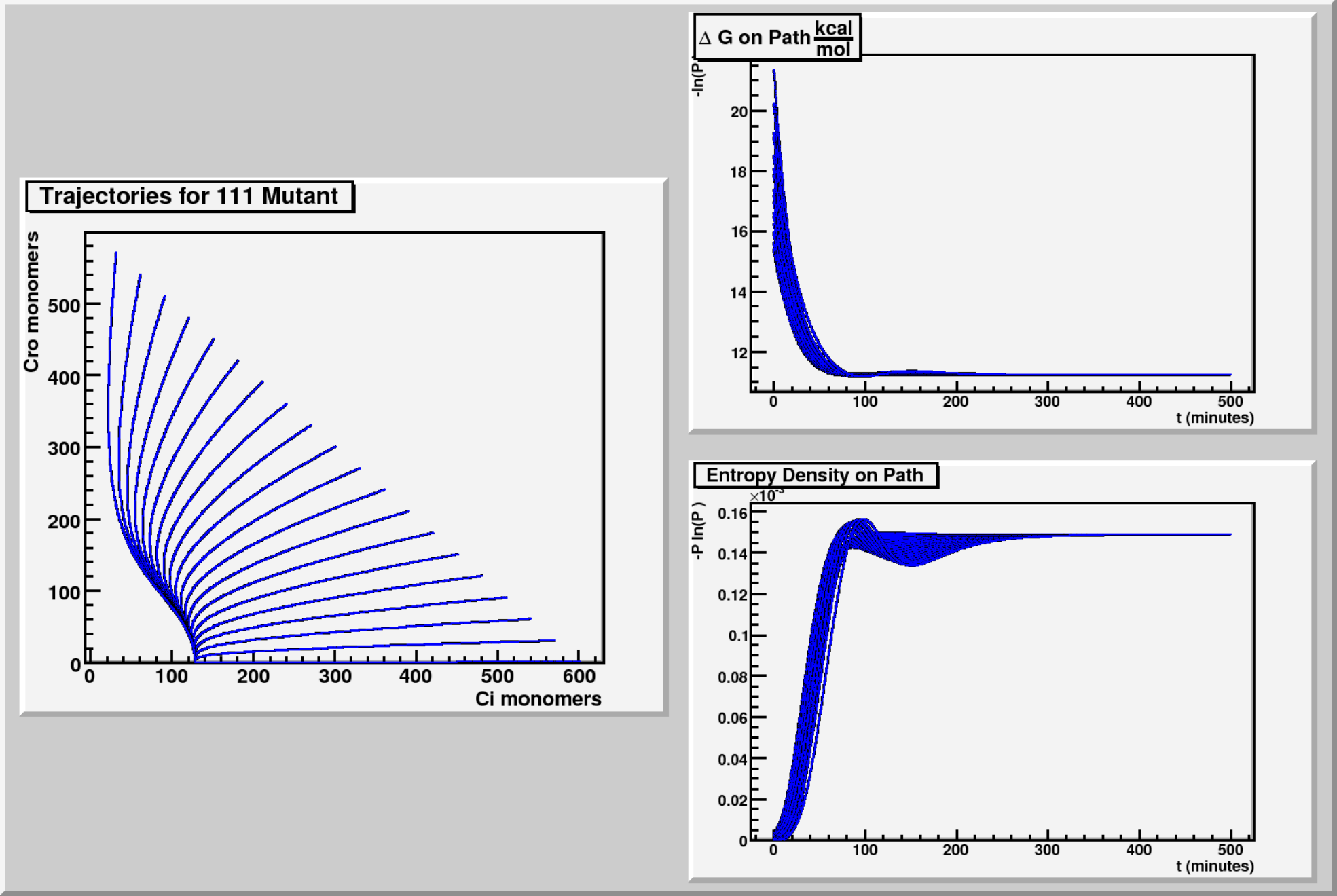}     
         &  
             \includegraphics[scale=.27]{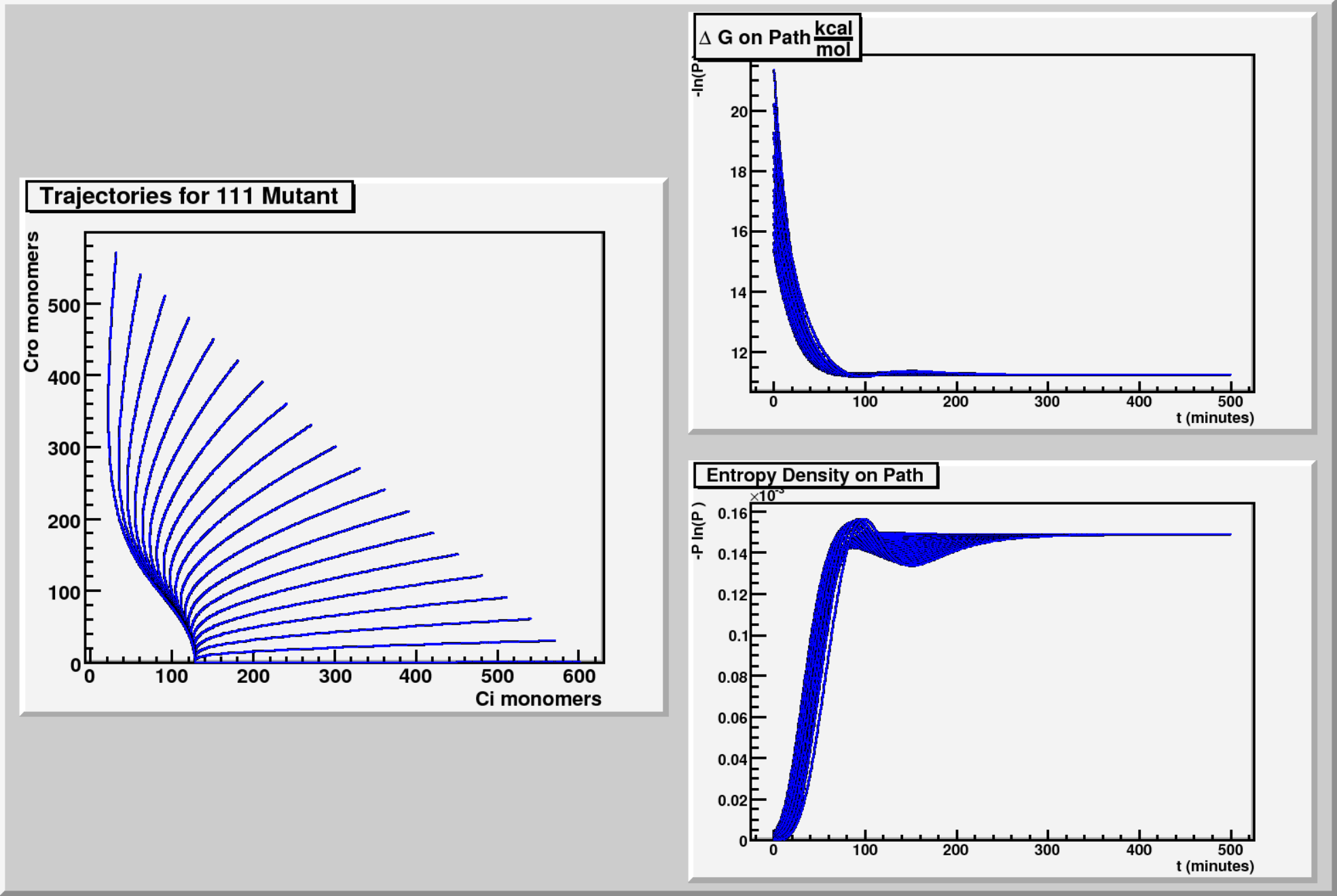}     
             &
             \includegraphics[scale=.27]{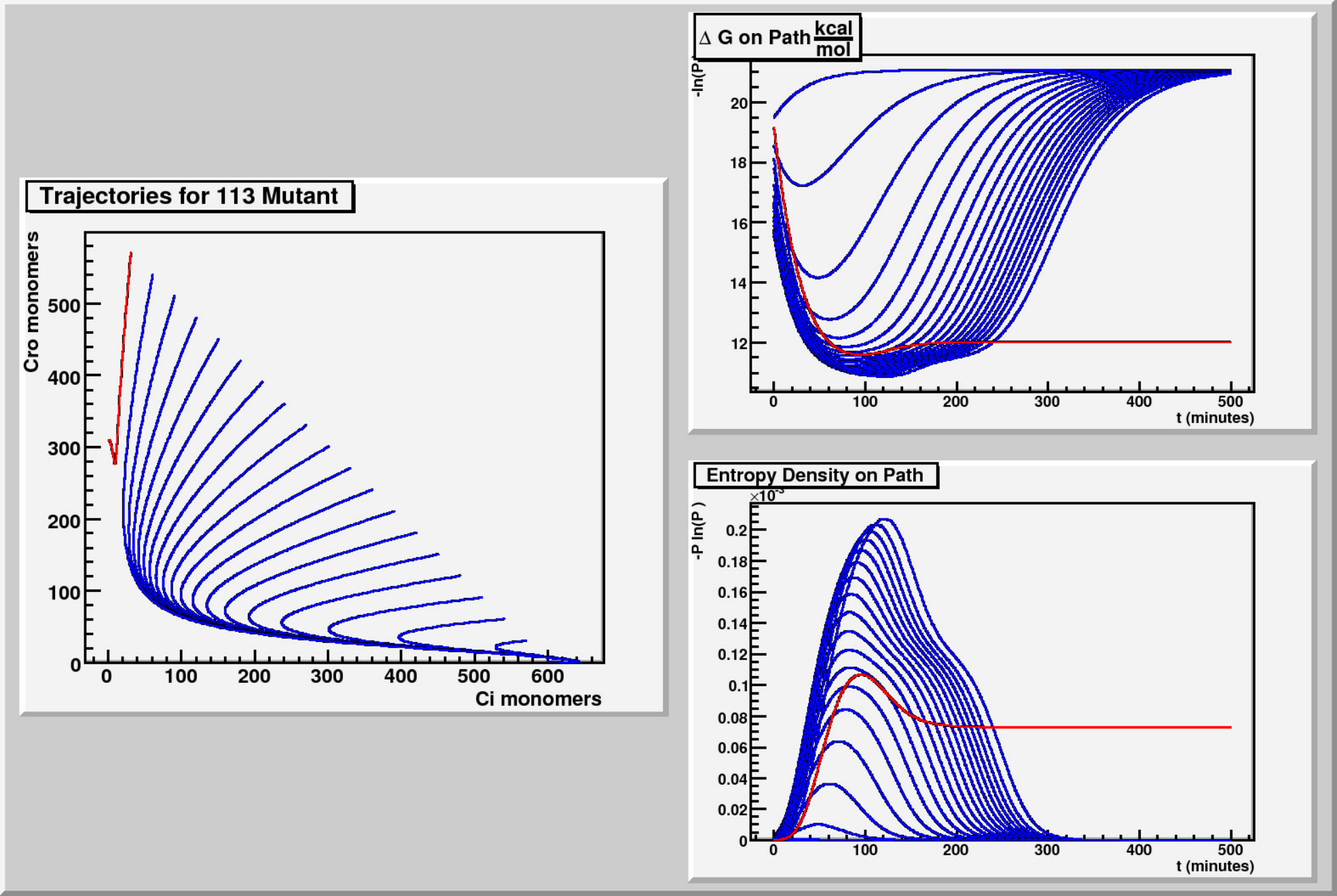}      

         \\ \hline \hline
             \includegraphics[scale=.27]{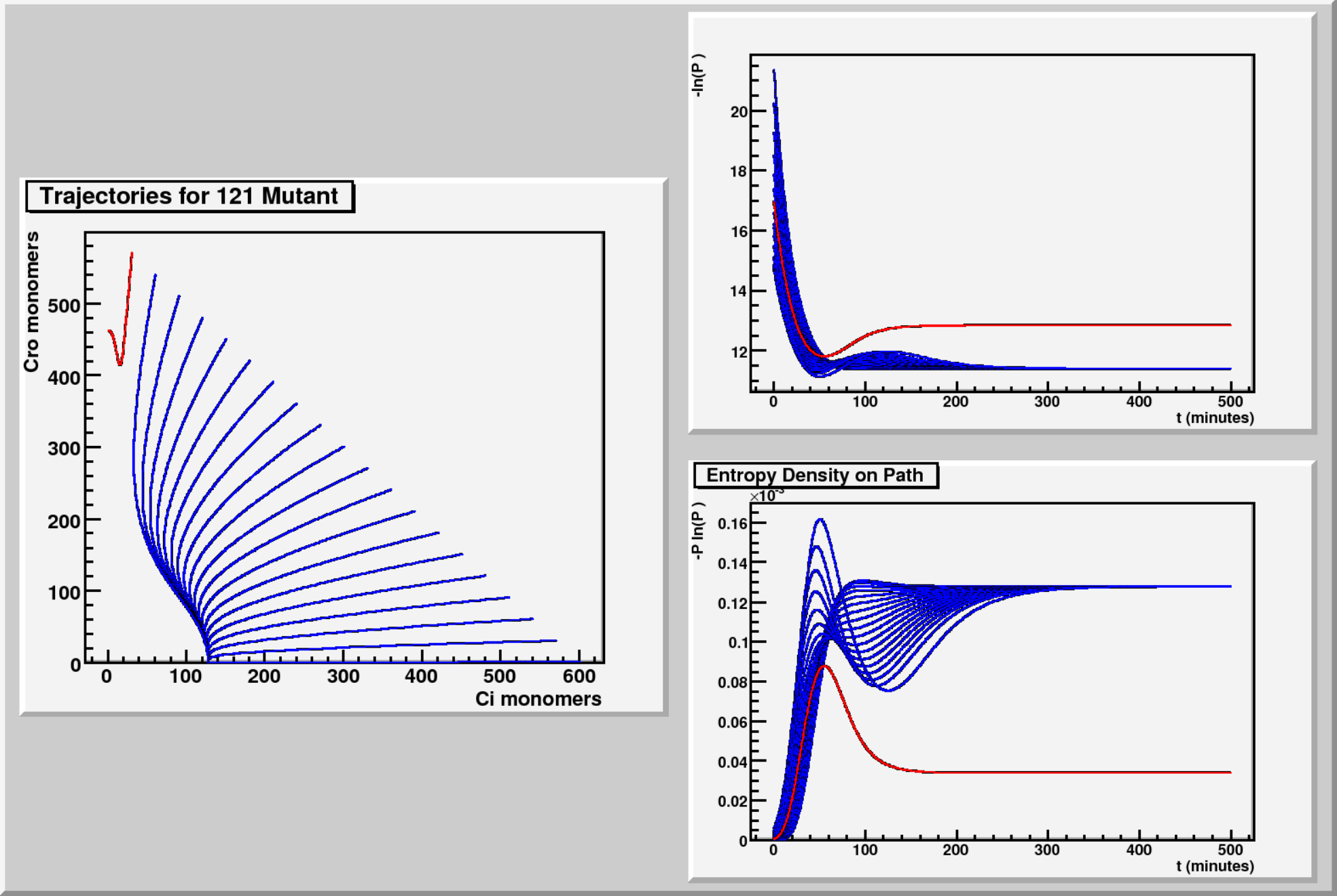}     
         &  
             \includegraphics[scale=.27]{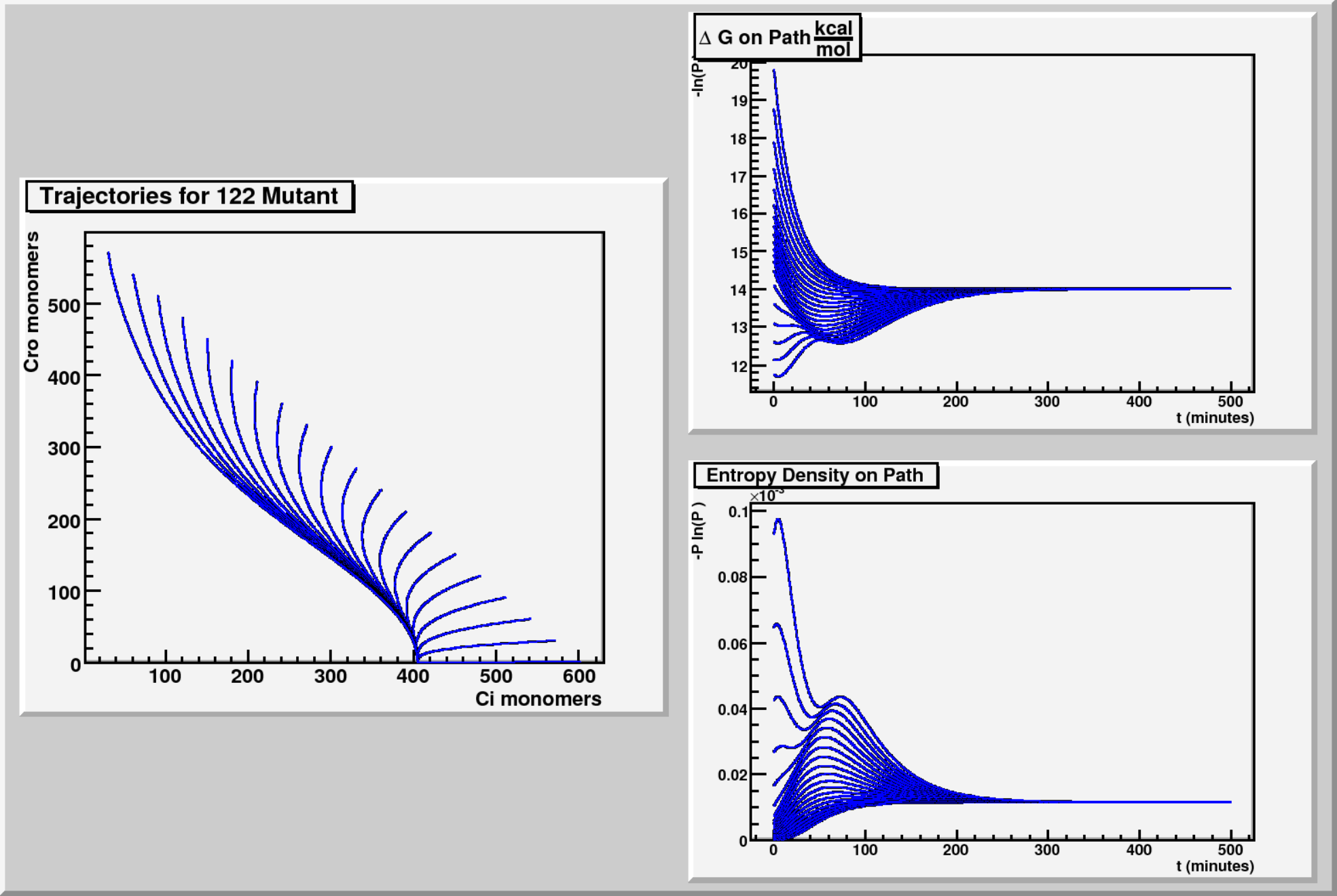}     
             &
             \includegraphics[scale=.27]{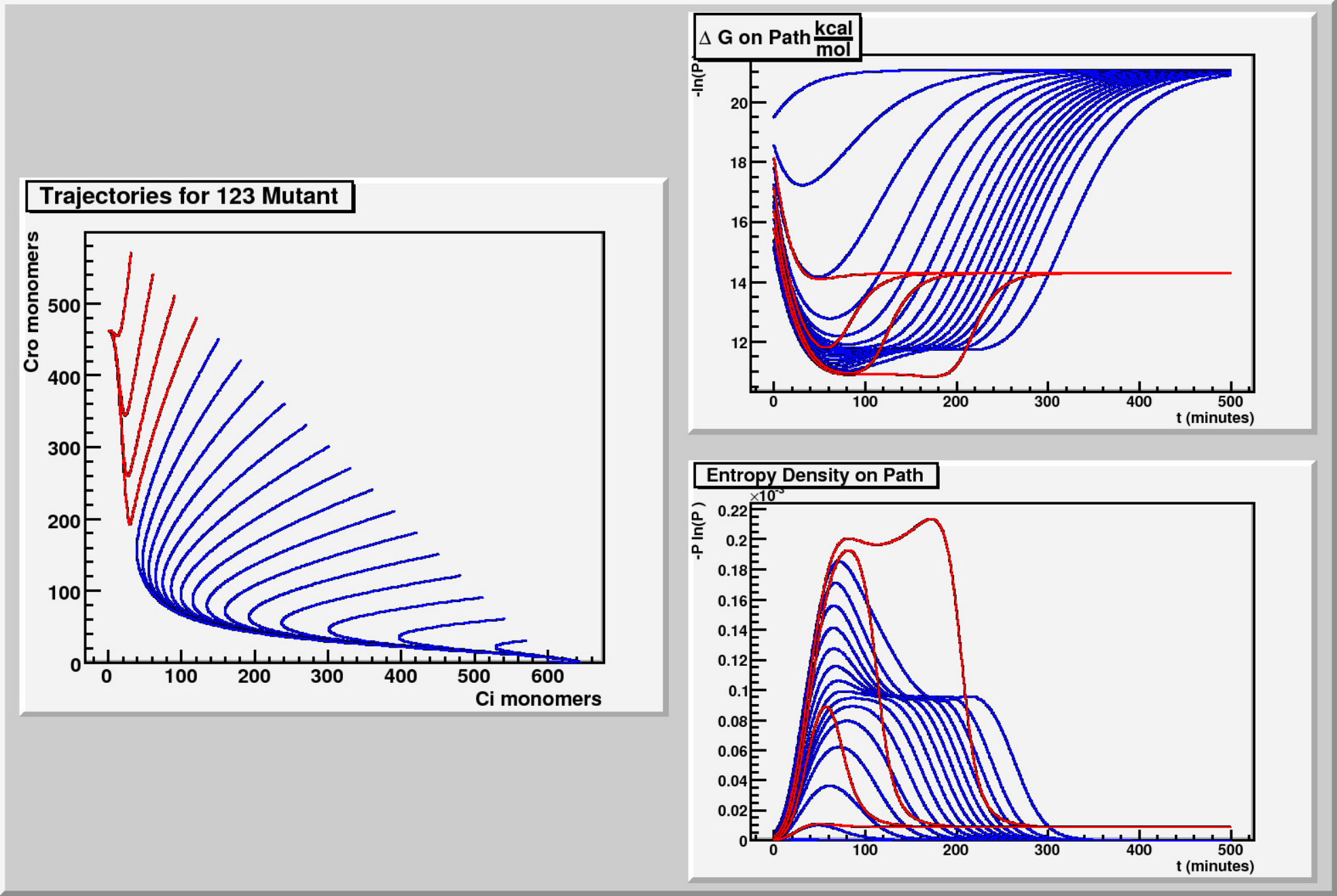}  

         \\ \hline \hline
             \includegraphics[scale=.27]{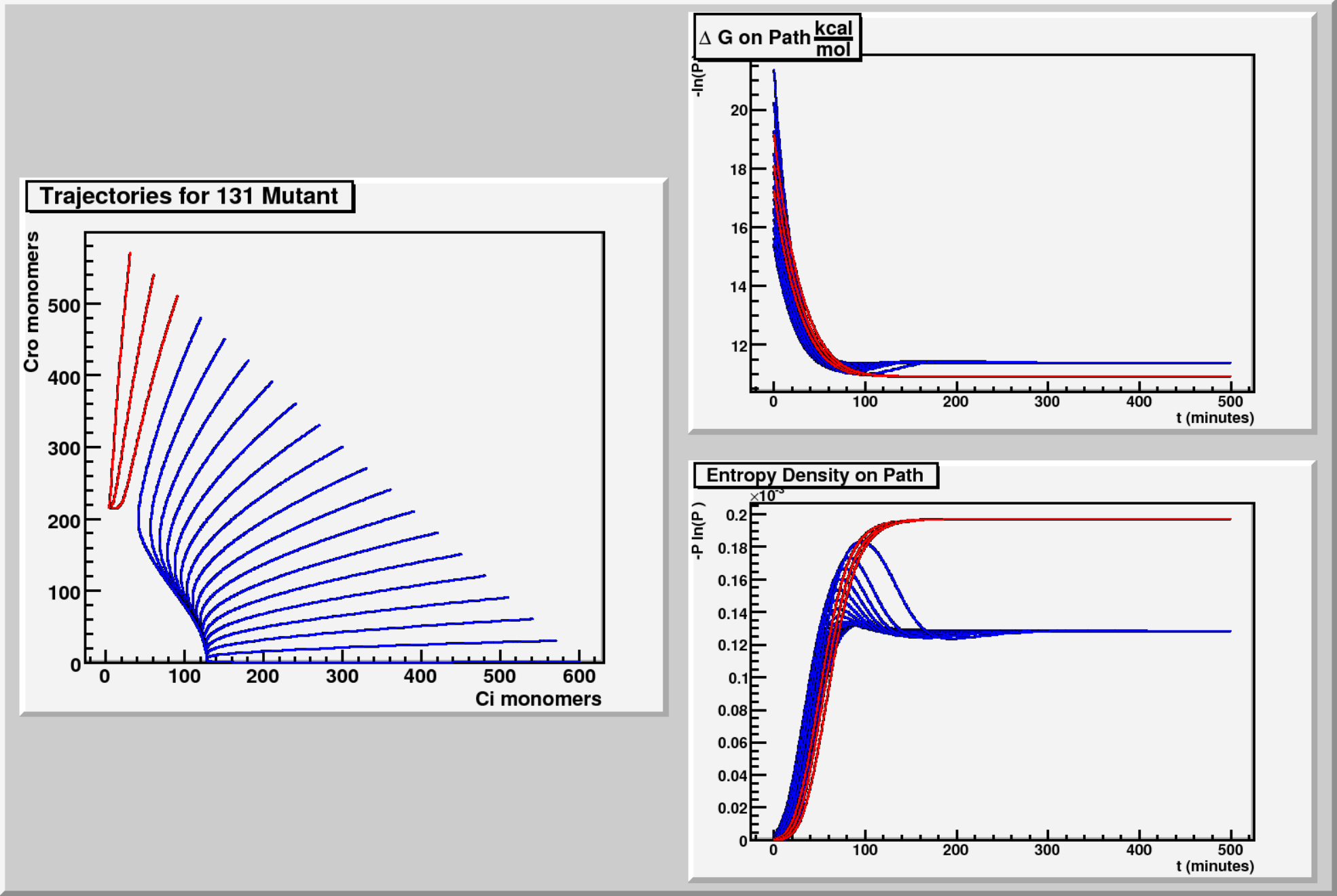}     
         &  
             \includegraphics[scale=.27]{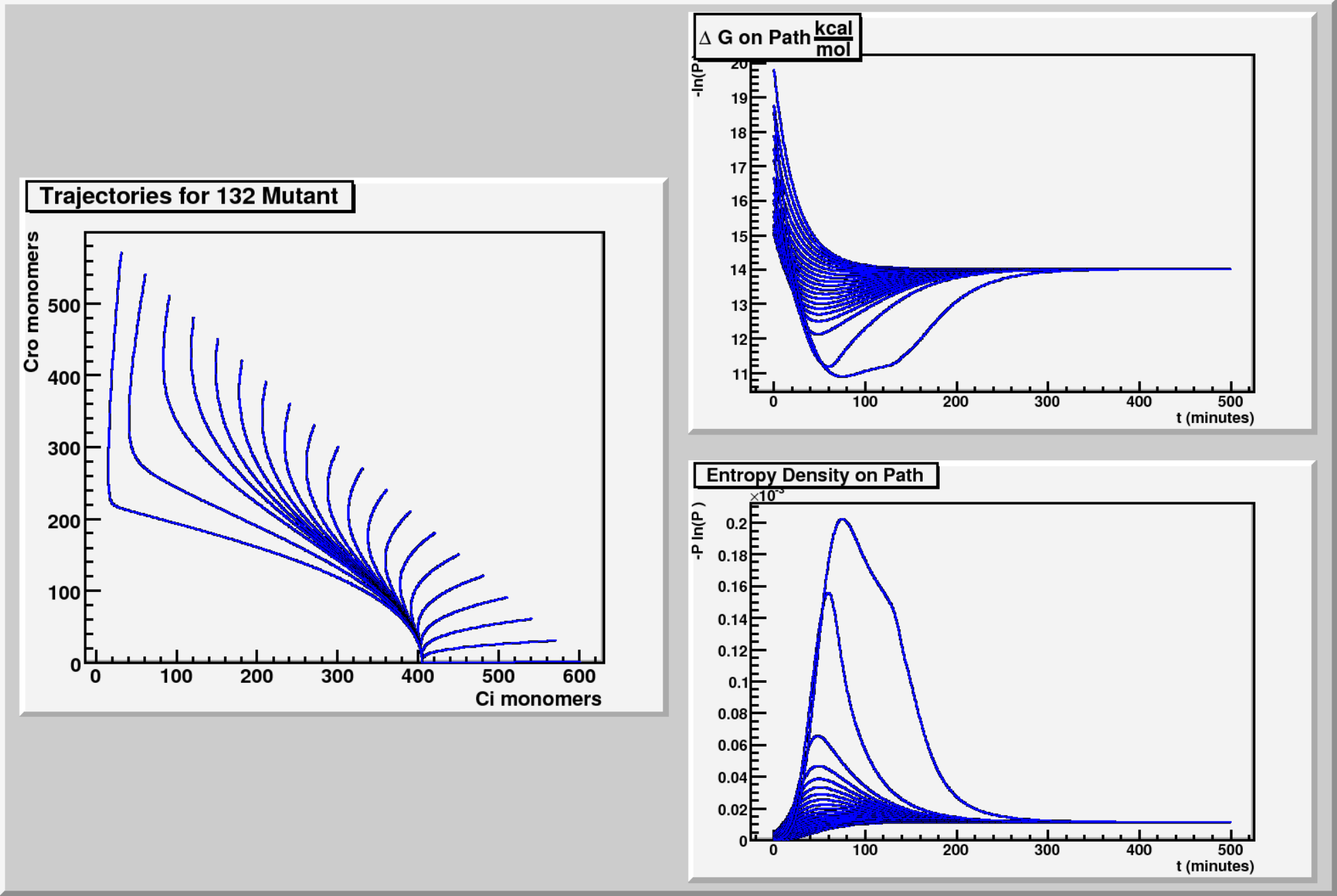}     
             &
             \includegraphics[scale=.27]{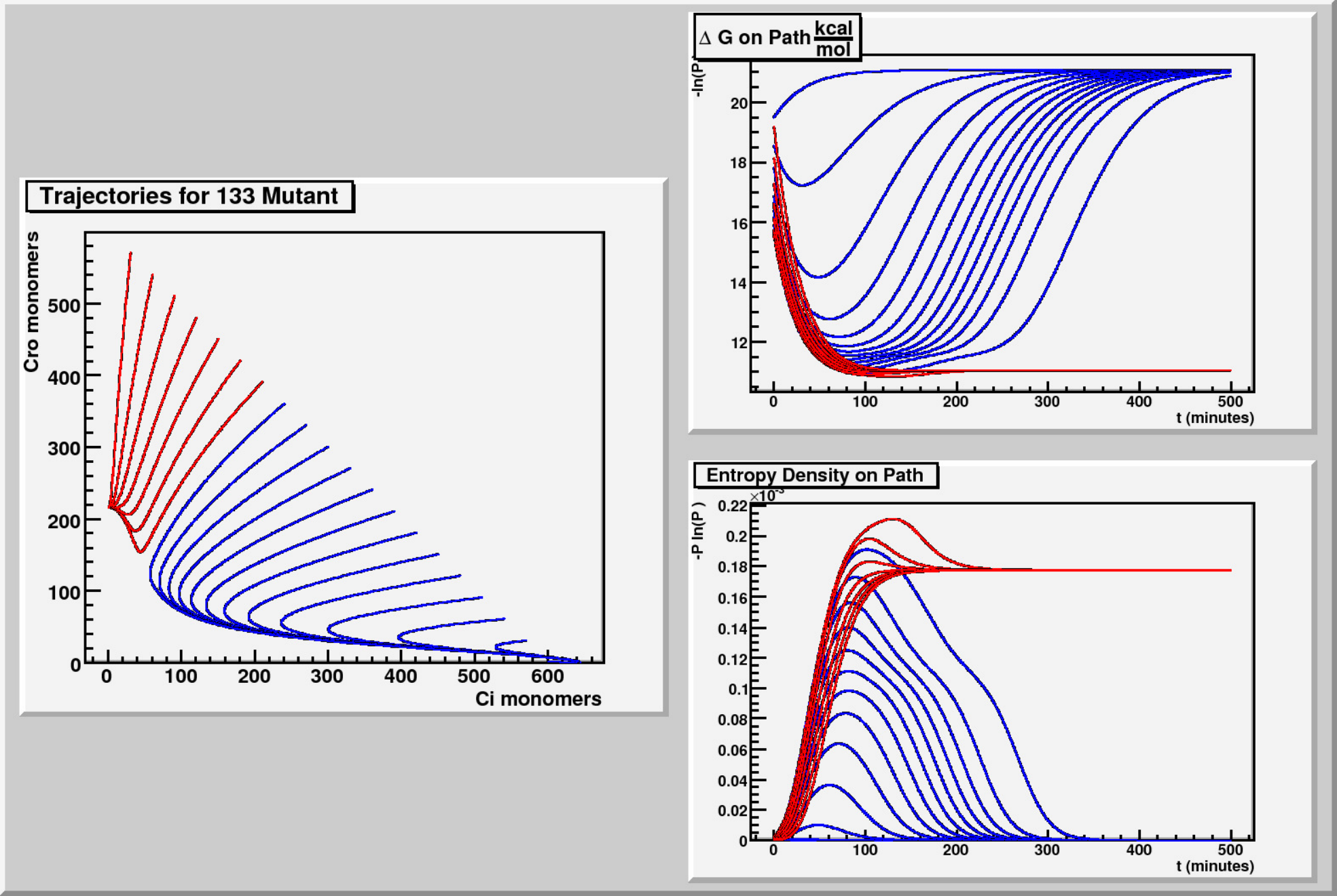}  

        \end{tabular}
}
    \caption{\bf {Entropy and Energy densities evaluated along example paths for mutations in $O_{R2,R3}$ with the $O_{R1}$ fixed.  Different trajectories can have different limiting values of entropy and energy density depending on the path taken.}}
        \label{tab:ltraj1}
    \end{table}%

\begin{table}[ht]
\noindent 
\resizebox{!}{5cm}{     
        \begin{tabular}{|c|c|c|}
          \hline
          \multicolumn{3}{|c|}{$\lambda_{2**}$ Path Entropy and Path Energy Densities}
          \\ \hline \hline
             \includegraphics[scale=.27]{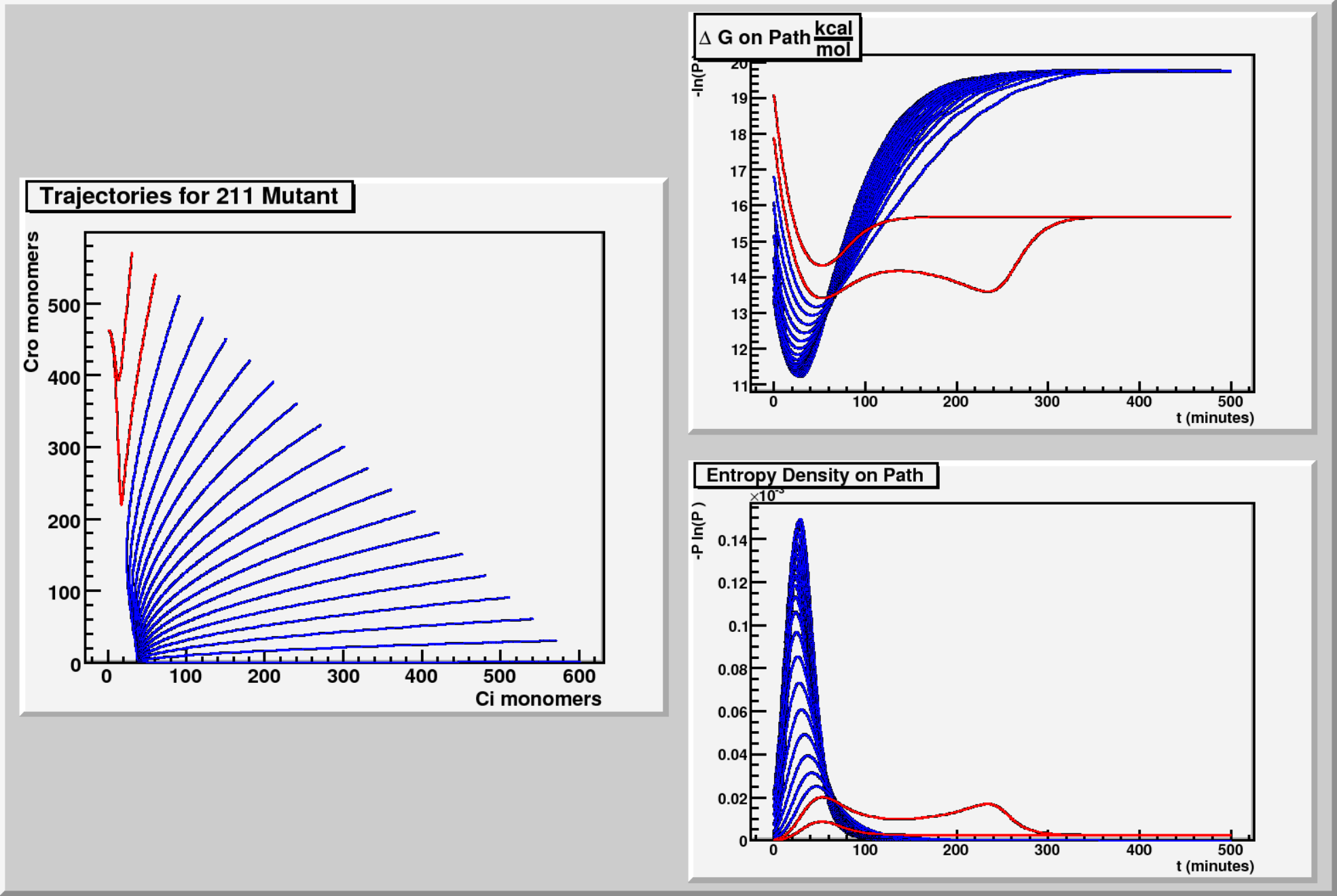}     
         &  
             \includegraphics[scale=.27]{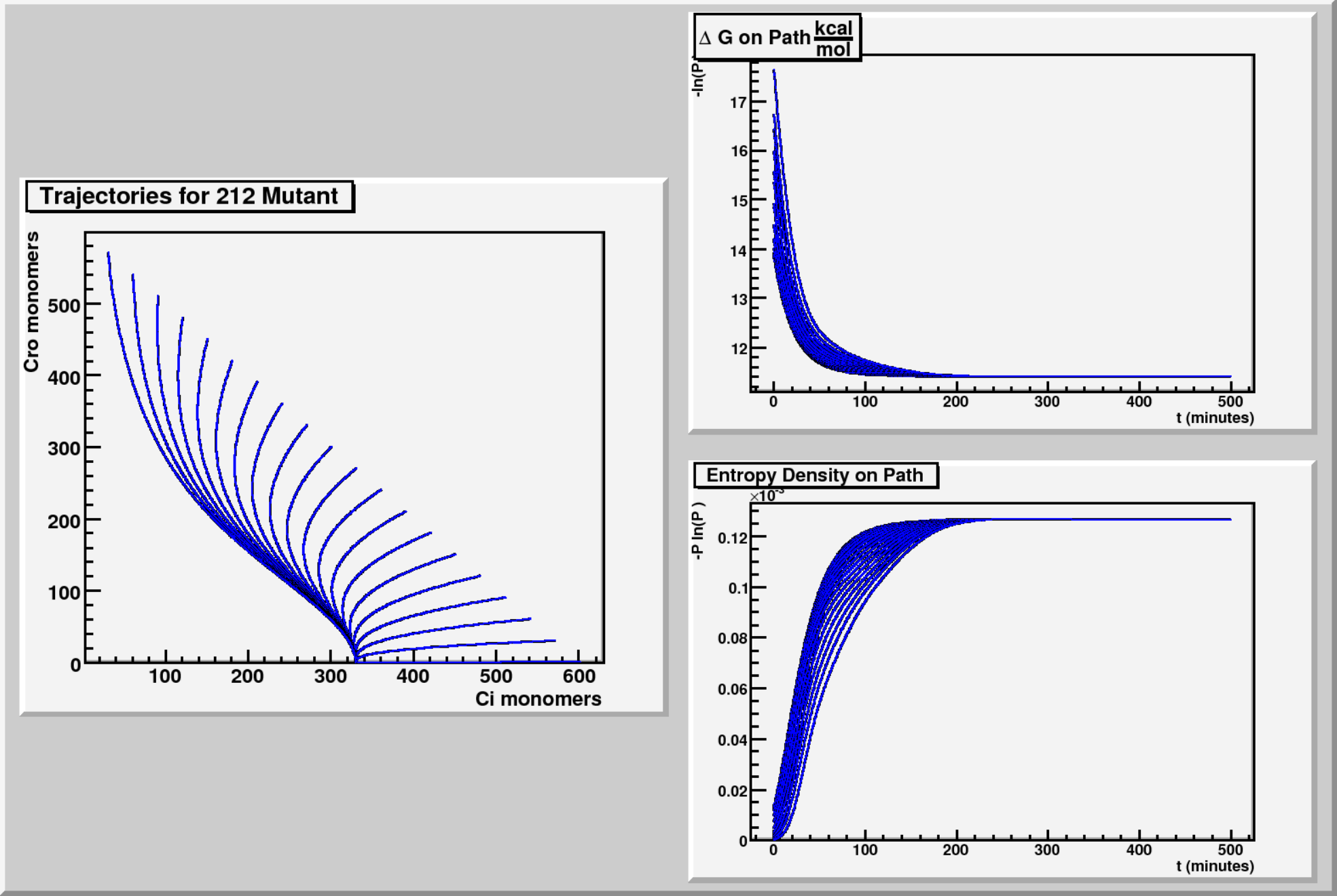}     
             &
             \includegraphics[scale=.27]{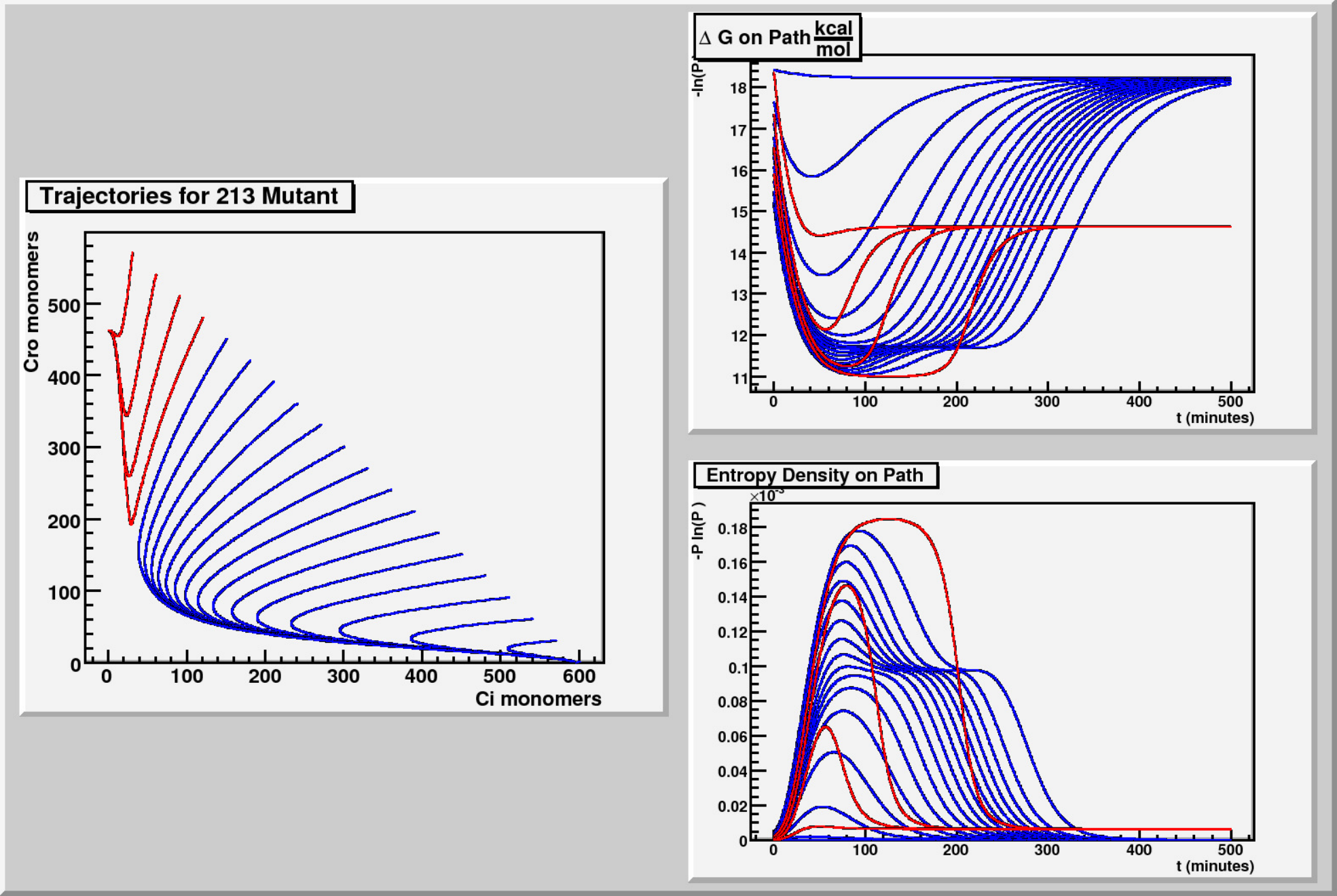}      

         \\ \hline \hline
             \includegraphics[scale=.27]{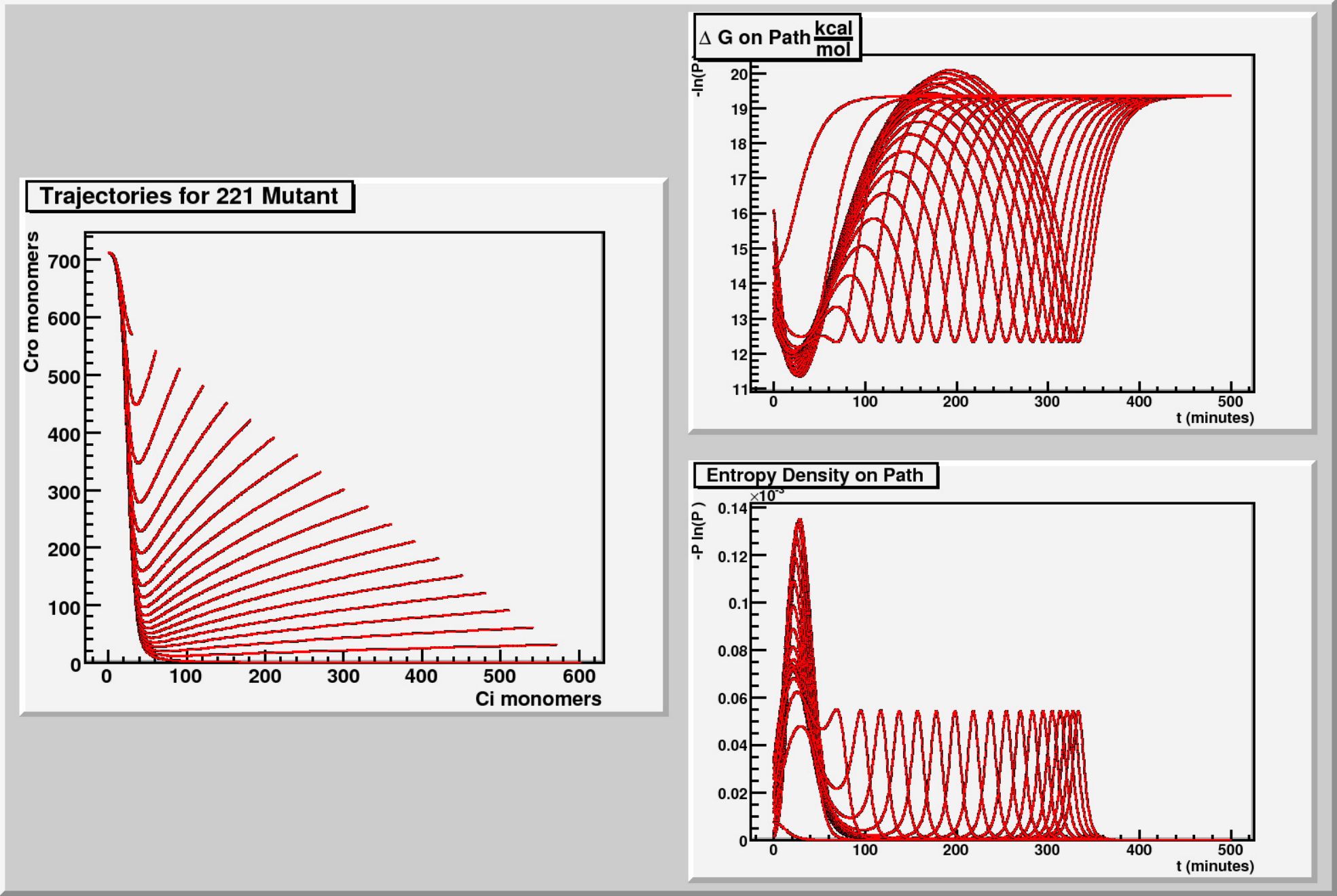}     
         &  
             \includegraphics[scale=.27]{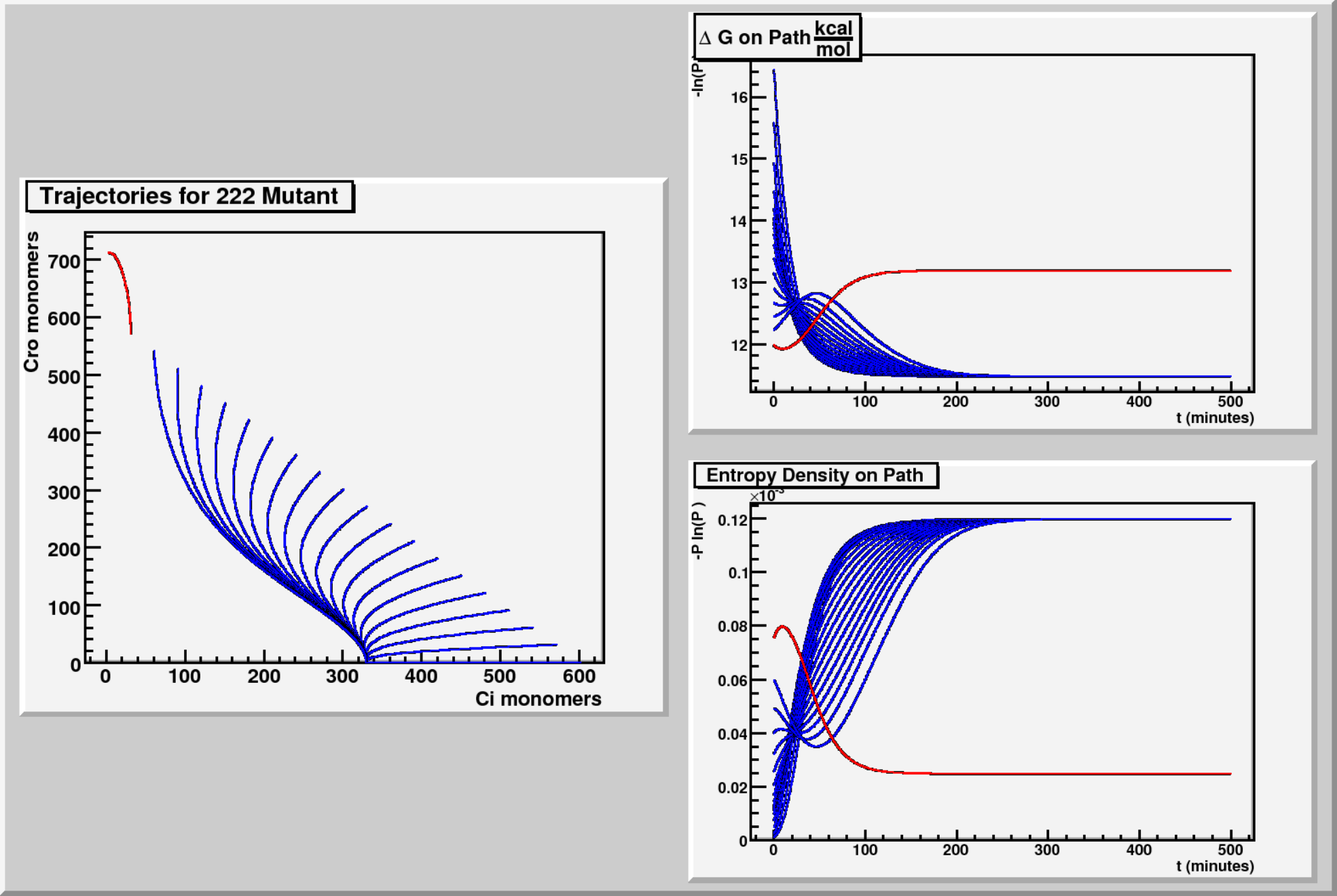}     
             &
             \includegraphics[scale=.27]{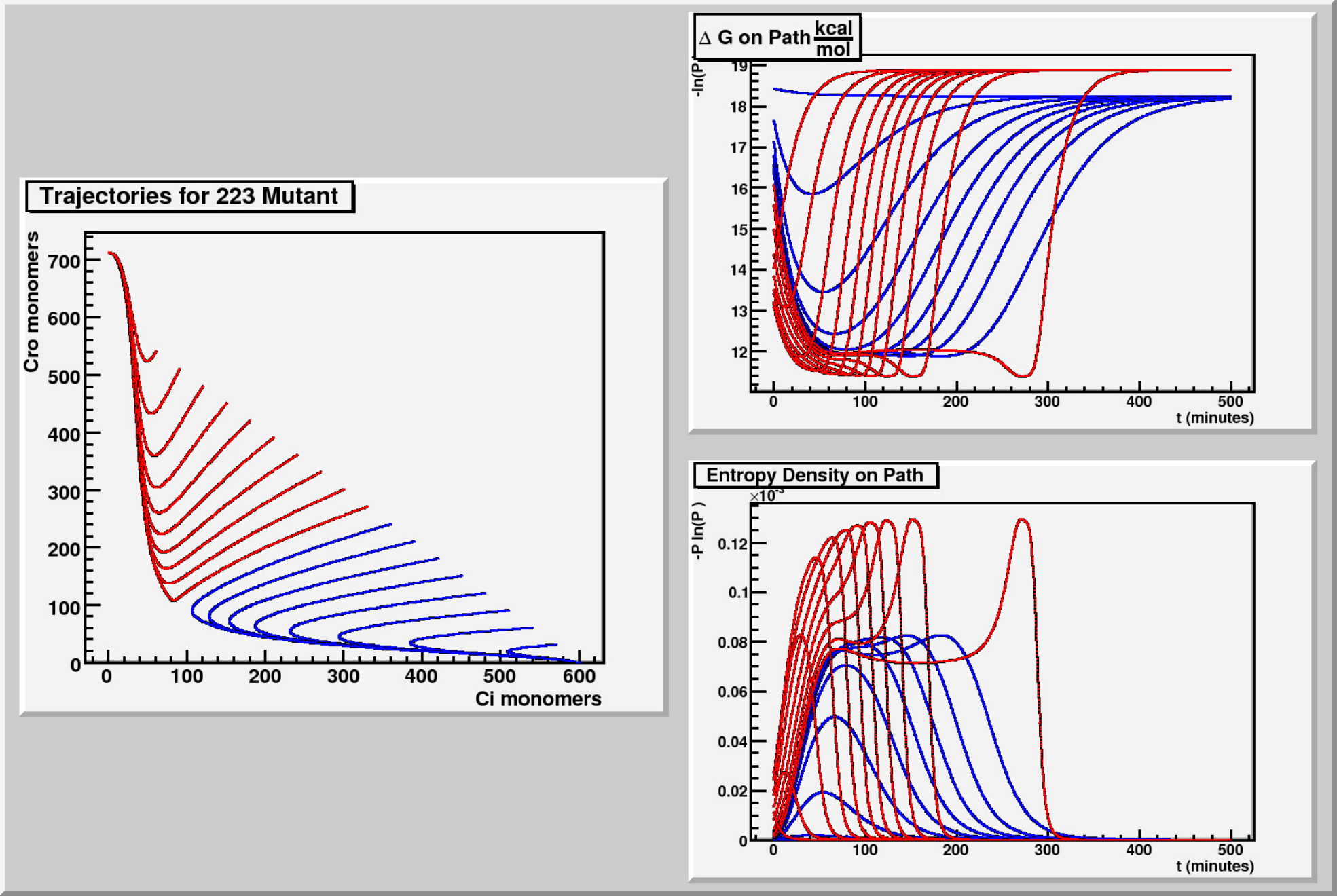}  

         \\ \hline \hline
             \includegraphics[scale=.27]{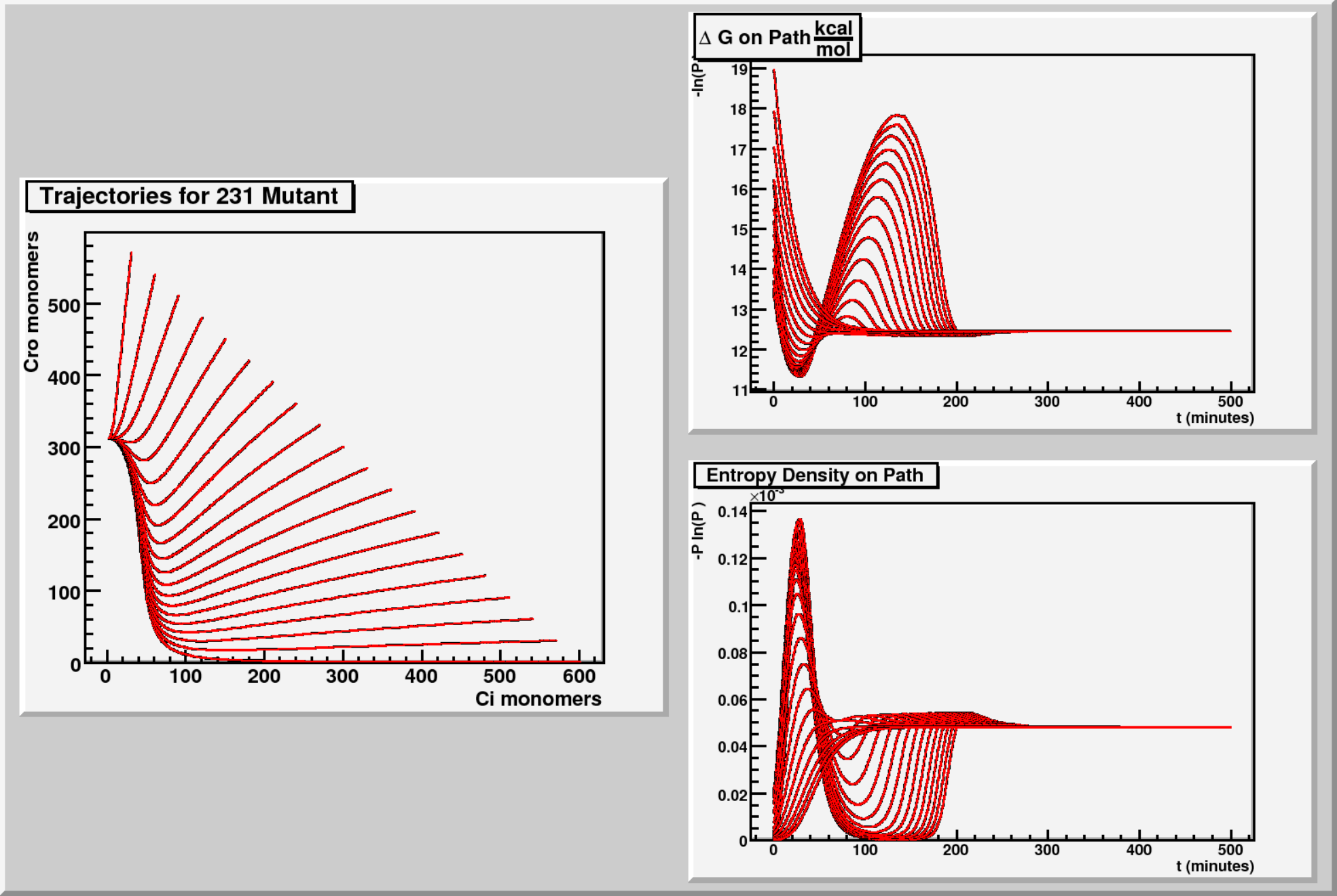}     
         &  
             \includegraphics[scale=.27]{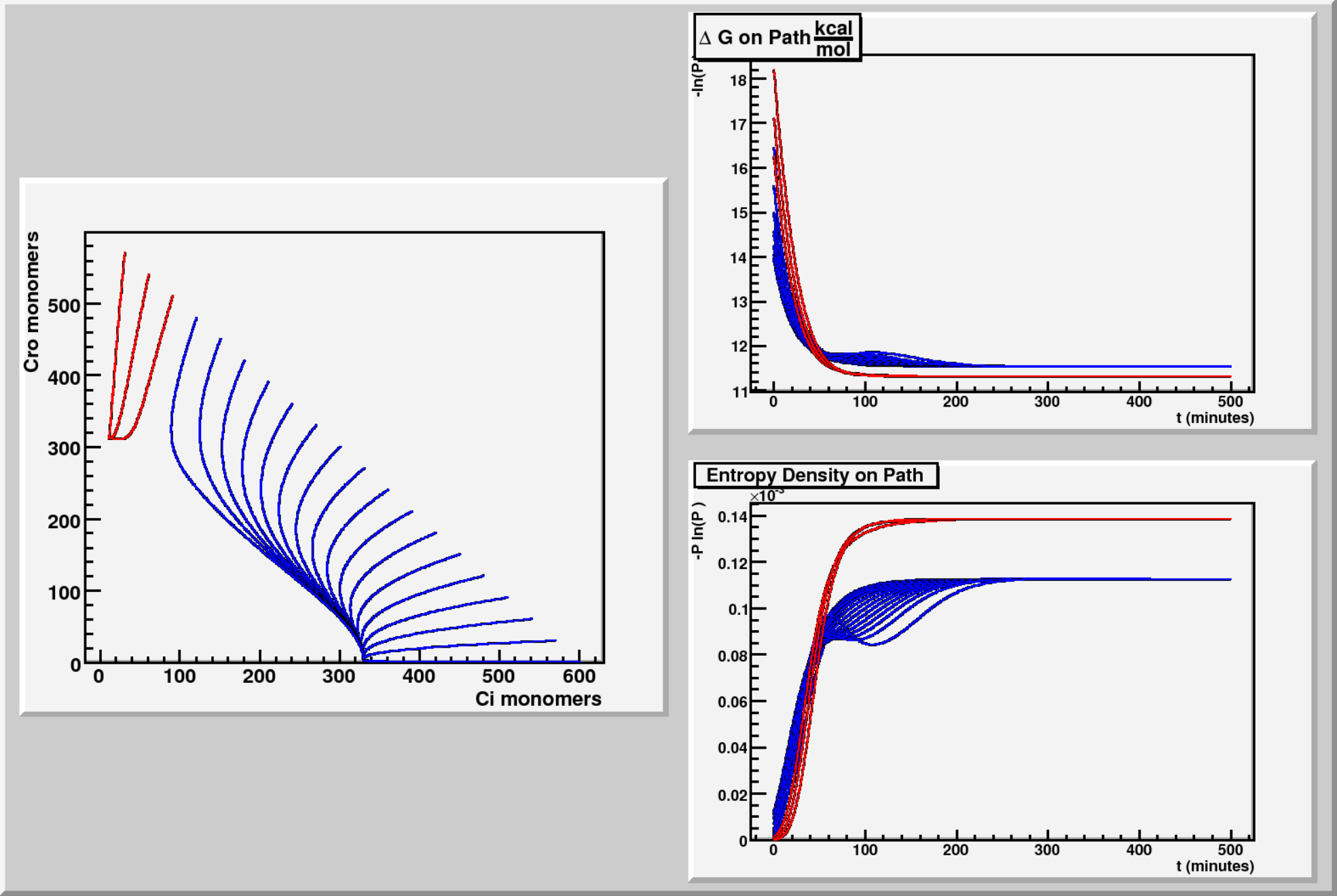}     
             &
             \includegraphics[scale=.27]{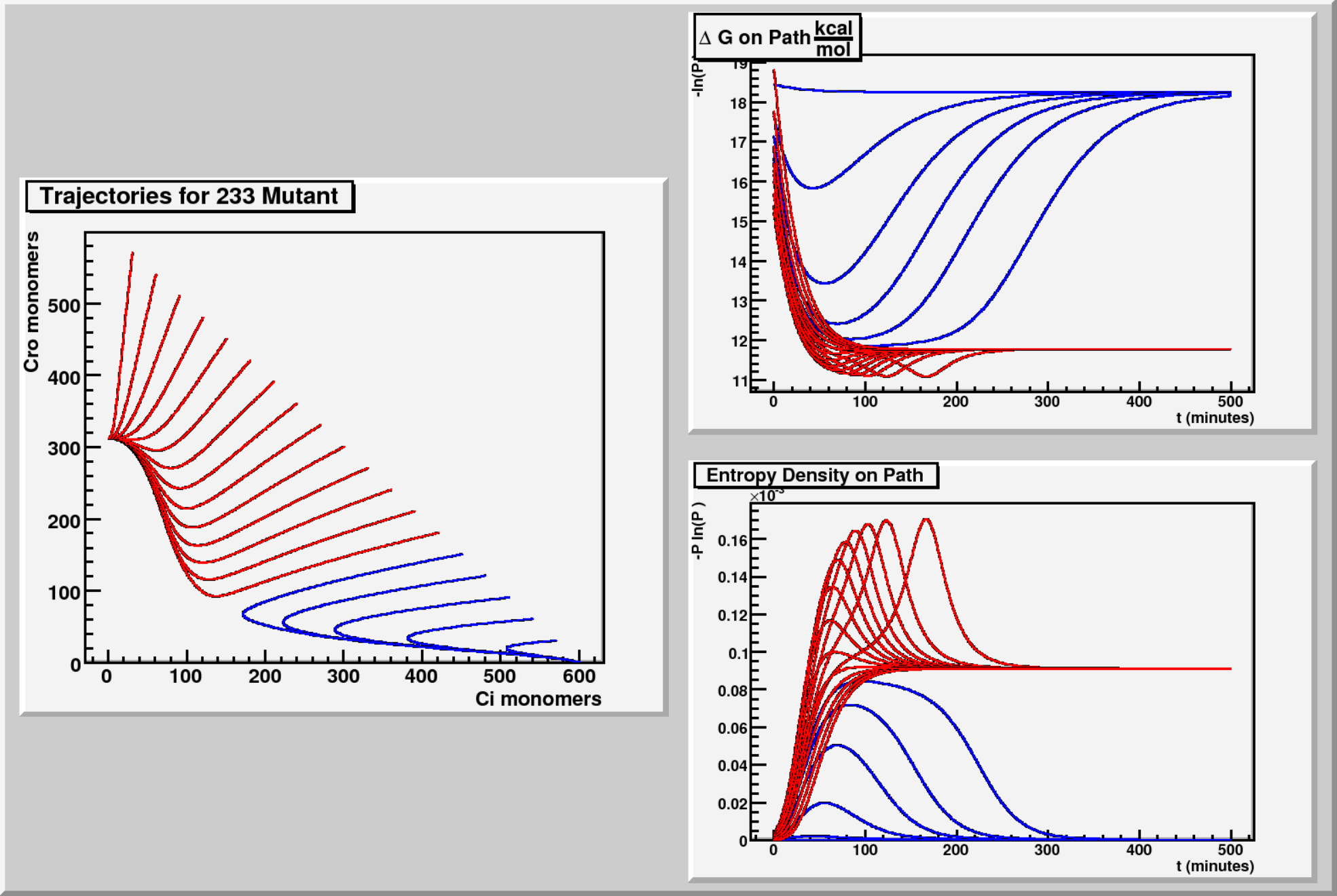}  

        \end{tabular}
}
    \caption{\bf {Entropy and Energy densities evaluated along example paths for mutations in $O_{R2,R3}$ with the $O_{R1}$ mutated into $O_{R2}$.  Different trajectories can have different limiting values of entropy and energy density depending on the path taken.}}
        \label{tab:ltraj2}
    \end{table}%

\begin{table}[ht]
\noindent 
\resizebox{!}{5cm}{     
        \begin{tabular}{|c|c|c|}
          \hline
          \multicolumn{3}{|c|}{$\lambda_{3**}$ Path Entropy and Path Energy Densities}
          \\ \hline \hline
             \includegraphics[scale=.27]{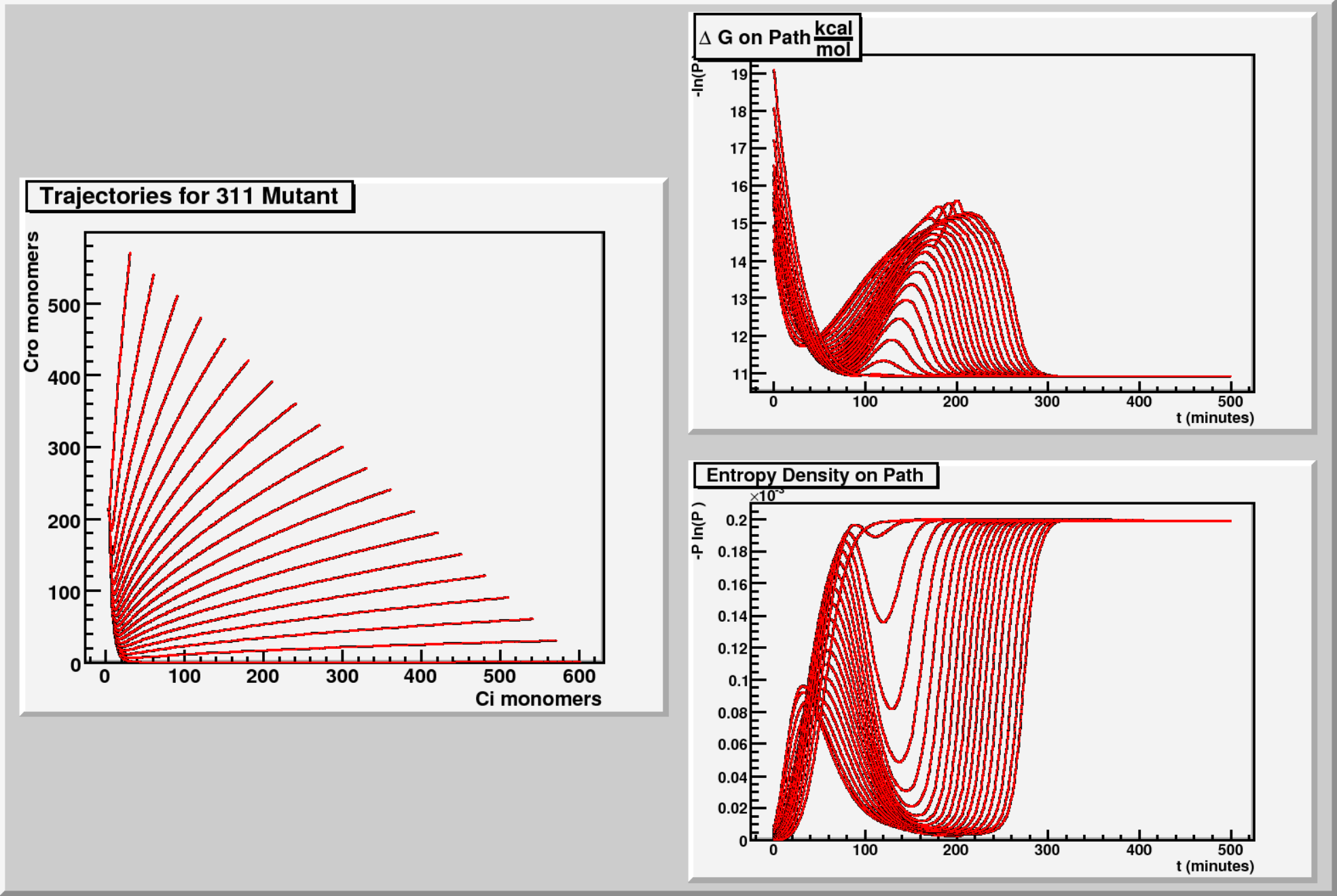}     
         &  
             \includegraphics[scale=.27]{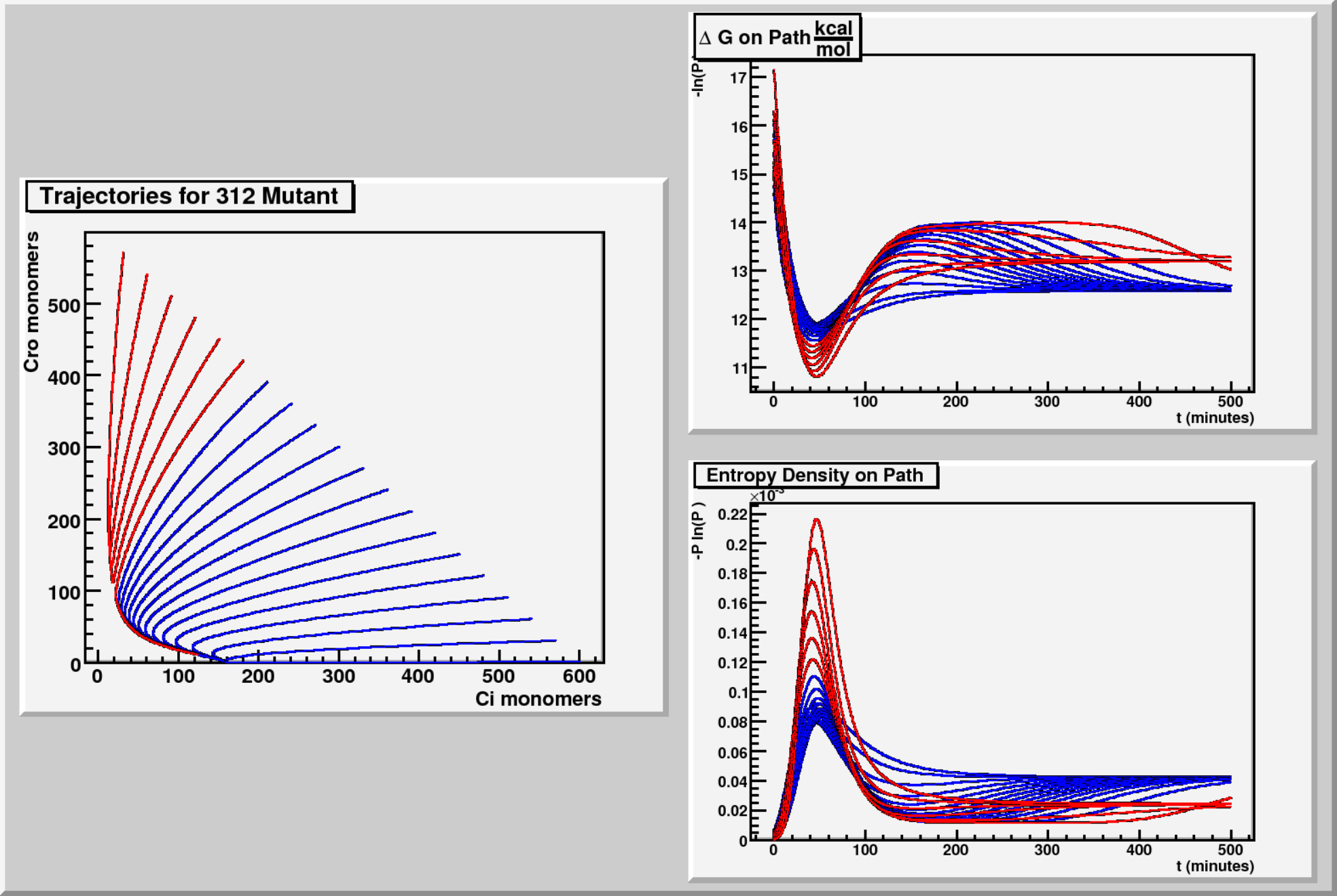}     
             &
             \includegraphics[scale=.27]{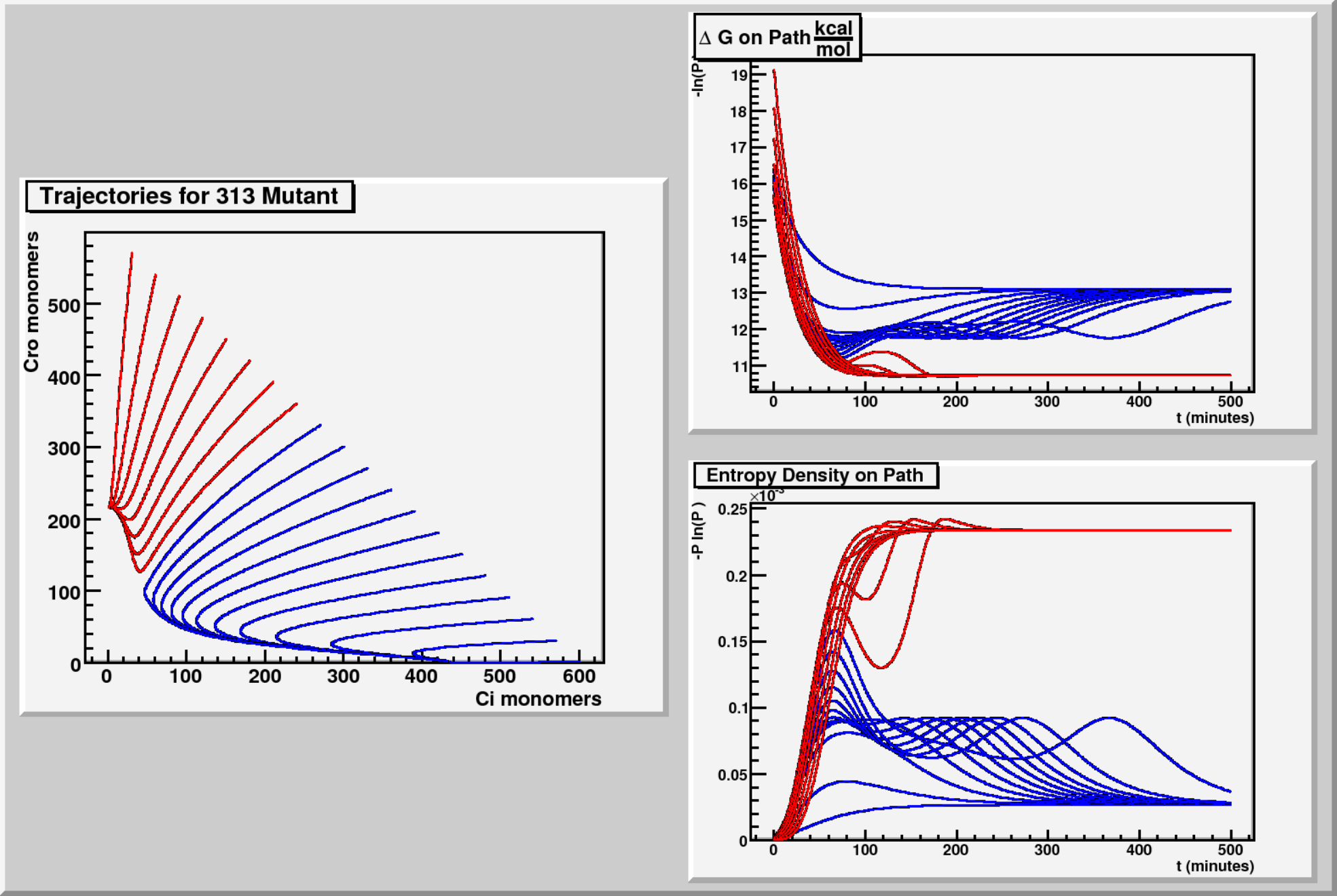}      

         \\ \hline \hline
             \includegraphics[scale=.27]{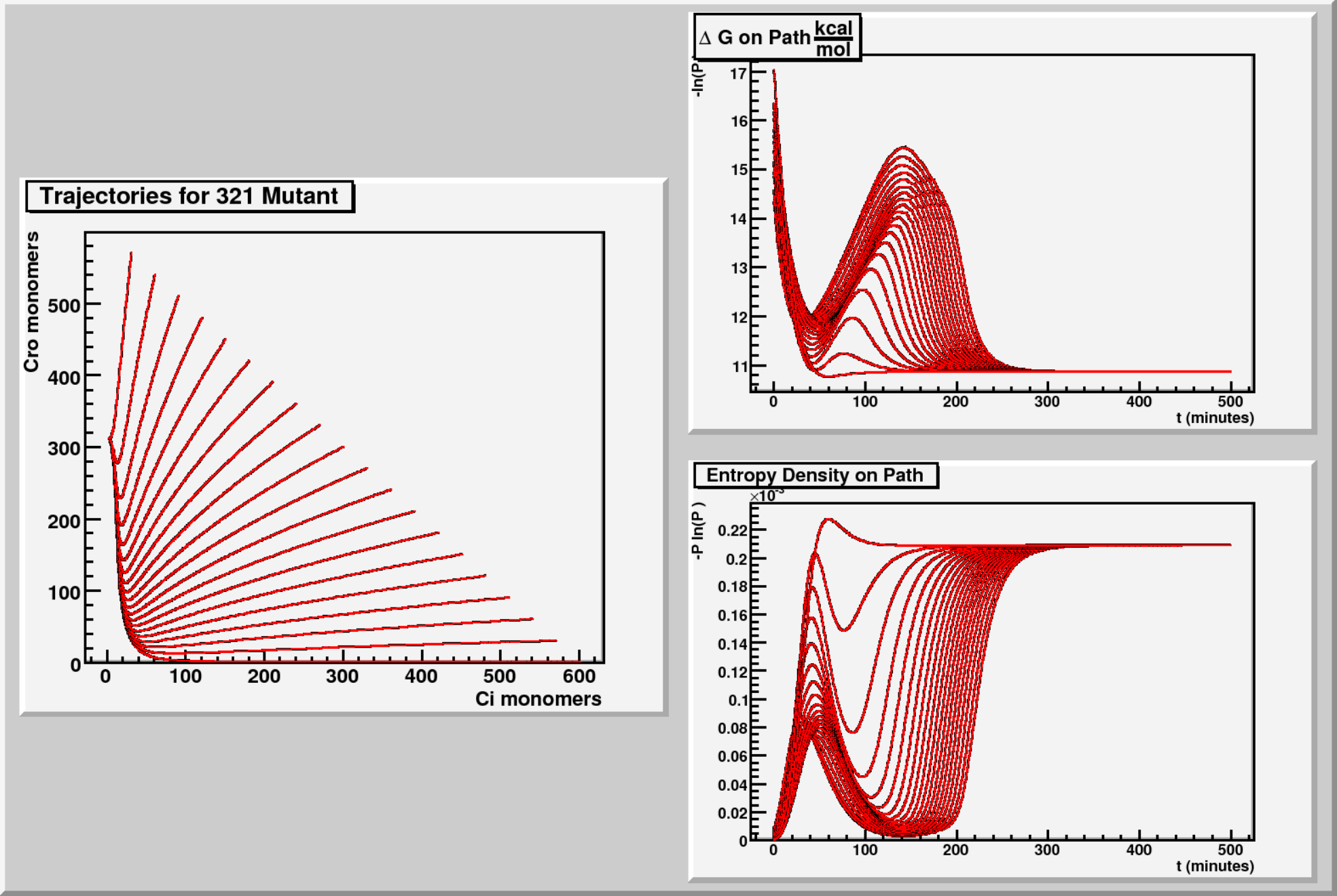}     
         &  
             \includegraphics[scale=.27]{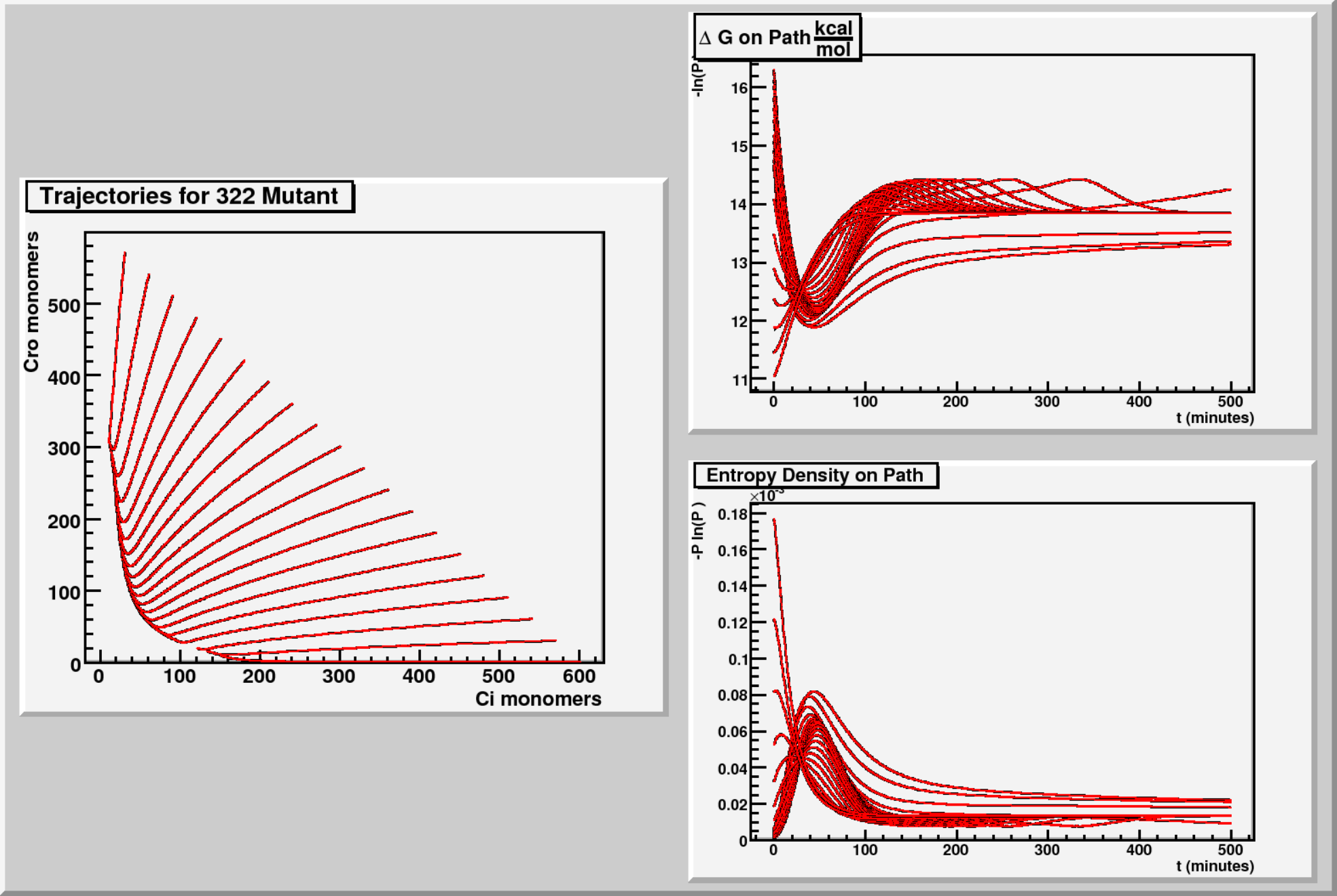}     
             &
             \includegraphics[scale=.27]{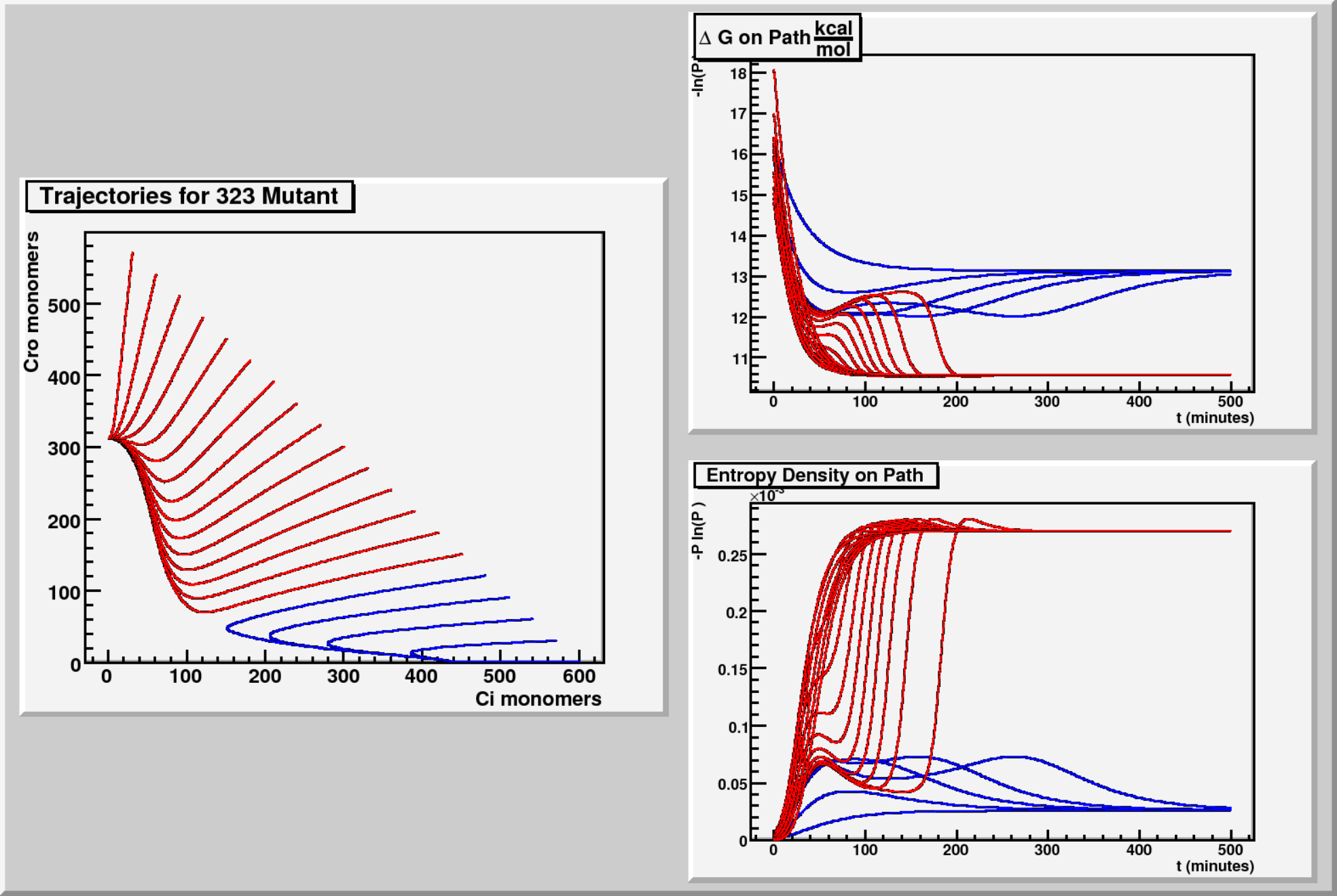}  

         \\ \hline \hline
             \includegraphics[scale=.27]{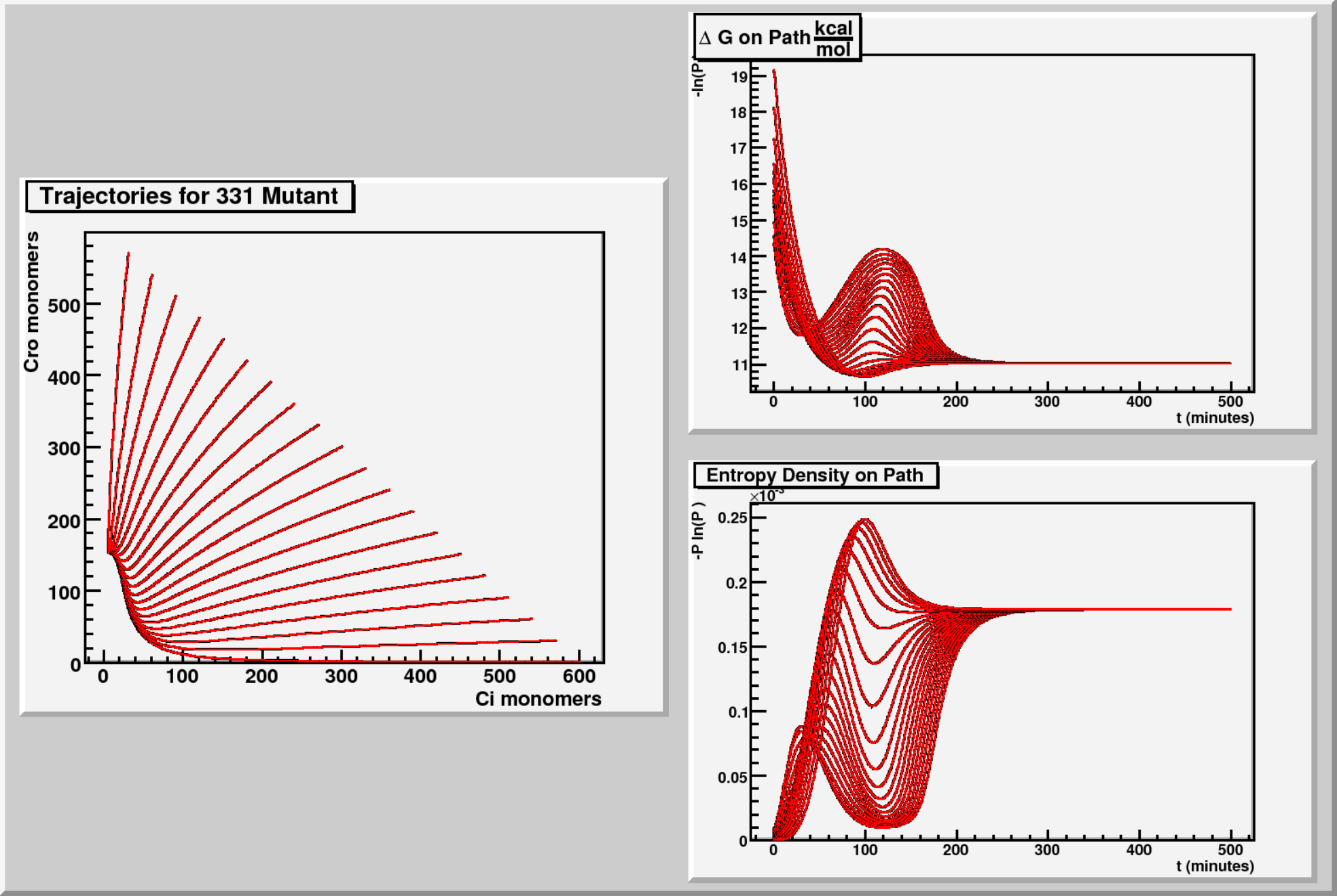}     
         &  
             \includegraphics[scale=.27]{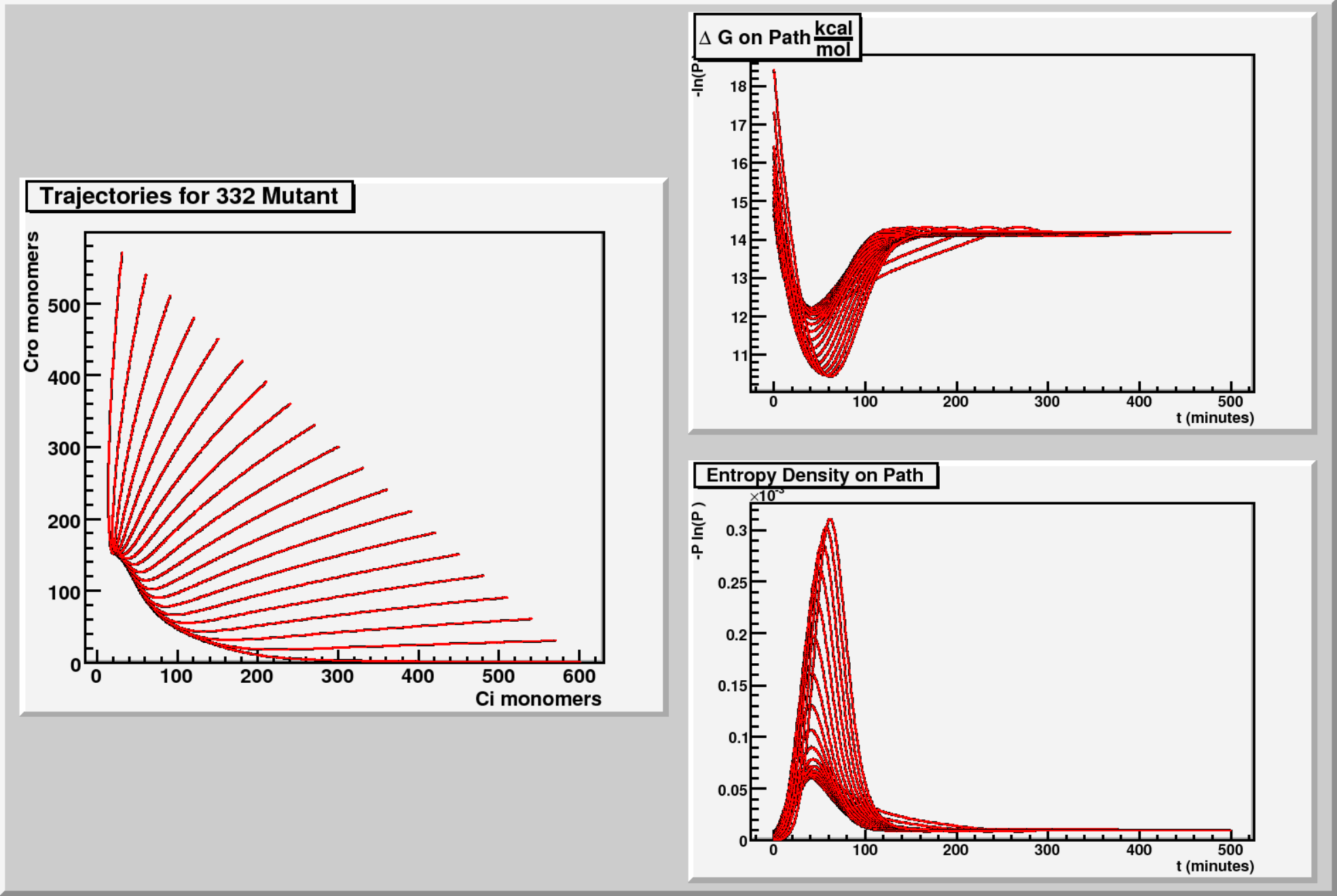}     
             &
             \includegraphics[scale=.27]{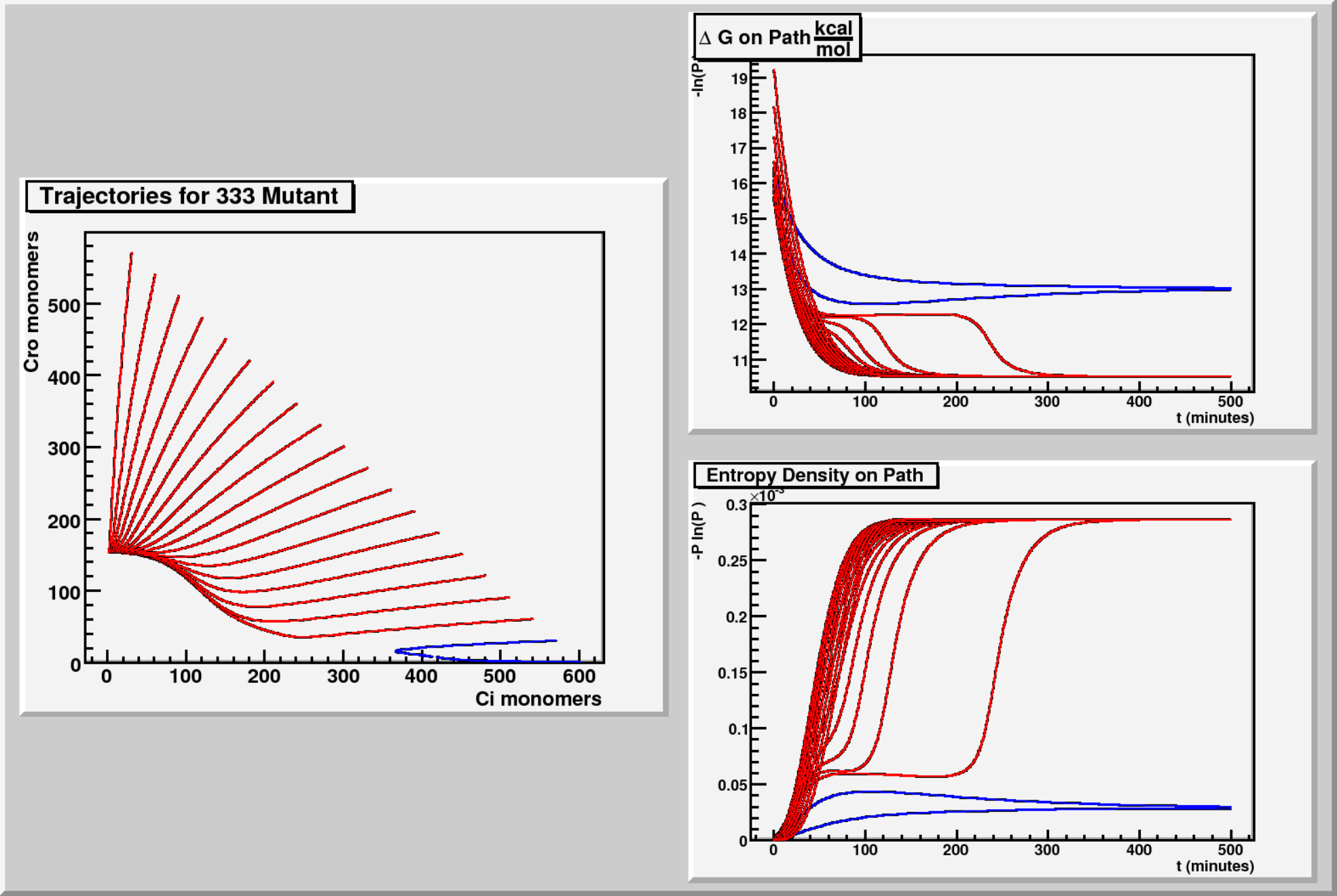}  

        \end{tabular}
}

    \caption{\bf {Entropy and Energy densities evaluated along example paths for mutations in $O_{R2,R3}$ with the $O_{R1}$ mutated into $O_{R3}$.  Different trajectories can have different limiting values of entropy and energy density depending on the path taken.}}
        \label{tab:ltraj3}
    \end{table}%

\clearpage
 
\clearpage

\pagestyle{prelim}
\begin{quotation}
\vspace{-4.4in}
\textit{Sometimes the glass unbreaks and falls to heaven.}

  \hspace{0.4in}Quickly approaching the zero, \textit{The Predicate}
\end{quotation}






\end{document}